\documentclass[pre, aps, twocolumn, longbibliography, showpacs, superscriptaddress, footnoteinbib]{revtex4-2}

\usepackage{graphicx}% Include figure files
\usepackage{dcolumn}% Align table columns on decimal point
\usepackage{bm}% bold math
\usepackage{amssymb,amsmath}
\usepackage{sidecap}
\usepackage{wrapfig}

\usepackage{graphicx}
\usepackage{leftidx}
\usepackage{color}

\usepackage{bbold} %This package allows to write the identity symbol

\usepackage{wasysym} % In order to introduce polygon symbols in the text

\usepackage{stackrel}
 
\usepackage{comment}

\usepackage{soul,xcolor}
\setstcolor{red}
\usepackage{mathtools}

\usepackage{mathrsfs}  % Some fancy cursive script

\usepackage{hyperref}

\newcommand{\be}{\begin{equation}}
\newcommand{\ee}{\end{equation}}
\newcommand{\ba}{\begin{align}}
\newcommand{\ea}{\end{align}}
\newcommand{\sysb}{\left\{\begin{array}}
\newcommand{\syse}{\end{array}\right.}
\newcommand{\baa}{\begin{array}}
\newcommand{\eaa}{\end{array}}
\newcommand{\bs}{\begin{split}}
\newcommand{\es}{\end{split}}

\newcommand{\matb}{\left(\begin{array}}
\newcommand{\mate}{\end{array}\right)}

\newcommand{\mal}{\mathcal}
\newcommand{\rmd}{{\rm{d}}}

\newcommand{\rme}[1]{{\rm{e}}^{#1}}

\newcommand{\id}{\mathbb{1}}
\newcommand{\trace}[1]{{\rm tr}\left\{ #1 \right\}}
\newcommand{\ptrace}[2]{{\rm tr}_{#1}\left\{ #2 \right\}}

\newcommand{\ha}{\frac{1}{2}}

\newcommand{\nnsum}[2]{\av{#1,\, #2}}

\newcommand{\lt}{\left(}
\newcommand{\rt}{\right)}
\newcommand{\lqq}{\left[}
\newcommand{\rqq}{\right]}
\newcommand{\lan}{\left\langle}
\newcommand{\ran}{\right\rangle}
\newcommand{\abs}[1]{\left| #1 \right|}
\newcommand{\eval}[1]{\left.\right|_{ #1 }}
\newcommand{\av}[1]{\lan #1 \ran}
\newcommand{\norm}[1]{\left\| #1 \right\|}
\newcommand{\set}[1]{\left\{  #1  \right\}}

\newcommand{\avl}[1]{\lan #1 \ran_{\mal{L}}}
\newcommand{\floor}[1]{\left\lfloor #1 \right\rfloor}

\newcommand{\R}{\mathbb{R}}
\newcommand{\N}{\mathbb{N}}

\newcommand{\C}{\mathbb{C}}

\newcommand{\ket}[1]{\left| #1 \ran}
\newcommand{\bra}[1]{\lan #1 \right|}
\newcommand{\bracket}[2]{\lan #1 \right| \!\left. #2 \ran}
\newcommand{\proj}[1]{\ket{#1} \bra{#1}}
\newcommand{\comm}[2]{\left[ #1, #2 \right]}
\newcommand{\acomm}[2]{\left\{ #1, #2 \right\}}

\newcommand{\rket}[1]{\left| #1 \rt}
\newcommand{\rbra}[1]{\lt #1 \right|}
\newcommand{\rbracket}[2]{\lt #1 \right| \!\left. #2 \rt}
\newcommand{\bigrbracket}[2]{\bigl( #1 \bigr| \bigl. #2 \bigr)}

\newcommand{\cosa}[1]{\cos \left(  #1 \right)}
\newcommand{\sina}[1]{\sin \left(  #1 \right)}

\newcommand{\cossa}[2]{\cos^{#1} \left(  #2 \right)}

\newcommand{\nol}{\nonumber \\}

\newcommand{\prodl}[2]{\prod\limits_{#1}^{#2}}
\newcommand{\suml}[2]{\sum\limits_{#1}^{#2}}

\newcommand{\liml}[1]{\lim\limits_{#1}}

\newcommand{\brak}[2]{\prescript{}{#2}{\bra{#1}}}

\usepackage[autostyle, english = american]{csquotes}
\MakeOuterQuote{"}

\begin{document}

\title{Effective Kinetic Monte Carlo for a Quantum Epidemic Process}
% ALTERNATIVE
% Effective Kinetic Monte Carlo for a Quantum Epidemic Process

\author{Alexander Sturges}
\affiliation{School of Physics and Astronomy, University of Nottingham, Nottingham, NG7 2RD, United Kingdom}
\affiliation{Faculty of Electrical Engineering, Mathematics and Computer Science, Delft University
of Technology, Mekelweg 4, Delft, 2628 CD, Zuid-Holland, The Netherlands}
\author{Hugo Smith}
\affiliation{School of Physics and Astronomy, University of Nottingham, Nottingham, NG7 2RD, United Kingdom}
\author{Matteo Marcuzzi}
\affiliation{School of Physics and Astronomy, University of Nottingham, Nottingham, NG7 2RD, United Kingdom}
\affiliation{Centre for the Mathematics and Theoretical Physics of Quantum Non-equilibrium Systems,
University of Nottingham, Nottingham NG7 2RD, UK}

\begin{abstract}
Inspired by previous works on epidemic-like processes in open quantum systems, we derive an elementary quantum epidemic model that is simple enough to be studied via Quantum Jump Monte Carlo simulations at reasonably large system sizes. We show how some weak symmetries of the Lindblad equation allow us to map the dynamics onto a classical Kinetic Monte Carlo; this simplified, effective dynamics can be described via local stochastic jumps coupled with a local deterministic component. Simulations are then used to reconstruct a phase diagram which displays stationary features completely equivalent to those of completely classical epidemic processes, but richer dynamics with multiple, recurrent waves of infection.  
\end{abstract}

\maketitle

\section{Introduction}
\label{Sec:Intro}

%Quantum physics is a field now about a century old and yet it still defies out efforts towards a comprehensive understanding of the quantum nature of reality. With its numerous counter-intuitive predictions and the inherent difficulties faced in fully reconciling it with other successful theories such as Relativity and Field Theory, it presented a challenge for many a generation of young physicists and still inspires several lines of active research.

Quantum physics is a field now about a century old and yet the quantum nature of reality still defies our understanding in many ways. With its numerous counter-intuitive predictions and the inherent difficulties faced in fully reconciling it with other successful theories such as Relativity and Field Theory, it presented a challenge for many a generation of young physicists and still inspires several lines of active research.

Our present work belongs and contributes, in our intentions, to the investigation of quantum many-body dynamics, a topic which has been addressed in a myriad different approaches; to name a few, quantum quenches \cite{PolkovnikovRMP, Mitra_quench} have played a prominent role for the past two decades, trying to probe quantum relaxation and the absence thereof in closed systems \cite{Kollath2007, Rossini2009, Rossini2010, Marino2014, Chanda2020}; with the support of cold atoms experiments \cite{Greiner2002, Kinoshita2006, Gring2012, AduSmith2013}, theoreticians have addressed the interplay between integrability, locality and thermalization \cite{Rigol2007, Rigol2008, Calabrese2012, Calabrese2012b, Sotiriadis2014, Bertini2015, Langen2016, Foini2017}; other works have instead tried to highlight and determine universal properties (in a statistical mechanics sense), or in other words general properties that do not strongly depend on the microscopic details of the system or of its dynamics \cite{Calabrese2006, Calabrese2007, Gambassi2012, Chiocchetta2015, Calabrese2016, Caux2016}. New instances of phase transitions, affecting the dynamical, rather than static, properties of quantum models have also been identified and described \cite{Heyl2013, Hickey2014, Heyl2018}. 

A different approach has been to look at "quantum generalizations" of selected classical models, namely stochastic ones whose dynamics is subject to so-called "kinetic constraints". To better frame the aims of such an approach, let us outline some basic concepts, together with some references the interested readers may find useful for seeking further details.
Kinetic constraints \cite{Ritort2003, Garrahan2011}) were first introduced in the classical theory of glasses to explain, among other aspects, the appearance of extremely long relaxation timescales \cite{Struik1976}. In simple terms, if we describe a stochastic dynamics by means of individual stochastic events (e.g.~a random walker jumping either left or right), we can see these "constraints" as conditions that must be satisfied for these events to take place (e.g.~a random walker can only jump on an empty position, but not on top of another, already present, walker).

Let us look at a simple example, which we will expand upon further below. We take an "infective" process where individuals (lattice sites, spins, ...) in a "healthy" state can switch to a second, "infected" one. However, they are only permitted to do so provided that they come into contact (e.g., be nearest neighbors) with another already "infected" individual. Constraints like this effectively hinder the dynamics and may, in some cases, lead to the emergence of absorbing states, i.e., states that can be reached in the course of the evolution, but cannot be left. 
Indeed, for any infective process of the kind just described, in the absence of "infected" individuals no infection dynamics can take place; in colloquial terms, one could say that once the last patient recovers (or dies) a pandemic is over. 

Because of the stark asymmetry between the possibility of evolving into, and the impossibility of leaving them, absorbing states are associated with a strong break of ergodicity which unavoidably leads the dynamics to either cease entirely or remain out of equilibrium \footnote{In general one could consider \emph{absorbing spaces} encompassing many non-absorbing states where an internal dynamics could still go on, but that would be beyond the scope of this work.}. A classical stochastic process with an absorbing state could not satisfy detailed balance \cite{Gardiner2004handbook}, nor an open quantum system microreversibility \cite{Agarwal1973}. Hence, phase transitions arising in the presence of absorbing state show a non-equilibrium character \cite{Hinrichsen2000}. For continuous transitions, in particular, this means that they lie in universality classes distinct from those of the more usual thermal \cite{Huang_book, Ma_book, Pelissetto2002, ZinnJustin_book, Mussardo_book} and quantum (zero-temperature) \cite{Mussardo_book, Sachdev_book} ones. 

The "robustness" of non-equilibrium features in the presence of absorbing states is potentially relevant for current research efforts on quantum non-equilibrium phase transitions: after all, microreversibility, much like other symmetries, can be recovered under coarse graining whenever the interaction terms that cause its breakdown at the microscopic level are irrelevant in a renormalization group sense \cite{DallaTorre2013, Sieberer2013} \footnote{A very simple instance is provided by the geometric symmetries of the microscopic lattice: a square lattice is only invariant under rotations by $\pi/2$, a honeycomb one under rotations by $\pi/3$, a triangular one under rotations by $2\pi/3$. Yet, an ordinary two-dimensional Ising model defined on any of these lattices will display the same isotropic statistical properties at large scales. The full $SO(2)$ symmetry under arbitrary rotations, broken at the microscopic level, is thus recovered under coarse-graining.}.
In other words, even if the microscopic rules obeyed by the individual constituents violate microreversibility (i.e., detailed balance if the dynamics is stochastic), the macroscopic properties of the system may equilibrate. Absorbing states prevent this from happening: at all scales a distinction can be made between active phases in which some dynamics takes place and inactive ones which feature no dynamics whatsoever \cite{Hinrichsen2000, Racz2002}.  

Quantum models with kinetic constraints thereby provide a reliable way for studying non-equilibrium phenomena in a quantum context and, as a consequence, they have attracted significant attention. At times, models featuring them were directly inspired by the corresponding classical ones, like in the case of the East model \cite{vanHorssen2015, Pancotti2020, Rose2022, Greissler2023, Causer2024, Brighi2024}, the contact process \cite{Buchhold2016, Carollo2019, Gillman2019, Jo2021, Makki2024} and the Domany-Kinzel cellular automaton \cite{Lesanovsky2019, Gillman2020, Nigmatullin2021, Gillman2022}. In other instances, these concepts have emerged as approximate descriptors of the dynamics of Rydberg atoms \cite{Mattioli2015, Marcuzzi2015, Marcuzzi2016, Valado2016, Gutierrez2017, Brady2023, Brady2024}; these strongly-interacting systems \cite{Gallagher_book} are known to feature epidemic-like (or avalanche-like) behavior \cite{Lesanovsky2014, Malossi2014, Klocke2019, Klocke2021} under conditions known as "facilitation" or "anti-blockade" \cite{Ates2007}. As we shall employ one of the latter models to draw comparisons with our own results, let us give a brief, intuitive overview of this facilitation mechanism: interactions between atoms in an excited state with high quantum number are much stronger than if either atom is in its ground state \cite{Gallagher_book}; as a consequence, in the neighborhood of an excited atom the energy price one has to pay to promote another atom to its excited state is effectively increased (decreased) by the strength of the repulsion (attraction) to the original excitation. In a more formal notation, if we call $\hbar \omega$ the energy gap between the excited and ground states, $\delta \omega$ the width of the corresponding spectral line, and $\Delta V (r)$ the strength of the interatomic interactions at some distance $r$, then in order to excite another atom at such a distance from a preexisting excitation one will have to provide an energy $\approx \hbar \omega + \Delta V$ (e.g., a photon of frequency $\omega + \Delta V / \hbar$). Think now of what happens if we shine a laser detuned from the ordinary atomic transition (say, of frequency $\omega + \Delta \omega$ with $\abs{\Delta \omega} \gg \abs{\delta \omega}$) on an atomic ensemble (a cloud, a tweezer lattice, \ldots). In the absence of excitations, because of the detuning $\Delta \omega$ the medium is essentially transparent to the laser. If, instead, a single excitation is prepared before the laser is switched on, then at a distance $r$ ($\hbar\Delta \omega = \Delta V(r)$) from said excitation resonance is achieved and other atoms can get excited. Furthermore, in the neighborhood of any novel excitation the same mechanism would be at work and new atoms would thus become sensitive to the radiation and therefore prone to excitation. From a single seed, one could progressively disseminate excitations throughout the entire medium in a process akin to an epidemic spreading from an initial patient.

We think it is important to improve our theoretical understanding of these processes, to identify their phases and to categorize their emergent phase transitions, keeping in mind that quantum fluctuations may lead to a broader spectrum of phenomena than stochastic ones. A model of particular relevance to the present work is a quantum epidemic process which was introduced a few years ago \cite{Espigares2017} and featured two very different parameter regimes: an effectively classical one, in which it underwent a continuous absorbing-state phase transition no different from the one appearing in epidemic classical models, and a quantum phase where instead multiple discontinuous transitions were predicted, associated to the emergence of multiple outbreaks from the initial seed. While the classical phase was described via a perturbative formalism valid under strong dephasing conditions \cite{Degenfeld2014, Marcuzzi2014}, the properties of the quantum one were derived from the analysis of the (non-uniform) mean-field equations of the model, meaning that quantum correlations between neighboring sites were disregarded. Because of this, it is difficult to establish whether the predicted phenomenology (a sequence of multiple waves of infection) is a feature of the system or rather an artefact of the approximation. Inspired by these questions, we introduce in this article a new quantum epidemic model which is simple enough to be simulated up to reasonable system sizes. We show that our numerical results support some of the predictions of the previous model.

This work is organized as follows: in Section \ref{Sec:Model} we briefly discuss the behavior of a classical epidemic model, we recall the main features of the previous quantum one and then introduce the current one. Following that, we explain in Section \ref{sec:theory} the peculiarities of our open quantum model that allow us to simulate it in two spatial dimensions up to sizes that would normally not be accessible. In Section \ref{sec:num} we present and discuss the numerical results from said simulations and provide our concluding remarks in Section \ref{sec:Concl}. In Section \ref{sec:outlook} we briefly comment on some potential ways to generalize our model. Several technical details are reported in various Appendices for the interested readers.

\section{The model}
\label{Sec:Model}

We start by specifying the general formalism we are going to adopt. To this end, note that to feature absorbing states, a dynamics need to be irreversible, i.e., it must lead to a (partial, at least) loss of information about the initial state over the course of the evolution. Closed quantum systems, being subject to a unitary (Hamiltonian-generated) evolution, are thus clearly prevented from showing an absorbing dynamics in the strict sense. It is thereby natural to shift our attention to open quantum systems. For simplicity, we shall describe our dynamics via a Lindblad equation \cite{Lindblad1976, Gorini1976, Breuer_book}, which is tantamount to saying we further require that the dynamics be Markovian.

To better illustrate how we constructed our model, we further divide this Section in three parts: in \ref{subsec:GEP}, for the unfamiliar reader, we report the main features of a very simple classical epidemic model; in \ref{subsec:RQEP} we recall the properties of the quantum epidemic model discussed in Ref.~\cite{Espigares2017} and point out the main differences which emerge with respect to the classical case. Lastly, in \ref{subsec:eQEP} we introduce the model constituting the main subject of this work. For brevity, we shall refer to the first as the \emph{General Epidemic Process} (GEP), to the second, being inspired by experiments on Rydberg atoms, as the \emph{Rydberg Quantum Epidemic Process} (RQEP) and finally to the third, due to its peculiar simplicity, as the \emph{elementary Quantum Epidemic Process} (eQEP). Additionally, unless explicitly stated otherwise, we will always assume that any dynamics we discuss starts at time $t = 0$ from an initial condition which includes one and only one "seed" placed in the center of the system, as illustrated in Fig.~\ref{fig:GEP1}. 

\subsection{The General Epidemic Process (GEP)}
\label{subsec:GEP}

\begin{figure}[h]
  \includegraphics[width=\columnwidth]{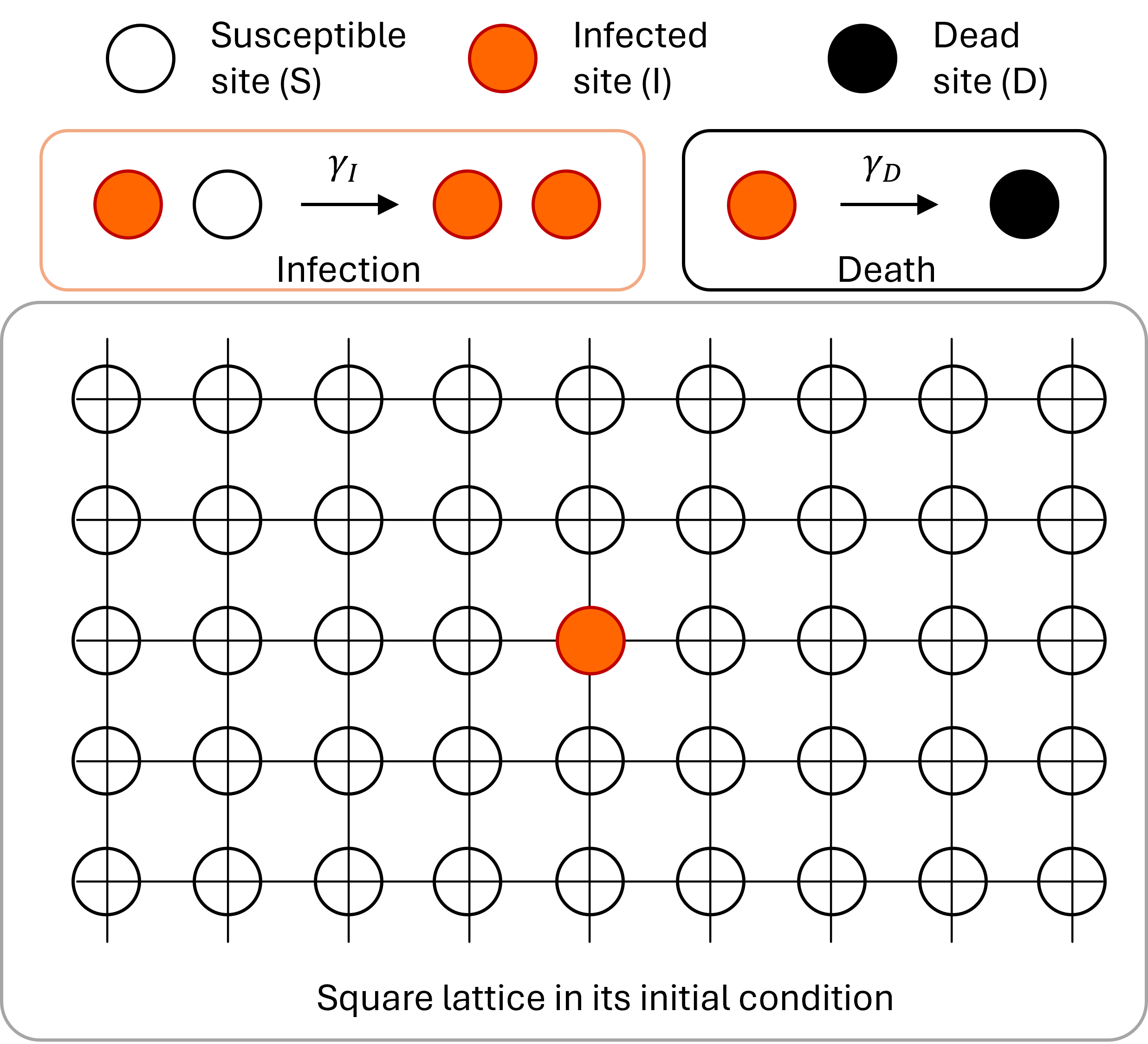}
  \caption{Structure of the GEP: at the top, the three possible values for the local state of a site are visualised as an empty circle for S (susceptible), an orange one for I (infected) and a black one for D (dead). The two basic processes are displayed one line below, where it is understood that the two sites shown under the "Infection" label are nearest neighbors; in the example provided, the leftmost one can infect its neighbor with rate $\gamma_I$. This rate is the same in all directions (the process is isotropic). Death, on the other hand, involves a single site and occurs at rate $\gamma_D$. The panel at the bottom displays the initial condition for a two-dimensional square lattice: all sites but the central one are in the S state; the one exception, i.e.~"the origin" of the infection, is set instead at time $t = 0$ in state I. }
\label{fig:GEP1}
\end{figure}

The General Epidemic Process (or GEP) is a continuous-time stochastic process on a lattice; each site is in one of three classical states, that we describe here as "susceptible" (S), "infected" (I) and "dead" (D). A schematic representation is provided in Fig.~\ref{fig:GEP1}, topmost row. On the second row we display the two distinct, competing local events that can take place, giving rise to the stochastic dynamics:
\begin{itemize}
\item{\underline{Infection:} an infected (I) site can infect a neighboring susceptible (S) site, i.e., promote it to state I, at a rate $\gamma_I$. This keeps the epidemics going, spreading the infection.}
\item{\underline{Death:} an infected (I) site can die, i.e., switch to state D, at a rate $\gamma_D$. Note how death hinders the growth of the infected area in two ways: first, by reducing the number of I sites which can pass the infection onward and, second, because D sites are inert, being unable to infect or be infected. The death of a site can be seen as analogous to its removal from the lattice.}
\end{itemize}
Note how both processes hinge on the presence of infected (I) sites. In the absence thereof, neither infection nor death can take place and the dynamics ceases. Hence, every system configuration made up of only S and D sites is an absorbing state. For a comprehensive discussion of the properties of the GEP we refer the reader to Ref.~\cite{Grassberger1983} and the references therein. Here we limit ourselves to a summary of its main features under the particular initial conditions mentioned above: a single starting "seed", i.e., a single infected (I) site in the center of an otherwise entirely susceptible (S) lattice (bottom panel of Fig.~\ref{fig:GEP1}). For brevity, we shall refer to said central site as "the origin". 

%
% FIGURE HERE
%

Taken separately, the two defining event types of the GEP lead to very different outcomes: while death by itself would simply deplete the system of any initially present infected sites, bringing the whole dynamics to a halt, infection without its competitor would eventually fill the entire system with infected sites (or continue spreading indefinitely if the system were unbounded). With both present, the relative strength of their rates $\gamma_{D/I}$ determines whether the long-time dynamics gets fully absorbed (absorbing phase) or not (active phase).

Unsurprisingly for a statistical physics model, one cannot neglect the role played by dimensionality. In one spatial dimension the process is invariably destined to fall into an absorbing state, no matter what values are chosen for the rates $\gamma_{D/I}$, with the only exception of the trivial case $\gamma_D = 0$, which we shall hereafter disregard. The one-dimensional GEP, in fact, is defined on a string of sites, each with at most one neighbor to its right and one to its left. Then, at any given time there cannot be more than two "extremal" I sites, the farthest one to the left and the farthest one to the right of the origin. In other words, the "front" of the infection, where new infections are possible, cannot consist of more than two sites. Because of this, there is always a finite probability ($\sim \gamma_D^2 / (\gamma_I + \gamma_D)^2$) of both dying before either can infect a susceptible site further away. 

This simple argument breaks down in higher dimensions: by spreading radially away from the origin the process generates a growing wavefront of infection (an "outbreak", borrowing the terminology of Ref.~\cite{Espigares2017}). This entails the presence of a growing number of infected sites in contact with susceptible ones further away; therefore, the probability of all the I sites dying before any could infect a new S site gets suppressed. In line with this intuition, the model exhibits a (continuous) phase transition already in two spatial dimensions (i.e., its lower critical dimension is $2$). This transition separates an inactive, absorbing phase for small $\gamma_I / \gamma_D$, with vanishing probability of the infection surviving at long times, from an active phase for large $\gamma_I / \gamma_D$ in which such a probability is instead finite. 

\begin{figure}[h]
  \includegraphics[width=\columnwidth]{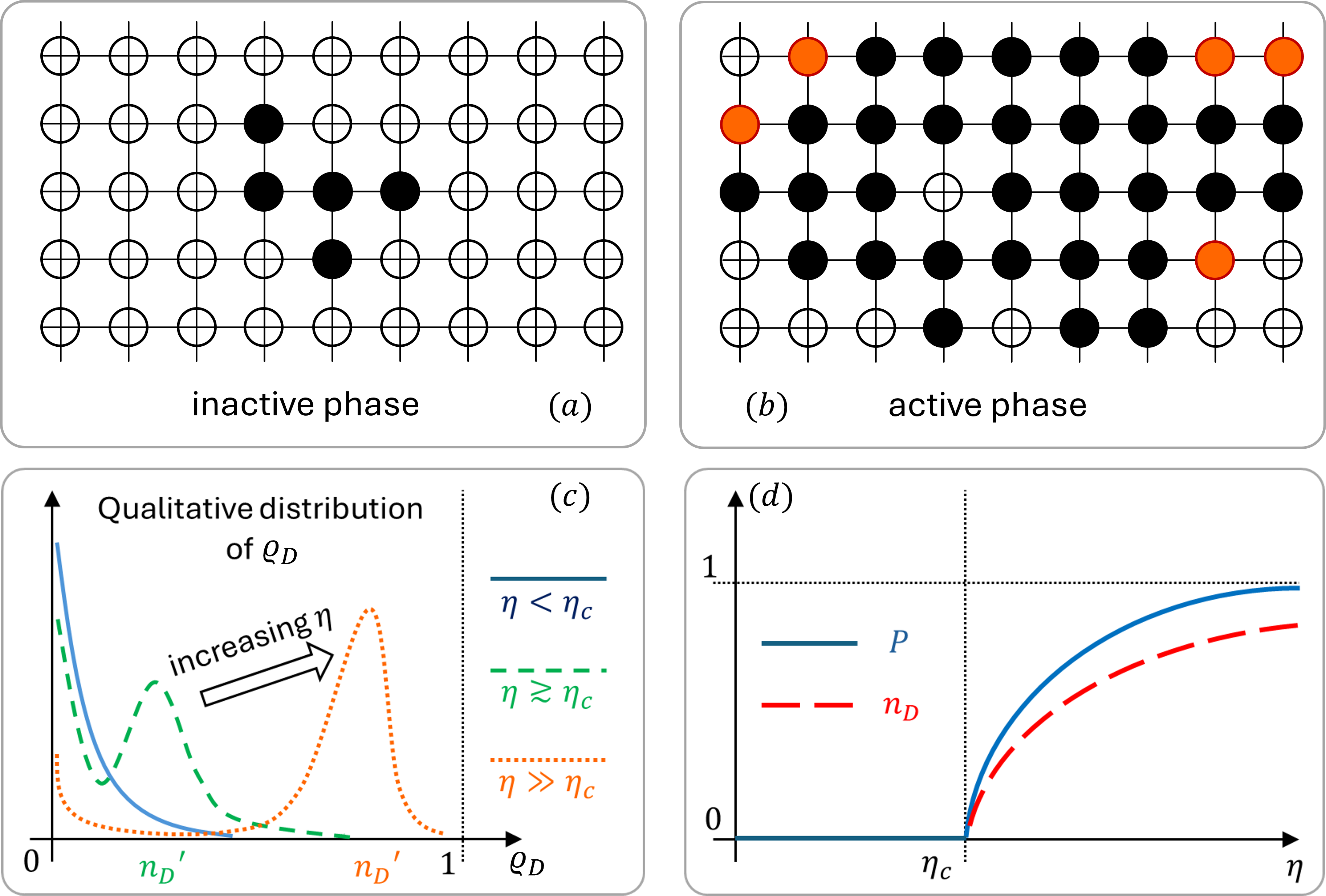}
  \caption{(a-b) A typical long-time configuration in (a) the inactive phase and (b) the active one. The color code for the sites is taken from Fig.~\ref{fig:GEP1}: empty or white (S), orange (I) and black (D). (c) Qualitative representation of the distribution of the final random DDS $\varrho_D$ in three different parameter regions: $\eta < \eta_c$ (blue, solid line), $\eta \gtrsim \eta_c$ (green, dashed line) and $\eta \gg \eta_c$ (orange, dotted line). The width of the peak around the origin is exaggerated for ease of visualization. The second peak progressively moves to the right as $\eta$ is increased, while also getting higher at the expense of the other one, as the probability of observing a dying trajectory also diminishes for growing $\eta$. Generally, the finite peaks are not centered around the averages $n_D$, but around larger values $n_D'$, which instead correspond to averages over the subset of "surviving" trajectories. (d) Qualitative behavior of the survival probability $P(\eta)$ (blue, solid line) and of the final average death density $n_D(\eta)$ (red, dashed line) across the transition. }
\label{fig:GEP2}
\end{figure}

To better illustrate how this transition manifests, we shall introduce two different stationary (i.e.~long-time) order parameters: first, we look at the probability of infection survival. To be more precise, let us denote by $N$ the system size (the number of lattice sites), by $\eta = \gamma_I / \gamma_D$ the ratio of the dynamical rates, by $\eta_c$ its critical value and by $N_I(t) \leq N$ the number of I sites in the system at time $t$ and define
\be
	P_N(t,\, \eta) = \mathbb{P} \lt \text{$N_I(t) > 0$ at size $N$, ratio $\eta$}   \rt
\label{eq:GEP_PN}
\ee
as the probability that at least one site out of $N$ is still infected at time $t$. Conceptually, for this quantity to be a meaningful order parameter, the thermodynamic limit must be approached as follows:
\be 
	P(\eta) = \liml{t \to +\infty} \liml{N \to +\infty} P_N(t,\, \eta),
\label{eq:GEP_PP}
\ee
where the order in which the limits are taken is crucial: at every finite size, in fact, the process is destined to eventually stop, as the system cannot support an indefinite radial growth of the outbreak; in mathematical terms, excluding again the trivial case $\gamma_D = 0$ ($\eta = +\infty$),
\be
	\liml{t \to +\infty} P_N(t,\, \eta) = 0 \,\,\, \forall\  N,\  \forall\  \eta.
\ee
Taking instead the limits in the order reported in Eq.~\eqref{eq:GEP_PP} yields a \emph{survival probability} $P(\eta)$ which is a continuous function of the ratio $\eta$, vanishes for $\eta \leq \eta_c$ and is $> 0$ for $\eta > \eta_c$, monotonically growing as $\eta$ is increased. A sketch of this behaviour is illustrated by the blue, solid line in Fig.~\ref{fig:GEP2}(d).

To further clarify what this implies, let us think in terms of Monte Carlo simulations of the GEP: provided the size of the lattice is large enough, deep in the inactive phase ($\eta < \eta_c$) no realization of the process (no "stochastic trajectory") spreads all the way to the boundaries, as exemplified by Fig.~\ref{fig:GEP2}(a); in the active phase ($\eta > \eta_c$), instead, a portion of trajectories $\approx P(\eta)$ will percolate through, producing at large times an extensive ($O(N)$) dead population, as shown in Fig.~\ref{fig:GEP2}(b).

A second, alternative order parameter for the GEP is the final density of dead sites (or DDS for short): since a site can only become dead (D) if it had previously been infected (I), the only way to have, at long times, a macroscopic portion of the system in the D state is by means of an infection outbreak occurring in the past; as we have mentioned, this takes place exclusively in the active phase, whereas for subcritical parameters $\eta < \eta_c$ only a small neighbourhood of the origin will feature a non-negligible D density. 

For a more precise formulation, let us now consider individual realizations (or stochastic trajectories) of the process, labeled by $\tau$; these trajectories appear according to a distribution $p_{st} (\tau)$ (see also our legend of terms and abbreviations in App.~\ref{App:Legend}).
For any given trajectory $\tau$, we denote by $N_D (t, \eta, \tau) \leq N$ the number of dead sites at time $t$; let then $\varrho_D(t, \eta, \tau) = N_D(t, \eta, \tau) \,/\, N \leq 1$ be the corresponding density. Given an observable $O$, its stochastic average can be expressed as
\be
	\overline{O} = \int \mal{D} \tau \, p_{st}(\tau)\, O(\tau).
    \label{eq:stoch_average}
\ee
Accordingly, we introduce the shorthand
\be
	n_D (t,\eta) = \overline{\varrho_D (t, \eta)}
	\label{eq:class_DDS}
\ee
for the average density of dead sites (again, DDS for short). Interestingly, to use this quantity as a meaningful order parameter the ordering of limits should be reversed with respect to Eq.~\eqref{eq:GEP_PP}: to understand why, consider that an outbreak spreads out from the origin at a finite speed, say $v$ \cite{Grassberger1983}. Hence, at any finite time $t$ the number $N_D$ must be finite: in two dimensions, for example, $N_D < \pi (vt)^2$; generally, $N_D < \omega_d (vt)^d$, where $\omega_d$ denotes the solid angle in $d$ dimensions. Thus,
\begin{align}
	\liml{N \to \infty}& n_D(t, \eta) = \liml{N \to \infty} \frac{\overline{N_D(t,\, \eta)}}{N} < \nol
    &< \liml{N \to \infty} \frac{\omega_d (vt)^d}{N} = 0 \,\,\, \forall\  t,\  \forall\  \eta.
\end{align}
Instead, one should take
\be
	n_D(\eta) = \liml{N \to \infty}\, \liml{t \to \infty} n_D(t, \eta) .
\ee  
With this ordering, the \emph{final average death density} $n_D(\eta)$ is a continuous quantity which vanishes in the subcritical regime $\eta < \eta_c$ and is positive in the supercritical one $\eta > \eta_c$. Again, to avoid confusion, we repeat that even in the active phase not all trajectories survive at long times; only a fraction $\approx P(\eta)$ manages to. Hence, the actual distribution of $\varrho_D$ for $\eta > \eta_c$ is typically bimodal, i.e.~it displays two separate peaks: one around $0$ accounting for the trajectories that fail to percolate through the system and one concentrated around a value $n_D' > n_D$ corresponding to the average DDS of the remaining ones, with 
\be 
	n_D (\eta) \approx \lt 1 - P(\eta) \rt 0 + P(\eta) n_D' (\eta) = P(\eta) n_D' (\eta).
	\label{eq:secondOP}
\ee  
For later convenience, we shall also mention that
\be
	\liml{\eta \to \eta_c} n_D' (\eta) = 0.
    \label{eq:vanishing_peak}
\ee
In other words, the second peak develops around $0$ and moves rightward (leftward) as the ratio $\eta = \gamma_I / \gamma_D$ is increased (decreased). This is qualitatively displayed in panel (c) of Fig.~\ref{fig:GEP2}, where the three lines represent sketches of the distribution of values of $\varrho_D$ for three choices of $\eta$: one in the inactive phase $\eta < \eta_c$ (blue, solid line), one in the near supercritical region $\eta \gtrsim \eta_c$ (green, dashed line) and one much further inside the supercritical region $\eta \gg \eta_c$ (orange, dotted line).  

%
% FIGURE HERE
%

Equation \eqref{eq:vanishing_peak} is not the only one compatible with the continuous behavior observed for $n_D (\eta)$ across the transition: let us assume for a moment that
\be
	\liml{\eta \to \eta_c} n_D' (\eta) = c_D' > 0.
    \label{eq:greater_than_0}
\ee

This would imply that, as the critical point $\eta_c$ is crossed from the inactive to the active phase, the finite peak would start developing around $c_D'$ instead of $0$. Yet, as one can glean from Eq.~\eqref{eq:secondOP}, the density $n_D (\eta)$ would still vanish at $\eta_c$ because $P(\eta)$ does. Yet, the physical behavior of the system would be observably different: in the near supercritical regime there would still be only a small fraction of surviving trajectories; those, however, would leave behind an extensive number of dead sites $\approx c_D' N$, no matter how close to the critical point. In this case, the scaling behavior $P(\eta) \sim (\eta - \eta_c)^\beta$ in the critical region would match that of $n_D(\eta)$, whereas in general $n_D(\eta) \sim (\eta - \eta_c)^{\beta'}$ and $P(\eta) \sim (\eta - \eta_c)^{\beta}$ with independent critical exponents $\beta' \neq \beta$ \cite{Grassberger1983}. If Eq.~\eqref{eq:greater_than_0} held, if one were to operate a post-selection of the surviving trajectories their average long-time D density $n_D'$ would experience a discontinuous jump $0 \to c_D'$ across $\eta_c$, effectively akin to a first-order phase transition.

\subsection{The Rydberg Quantum Epidemic Process (RQEP)}
\label{subsec:RQEP}

Our model, which we will introduce further below in Sec.~\ref{subsec:eQEP}, takes inspiration in some measure from the one of Ref.~\cite{Espigares2017}. In this Section, we shall summarize some relevant definitions and results from the previous work, reformulating them into our current notation whenever necessary.

%For the sake of illustrating how our model, which will be introduced further below in Sec.~\ref{subsec:eQEP}, was conceived and in what sense it constitutes a simplification with respect to the one of Ref.~\cite{Espigares2017}, we shall summarize some relevant definitions and results from that previous work, translating them into our current notation whenever necessary.

As already mentioned, a closed quantum system defined via a Hermitian, time-independent Hamiltonian cannot feature any absorbing states: its evolution is in fact unitary and thus, if a state $\ket{f}$ can be reached through evolution from a state $\ket{i}$, then the reverse is also equally possible. Formally, for finite systems, if
\be
	A_{i\to f} (t) = \bra{f} \rme{-iHt / \hbar} \ket{i} \neq 0
	\label{eq:conn_under_evol}
\ee
for some $t > 0$ then $\exists \,\, t' > 0$ such that
\be
	A_{f\to i} (t') = \bra{i} \rme{-iHt' / \hbar} \ket{f} \neq 0.
\ee
Hence, there is no state $\ket{f}$ which can be reached but not left.

One could try with either non-Hermitian or time-dependent Hamiltonians; we are not aware of any attempts to define an absorbing dynamics with the former. As for the latter, we do not recall any examples with a continuous time-dependence. There are instead examples of absorbing states and phase transitions in quantum cellular automata \cite{Lesanovsky2019, Gillman2020, Nigmatullin2021, Gillman2022}; these systems are evolved via the application of a sequence of unitary transformations in a procedure reminiscent of quantum computation; as these unitary transformations usually change from step to step, the whole procedure can be regarded as an evolution under a Hailtonian subject to sharp changes at discrete intervals.

To our knowledge, all other attempts to construct absorbing quantum systems involve some coupling to an external environment, i.e., consider \emph{open systems}; although the whole (system plus environment) evolves unitarily under an extended Hamiltonian, the partial trace over the external degrees of freedom introduces loss of information, making the reduced evolution of the system non-unitary in nature. The simplest framework to treat open quantum system is the Lindblad formalism \cite{Lindblad1976, Gorini1976, Breuer_book}, which adds the extra requirement of Markovianity on the reduced system dynamics.    

% It is thus unsurprising that most attempts made thus far to define an absorbing quantum dynamics --- to our knowledge, with the only exceptions being some quantum cellular automata \cite{Lesanovsky2019, Gillman2020, Nigmatullin2021, Gillman2022}, which are "evolved" via the application of a sequence of unitary transformations in a procedure reminiscent of quantum computation --- have done so via the Lindblad formalism, which constitutes the most general framework describing a time-independent, Markovian dynamics preserving the fundamental properties of a physical state (i.e., safeguarding conservation of total probability and of complete positivity) \cite{Lindblad1976, Gorini1976, Breuer_book}.

Having to deal with open quantum systems, we need to account for statistical mixtures; in general, therefore, the state at some time $t$ will be described by a density matrix $\rho(t)$. Its evolution is captured by the Lindblad equation, which in turn can be expressed as a linear operation $\mal{L}$ as follows:
%Denoting by $\rho(t)$ the density matrix describing the state of the system at time $t$, the Lindblad equation can be expressed as a linear operation $\mal{L}$ such that
\begin{align}
	\dot\rho(t) & = \mal{L} \rho(t) =   \nonumber  \\  
	 & -\frac{i}{\hbar} \comm{H}{\rho(t)} + \sum_\alpha \lqq L_\alpha \rho L_\alpha^\dag - \ha \acomm{L_\alpha^\dag L_\alpha}{\rho(t)} \rqq 
     \label{eq:Lindblad}
\end{align}
where $\comm{A}{B} = AB - BA$ stands for commutation, $\acomm{A}{B} = AB + BA$ for anticommutation, $H$ represents a Hamiltonian term (which includes the Hamiltonian of the closed system plus possible corrections to the energy levels induced by the coupling to the environment), whereas the final term encodes the non-reversible interactions with the outside world and is often dubbed the "dissipator" of the equation. The dissipator is a sum which runs over all distinct external "events" that may affect the system (say, the emission of a photon); the action of each such event is captured by the corresponding "jump operator" $L_\alpha$, with $L_\alpha^\dag$ denoting its Hermitian conjugate. 

Equation \eqref{eq:Lindblad} represents the most general form that a linear, Markovian equation can take under the requirement that it preserve the fundamental properties of a physical state, i.e.~that it conserve total probability and complete positivity \cite{Lindblad1976, Gorini1976}.

The average for an observable $O$ on a state $\rho(t)$ --- evolved under the Lindblad equation from an initial one $\rho_0$ at time $t_0$ --- corresponds to the trace
\be
	\avl{O(t)} = \trace{O \rho(t)} = \trace{O \lt \rme{\mal{L} \lt t - t_0 \rt }  \rho_0 \rt}.
    \label{eq:Lind_average}
\ee
Note that, while operators such as $H$ or $L_\alpha$ act on the usual Hilbert space $\mal{H}$ ($H \ket{\psi} \in \mal{H}$ for any $\ket{\psi} \in \mal{H}$), the Lindbladian $\mal{L}$ constitutes instead an operator over the Liouville space $\sim \mal{H} \otimes \mal{H}^\ast$\footnote{Here $\mal{H}^\ast$ denotes the adjoint Hilbert space, or, in Dirac notation, the space of ``bra'' vectors $\bra{\psi}$. }, which we can regard as a "space of density matrices"; to avoid confusion, one usually refers to linear operations on Liouville space objects as "superoperators", with the prefix "super-" just being a reminder that they act on a space of greater dimension than $\mal{H}$. 

Looking at Eq.~\eqref{eq:Lindblad}, it is clear that to fully define a Markovian open quantum model it is sufficient to define the Hilbert space $\mal{H}$ (or the corresponding Liouville space) and specify the Hamiltonian $H$ and all jump operators $L_\alpha$. This is how we are going to introduce the Rydberg Quantum Epidemic Process, or RQEP. For simplicity, from this point onward we will be working in natural units ($\hbar = 1$).

% Within the Lindblad formalism, fully defining a (Markovian open quantum) model can be done by specifying the Hilbert space $\mal{H}$ (or the corresponding Liouville space), the Hamiltonian $H$ and all jump operators $L_\alpha$. For simplicity, from this point onward we will be working in natural units ($\hbar = 1$).

\begin{figure}[h]
  \includegraphics[width=\columnwidth]{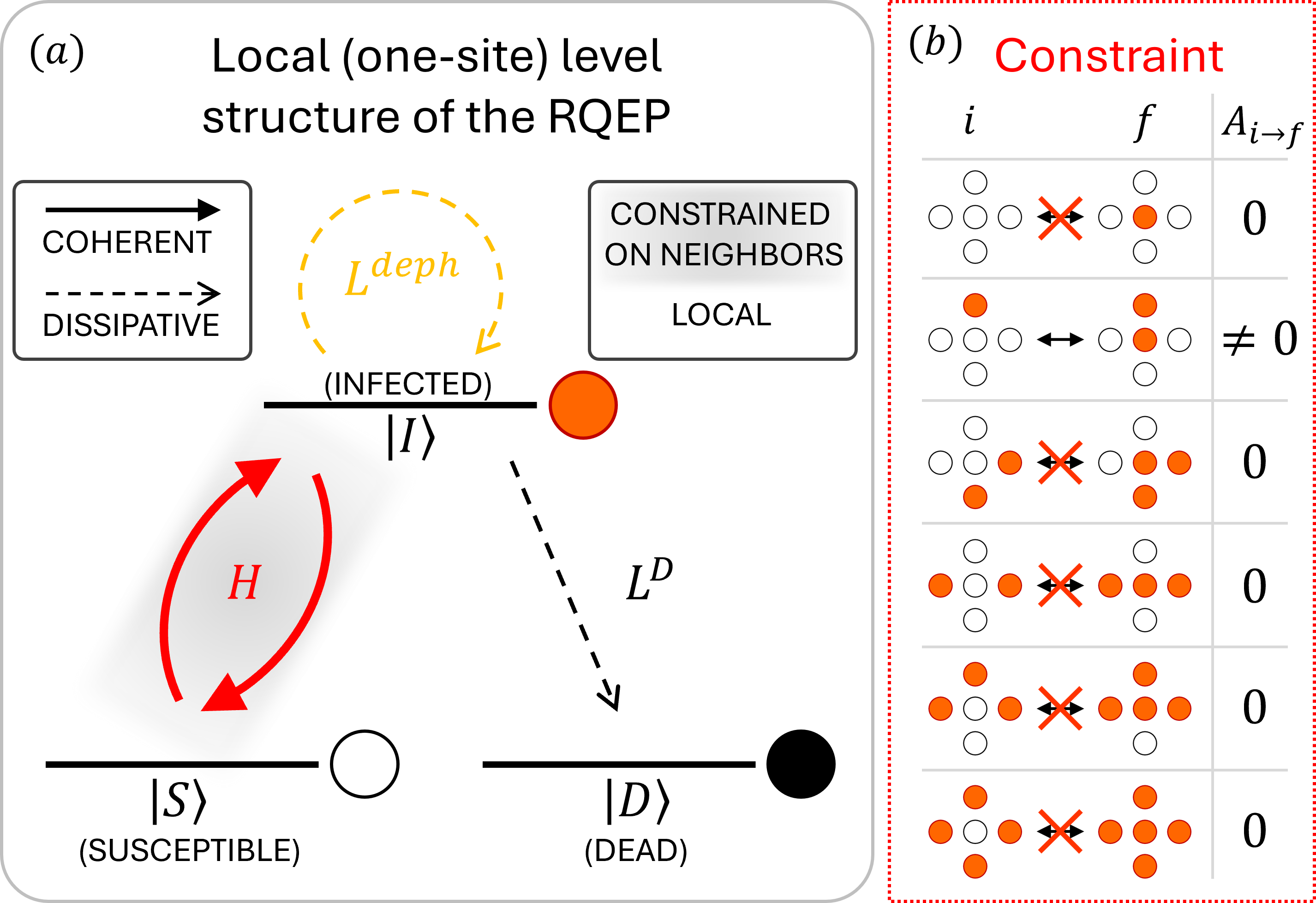}
  \caption{Visualization of the RQEP's structure: (a) level structure of a single site. The three GEP-like states $\ket{S}$, $\ket{I}$ and $\ket{D}$ are displayed as three black horizontal lines, each with its corresponding ket directly below. To the right of each, the corresponding classical state is reported in the style of Figs.~\ref{fig:GEP1} and \ref{fig:GEP2}. Arrows represent dynamical processes; solid ones denote Hamiltonian terms, whereas dashed ones dissipative processes associated to jump operators. From left to right, we have "infection" (red), "dephasing" (dark yellow) and "death" (black). The shaded area highlights the only process, infection, which involves more than one site and is thus capable of producing spatial correlations. The kinetic constraint of the RQEP is illustrated in panel (b), where we use as an example five sites in a cross configuration. Due to the isotropy of the model, configurations rotated by multiples of $\pi/2$ obey analogous rules. Every case (plus rotations thereof) with a crossed-out arrow is prohibited by the constraint. The second case from the top (plus rotations thereof) is the only allowed transformation among those drawn.
The rightmost column, labeled $A_{i \to f}$, shows whether or not states $\ket{i}$ and $\ket{f}$ are connected under the infection Hamiltonian \eqref{eq:conn_under_evol}. The second entry from the top should be read as $\exists\,\, t > 0$ such that $A_{i \to f} (t) \neq 0$. } 
\label{fig:RQEP1}
\end{figure}

The RQEP is a many-body system defined on a two-dimensional square lattice of size $n \,\times \,n = N$ with open boundary conditions. The level structure for a single site consists of three local orthonormal states, which we dub $\ket{S}$ ("susceptible"), $\ket{I}$ ("infected") and $\ket{D}$ ("dead") in analogy with the GEP presented earlier. In the following, we shall refer to them as "GEP-like states" and, all three together, as the "local classical basis".
A sketch of this level structure can be found in Fig.~\ref{fig:RQEP1}(a). 

%
% FIGURE HERE
%

We shall use Roman lowercase letters ($j$, $k$, $l$) for the lattice indices, so that for instance $\ket{D}_k$ indicates that the $k$-th site is in state $\ket{D}$. For later convenience, we also introduce the elementary local matrices
\be
	\sigma_k^{\mu \nu} =  \ket{\mu}_k   \brak{\nu}{k}
	\label{eq:supPauli}
\ee
with $\mu, \, \nu \in \left\{ S,\, I,\, D   \right\}$, which satisfy
\begin{subequations}
	\label{eqs:sigmas}
\begin{align}
	&\sigma_k^{\mu \nu} \ket{\nu'}_k = \delta_{\nu \nu'} \ket{\mu}_k, \label{subeq:matrixvec}\\
	&\sigma_k^{\mu \nu} \sigma_k^{\nu' \mu'} = \delta_{\nu \nu'}\, \sigma_k^{\mu \mu'}  \label{subeq:matmat} \\
	&\lt \sigma_k^{\mu \nu} \rt^\dag = \sigma_k^{\nu \mu}.
\end{align}
\end{subequations}
We also point out that $\sigma_k^{\mu\mu}$ acts locally as the projector over the state $\ket{\mu}_k$.

To avoid making the notation too cumbersome, we shall use the same symbols to describe these matrices once trivially extended over the whole Hilbert space:
\be
	\sigma_k^{\mu\nu} = \lqq \bigotimes_{j = 1}^{k-1} \id_j \rqq \otimes \lt  \ket{\mu}_k   \brak{\nu}{k} \rt  \otimes \lqq \bigotimes_{j=k+1}^{N} \id_j \rqq,
	\label{eq:extended} 
\ee
with $\id_j = \sum_{\mu} \ket{\mu}_j \brak{\mu}{j} = \sum_\mu \sigma_j^{\mu\mu}$ the $3\times 3$ local identity operator. Properties \eqref{eqs:sigmas} remain essentially the same with definition \eqref{eq:extended}. This modest abuse of notation allows us to write the two-site commutation relations
\be
	\comm{\sigma_k^{\mu \nu}}{ \sigma_j^{\mu' \nu'}} = 0 \ \ \ \forall j \neq k.
    \label{eq:sigma_comm}
\ee
As we have done for a single site, we introduce a "global classical basis" as the set of all product states $\ket{\vec{\mu}} = \bigotimes_k \ket{\mu_k}_k$ with $\mu_k \in \left\{ S,\, I,\, D   \right\}$, i.e., in which each factor is a local GEP-like state; the vector $\vec{\mu}$ is in this case simply a shorthand for the list of labels $\vec{\mu} = \lt \mu_1, \, \mu_2 , \,\ldots,\, \mu_N  \rt$. Having three local states per site and $N$ sites in the lattice, the dimension of the whole Hilbert space is $\dim \mal{H} = 3^N$, corresponding to all the possible ways in which the labels in $\vec{\mu}$ can be fixed independently.

The RQEP's dynamics takes place under the action of three distinct processes, also reported in Fig.~\ref{fig:RQEP1}(a):
\begin{itemize}
\item{\underline{Infection:} as in the GEP's case, an infected site here has the potential to infect a susceptible neighbor. Differently from before, however, the RQEP features a stricter constraint: the excitation $\ket{S} \to \ket{I}$ is only allowed in the presence of one, \emph{and only one}, infected neighbor. This peculiar restriction has its origin in the "facilitation" physics of Rydberg atoms, which the RQEP takes inspiration from: as we have briefly explained in Sec.~\ref{Sec:Intro}, in these atomic systems an excitation can, by virtue of its strong interactions $\Delta V$ with its neighbors, shift the transition frequency $\omega_{atom}$ of a nearby atom from its ordinary value $\omega$ to $\omega + \Delta V$. Assuming that the laser's frequency is also $\omega_{las} = \omega + \Delta V$, the interatomic interactions would bring its neighbor into resonance with the radiation. However, having multiple (say, $m$) excitations nearby would produce a further shift and $\omega_{atom} \approx \omega + m\Delta V$ would be again off resonant, rendering the neighborhood transparent (and thus not excitable).   
 
Infection in the RQEP is therefore modeled via the Hamiltonian  
\be 
	H = \Omega_I \sum_k \Pi_k \lt \sigma_k^{SI} + \sigma_k^{IS} \rt  ,
	\label{eq:RQEP_H}
\ee 
where the matrix within round brackets, written in the notation of Eq.~\eqref{eq:supPauli}, is essentially a Pauli matrix $\sigma^x$ in the $\ket{S}$, $\ket{I}$ subspace; it thus induces a form of precession between these two states. The term $\Pi_k$ is instead a projector meant to enforce the kinetic constraint. As such, it has to project over all states including a single excitation among the nearest neighbors of site $k$. To write it in a compact form, let us denote by $\Lambda_k$ the set of said nearest neighbors; then,
\be
	\Pi_k = \sum_{j \in \Lambda_k} \lqq \sigma_j^{II} \prodl{l \in \Lambda_k/\left\{ j \right\}}{} \lt 1 - \sigma_l^{II} \rt \rqq.
	\label{eq:RQEP_proj}
\ee
The strength (and speed) of infection is controlled via the parameter $\Omega_I$ (a sort of Rabi frequency).
}
\item{\underline{Death:} as it does in the GEP, death locally induces a decay $\ket{I} \to \ket{D}$ independently on each site. It is described via $N$ jump operators (one per site):
\be
	L^D_{k} = \sqrt{\gamma_D} \, \sigma_k^{DI}
    \label{eq:RQEP L_D}
\ee 
with $\gamma_D$ the corresponding rate. By applying Eq.~\eqref{subeq:matrixvec} it is straightforward to find
\be
	\sigma_k^{DI} \ket{I}_k = \ket{D}_k
    \label{eq:Death on ket{D}}
\ee 
and
\be
\sigma_k^{DI} \ket{S}_k = \sigma_k^{DI} \ket{D}_k = 0 ,
\label{eq:Death zero eig}
\ee
showing that this jump operator indeed acts as intended.
}
\item{\underline{Dephasing:} a RQEP process without a GEP equivalent, dephasing induces decoherence independently on each site; its action induces a decay in all quantum superpositions involving GEP-like state $\ket{I}$. To more clearly explain its effects, consider a local density matrix $\rho_k$ represented in the local classical basis: dephasing would leave its diagonal elements untouched, while killing the off-diagonal ones exponentially fast in time. If we let dephasing acting on its own for long enough, we would eventually get a diagonal matrix
\be
	\rho_k^{(deph)} = \sum_\mu \brak{\mu}{k} \rho_k \ket{\mu}_k  \sigma_k^{\mu\mu}.
\ee
This type of action is achieved via another set of $N$ jump operators, one per site, reading
\be
	L^{deph}_{k} = \sqrt{\gamma_{deph}}\, \sigma_k^{II}
\ee
with $\gamma_{deph}$ the corresponding rate. This process controls the degree to which quantum coherences are allowed to develop in the course of the dynamics. The higher the dephasing rate, the smaller and shorter-lived quantum effects (quantum correlations, for instance) will be in the system.
}
\end{itemize}

With the RQEP fully defined, we are now in the position to contrast its main features with those of the GEP. We start by considering what an absorbing state is for an open quantum system: the obvious requirement $\mal{L} \rho = 0$ is, albeit necessary, not sufficient. It, in fact, merely identifies a \emph{stationary} state and every well-defined Lindblad equation $\mal{L}$ (and in particular any Lindblad equation acting on a finite-dimensional space of states) features at least one such $\rho$. 
Instead, we define a density matrix $\rho^{(a)}$ to be absorbing if
\begin{itemize}
\item[(i)] it is a pure state, $\rho^{(a)} = \ket{\psi^{(a)}} \bra{\psi^{(a)}}$, and $\ket{\psi^{(a)}}$ is a \emph{dark state}, i.e., is
\item[(ii)] an eigenvector of the Hamiltonian $H\ket{\psi^{(a)}} = \lambda \ket{\psi^{(a)}}$ and
\item[(iii)] annihilated by (i.e.~a kernel eigenvector of) all jump operators: $L_\alpha \ket{\psi^{(a)}} = 0 \,\,\, \forall\,\, \alpha$.
\end{itemize}
%If $\ket{\psi^{(a)}} = \ket{\vec{\mu}}$ for some classical basis vector, we shall call it a "classical absorbing state". 
Our choice of conditions (i)-(iii) is expanded upon in Appendix \ref{App:abs_dark}; it should also become more transparent once we introduce the stochastic unraveling of $\mal{L}$ in Sec.~\ref{sec:theory}; here we simply take it as a working definition.

Second, guided by the similarities with the GEP, it is not too difficult to prove that all GEP-like states $\ket{\vec{\mu}}$ whose components $\mu_k$ are either $S$ or $D$ (in any given combination) are absorbing states for the RQEP. This is because each of the three processes (Infection, Death, Dephasing) introduced in Ref.~\cite{Espigares2017} includes an operator ($\Pi_k$, $\sigma_k^{DI}$, $\sigma_k^{II}$) which annihilates $S$ and $D$ components somewhere:
\be
    \sigma_k^{DI} \ket{S}_k = \sigma_k^{DI} \ket{D}_k = \sigma_k^{II} \ket{S}_k = \sigma_k^{II} \ket{D}_k = 0
\ee
and
\be
    \Pi_k \ket{S}_j = \Pi_k \ket{D}_j = 0
\ee
whenever $j$ is a neighbor of $k$ ($j \in \Lambda_k$). We thus recover one of the most basic rules of the GEP: any state devoid of infected sites is an absorbing state and constitutes an end for the epidemic.

Moving now to the most visible differences with respect the GEP, we first highlight the fact that the constraint imposed over the infection process is stricter: infection is only allowed with one --- and only one --- infected neighbor. A pictorial representation of this restriction is provided in Fig.~\ref{fig:RQEP1}(b) (see also App.~\ref{App:Afi}). However, this should not yield major discrepancies in the qualitative features of the phase transition: after all, close to the critical point the average local infection density $\av{\sigma_j^{II}}$ is expected to remain low throughout the dynamics. If we expand the projector \eqref{eq:RQEP_proj} to linear order in the $\sigma^{II}$s, we find $\Pi_k \approx \sum_{j \in \Lambda_k} \sigma_j^{II}$, i.e., a GEP-like constraint where each infected neighbor increases the likelihood of infection at $k$. Hence, at low local infected density (and consequently in the proximity of the critical point or at lower $\gamma_I$) we should not expect major qualitative differences to arise.

More significantly, the RQEP features an entirely new process, dephasing. This was used in Ref.~\cite{Espigares2017} to control the relevance of quantum effects. Under very strong dephasing ($\gamma_{deph} \gg \gamma_D,\, \Omega_I$), a perturbative procedure \cite{Degenfeld2014, Marcuzzi2014} approximately maps the RQEP onto a classical stochastic process very similar to the GEP with an effective infection rate $\gamma_I = 4\Omega_I^2 / \gamma_{deph}$, an effective death rate of $\gamma_D$, but with the stricter constraint just presented.

Unsurprisingly, then, in the strong dephasing regime the RQEP undergoes a continuous phase transition from an absorbing phase (where the infection does not propagate far from the origin) to an active one (featuring an outbreak propagating from the center outward), much like the GEP. Qualitatively, the behavior is similar to the one displayed in Fig.~\ref{fig:GEP1}(d), with the local DDS at time $t$ now being defined as $\avl{\sigma_k^{DD}(t)}$ and its final average reading
\be 
	n_D = \liml{t \to +\infty} n_D (t) = \liml{t \to +\infty} \frac{1}{N} \sum_k \avl{\sigma_k^{DD}(t)}.
	\label{eq:qDDS}
\ee

\begin{figure}[h]
  \includegraphics[width=\columnwidth]{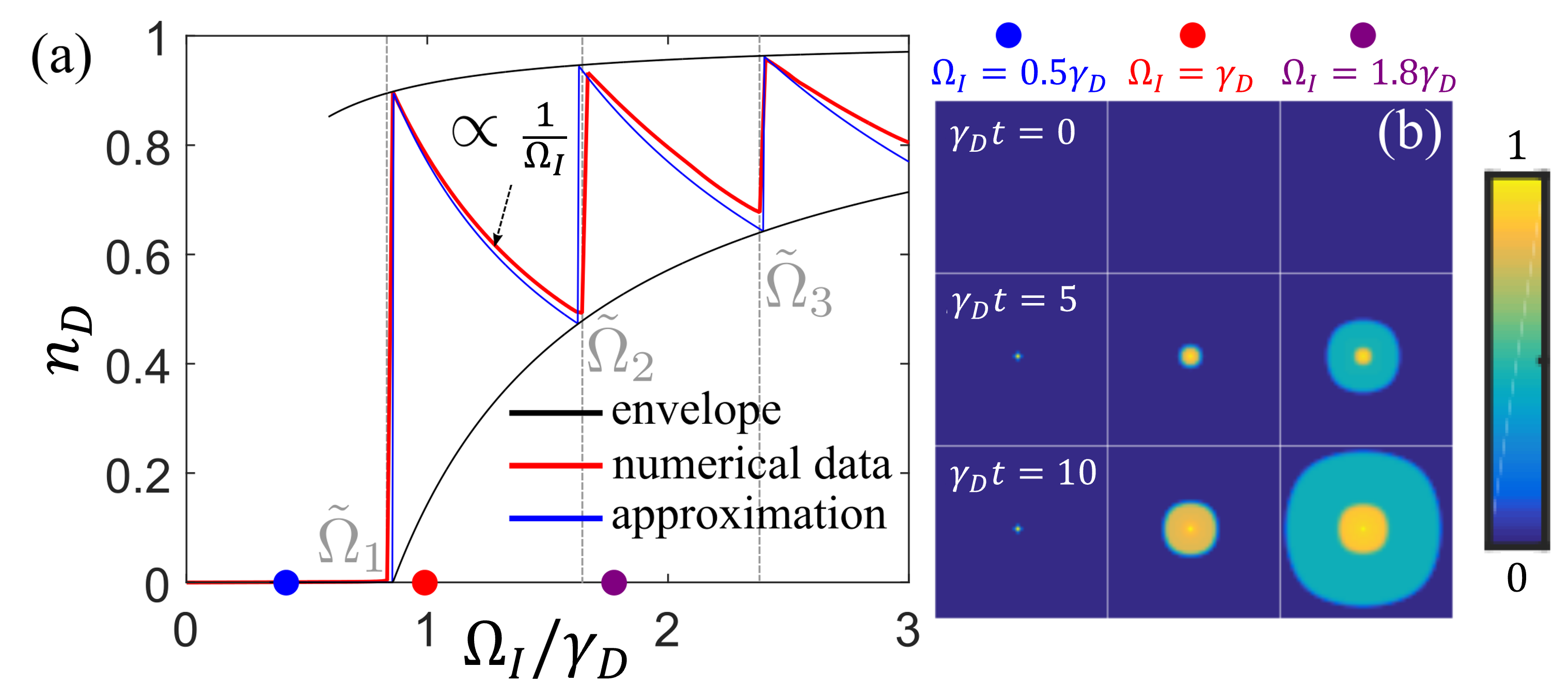}
  \caption{This figure has been adapted from Ref.~\cite{Espigares2017}; the labels have been modified to match our present notation. Panel (a) displays the final average DDS $n_D$, defined in Eq.~\eqref{eq:qDDS}, as a function of the ratio $\Omega_I / \gamma_D$ for $\gamma_{deph} = 0$. The red line illustrates the numerical solution of the inhomogeneous mean-field equations; the blue and black lines constitute predictions obtained via a scaling argument (not reported here) and can for our present purposes be safely ignored. The repeating structure of jumps followed by a continuous decay in the final DDS is here apparent; from the same scaling argument the jumps across the thresholds $\tilde{\Omega}_i$ $(i = 1,\,2,\,3,\,\ldots)$ are expected to be discontinuous. Panel (b) shows the corresponding patterns appearing in the dynamics for (moving horizontally) three reference values of the ratio $\Omega_I / \gamma_D$ and (moving vertically) three different times (measured in units of $\gamma_D^{-1}$). Each of the nine subpanels represents a map of the local DDSs $\av{\sigma_k^{DD} (t)}$ over the entire $51\times 51$ lattice. Reading each column from left to right we observe the presence of zero, one and two outbreaks, respectively. The three colored dots at the top are used in panel (a) as a guide to the eye on where the three chosen parameter values lie on the horizontal axis.  }
\label{fig:old_one}
\end{figure}

The predictions of Ref.~\cite{Espigares2017}, however, significantly diverge from the GEP's behavior in the opposite limit, i.e., when $\gamma_{deph}$ is small or vanishes entirely. According to a non-uniform mean-field analysis, letting quantum fluctuations survive makes the transition discontinuous, as shown in Fig.~\ref{fig:old_one}(a): crossing a threshold value $\tilde{\Omega}_1$, the final average DDS leaps from $0$ to a finite, large value $\gtrsim 0.8$.  
Counter-intuitively, upon further increasing the infection parameter $\Omega_I$ the average DDS decreases. Even more strangely, after hitting a second threshold $\tilde{\Omega}_2$ we encounter another upwards jump, beyond which $n_D$ starts decreasing once again. In a similar fashion, as $\Omega_I$ is increased even further, a third discontinuity is encountered, then a fourth, a fifth and so on (only a few are displayed in the figure), in a structure we could describe as a sequence of first-order transitions.

Interestingly, this odd behavior of the stationary average DDS is accompanied by the emergence of an equally peculiar dynamical pattern, illustrated by the dynamical, site-resolved snapshots of the dynamics in Fig.~\ref{fig:old_one}(b). Each of the jumps of $n_D$ shown in panel (a) coincides, in fact, with the occurrence of a new outbreak at some point in time. Each column of the $3\times 3$ grid corresponds to a different value of $\Omega_I$ and displays, from top to bottom, the average local DDS over the entire ($51\times 51$) lattice at three different times $\gamma_D t = 0$, $5$ and $10$. Below the first threshold ($\Omega_I < \tilde{\Omega}_1$) the system is in the inactive phase; there are no outbreaks and only a small neighborhood of the origin develops a non-negligible DDS (bottom-left subpanel). 

Increasing the infection parameter $\Omega_I$ beyond the first threshold, but keeping it below the second one, $\Omega_I \in \lqq \tilde{\Omega}_1, \, \tilde{\Omega}_2  \rqq$, the dynamics features an outbreak, i.e., a wave of infection travels throughout the system eventually covering the entire lattice with some finite DDS. This behavior looks analogous to the one observed in the active phase of the GEP, if not for the fact that the outbreak now covers the system with a large, finite value of DDS no matter how close to the threshold point $\tilde{\Omega}_1$ the parameters are. 

After crossing the second threshold $\tilde{\Omega}_2$, something different takes place: as the propagation of the original outbreak is underway, a second one arises from the origin and starts radiating outwards. In a sense, where this wave passes some further DDS is "deposited on top" of the one that the first outbreak left behind. Should the third threshold be crossed, another wave of infection will develop following the first and second ones, adding its own contribution to the final DDS. Each jump in $n_D$ seen in Fig.~\ref{fig:old_one}(a) is the result of a new outbreak appearing in the dynamics and a new "layer" of dead density being added on top of the previous ones.

This surprising departure from the simple, intuitive behavior of the GEP hinges on the validity of the inhomogeneous mean-field approximation adopted in \cite{Espigares2017}, which neglects correlations between different sites, i.e., performs the substitution
\be
	\avl{\sigma_j^{\mu \nu} \sigma_{k}^{\mu' \nu'}} \to \avl{\sigma_j^{\mu \nu}} \avl{ \sigma_{k}^{\mu' \nu'}} \ \ \ \forall \,\, k \neq j
\ee
whenever such a combination appears in the Heisenberg equations of motion for the local observables $\sigma_k^{\mu\nu}$.

Generally, mean-field approaches of this kind succeed in predicting the qualitative behavior of the system as long as the correlation length (the characteristic distance over which correlations between statistical fluctuations dissipate) is not much larger than the lattice spacing. This can be particularly problematic in the presence of continuous phase transitions, as the order parameter's correlation length diverges at the critical point \cite{ZinnJustin_book}. 
Since the new physics predicted for the RQEP involves discontinuous transitions instead, and since local processes like death and dephasing curtail the spread of correlations, it is not completely unreasonable to think that the mean-field equations may be capturing some actual features of the dynamics. However, as we lack an estimate of the correlation length, it is impossible to draw a conclusion either way.

Part of the difficulty in deciding one way or the other lies in the fact that the RQEP cannot be simulated via standard methods: exact numerical diagonalisation of the superoperator $\mal{L}$ is, in the absence of obvious symmetries, stifled by the extreme growth of the vector space dimension with the system size $N$ (with three states per site the full Hilbert space has dimension $3^N$, the Liouville space further gets that squared, $3^{2N}$, and thus $\mal{L}$ is a $3^{2N} \times 3^{2N}$ matrix). To illustrate the scales at play, consider that, without recourse to high-performance computing, it is already quite difficult to reliably diagonalize matrices larger than $\sim 10^6 \times 10^6$. For the RQEP, $\mal{L}$ reaches this size at $N \approx 6$ sites, way too small to provide any meaningful insight.

Stochastic methods such as Quantum Jump Monte Carlo \cite{Qjumps1, Qjumps2, Breuer_book} have the advantage of working with vectors $\ket{\psi}$ from the original Hilbert space $\mal{H}$; the density matrix is then reconstructed by taking an average over a (sufficiently large) set of stochastic trajectories $\rho = \overline{\ket{\psi} \bra{\psi}}$. Bringing back the problem to a formulation over $\mal{H}$ means that the typical matrices one works with are now $3^N \times 3^N$ and one can, in principle, simulate twice the system size. Still, $N\approx 12$ sites are way too few to even distinguish between the inactive and active phases. This is made even worse by having to study the RQEP in (at least) two dimensions. If we think, for instance, of a RQEP on a $3 \times 4$ lattice, then once the infection leaves the origin it is already at the edges. 

As discussed in Sec.~\ref{subsec:GEP}, in one spatial dimension any (short-range) epidemic dynamics becomes trivial. More advanced methods based on the Density Matrix Renormalization Group, or DMRG \cite{DMRG1, DMRG2}, are likely not applicable to the RQEP as they are specialized in dealing with one-dimensional cases.

Our approach in this work is to look for a simpler epidemic model which is amenable to numerical simulation up to a reasonable system size, while still retaining some of the basic properties that make the RQEP different from the GEP. We shall elucidate our thought process in the next Section. Before bringing our account of the RQEP to a close, however, we recall one last qualitative argument from Ref.~\cite{Espigares2017}, namely the fact that the appearance of multiple outbreaks could be due to the "precession-like" nature of the Hamiltonian term (see our discussion below Eq.~\eqref{eq:RQEP_H}). Let us assume that, after an initial transient, a small region of infected sites forms around the origin oscillating between $I$ and $S$ states under the action of $H$. They are able to do so by mutually satisfying the constraint for each other, e.g.. ~the origin permits the oscillation of its neighbors, which then acquire some component over $\ket{I}$, which then permits the oscillation of the origin. This region then acts as a pulsating generator of outbreaks: at times when its distal sites have large infected (I) components we find the highest probability of infection for sites immediately outside this region; these times therefore correspond to the moments at which outbreaks are spawned. Conversely, when the central region is mostly susceptible (S) the probability of starting another outbreak is strongly suppressed. Thus, it is presumably this oscillation of the central region in and out of infection that creates periodic opportunities for outbreaks to start and gives rise to the recurrent structure just described.

\subsection{The current model: the elementary Quantum Epidemic Process (eQEP)}
\label{subsec:eQEP}

We devote this Section to introducing our own quantum epidemic model, which we dub the "elementary Quantum Epidemic Process", or eQEP for short. We subdivide our discussion in two parts: in the former we try, for the interested reader, to elucidate our reasoning. In the latter, we directly define the model, summarizing its most relevant features.

\subsubsection{Our reasoning}

We are looking for the simplest quantum epidemic model we can find which satisfies a certain set of requirements; let us summarize and discuss them one by one:
\begin{itemize}
    \item[(I)] it is a Markovian open quantum model.
\end{itemize}
As discussed earlier, closed quantum models evolve in a reversible fashion. We need dissipation in order to have absorbing states. The assumption of Markovianity, albeit not strictly necessary, is a simplification that allows us to build the model as a Lindblad equation. This inclination towards choosing the simplest option will be recurring throughout this Section and it, in substance, the reason behind calling the model "elementary".

In order to fully define the eQEP within the Lindblad framework, we need to specify a space of states, a Hamiltonian and a set of jump operators, which will be the aim of the rest of this Section. Besides being dissipative and quantum, we ask that the eQEP be
\begin{itemize}
    \item[(II)] an epidemic model, i.e., featuring
            \begin{itemize}
                \item[(IIa)] a constrained infection and
                \item[(IIb)] an irreversible death.
            \end{itemize}
\end{itemize}
In order to even make sense of these terms, we imagine a system on a lattice in (at least) two dimensions, where we need to have at least three local levels per site, a susceptible $\ket{S}_k$, an infected $\ket{I}_k$, and a dead one $\ket{D}_k$, with $k$ the lattice index. Infection must evolve $\ket{S}_k$ into $\ket{I}_k$, death $\ket{I}_k$ into $\ket{D}_k$. By analogy with the GEP, we further ask that only susceptible sites may be infected and only infected ones may die. Requirement (IIb) forces death to be a dissipative process; since we cannot think of any reason to make death of an infected dependent on sites other than the one affected, we simply copy it from the RQEP:
\be
	L^D_{k} = \sqrt{\gamma_D} \, \sigma_k^{DI}.
    \label{eq:eQEP L_D_0}
\ee 

From the RQEP we also take another ingredient; more precisely, we extract it from the interpretation we summarized at the end of Sec.~\ref{subsec:RQEP}, that sees the RQEP Hamiltonian as responsible of local oscillations which "activate" and "deactivate" sites. We ask that, in the eQEP,
\begin{itemize}
    \item[(III)] the Hamiltonian induces oscillations between the infected state $\ket{I}_k$ and some other, inactive state. By "inactive" we mean that it does not participate in the epidemic dynamics: it can neither infect others nor die.
\end{itemize}
Note that, whatever this state may be, the Hamiltonian will connect it to $\ket{I}_k$, implying that it will generally be able to oscillate back to being infected- Therefore, this other state cannot be $\ket{D}_k$, or we would violate death irreversibility, (IIb). It can either be $\ket{S}_k$, as in the RQEP, or a new local state. In the former case, the Hamiltonian would be responsible for infection (at least partly) and it would need to be constrained, lest we violate (IIa). Then, it would look like Eq.~\eqref{eq:RQEP_H}, with some appropriate projector $\Pi_k$ enforcing the constraint. This projector has support (i.e., acts non-trivially) not on site $k$ directly, but on its neighbors. A Hamiltonian in this form, therefore, generates correlations between different sites. We identify this as the major obstacle in simulating the RQEP and wish to replace Eq.~\eqref{eq:RQEP_H} with a sum of local terms. 

We thus exclude $\ket{S}_k$ as well and are forced to introduce a new local state, which, to remain more or less faithful to the epidemic lexicon of the previous models, we call "bedridden" and denote by $\ket{B}_k$. The matrices $\sigma_k^{\mu \nu}$ can still be defined via Eq.~\eqref{eq:supPauli}, with the obvious caveat that now $\mu$, $\nu \in \set{S, I, B, D}$ with the addition of the new label $B$. The properties \eqref{eqs:sigmas} and the commutation relations \eqref{eq:sigma_comm} still hold in the same form.

\begin{figure}[h]
  \includegraphics[width=\columnwidth]{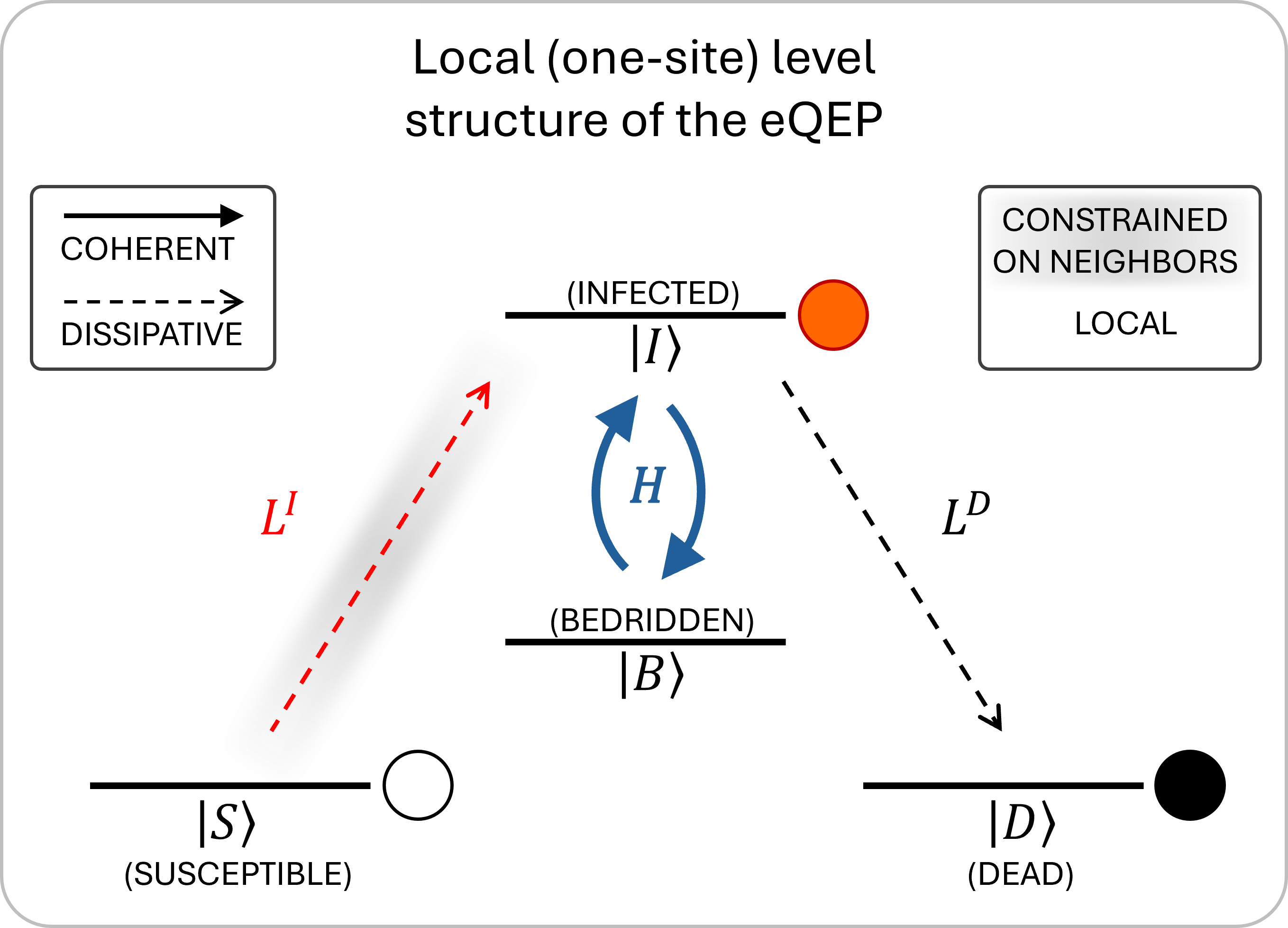}
  \caption{Pictorial representation of the eQEP: each site features $4$ levels, $\ket{S}$ (susceptible), $\ket{I}$ (infected), $\ket{B}$ (bedridden) and $\ket{D}$ (dead). Except for the newly-introduced bedridden state, there is a straightforward relation to the classical ones of the GEP, shown here as colored circles with the same style of Fig.~\ref{fig:GEP1}
  We adopt here some of the conventions used for Fig.~\ref{fig:RQEP1}: dashed lines denote dissipative processes, solid lines coherent (Hamiltonian) terms. There are two of the former, infection (red) and death (black), and one of the latter (blue), which is the only one connecting $\ket{B}$ to the rest. Shaded areas show where the constraint applies, i.e.~just to infection. Both the Hamiltonian and death jump operators are collections of one-site, local terms. Note that in the picture above there are no arrows going left: once you move right you can never go back. This is a visualization of the $S \to I \to D$ "directionality" of the GEP, which is recovered in the eQEP.
}
\label{fig:eQEP1}
\end{figure}

The simplest Hamiltonian obeying (III) is then
\be
    H = \sum_{k}\Omega \lt \sigma_k^{IB} + \sigma_k^{BI} \rt. 
    \label{eq:eQEP_H_0}
\ee
In the IB subspace of any given site $k$, this Hamiltonian acts as the Pauli operator $\sim \sigma^x$ and generates a precession with frequency $\propto \Omega$. The new four-level structure is depicted in Fig.~\ref{fig:eQEP1}.

We are still left with infection. For the same reason stated above, we do not want it to be described by a Hamiltonian term, which leaves us with the only choice of making it part of the dissipator. Since we do not expect fundamental qualitative changes in behavior from making the infection constraint laxer (as in the GEP) or stricter (as in the RQEP), we adopt here the GEP convention, i.e., we allow susceptible sites in the eQEP to be infected at a faster and faster rate the more infected neighbors they have. 

To achieve this, we introduce a set of jump operators
\be
    L^{I}_{kj} = \sqrt{\gamma_I}\, \sigma_k^{IS} \sigma_j^{II},
    \label{eq:eQEP_LI_0}
\ee
with the understanding that $k$ and $j$ are nearest neighboring sites, i.e., $j \in \Lambda_k$ (and $k \in \Lambda_j$). The jump above displays the desired behavior on the classical basis:
\be
    L_{kj}^I \ket{\mu}_k \ket{\nu}_j = \sqrt{\gamma_I}\, \delta_{\mu S}\, \delta_{\nu I}\,  \ket{I}_k \ket{I}_j,
\ee
i.e., infects site $k$ provided that $k$ is susceptible ($\mu = S$) and $j$ infected ($\nu = I$), annihilating any other $(\mu,\, \nu)$ combination. 

On a large square lattice (size $N = n \times n$ with $n \gg 1$) there would be $\approx 4N$ distinct infection jump operators, up to subleading boundary corrections; there would be, in fact, $\approx 2N$ "bonds" in the lattice, with each bond supporting two of these operators: $L_{kj}^I$ ($j$ infects $k$) and $L_{jk}^I$ ($k$ infects $j$).

Some observations are now in order: first, from our definitions we recover one of the properties of the GEP that was lost in the RQEP, namely its "directionality" $S \to I \to D$. In the GEP, once a site gets infected, it can never become susceptible again; once it dies, it can never be re-infected. The bedridden state acts as a shelving state for infection that does not take part in the original epidemic dynamics. In Fig.~\ref{fig:eQEP1} the directionality translates into the absence of arrows pointing left.

Second, all the absorbing states of the RQEP, $\ket{\vec{\mu}}$ with $\mu_k \in \set{S,\, D}$ $\forall \,k$, also annihilate the terms \eqref{eq:eQEP L_D_0}, \eqref{eq:eQEP_H_0}, \eqref{eq:eQEP_LI_0} we have just introduced: this can be again demonstrated by realizing that each of them contains at least one operator which annihilates both $\ket{S}$ and $\ket{D}$:
\begin{subequations}
\begin{align}
    &\lt \sigma_k^{IB}  + \sigma_k^{BI}  \rt \, \ket{S(D)}_k = 0, \\[0.2cm]
    &L_k^D \, \ket{S(D)}_k = L_{jk}^I \, \ket{S(D)}_k = 0.
\end{align}
\end{subequations}
Hence, we have a model that features a space of absorbing states which includes those of the RQEP, which in turn matches the classical GEP space. In fact, it can be shown (see App.~\ref{app:eQEP_absorbing}) that for $\Omega \neq 0$, the GEP-like absorbing states are \emph{all} the absorbing states present.

Third, for $\Omega = 0$ we recover, for our choice of initial conditions, the classical GEP dynamics. A proof can be found in App.~\ref{app:eQEP_to_GEP}. Here, instead, we offer a more intuitive argument: setting the frequency $\Omega$ to $0$ is equivalent to "turning off" the Hamiltonian and decoupling the $\ket{B}$ state from all the others, so that it can be safely ignored. The remaining terms are purely dissipative, plus the classical basis is closed under their action, i.e.~they map it onto itself. Furthermore, their action produces exactly the same transformations that infection and death would produce in the GEP.
This also implies that a comparison with the classical case can be done by simply setting $\Omega = 0$ and our model does not need to rely on dephasing as the RQEP.

In conclusion, the Lindblad dynamics we have hereby defined not only satisfies our basic requirements (I)-(III), but also matches the GEP's absorbing space structure and provides an easy way to control quantum fluctuations, which can be turned off entirely without affecting the epidemic nature of the dynamics. In line with our ever-present intention to keep the model as simple as possible, we do not see any reason at this stage to add more terms or further complicate any of the existing ones.

\subsubsection{The eQEP}

We define the eQEP as an open quantum model on a lattice of $N$ sites. Each site is associated to a four-dimensional local Hilbert subspace $\mal{H}_k = \mathrm{Span} \set{\ket{S}_k, \ket{I}_k, \ket{B}_k, \ket{D}_k} $, which adds a new, "bedridden" state $\ket{B}_k$ to the RQEP levels, see Fig.~\ref{fig:eQEP1}. The dynamics is described by a Lindblad equation including three distinct processes:
\begin{itemize}
\item \underline{Infection:} a dissipative process between nearest neighbors, in which an infected site $j$ can infect a susceptible one $k$. It is described by the jump operators
\be
    L^{I}_{kj} = \sqrt{\gamma_I}\, \sigma_k^{IS} \sigma_j^{II},
    \label{eq:eQEP_LI}
\ee
two per lattice bond ($j$ infecting $k$ and $k$ infecting $j$). The rate $\gamma_I$ measures the prominence of this term. The constraint is imposed according to the GEP prescription, i.e., a susceptible site with multiple infected neighbors has a higher chance of being infected.
\item \underline{Death:} another dissipative process, taking place independently on each site via 
$N$ jump operators
\be
    L^{D}_k = \sqrt{\gamma_D} \, \sigma_k^{DI}.
    \label{eq:eQEP L_D}
\ee
Formally, this is equivalent to death in the RQEP, see Eq.~\eqref{eq:RQEP L_D}.
\item \underline{(De)activation:} a coherent term described by a decoupled Hamiltonian 
\be
    H = \sum_k \Omega \lt \sigma_k^{IB} + \sigma_k^{BI} \rt,
    \label{eq:eQEP_H}
\ee
and the only term involving the new bedridden state. Its role is to reproduce a prominent feature of the RQEP, the fact that infected sites can be "deactivated", i.e., transformed into states that cannot infect and cannot die. In the RQEP, this took place via oscillations between $\ket{I}$ and $\ket{S}$ with deactivated sites being identified with susceptible ones. In the eQEP case, instead, deactivation occurs via oscillations to the $\ket{B}$ state which does not participate in the pure epidemic dynamics (thus the name "bedridden", since it cannot infect nor can it die, but can still become dangerous again).
\end{itemize}

As we discussed in Sec.~\ref{subsec:eQEP}, the eQEP recovers the $S \to I \to D$ "directionality" of the GEP, which was lost in the RQEP as infected sites could evolve into susceptible ones under Hamiltonian \eqref{eq:RQEP_H}. Furthermore, as shown in App.~\ref{app:eQEP_absorbing}, except for the special case $\Omega = 0$ the eQEP shares the same absorbing state of the RQEP, i.e. all $\ket{\vec{\mu}}$ with only susceptible or dead components $\mu_k$. Finally, as we prove in App.~\ref{app:eQEP_to_GEP}, for $\Omega = 0$ and with our initial conditions, the eQEP reduces exactly to the classical GEP.

\section{Reduction to a simpler stochastic model}
\label{sec:theory}

Numerical simulation of quantum systems is notoriously difficult due to the exponential growth of the size of the Hilbert space $\mal{H}$ with the system size. For the eQEP, each site has a four-level inner structure, so for a lattice of $N$ sites the Hilbert space's dimension is $4^N$. As already highlighted in the Sec.~\ref{subsec:RQEP}, this problem is exacerbated by the fact that the Lindblad equation acts on density matrices, which are in turn elements of the Liouville space $\sim \mal{H} \otimes \mal{H}^\ast$, whose dimension is the square of the Hilbert space's: $\mathrm{dim} \lt \mal{H} \otimes \mal{H}^\ast \rt = \lt \mathrm{dim} \mal{H} \rt^2$. Solving the dynamics of the eQEP, for instance, would mean facing the diagonalization of $\mal{L}$ as a $4^{2N} \times 4^{2N}$ matrix, which would be exceedingly difficult to perform numerically already for $N \approx 10$.

An alternative to studying the evolution in Liouville space is to perform a "stochastic unraveling" of the Lindblad equation. This consists in defining an auxiliary dynamical process acting on the original Hilbert space $\mal{H}$, which combines a deterministic evolution with a stochastic process. The random events produced under the latter are in a one-to-one correspondence with the Lindblad jump operators $L_\alpha$ and are often referred to as "quantum jumps" (or "jumps" for short).

The full Lindblad evolution of the density matrix is recovered by averaging over different realizations of the stochastic process. Note that this means that the price to pay for the reduction in dimensionality (from $4^{2N}$ to $4^N$ in our case) is that one must now average over the distribution of stochastic trajectories. However, this average can usually be efficiently approximated via Monte Carlo sampling, which is why the method takes the name of Quantum Jump Monte Carlo (QJMC for short) or Monte Carlo Wave Function (MCWF). 

We refer the unfamiliar reader to Refs.~\cite{Qjumps1, Qjumps2, Breuer_book} and to App.~\ref{App:QJMC}, where we have left our own description of the method. In the next Section we will merely list the fundamental instructions to implement a QJMC algorithm. This should be enough for us to then discuss the "symmetries" of the model that are going to make our numerical simulations much simpler.

\subsection{The Quantum Jump Monte Carlo algorithm}
\label{subsec:QJMC_algorithm}

For later convenience we introduce the so-called "effective Hamiltonian"
\be
	H_{eff} = H - \frac{i}{2} \sum_\alpha L_\alpha^\dag L_\alpha
	\label{eq:Heff}
\ee
and stress once again that the QJMC approach performs an evolution over the Hilbert space, i.e.~it acts on vectors $\ket{\psi} \in \mal{H}$ (we shall try to avoid using the term "state" for them as, without further assumptions, the \emph{actual state of the system} is instead a density matrix obtained from these vectors via an averaging process). This means that in any individual realization of the process, at all simulated times $t$ the algorithm is working with some vector $\ket{\psi(t)}$. In the following, we are often going to refer to the partial evolution under $H_{eff}$ as the \emph{deterministic} part of the evolution.

Consider the following task: we fix a time $t_{end} \geq 0$ and wish to reconstruct the state (density matrix) $\rho\lt t_{end} \rt$ evolved from some previous initial state fixed at $t = 0$. Let us focus on an individual stochastic trajectory; say that, after a number $z \geq 0$ of jumps this trajectory has brought the process to time $t_z < t_{end}$ and (normalized) vector $\ket{\psi_z}$. Then, the basic QJMC step consists of the following operations:
\begin{itemize}
\item[(I)] \emph{\underline{Determine the time of the next jump:}} with the shorthand $TFJ$ for "time of first jump" after $t_z$, we define the \emph{survival probability} $\mathbb{P} (TFJ \geq t) = \norm{\rme{-i H_{eff} (t-t_z)} \ket{\psi_z} }^2 $ as the squared norm of the deterministically evolved vector $\ket{\psi_z}$, which measures the probability that the first jump after $t_z$ occurs later than $t$ (i.e., after an interval of at least $t-t_z$ has elapsed). This $\mathbb{P}$ can be equivalently seen as the probability that no jumps take place between $t_z$ and $t$. We now uniformly pick a random real number $u \in \lqq 0,\, 1 \rqq$ and set $t_{z+1} = \mathbb{P}^{-1} (u)$ as the new jump time. 
%This condition is of course the same as $\mathbb{P}(TFJ \geq t_{z+1} ) = u$. The time $t_{z+1}$ is unambiguously defined since $\mathbb{P}$ is a non-increasing function of $t$; 
if $\mathbb{P} \equiv 1$ is constant, we set $t_{z+1} = +\infty$. If $t_{z+1} \leq t_{end}$ we move to instruction (II) below, otherwise we skip (II) and (IIIa) and leap further down to (IIIb).
\item[(II)] \emph{\underline{Find out which jump takes place:}} with the shorthand $\ket{\psi_{eff}(t_{z+1})} = \rme{-i H_{eff} (t_{z+1}-t_z)} \ket{\psi_z}$ we define $Q_\alpha = \norm{L_\alpha \ket{\psi_{eff}(t_{z+1})\,} \,}^2 \geq 0$ and $q_\alpha = Q_\alpha / \lt \sum_\beta Q_\beta \rt$, where the sum runs over all Lindblad jump operators. By construction, $\sum_\alpha q_\alpha = 1$ and we can interpret $\left\{ q_\alpha \right\}_{\alpha}$ as a set of probabilities. We then randomly pick a label $\overline{\alpha}$ with probability $q_{\overline{\alpha}}$. Note that if $L_\alpha \ket{\psi_{eff}(t_{z+1})} = 0$ then $\alpha$ has zero probability of being selected.
\item[(IIIa)] \emph{\underline{Apply the selected jump:}} adopting the notation just introduced in (II), the new vector at time $t_{z+1}$ is $\ket{\psi_{z+1}} \equiv L_{\overline{\alpha}} \ket{\psi_{eff}(t_{z+1})} / \sqrt{Q_{\overline{\alpha}}}$. We now have a new unit vector at a new time and can thus repeat the same steps starting back at (I). 
\item[(IIIb)] \emph{\underline{Apply the last stretch of deterministic evolution:}} we set the vector at $t_{end}$ as $\ket{\psi_{end}} = \rme{-i H_{eff} (t_{end}-t_z)} \ket{\psi_z} / \sqrt{\mathbb{P} \lt TFJ \geq t_{end} \rt}$, where the denominator once again ensures that $\ket{\psi_{end}}$ is normalized to $1$. The trajectory is now over and $\ket{\psi_{end}}$ is returned. 
\end{itemize}   
Steps (II) and (III) hold all the necessary bits of information to reconstruct the stochastic probability density: for a generic trajectory with $Z$ jumps
\be
    \tau = \left\{ \lt t_1, \alpha_1 \rt, \, \lt t_2, \alpha_2 \rt, \ldots, \, \lt t_Z, \alpha_Z \rt \right\}
\ee
we would simply have
\begin{align}
    p_{st} & (\tau) = \prod_{z=1}^Z  \Bigl( q_{\alpha_z}\, p_{TFJ} \lt t_z \, | \, t_{z-1} \rt  \Bigr) \times \nol
    &\times \mathbb{P} \lt TFJ \geq t_{end} \,|\, t_Z \rt. 
    \label{eq:pst_factors}
\end{align}
with $p_{TFJ}(t\,|\,t') = -\partial_t \mathbb{P}\lt TFJ \geq t \,|\, t' \rt$ the probability density for the time of first jump (TFJ) and $t_0 = 0$.
Analogously to the purely classical case (see Eq.~\eqref{eq:stoch_average}), we can introduce the stochastic averaging
\be
    \bar{\mal{A}} = \int \mal{D} \tau \, p_{st} (\tau) \, \mal{A}(\tau)
    \label{eq:stoch_av_quantum}
\ee
for any trajectory-dependent quantity $\mal{A}$. This allows us to recover the the Lindblad-evolved density matrix $\rho(t)$ from the vectors $\ket{\psi(t, \tau}$ obtained via the QJMC algorithm:
\be
    \rho(t) = \overline{\bigl. \ket{\psi(t, \tau)} \bra{\psi(t, \tau)} \bigr.}.
    \label{eq:stoch_rho}
\ee
Combining this expression with Eq.~\eqref{eq:Lind_average} we see that for any observable $O$ at time $t$ we have
\be
    \av{O(t)}_{\mal{L}} = \trace{O \rho(t)} = \overline{  \bra{\psi(t, \tau)} O \ket{\psi(t, \tau)} }.
\ee

Numerical simulations produce a finite number of trajectories $\tau_m$, $m = 1, \ldots, M_{tr}$; the advantage of Monte Carlo methods is that these are produced already weighed according to the corresponding stochastic distribution: in practical terms, since simulations can only produce trajectories from a finite ensemble, we can also think in these terms: if we have two trajectories $\tau$ and $\tau'$ such that $p_{st}(\tau) = 2 p_{st}(\tau')$ then in the limit $M_{tr} \to \infty$ $\tau$ will appear twice as frequently as $\tau'$. Hence, the observable average above is (approximately) reconstructed via the arithmetic average
\be
    \av{O(t)}_{\mal{L}} \approx \frac{1}{M_{tr}} \sum_m \, \bra{\psi(t, \tau_m)} O \ket{\psi(t, \tau_m)}.
\ee

We are now in a position to make more sense of the definition of "absorbing state" given in Sec.~\ref{subsec:RQEP} (see also App.~\ref{App:abs_dark}): we are not just asking for $\partial_t \,\rho(t) = 0$, but also for the QJMC dynamics to stop completely. To see this, let $\ket{\psi^{(a)}}$ again denote a dark state; then, $L_\alpha \ket{\psi^{(a)}} = 0$ $\forall\, \alpha$ implies $q_\alpha = 0$ $\forall\, \alpha$, i.e., no jump ever occurs. Furthermore, $H\ket{\psi^{(a)}} = \lambda \ket{\psi^{(a)}}$ implies
\begin{subequations}
\label{eqs:glob_phase}
\begin{align}
    &H_{eff} \ket{\psi^{(a)}} = \lambda \ket{\psi^{(a)}} \ \ \text{ and } \label{subeq:abs_Heff}\\ 
    &\rme{-iH_{eff}t} \ket{\psi^{(a)}} = \rme{-i\lambda t} \ket{\psi^{(a)}} \label{subeq:abs_exp}
\end{align}
\end{subequations}
with $\lambda \in \R$ (being an eigenvalue of the actual Hamiltonian $H$). In other words, over the course of the deterministic evolution a dark state merely picks up an immaterial global phase. For all intents and purposes, the trajectory is over.

\subsection{Symmetries of the eQEP and eigenspace structure}
\label{subsec:weak_sym}

Despite the noticeable reduction of computational complexity achieved by moving the problem from Liouville space to the original Hilbert space, QJMC simulations remain, in general, a costly endeavor. While the Monte Carlo sampling can typically be run very efficiently, the deterministic part of the evolution, generated by the effective Hamiltonian \eqref{eq:Heff}, is as complex to evaluate as the evolution of a general closed quantum system under its Hamiltonian, and constitutes the major bottleneck capping the achievable system size $N$. 

Whenever possible, it is thus important to identify any underlying symmetries and conserved quantities of the system, as this will yield a fragmentation of the space and thereby an effective reduction of the dimensionality of the problem. This will be the main aim of this Section.

The effective Hamiltonian defined in Eq.~\eqref{eq:Heff} for the eQEP reads
\begin{align}
    H_{eff} & = \Omega \sum_{k} \lt \sigma_k^{IB} + \sigma_k^{BI}  \rt - \frac{i}{2} \gamma_I \sum_{\nnsum{k}{j}} \sigma_k^{SS} \sigma_j^{II} +  \nol
    & - \frac{i}{2} \gamma_D \sum_{k} \sigma_k^{II},
    \label{eq:eQEP_Heff}
\end{align}
where, as in Sec.~\ref{subsec:eQEP}, $\nnsum{k}{j}$ denotes summation over all nearest neighboring site pairs. Exploiting the same-site orthogonality \eqref{subeq:matmat} and different-site independence \eqref{eq:sigma_comm} of the $\sigma^{\mu \nu}$ matrices one can see that, for any given site $k$, 
\be
    \comm{H_{eff}}{\,\sigma_{k}^{SS}} = \comm{H_{eff}}{\,\sigma_{k}^{DD}} = 0,
\ee
i.e.~the local susceptible and dead density operators \emph{on every site} are preserved under the action of $H_{eff}$. Note that conservation laws valid for $H_{eff}$ are not necessarily replicated in the full Lindblad evolution. Indeed, if that were the case, no initially susceptible site could be infected, nor any "alive" site could die. Within a QJMC approach, if changes in these densities cannot be effected by the deterministic evolution, they must come as a consequence of jumps.

We mention here that the unitary operators $U^\mu_k \lt \theta \rt = \rme{i \theta \sigma_k^{\mu\mu}}$ for $\mu \in \left\{ \text{S, D} \right\}$ and $\theta \in \R$ are, in fact, symmetries of the Lindblad equation, i.e., they leave it invariant in the following sense: adopting for a moment the notation $\mal{L} \lqq H, \left\{ L_\alpha \right\}_\alpha \rqq$ for a Lindblad equation constructed via a Hamitonian $H$ and jump operators $L_\alpha$, we say that the associated superoperator is invariant under a unitary transformation $U$ if 
\be
    \mal{L}  \lqq H, \left\{ L_\alpha \right\}_\alpha \rqq = \mal{L}  \lqq U^\dag H U, \left\{ U^\dag L_\alpha U \right\}_\alpha \rqq.
    \label{eq:Uinvariance}
\ee
Yet, none of the corresponding $2N$ densities is preserved. The unfamiliar reader may find it surprising to see symmetries $U$ associated to non-conserved observables $\sigma$, but this is a well-known occurrence in Lindblad physics. Symmetries of this kind are often dubbed "weak" to distinguish them from the "strong" ones which instead feature conserved generators. We refer the interested reader to Refs.~\cite{Buca2012, Albert2014} or, for a brief summary of the main concepts, to our own App.~\ref{app:strong_weak}.

In order to elucidate in what way the jump operators defy the conservation laws one could naively expect, we need to discuss the relevant eigenspaces: both $\sigma_k^{SS}$ and $\sigma_k^{DD}$ are (orthogonal) projectors with support over a single site. By virtue of being projectors, they admit eigenvalues $0$ and $1$ only. Because all these operators commute, as can be seen from Eqs.~\eqref{eq:sigma_comm} and \eqref{eqs:sigmas}, eigenvalues relating to different sites can be chosen independently and global eigenspaces can be constructed as a tensor product of local ones. 

\begin{figure}[h]
  \vspace{0cm}
  \includegraphics[width=0.9\columnwidth]{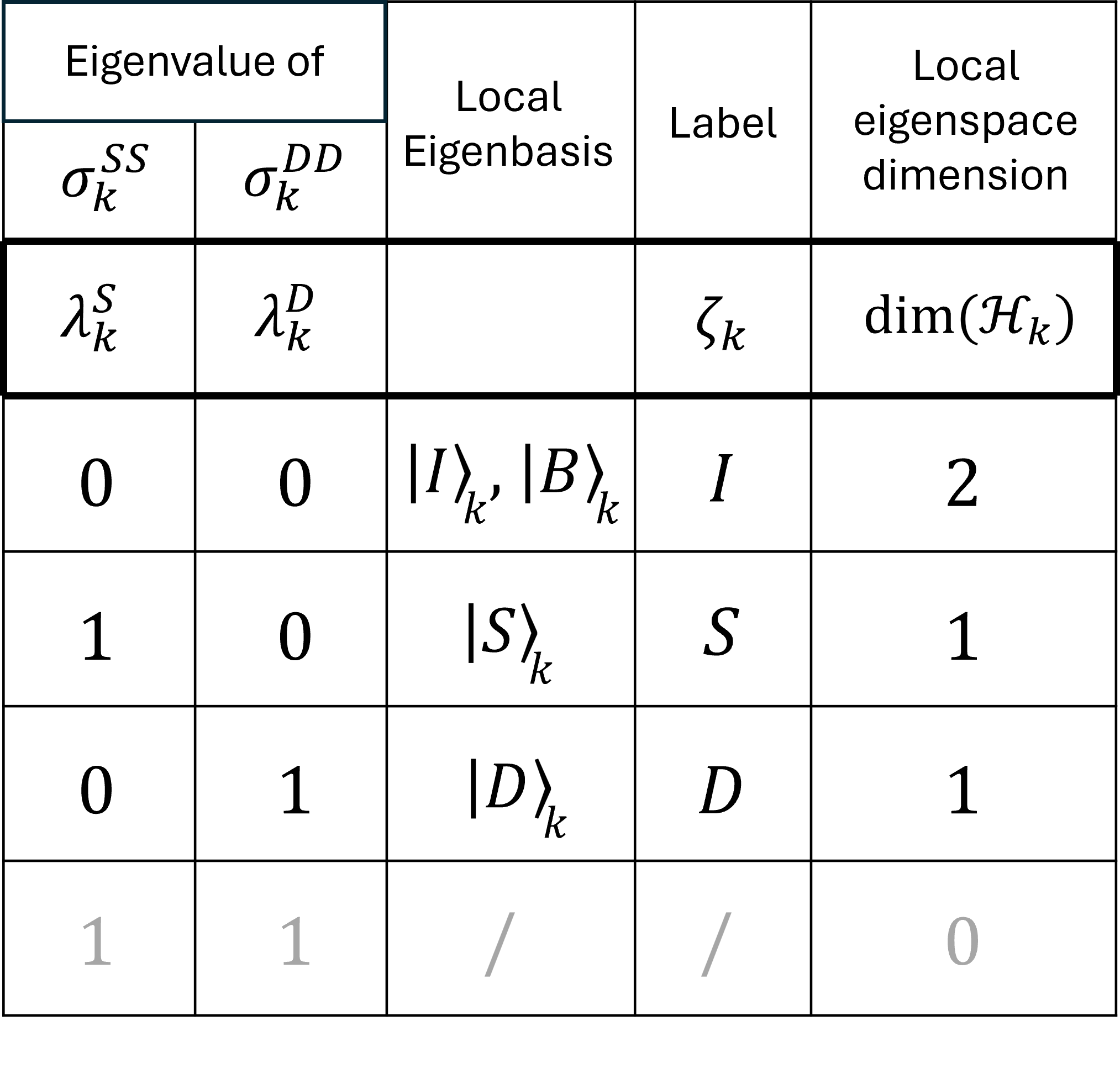}
  \caption{Eigenvalues and eigenspaces of the S and D local density operators. The second row, highlighted with thicker borders, displays, where possible, the notation used in the main text for the quantities considered in each column. The first two columns display all possible choices for the eigenvalues; the third column shows a basis for the corresponding eigenspaces, whose dimensions are reported in the fifth one. The fourth column reports the labels we introduce for each of the three relevant choices. The last row is grayed out to stress the fact that no non-zero vectors are associated to it.  } 
\label{fig:eigentable}
\end{figure}

To better formalize these considerations, let us consider a single site $k$; its degrees of freedom lie in a four-dimensional (local) Hilbert space
\be
    \mal{H}_k  = \mathrm{Span} \left\{ \ket{S}_k, \, \ket{I}_k, \, \ket{B}_k, \, \ket{D}_k \right\};
\ee 
the two eigenvalues $\lt \lambda_k^{S} ,\,  \lambda_k^{D}\rt$ of $\sigma_k^{SS}$ and $\sigma_k^{DD}$ can be chosen, in principle, in $4$ different ways, as we illustrate in Fig.~\ref{fig:eigentable}: 
%$(0, 0)$ corresponds to a local two-dimensional eigenspace $\mal{H}_k^{ \lt 0,0 \rt} = \mathrm{Span} \left\{ \ket{I}_k, \, \ket{B}_k \right\}$. The choices $\lt 1, 0 \rt$ and $\lt 0, 1 \rt$ lead to the one-dimensional $\mal{H}_k^{ \lt 1,0 \rt} = \mathrm{Span} \left\{ \ket{S}_k \right\}$ and $\mal{H}_k^{ \lt 0,1 \rt} = \mathrm{Span} \left\{ \ket{D}_k \right\}$, respectively. The final possibility $\lt 1, 1 \rt$ is actually factitious as a site cannot be susceptible and dead at the same time (or, equivalently, there is no non-trivial intersection between the latter two eigenspaces). 
%
$(0, 0)$ corresponds to a local two-dimensional eigenspace $\mal{H}_k^{ \lt I \rt} = \mathrm{Span} \left\{ \ket{I}_k, \, \ket{B}_k \right\}$. The choices $\lt 1, 0 \rt$ and $\lt 0, 1 \rt$ lead to the one-dimensional $\mal{H}_k^{ \lt S \rt} = \mathrm{Span} \left\{ \ket{S}_k \right\}$ and $\mal{H}_k^{ \lt D \rt} = \mathrm{Span} \left\{ \ket{D}_k \right\}$, respectively. The final possibility $\lt 1, 1 \rt$ is actually fictitious as a site cannot be susceptible and dead at the same time (or, equivalently, there is no non-trivial subspace of both $\mal{H}_k^{ \lt S \rt} $ and $\mal{H}_k^{ \lt D \rt} $). Therefore, 
\be
\mal{H}_k = \mal{H}_k^{ \lt S \rt}  \oplus \mal{H}_k^{ \lt I \rt} \oplus \mal{H}_k^{ \lt D \rt} \ \ \forall \ k .
\ee

To each of the three meaningful choices for $\lt \lambda_k^{S} ,\,  \lambda_k^{D}\rt$ we have associated a label $\zeta_k \in \left\{ S, I, D  \right\}$; the correspondence is illustrated in Fig.~\ref{fig:eigentable} (first, second and fourth columns in particular). We shall in the following use these labels to specify the eigenspaces we work with; as locally fixing $\zeta_k$ yields a reduction
\be
    \mal{H}_k \to \mal{H}_k^{(\zeta_k)},
\ee
a list of $N$ such choices $\vec{\zeta} = \left\{ \zeta_1,\, \zeta_2,\, \zeta_3,\, \ldots \zeta_N  \right\}$, $\zeta_k \in \left\{ S, I, D \right\}$, one per site, we restrict the full Hilbert space
\be
    \mal{H} = \bigotimes_{k} \mal{H}_{k}
\ee
to the collective eigenspace
\be
    \mal{H}^{(\vec{\zeta})} = \bigotimes_{k} \mal{H}_{k}^{(\zeta_k)}
\ee
whose dimension, calculated from the rightmost column of Fig.~\ref{fig:eigentable}, can be expressed as
\be
    \mathrm{dim} \lt \mal{H}^{(\vec{\zeta})} \rt = \prod_k \mathrm{dim} \lt \mal{H}_{k}^{(\zeta_k)} \rt = 2^{\chi_I(\vec{\zeta})},
    \label{eq:eigensp_dim}
\ee
where
\be
    \chi_I(\vec{\zeta}) = \text{number of "$I$"s in $\vec{\zeta}$}
    \label{eq:chi_I}
\ee
is the characteristic function which counts the number of "$I$" components in the string $\vec{\zeta}$. 

To clarify, let us look at some $N=3$ examples: first, let us take $\vec{\zeta} = \lt S, D, I \rt$, or equivalently $\zeta_1 = S$, $\zeta_2 = D$, $\zeta_3 = I$; this identifies the global eigenspace
\be
    \mal{H}^{((S,D,I))} = \mal{H}^{(S)}_1 \otimes \mal{H}^{(D)}_2 \otimes \mal{H}^{(I)}_3.
\ee
Of the three subspaces in the tensor product above, the first and second are one-dimensional, whereas the latter is bi-dimensional. Hence, the dimension of the collective eigenspace is $2^0 \times 2^0 \times 2^1 = 2^{0 + 0 + 1}$,
% \be
%     2^0 \times 2^0 \times 2^1 = 2^{0 + 0 + 1},
% \ee
where the exponent is just reflecting the presence of a single "$I$" component, the one for the third site. Counting how many "$I$" components there are in $\vec{\zeta}$ is exactly what $\chi_I$ does and in this case $\chi_I ( \vec{\zeta} ) = 1$. 

Similarly, $\vec{\zeta} = \lt I, I, I \rt$ has three "$I$" labels, meaning that $\chi_I ( \vec{\zeta} ) = 3$; the collective eigenspace is
\be
    \mal{H}^{((I,I,I))} = \mal{H}^{(I)}_1 \otimes \mal{H}^{(I)}_2 \otimes \mal{H}^{(I)}_3,
\ee
of dimension is $2^1 \times 2^1 \times 2^1 = 2^3 = 2^{\chi_I ( \vec{\zeta} )}$.

Turning again to the general $N$ case, on any collective eigenspace $\mal{H}^{ ( \vec{\zeta} )}$ the action of generic S or D densities is indistinguishable from simple multiplication by constants:
\begin{subequations}
\label{eqs:as_const}
\begin{align}
    &\sigma_k^{SS} \,\, \mal{H}^{(\vec{\zeta})} \equiv \lt \lambda_k^{S} \id \rt \, \mal{H}^{(\vec{\zeta})}, \\
    &\sigma_j^{DD} \,\, \mal{H}^{(\vec{\zeta})} \equiv \lt \lambda_j^{D} \id \rt \, \mal{H}^{(\vec{\zeta})},
\end{align}
\end{subequations}
with $\lt \lambda_k^{S},\, \lambda_k^{D} \rt$ uniquely identified by the labels $\vec{\zeta}$. For brevity, we shall refer to the collective eigenspaces $\mal{H}^{(\vec{\zeta})}$ as "sectors".

With three choices per site, the full Hilbert space is thus fragmented in $3^N$ different sectors, which we can further group by dimension: for each $i \in \left\{0, 1, \ldots N\right\}$ there are $\binom{N}{i} 2^{N-i}$ such subspaces of dimension $2^i$. The largest among them ($i = N$) is unique, corresponds to $\zeta_k = I \,\, \forall \, k$, and its dimension is $2^N$. The $2^N$ smallest ones ($i = 0)$ are one-dimensional, correspond to all possible choices of $\vec{\zeta}$ with $\zeta_k \neq I \ \,\forall\, k$, and are in fact all the absorbing GEP-like configurations made up of only susceptible and dead sites (see App.~\ref{app:eQEP_absorbing}).

In spite of the considerable reduction in complexity from $\mathrm{dim} \lt \mal{H} \rt = 4^N$ we are still dealing with an exponentially large dimension in the system size. Thankfully, as we shall show in the following Section, the eQEP features some additional, decisive simplifications.

\subsection{The action of jump operators on sectors}
\label{ssec:jump_action}

We now turn to consider the action of jumps on the eigenspace structure. We start with those that encode infection processes; we have 
\be
    \comm{L^I_{kj}}{\,\sigma_l^{DD}} = 0 \,\,\, \forall \,\, k,\, j,\, l
\ee
implying that infection processes leave the $\sigma^{DD}$ eigenspaces invariant. This is not particularly surprising as it tells us that infection is unable to change the state of a dead site. By the same intuitive approach, we should expect the same not to hold for all $\sigma^{SS}$s. We have in this latter case
\be
    \comm{L^I_{kj}}{\,\sigma_l^{SS}} = 0 \,\,\, \forall \,\, k,\, j,\, l \neq k,
\ee
but not for $l = k$. Instead of expressing the commutator, however, let us apply properties \eqref{eqs:sigmas} and \eqref{eq:sigma_comm} to establish the following equalities:
\begin{subequations}
\label{eqs:LI_rules}
\begin{align}
    &L^I_{kj} = \sqrt{\gamma_I} \sigma_k^{IS} \sigma_j^{II} = \sqrt{\gamma_I} \sigma_k^{IS} \sigma_k^{SS} \sigma_j^{II}  = L^I_{kj} \sigma_k^{SS} , \label{subeq:LI_sigma} \\
    &L^I_{kj} = \sqrt{\gamma_I} \sigma_k^{IS} \sigma_j^{II} = \sqrt{\gamma_I} \sigma_k^{II} \sigma_k^{IS}  \sigma_j^{II}  = \sigma_k^{II} L^I_{kj}. \label{subeq:sigma_LI} \\
    &L^I_{kj} = \sqrt{\gamma_I} \sigma_k^{IS} \sigma_j^{II} = \sqrt{\gamma_I} \sigma_k^{IS}  \lt \sigma_j^{II} \rt^2 = \sigma_j^{II} L^I_{kj} = \nol
    & = L^I_{kj} \sigma_j^{II} . \label{subeq:sigma_LI2}
\end{align}
\end{subequations}
Note now that Eq.~\eqref{subeq:LI_sigma} is telling us that any vector $\ket{\psi}$ annihilated by $\sigma_k^{SS}$ is necessarily annihilated by $L_{kj}^I$ as well; in other words,
\be
    L_{kj}^I \, \mal{H}^{(\vec{\zeta})} = 0 \text{ whenever } \zeta_k \neq S.
\ee
This is actually a reiteration of the fact that infection cannot take place on non-susceptible sites. We recall from rule (II) of the QJMC algorithm that if a jump operator annihilates a vector then it will never be selected to act upon it, or equivalently its associated probability vanishes. In a similar fashion, we can conclude from the second line of Eq.~\eqref{subeq:sigma_LI2} that 
\be
    L_{kj}^I \, \mal{H}^{(\vec{\zeta})} = 0 \text{ whenever } \zeta_j \neq I,
\ee
which reminds us that neighbor $j$ must be infected in order to pass on the illness onto susceptible site $k$.

Combining Eqs.~\eqref{subeq:sigma_LI} and \eqref{subeq:sigma_LI2} we have instead that whatever vector $\ket{\psi}$ the jump operator $L_{kj}^I$ is acting upon will have its $k$-th and $j$-th components transformed into $\ket{I}$. From the eigenspace viewpoint, we can write
\be
    L_{kj}^I \, \mal{H} \subseteq \mal{H}_1 \otimes \mal{H}_2 \otimes \ldots \otimes \mal{H}^{(I)}_j \otimes \ldots \otimes \mal{H}^{(I)}_k \otimes \ldots \otimes \mal{H}_N. 
\ee
Specializing this statement to when we start already from a sector of label $\vec{\zeta}$ we can write
\be
    L_{kj}^I \, \mal{H}^{(\vec{\zeta})} \subseteq \mal{H}^{(\vec{\zeta}')} 
    \label{eq:LI_esector_map}
\ee
with $\zeta'_k = \zeta'_j = I$ and $\zeta'_l = \zeta_l \ \ \forall \, l\neq j, \,k$.

Equation \eqref{eq:LI_esector_map} tells us that $L_{kj}^I$ always maps a sector onto another one and we can therefore describe its action between sectors in terms of label transformations: combining all the previous results, we have that if $\zeta_k = S$ and $\zeta_j = I$ then $\zeta'_k = \zeta'_j = I$. In all other cases $L_{kj}^I$ annihilates the vector $\ket{\psi} \in \mal{H}^{(\vec{\zeta})}$ and is thus never picked in the course of the QJMC algorithm.
Recalling that, by definition \eqref{eq:chi_I}, $\chi_I (\vec{\zeta'}) = \chi_I ( \vec{\zeta}) + 1$ since the "$I$" count goes up by $1$ over the jump, we also see that every infection event doubles the dimension of the sector the dynamics is restricted to. 

%either does not act at all or maps a sector \emph{precisely} onto another one; in the latter case, the mapping can be illustrated via a change of label $\vec{\zeta} \to \vec{\zeta'}$ where $\zeta_{k'} = \zeta'_{k'}$ $\forall \, k' \neq k$, while $\zeta_k = S$ and $\zeta'_k = I$. Recalling that, by definition \eqref{eq:chi_I}, $\chi_I (\vec{\zeta'}) = \chi_I ( \vec{\zeta}) + 1$ since the "$I$" count goes up by $1$ over the jump, we also see that every infection event doubles the dimension of the sector the dynamics is restricted to. 

We now turn to death events: the jump operator $L^D_k$ can be treated in a completely analogous manner: it commutes with $\sigma_l^{SS}$ $\forall \, l$ and with $\sigma_l^{DD}$ $\forall \, l \neq k$. On site $k$ we find
\begin{subequations}
\label{eqs:LD_rules}
\begin{align}
    &L^D_{k} =  L^D_{k} \sigma_k^{II} \text{  and} \label{subeq:LD_sigma} \\
    &L^D_{k} = \sigma_k^{DD} L^D_{k}. \label{subeq:sigma_LD}
\end{align}
\end{subequations}
Following the same reasoning applied to $L^I_{kj}$ we derive that $L_k^D$ either does not act or maps a sector onto a single other one, with a label change $\vec{\zeta} \to \vec{\zeta'}$, where $\zeta_k = I$, $\zeta'_k = D$ and $\zeta_{k'} = \zeta'_{k'}$ $\forall \, k' \neq k$. In this case the "$I$" count decreases by $1$, i.e.~$\chi_I (\vec{\zeta'}) = \chi_I ( \vec{\zeta}) - 1$ and the sector dimension is halved.

\begin{figure}[h]
  \vspace{0cm}
  \includegraphics[width=\columnwidth]{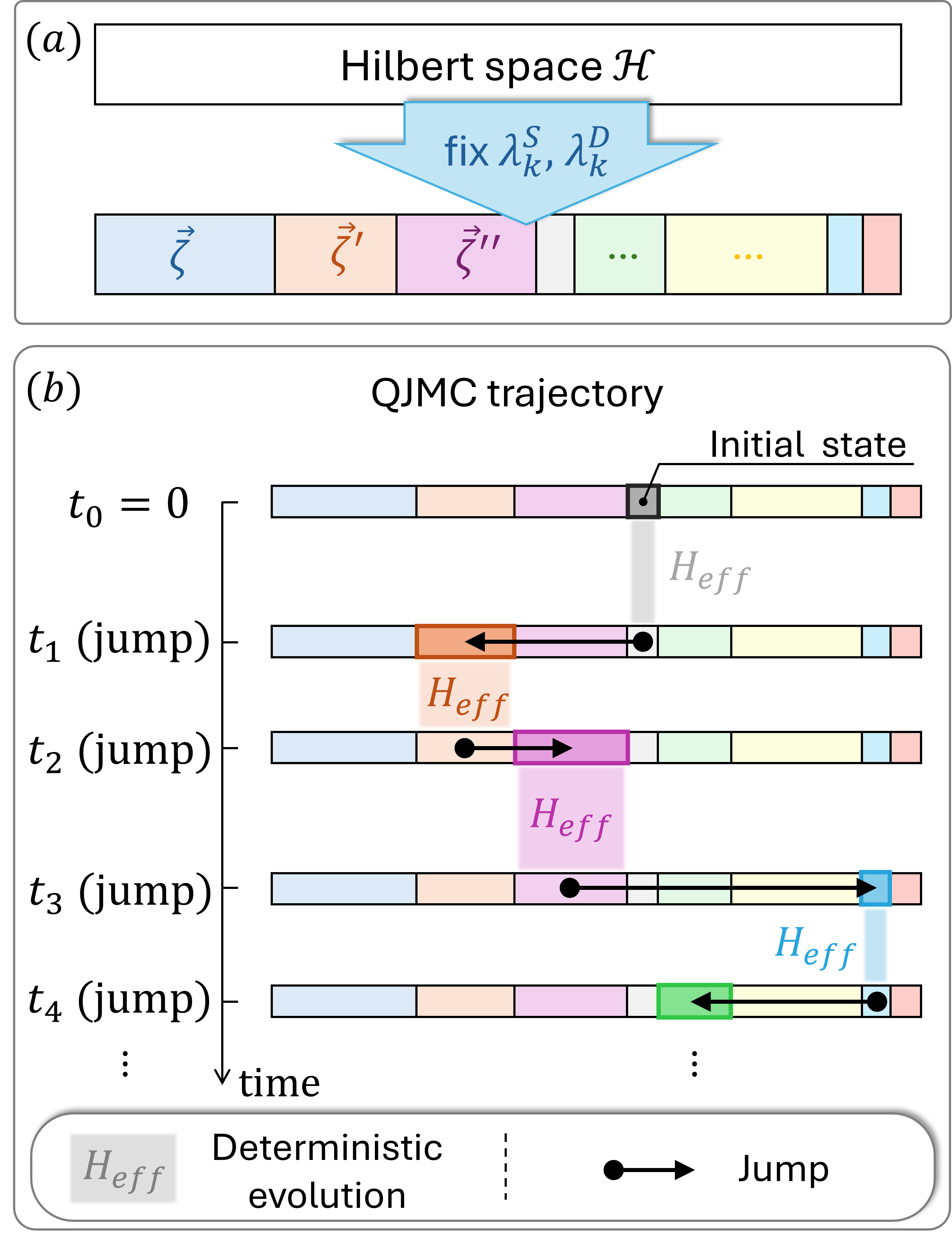}
  \caption{Effect of the symmetries of $H_{eff}$ on the QJMC dynamics: (a) The Hilbert space $\mal{H}$ can be fragmented in $3^N$ distinct sectors $\mal{H}^{(\vec{\zeta})}$, each labeled by a string $\vec{\zeta}$ which uniquely identifies the chosen eigenvalues for every $\sigma_k^{SS}$ and $\sigma_k^{DD}$. Here, different sectors are drawn in different colors. (b) Sketch of a QJMC trajectory: the colored horizontal strips represent the fragmented Hilbert space, displaying the same pattern of panel (a). The initial state lies entirely within one highlighted sector. The deterministic evolution generated by the effectve Hamiltonian, which takes place between jumps, leaves each sector invariant and is here displayed as a shaded area between one $\mal{H}$ strip and the next. At each of the displayed times $t_0$, $t_1$ etc. a jump occurs which shifts the dynamics from the previous sector to a new (highlighted) one, illustrated here with a black horizontal arrow. The overall effect is that the dynamics is at all times restricted within a single sector, though the sector itself changes at every jump.} 
\label{fig:QJMC_traj}
\end{figure}

Next, we recall that our initial state is a GEP-like configuration in which all sites but one are susceptible, with the remaining one (the origin) being infected. The corresponding QJMC initial vector $\ket{\psi_0}$, therefore, belongs to the sector $\vec{\zeta}^{(in)}$, where, denoting by $k_o$ the origin's site index,
\be
    \zeta^{(in)}_k = S \,\,\, \forall \,\, k \neq k_o \  \text{ and } \  \zeta^{(in)}_{k_o} = I.
    \label{eq:zeta_in}
\ee

Collecting all the arguments above, we can write the following rules for the QJMC dynamics of the eQEP:
\begin{itemize}
    \item[(a)] The initial state belongs to the specific sector just described: $\ket{\psi_0} \in \mal{H}^{\lt \vec{\zeta}^{(in)} \rt}$.
    \item[(b)] For any given $\vec{\zeta}$ and any vector $\ket{\psi} \in \mal{H}^{(\vec{\zeta})}$, the deterministic evolution never leaves the sector: $\rme{-iH_{eff} t} \ket{\psi} \in \mal{H}^{(\vec{\zeta})}$ $\forall \, t$, i.e.~every sector is closed under the evolution generated by the effective Hamiltonian \eqref{eq:eQEP_Heff}.
    \item[(c)] For any given $\vec{\zeta}$ and any vector $\ket{\psi} \in \mal{H}^{(\vec{\zeta})}$, if $L_\alpha$ is a jump operator with a non-zero probability of acting upon $\ket{\psi}$ then $L_\alpha \ket{\psi} \in \mal{H}^{(\vec{\zeta}')}$ for some $\vec{\zeta}'$ which differs from $\vec{\zeta}$ by a single entry.
    \item[(d)] A jump generated by $L^I_{kj}$ induces the change $\zeta_k = S \to \zeta'_k = I$.
    \item[(e)] One generated by $L^D_k$, instead, changes $\zeta_k = I \to \zeta'_k = D$.
\end{itemize}
The general rules (a)-(c) imply that the QJMC dynamics of the eQEP is always restricted within one, and one only, of the many sectors $\mal{H}^{(\vec{\zeta})}$ the Hilbert space is subdivided in. This sector changes at every jump. An illustration of this is provided in Fig.~\ref{fig:QJMC_traj}.

\subsection{Decoupling of $H_{eff}$}
\label{subsec:decoupling}

Combining the closing remarks of the previous Section with Eqs.~\eqref{eqs:as_const} we see that between jumps we can effectively treat the conserved densities as constants:
\be
    \sigma_k^{SS} \to \lambda^S_k \ \text{ and } \sigma_k^{DD} \to \lambda^D_k.
    \label{eq:subst}
\ee
Performing this substitution in the effective Hamiltonian yields
\begin{align}
    H_{eff} & = \Omega \sum_{k} \lt \sigma_k^{IB} + \sigma_k^{BI}  \rt - \frac{i}{2} \gamma_I \sum_{\nnsum{k}{j}} \lambda_k^{S} \sigma_j^{II} +  \nol
    & - \frac{i}{2} \gamma_D \sum_{k} \sigma_k^{II}.
\end{align}
Up to a trivial renaming of the summation indices $j \leftrightarrow k$ this can be recast as $H_{eff} = \sum_k h_k$ with
\be
    h_k = \Omega  \lt \sigma_k^{IB} + \sigma_k^{BI}  \rt - \frac{i}{2}  \lt \gamma_D + \mal{N}^S_k \gamma_I \rt \sigma_k^{II}
    \label{eq:hk}
\ee
a collection of $N$ local terms, each with support over a single site, having introduced the shorthand
\be
    \mal{N}_k^S = \sum_{j \in \Lambda_k} \lambda_j^S .
\ee
This $\mal{N}_k^S$ is, by definition, the number of sites, among the nearest neighbors of $k$, that are restricted to the local eigenspace $\mal{H}_j^{(S)}$; in simpler terms, we can say it counts the number of neighboring susceptible sites. For a square lattice (except for boundary sites) $\mal{N}^S \in \left\{ 0, 1, \ldots 4  \right\}$. 

At a first glance, Eq.~\eqref{eq:hk} may seem confusing: up to this point, we have expressed the constraint in a "passive" sense, i.e., we have looked at \emph{when a susceptible site can be infected by an ill neighbor}. Equation \eqref{eq:hk}, on the other hand, is more easily interpreted in an "active" sense, i.e., by asking ourselves \emph{when infected site $k$ can infect its susceptible neighbors}. Additionally, in the same way having multiple infected neighbors amplifies the likelihood of infection for a susceptible site, having multiple susceptibles nearby increases the infection opportunities for an infected, which is why we see the individual infection rate $\gamma_I$ multiplied by $\mal{N}^S$. Of course, these two formulations are formally completely equivalent ($j$ is infected by $k$ $\Leftrightarrow$ $k$ infects $j$). 

% To avoid confusion going forward, we remark that Eq.~\eqref{eq:hk}, in a sense, reformulates the constraint in an "active" sense: instead of looking at whether an S site \emph{can be infected} by an I neighbor, it is now easier to consider whether an I site \emph{can infect} one of its S neighbors. Additionally, in the same way having multiple infected neighbors amplifies the likelihood of infection for a susceptible site, having multiple susceptibles nearby increases the infection opportunities for an infected, which is why we see the individual infection rate $\gamma_I$ multiplied by $\mal{N}^S$. 

Note now that, with the substitutions \eqref{eq:subst} we have decoupled different sites in the effective Hamiltonian, which is now reduced to a sum of local terms $h_k$. The forms these local terms take are, however, still dependent upon the sector the dynamics is restricted within:
\be
    h_k = h_k \lt  \vec{\zeta} \rt.
\ee
% In general, $h_k ( \vec{\zeta} ) \neq h_k (  \vec{\zeta}' )$ if $\vec{\zeta} \neq \vec{\zeta}'$

Hence, if a QJMC vector $\ket{\psi}$ (A) lies entirely within a sector $\vec{\zeta}$ and (B) is factorized ($\ket{\psi} = \bigotimes_k \ket{\psi_k}_k$), then $\ket{\psi_{eff}(t)} = \rme{-iH_{eff}t} \ket{\psi}$ remains factorized at all times:
\be
    \ket{\psi_{eff}(t)} = \bigotimes_k \lt \rme{-ih_k ( \vec{\zeta} ) t} \ket{\psi_k}_k \rt.
\ee
For brevity, in the following we will usually drop the explicit dependence of $h_k$ on the sector label, which remains in any case implicitly understood.

The factorization just discussed is clearly a great simplification, as whenever conditions (A) and (B) are satisfied it allows us to treat the sites as being effectively independent. Examining the initial condition
\be
    \ket{\psi^{(in)}} = \lt \bigotimes_{k < k_o} \ket{S}_k \rt \otimes \ket{I}_{k_o} \otimes \lt \bigotimes_{k > k_o} \ket{S}_k \rt,
    \label{eq:init_vec}
\ee
where again we denote by $k_o$ the site index of the origin (the center of the lattice), we see that it is factorized, satisfying (B). We have already seen it satisfies (A), i.e., it belongs to the sector defined in Eq.~\eqref{eq:zeta_in}. Thus, up to the first jump, we are always dealing with a factorized vector.

Let us see then what happens at jumps. We know what their effect is among sectors; we need to understand, then, what they do to factorized vectors $\ket{\psi} = \otimes_k \ket{\psi_k}_k$.
Death is particularly simple: by definition, we have, up to normalization factors,
\be
    L_k^D \ket{\psi} = \lt \bigotimes_{k' < k} \ket{\psi_{k'}}_{k'} \rt \otimes L_k^D \ket{\psi_k}_{k} \otimes \lt \bigotimes_{k' > k} \ket{\psi_{k'}}_{k'} \rt.    
\ee
The infection case is equally straightforward, albeit a bit more convoluted to write down: by recalling our definition $L^I_{kj} = \sqrt{\gamma_I}\, \sigma_k^{IS} \otimes \sigma_j^{II}$ ($k<j$) one finds
\begin{align}
    L^I_{kj} \ket{\psi} &= \sqrt{\gamma_I}  \lt \bigotimes_{k' < k} \ket{\psi_{k'}}_{k'} \rt \otimes \sigma_k^{IS} \ket{\psi_k}_k \otimes \nol
    & \otimes \lt \bigotimes_{k < k' < j} \ket{\psi_{k'}}_{k'} \rt \otimes \sigma_j^{II} \ket{\psi_j}_j \otimes \nol
    & \otimes \lt \bigotimes_{k' > j} \ket{\psi_{k'}}_{k'} \rt,
\end{align}
which is still in a factorized form (essentially because $L^I_{kj}$ is too). The case $L^I_{kj} = \sqrt{\gamma_I}\, \sigma_j^{II} \otimes \sigma_k^{IS}$ ($k > j$) is analogous.

% Finally, the initial state itself is a factorized vector:
% \be
%     \ket{\psi^{(in)}} = \lt \bigotimes_{k < k_o} \ket{S}_k \rt \otimes \ket{I}_{k_o} \otimes \lt \bigotimes_{k > k_o} \ket{S}_k \rt,
%     \label{eq:init_vec}
% \ee
% where again we denote by $k_o$ the site index of the origin (the center of the lattice).

We can thus conclude that over the course of a QJMC trajectory the vector $\ket{\psi}$ remains factorized at all times. Hence, there is no entanglement created by the QJMC algorithm and correlations between different sites arise exclusively from the stochastic averaging \eqref{eq:stoch_rho}.

Unsurprisingly, this produces further simplifications down the line, and in particular in the QJMC algorithm: now the calculation of the TFJ (time of first jump) cumulative probability can be done independently on each site, facing a local Hilbert subspace dimension of $2$ (at most). To see how this works more precisely, we start by recalling the definition
\be
    \mathbb{P} \bigl( TFJ \geq t  \,\,|\, \ket{\psi}, t' \bigr) = \norm{\rme{-iH_{eff} (t-t')} \ket{\psi}}^2,
\ee
having made the conditional structure of $\mathbb{P}$ explicit. Exploiting the fact that in our QJMC trajectories we can always decouple $H_{eff}$ and $\ket{\psi}$ always remains factorized, we can equivalently write
\begin{align}
    \mathbb{P} &\bigl( TFJ \geq t  \,\,|\, \ket{\psi}, t' \bigr)  = \prod_k \norm{\rme{-ih_k (t-t')} \ket{\psi_k}_k}^2 = \nol
    & = \prod_k \mathbb{P} \bigl( TFJ_k \geq t  \,\,|\, \ket{\psi_k}_k, t' \bigr) \equiv \prod_k P_k(t \,|\, t'),
    \label{eq:Pfact}
\end{align}
where, in the last line, each factor expresses the effective probability of a site (say, $k$) to jump for the first time after $t$, under the action of its local $h_k$, given that it was in local state $\ket{\psi_k}_k$ at time $t'$. The last equality introduces a shorthand $P_k$ that will be often employed in the following.

Now, $h_k \ket{S}_k = h_k \ket{D}_k = 0$, so, whenever $\ket{\psi_k}_k$ is either purely susceptible or dead, $P_k = 1$ and can be ignored for the calculation of the whole probability. We are left with all cases $\ket{\psi_k}_k \in \mal{H}^{(I)}_k$ (i.e., cases that have components exclusively on $\ket{I}_k$ and $\ket{B}_k$). On this two-dimensional subspace the effective Hamiltonian component can be recast as a $2 \times 2$ matrix
\be
    h_k = \matb{cc} - i \frac{\gamma_{eff}}{2} & \Omega \\ \Omega & 0    \mate,
    \label{eq:hk_matrix}
\ee
with the shorthand
\be
    \gamma_{eff} = \gamma_D + \mal{N}_k^S \gamma_I.
    \label{eq:gamma_eff}
\ee
Note that the dependence on the site $k$ is retained solely through the number of susceptible neighbors $\mal{N}^S_k$. 

The local effective Hamiltonian $h_k$ can be diagonalized analytically; its eigenvalues and eigenvectors can be found in App.~\ref{app:eigen}. From those we have then been able to derive a closed expression (a sum of complex exponentials) for each $P_k$ (also in App.~\ref{app:eigen}). However, rule (I) of the QJMC algorithm requires us to invert $\mathbb{P}$ in order to extract a jump time and we have been unable to obtain an analytic form for $\mathbb{P}^{-1}$. In principle, since $\mathbb{P}$ is a monotonically-decreasing function of $t$, we could estimate the jump time $t_{TFJ}$ by applying a bisection procedure to the equation $\mathbb{P}(t) = u$. However, considering its cumbersome structure as a product of, potentially, $O(N)$ terms, we have found it more convenient to take a different numerical approach. 

To elucidate how it works, we need to go back to Eq.~\eqref{eq:Pfact}: we interpret the product in the second line as a product of probabilities of independent random events. Note that we are \emph{not} stating that jump times on different sites are independent (see App.~\ref{app:not_indep}); we are merely using the proposed interpretation as a convenient tool to extract the global jump time.
Applying the general rule $\mathbb{P}(A) \, \mathbb{P}(B) = \mathbb{P}(A \text{ and } B)$ for any independent $A$ and $B$ we can re-express
\be
    \prod_k P_k(t\,|\, t') = \mathbb{P} \bigl( TFJ_k \geq t  \,\, \forall \, k\,\,|\, \ket{\psi}, t'  \bigr) 
\ee
as the probability that \emph{all} sites experience their first jump after $t$. This is equivalent to the probability that the minimum among all those times is greater than $t$, i.e.,
\be
    \prod_k P_k(t \,|\, t') = \mathbb{P} \bigl( \min_k \lt TFJ_k \rt \geq t  \,\,|\, \ket{\psi}, t'  \bigr). 
\ee
Our strategy is then very straightforward: for each site in the local $I$ eigenspace we extract a random number $u_k \in \lqq 0, 1 \rqq$ and then numerically solve
\be
    P_k(t\,|\, t') = u_k
\ee
via a bisection method. In doing so, we extract a time $t_k$. Finally, we select the smallest one as the jump time ($t_{TFJ} = \min_k \left\{ t_k \right\}$), discarding all the others. Times extracted in this fashion have the same distribution as those extracted by inverting $\mathbb{P}$. We refer the unconvinced reader to App.~\ref{app:not_indep}. 
%This has allowed us to work with the more compact expression of the individual $P_k$s and, more importantly, it also singles out the site $k$ the jump originates from.

A subtle consequence of this approach is that the extraction of the jump time now singles out a site $k$ too. In doing so, it is in fact restricting the available jump types to those that originate from $k$. Effectively, this means that our procedure is a "mix" of steps (I) and (II) of the QJMC algorithm, covering the former and part of the latter. We need now to discuss how to complete the jump selection: once site $k$ has been picked, there are $1 + \mal{N}_k^S$ (between one and five) different possibilities: death of $k$ plus all possible infections of one of its susceptible neighbors. 

To derive the relative probabilities, we introduce the notation
\be
    \ket{\psi_k}_k = \ket{a,\, b}_k = a \ket{I}_k + ib\ket{B}_k 
\ee
for component $k$ of the QJMC vector, where the r.h.s.~is a generic parametrization for a vector $\in \mal{H}_k^{(I)}$ with $a$, $b$ $\in \C$. Assuming $j$ is one of the susceptible sites around $k$, we find
\begin{align}
    L_{jk}^I & \ket{a,\,b}_k \ket{S}_j  = \sqrt{\gamma_I}\, \sigma_k^{II} \sigma_j^{IS} \ket{a,\,b}_k \ket{S}_j = \nol
    & = \sqrt{\gamma_I} \lqq \sigma_k^{II} \bigl( a \ket{I}_k + ib\ket{B}_k \bigr) \rqq \lt \sigma_j^{IS} \ket{S}_j \rt = \nol
    & =\sqrt{\gamma_I} \, a \, \ket{I}_k \ket{I}_j
\end{align}
which in turn implies, according to rule (II) of the QJMC algorithm in Sec.~\ref{subsec:QJMC_algorithm},
\be
    Q_{jk}^I = \norm{L_{jk}^I  \ket{a,\,b}_k \ket{S}_j}^2 = \gamma_I \abs{a}^2.
    \label{eq:QI_for_k}
\ee
For the potential death of $k$ we find instead
\begin{align}
    L_{k}^D & \ket{a,\,b}_k = \sqrt{\gamma_D}\, \sigma_k^{DI} \bigl( a \ket{I}_k + ib\ket{B}_k  \bigr) = \nol
    &= \sqrt{\gamma_D}\,a \,\ket{D}_k
\end{align}
and the corresponding statistical weight is
\be
    Q_k^D = \norm{L_{k}^D  \ket{a,\,b}_k}^2 = \gamma_D \, \abs{a}^2,
\ee
independently of any neighbors.

Note now that Eq.~\eqref{eq:QI_for_k} is valid provided that site $j$ is, in fact, susceptible, otherwise $Q_{jk}^I = 0$. Recalling that we have called $\mal{N}_k^S$ the number of susceptible neighbors of site $k$ we can calculate the associated probability
%Numbering for simplicity the susceptible neighbors of $k$ from $1$ to $\mal{N}_k^S$, the probability of selecting an infection is thus
% \begin{align}
%     %q_{jk}^I &= \frac{\gamma_I \abs{a}^2}{ \gamma_D \abs{a}^2 + \sum_{j' = 1}^{\mal{N}_k^S} \gamma_I \abs{a}^2  } = \nol
%      q_{jk}^I = \frac{\gamma_I}{\gamma_D + \mal{N}_k^S \gamma_I} = \frac{\gamma_I}{\gamma_{eff}},
%     \label{eq:qI}
% \end{align}
\be
    q_{jk}^I = \sysb{lcr} \frac{\gamma_I}{\gamma_D + \mal{N}_k^S \gamma_I} = \frac{\gamma_I}{\gamma_{eff}} & \ & \text{if } \zeta_j = S, \\
    0 & \ & \text{otherwise.} \syse
    \label{eq:qI}
\ee
The probability of $k$ dying, instead, reads
\be
    q_k^D =  \frac{\gamma_D}{\gamma_D + \mal{N}_k^S \gamma_I} = \frac{\gamma_D}{\gamma_{eff}}.
    \label{eq:qD}
\ee
As a side remark, in removing $\abs{a}^2$ from our fractions we have implicitly assumed $a \neq 0$. This can be done without loss of generality; we refer the interested reader to App.~\ref{app:eigen} for the proof. Here we limit ourselves to an intuitive argument: $a = 0$ means that $\ket{\psi_k}_k \propto \ket{B}_k$; this vector annihilates the anti-Hermitian part of the local effective Hamiltonian \eqref{eq:hk}, i.e., the part responsible for the decline of the survival probability $\mathbb{P}\lt TFJ_k \geq t \rt$ in Eq.~\eqref{eq:Pfact}. Let $\bar{t}$ be a time at which $a = 0$; if we accept that the probability $\mathbb{P}$ locally flattens out at $\bar{t}$ we can derive, for the distribution of $TFJ_k$, that 
\be
    p_{TFJ_k} \lt \bar{t} \rt = -\partial_t \bigl( \mathbb{P} \lt TFJ_k \geq t \rt  \bigr) \eval{t = \bar{t}} = 0.
\ee
Hence, times like $\bar{t}$ are \emph{never} selected for a jump and we never have to face the case $a = 0$ in Eqs.~\eqref{eq:qI} and \eqref{eq:qD}.

In conclusion, the relative probabilities of infection and death events, once the jumping time and site have been determined, only depend on the values of $\gamma_D$, $\gamma_I$ and $\mal{N}_k^S$ and are on the contrary independent from the instantaneous vector (and its components $a$ and $b$). Thus, the choice of jump is essentially made on the basis of the relative strength between the rates $\gamma_D$ and $\gamma_I$, accounting of course for the fewer or greater opportunities for infection in the neighborhood.  

once the time and sites have been determined the remaining choice of jump is independent of the instantaneous vector and is essentially made based on the relative strength between $\gamma_D$ and $\gamma_I$.

\subsection{The eQEP algorithm}
\label{subsec:eQEP_algo}

For the interested reader, we hereafter provide a breakdown of the QJMC algorithm we have used to simulate the eQEP, which exploits all the various simplifications listed up to this point throughout Sec.~\ref{sec:theory}, and which we refer to as the "eQEP algorithm" or "our algorithm" for brevity.

For simplicity, we choose again, as the algorithm's aim, the (approximate) reconstruction of the state $\rho$ at some fixed time $t_{end} > 0$. The initial time is $t = 0$ and the initial state $\rho_0 = \ket{\psi_0} \bra{\psi_0}$ with $\ket{\psi_0}$ equal to $\ket{\psi^{(in)}}$ of Eq.~\eqref{eq:init_vec}. This ket acts as the initial vector for \emph{all} trajectories. When applying the jumps, we will keep the vector normalized to $1$, as we did in step (IIIa) of the general algorithm. With all these assumptions, for a trajectory which has reached state $\ket{\psi_z}$ at time $0 \leq t_z \leq t_{end}$ after experiencing $z \geq 0$ jumps, we
\begin{itemize}
    \item[(I)] \emph{\underline{Find all sites with $\zeta_k = I$:}} knowing that $\ket{\psi_z} = \otimes_k \ket{\psi^{z}_k}_k$ we look for all components satisfying $\ket{\psi^z_k}_k \in \mal{H}_k^{(I)}$. These will be the only ones experiencing a non-trivial deterministic evolution and, thus, capable of jumping. For each such site, we also record the number $\mal{N}_k^S$ and position of susceptible neighbors. Should there be no vector coponents in $(I)$ subspaces, an absorbing state has been reached; in this case, the procedure ends and the vector $\ket{\psi_{end}} = \ket{\psi_z}$ is returned.
    \item[(II)] \emph{\underline{Calculate all effective local jump times:}} for each component isolated in step (I) we uniformly extract a random number $u_k \in \lqq 0, 1 \rqq$. We then solve $\overline{P_k}(t) = u_k$ numerically by bisection, determining in this way a time $t_k^{(z+1)}$ per site. Here, $\overline{P_k}(t) = P_k(t+ t_z\,|\,t_z)$, with $P_k$ defined in Eq.~\eqref{eq:Pfact}.
    \item[(III)] \emph{\underline{Select jump time and site:}} we fix the next jump time to be $t_{z+1} = \min_k \left\{ t_k^{(z+1)}  \right\}$. We call $\bar{k}$ the site which achieves the minimal jump time, so that $t_{z+1} = t_{\bar{k}}^{(z+1)}$ for some $\bar{k}$, since we are working with a finite system. Note that in this way we are extracting both a time $t_{z+1}$ and a site $\bar{k}$, determining when \emph{and where} a jump takes place. If $t_{z+1} > t_{end}$ we skip the following steps and leap to (VII). Otherwise, we proceed to step (IV).
    \item[(IV)] \emph{\underline{Apply deterministic evolution locally up to $t_{z+1}$:}} all susceptible or dead sites in the lattice are not subject to variation. For all sites singled out in step (I) we have $\ket{\psi_k^z}_k = a_k (t_z) \ket{I}_k + ib_k (t_z) \ket{B}_k$ for some coefficients $a_k(t_z)$, $b_k(t_z)$ such that $\abs{a_k(t_z)}^2 + \abs{b_k(t_z)}^2 = 1$. Via our diagonalization of $h_k$ we update the coefficients $a_k (t_z) \to a_k (t_{z+1})$, $b_k (t_z) \to b_k (t_{z+1})$, yielding a normalized pre-jump vector $\ket{\varphi_k^{z+1}}_k $, which represents the QJMC vector at time $t_{z+1}^-$ (see App.~\ref{app:eigen} for the actual expressions).
    \item[(V)] \emph{\underline{Find out which jump takes place on $\bar{k}$:}} we select death with probability $\gamma_D / \gamma_{eff}$, infection with probability $\mal{N}_k^S \gamma_I / \gamma_{eff}$ with $\gamma_{eff}$ as in Eq.~\eqref{eq:gamma_eff}. In the former case, proceed to (VI-d), in the latter to (VI-i). 
    %In the latter case, we randomly select, with equal probability, one of the susceptible neighbors of $\bar{k}$ (say, $\bar{j}$).
    %
    \item[(VI-d)] \emph{\underline{Apply death jump:}} we perform the substitution 
    \be
        \ket{\varphi_{\bar{k}}^{z+1}}_{\bar{k}} \to \ket{\psi_{\bar{k}}^{z+1}}_{\bar{k}} = \ket{D}_{\bar{k}}.
    \ee
    For all other sites $k \neq \bar{k}$, $\ket{\psi_{k}^{z+1}}_{k} = \ket{\varphi_{k}^{z+1}}_{k}$, i.e., the pre-jump and post-jump vector components coincide. With the (global) vector $\ket{\psi_{z+1}}$ thus obtained we start again from (I). 
    \item[(VI-i)] \emph{\underline{Apply infection jump:}} we select, with equal probability, one of the suceptible neighbors of $\bar{k}$ (say, $\bar{j}$). We perform the susbtitutions 
    \be
        \ket{S}_{\bar{j}} \to \ket{\psi^{z+1}_{\bar{j}}}_{\bar{j}} = \ket{I}_{\bar{j}}
    \ee
    and
    \be
        \ket{\varphi_{\bar{k}}^{z+1}}_{\bar{k}} \to \ket{\psi_{\bar{k}}^{z+1}}_{\bar{k}} = \ket{I}_{\bar{k}}.
    \ee
    Note that, despite all major simplifications encountered, there are still traces of the quantum nature of the problem: here we see that the constraint does not only affect the site being infected ($\bar{j}$, in this case), but the infecting site $\bar{k}$ as well through the action of the projector $\sigma_{\bar{k}}^{II}$ (and subsequent normalization). For all other sites $k$, $\ket{\psi_{k}^{z+1}}_{k} = \ket{\varphi_{k}^{z+1}}_{k}$, i.e., their pre-jump and post-jump expressions are the same. With the newly-obtained global vector $\ket{\psi_{z+1}}$ we go back to (I).
    \item[(VII)] \emph{\underline{Apply the last stretch of deterministic evolution:}}  all susceptible or dead sites are kept untouched. For all sites identified in step (I) we have $\ket{\psi_k^z}_k = a_k (t_z) \ket{I}_k + ib_k (t_z) \ket{B}_k$. Via our diagonalization of $h_k$ we update the coefficients $a_k (t_z) \to a_k (t_{end})$, $b_k (t_z) \to b_k (t_{end})$ to the final time $t_{end}$. The corresponding vector $\ket{\psi_{end}}$ is then normalized and returned, halting the procedure.
  
\end{itemize}

Taking into account our analytical approach to the deterministic part of the evolution, we have via the algorithm above effectively reduced the eQEP dynamics to a classical stochastic simulation. There are, of course, extra complications arising with respect to common classical processes (such as the GEP): first, the configurations of the system are unit vectors in the space 
\be
	\widetilde{\mal{H}} = \left\{ \ket{\psi} \in \mal{H} \, | \, \bracket{\psi}{\psi} = 1    \right\}.
\ee
which can therefore vary continuously. This is captured by the continuous nature of the two components $a$ and $b$ in each $\mal{H}^{(I)}$ subspace. Second, instead of being defined in terms of classical rates, the dynamics stems from the survival, or "no-jump", probability $\mathbb{P}$ coupled with the selection probabilities $q_\alpha$. As we have been unable to invert this function $\mathbb{P}$ analytically, we have been forced to look for solutions numerically, which made the procedure more computationally demanding.
%As discussed in App.~\ref{app:KMC}, this is actually equivalent to working with effective time-dependent rates 
% \be
%     \gamma_\alpha  \lt t  \rt = q_\alpha  \bigl[ -\partial_t \ln \mathbb{P} \bigl( TFJ \geq t \,|\, \ket{\psi}^z, t_z \bigr)  \bigr].
% \ee

\section{Numerical results}
\label{sec:num}

We have applied the algorithm outlined in the previous Section to simulate the eQEP on a two-dimensional square lattice of size $N = 101 \times 101 = 10201$ with open boundary conditions. In terms of computational time, the most costly steps have proved to be (II) and (III), i.e.~the determination of the jump times. This is not too surprising, as they are in fact the only wasteful passages, in the sense that while we numerically extract a time $t_k$ for every infected site present, we only retain the smallest and discard all the others. Thereby, the more infected sites are present at any given time in a trajectory, the longer it takes to establish the time of the next jump. In accordance with this very simple intuition, the simulations take longer and longer the more the infection rate $\gamma_I$ is increased. Still, we were able to produce some statistically meaningful results on an average computer, without making use of high-performance facilities, and thus showing the advantage of all the simplifications previously listed.

\begin{figure}[h]
  \includegraphics[width=\columnwidth]{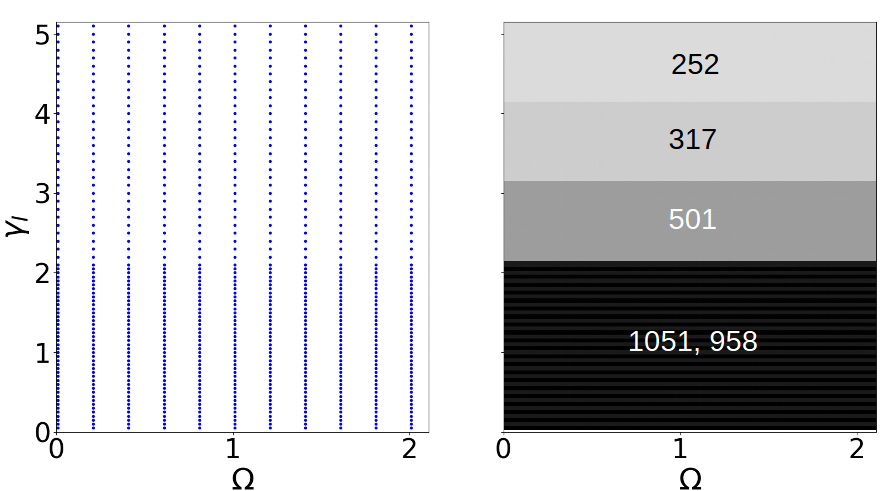}
  \caption{(Left) Simulated parameter pairs $(\Omega, \gamma_I)$. Each blue dot identifies one such pair. While the same $\Omega$s consistently repeat throughout (the horizontal spacing is uniform throughout), the density of points doubles for $\gamma_I \leq 2.1$ with respect to $\gamma_I >2.1$, i.e., the vertical spacing is $0.05$ up to the threshold and $0.1$ above it. (Right) Number of trajectories taken for each portion of parameter space. Here, lighter shades refer to lower trajectory counts and vice versa. The actual numbers are displayed close to the center of each region. In the lowermost part, up to $\gamma_I = 2.1$, there actually are two alternating subsets with a similar, but not exactly equal, trajectory count.  } 
\label{fig:cells_and_trajs}
\end{figure}

In all our numerical simulations we have set $\gamma_D = 1$, or in other words we measure all times in units of $\gamma_D^{-1}$ and all frequencies and energies in units of $\gamma_D$. As already stated, all simulations started from the same initial state: a GEP-like configuration of susceptible sites with a single infected one in the very center, site $\lt 51,\, 51 \rt$. We have run trajectories for the following values of the dynamical parameters:
\begin{subequations}
\label{eqs:sim_parameters}
\begin{align}
    &\Omega = 0.01 + 0.2 \,c_{\Omega}, &c_\Omega = 0,\, 1,\, 2,\, \ldots ,\, 10, \label{subeq:taken_Omegas} \\
    &\gamma_I = 0.05 \, c_{I}, & c_{I} = 1,\, 2,\, \ldots ,\, 42, \\
    &\gamma_I = 2.1 + 0.1 \,c_I', & c_I' = 1,\, 2,\, \ldots,\, 30.
\end{align}
\end{subequations}
% A depiction of the parameter region covered by our simulations can be found in Fig.~\ref{fig:cells_and_trajs}, where it is shown divided in smaller boxes, each corresponding to a single parameter pair $(\Omega, \gamma_I)$.
A depiction of the parameter region covered by our simulations is provided in Fig.~\ref{fig:cells_and_trajs}. Each blue dot in the left panel corresponds to one of the parameter pairs $(\Omega, \gamma_I)$ that can be constructed from Eqs.~\eqref{eqs:sim_parameters}.

\begin{figure}[h]
  \includegraphics[width=\columnwidth]{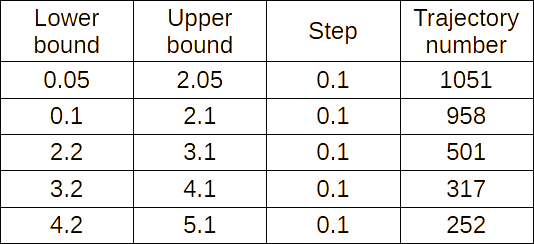}
  \caption{Sets of $\gamma_I$ simulated. Each row in this table indicates the smallest and largest values included. Intermediate values between these bounds are taken at regular steps of $0.1$ in all cases. For each $\gamma_I$ subset the rightmost column reports the actual number of stochastic trajectories simulated. To avoid misunderstandings, this number is per individual $\gamma_I$ value \emph{and} $\Omega$ value. For instance, $1051$ trajectories have been collected for $(\Omega = 0.01,\, \gamma_I = 0.05)$, \emph{another} $1051$ for $(\Omega = 0.21,\, \gamma_I = 0.05)$ and so on for each of the $11$ values of $\Omega$ in Eq.~\eqref{subeq:taken_Omegas}.  } 
\label{fig:tab_trajectories}
\end{figure}

As mentioned above, at higher values of $\gamma_I$ the simulations become more time-consuming; we have thus opted for a lower density of points together with a smaller number of trajectories as we increased $\gamma_I$. The right panel in Fig.~\ref{fig:cells_and_trajs} shows how many stochastic trajectories have been taken for points in each region. The black area at the very bottom is associated to two different values, $1051$ and $958$, because it contains interspersed points associated to either. A more detailed breakdown can be found in Fig.~\ref{fig:tab_trajectories}.
%
%
%
%A more detailed breakdown of these associations is reported, for the interested reader, in Fig.~\ref{fig:tab_trajectories}. 

Most of the results presented in this Section come from the analysis of the aforementioned dataset. Some additional trajectories have been collected for selected parameter values; the corresponding choices will be specified separately for each such case. 

Finally, as a basic reliability check we have compared our algorithm's results against exact numerical diagonalization of the Lindblad superoperator $\mal{L}$ for a tiny system ($N = 2$) and observed agreement between the two approaches as long as a large enough number of stochastic trajectories is taken. An example of this can be seen in App.~\ref{app:comparison}.

\subsection{Stationary properties}
\label{subsec:stat}

We start by discussing the eQEP's stationary state. In order to collect stationary data, we have run each trajectory until it hit an absorbing GEP-like configuration. As we stated in Sec.~\ref{subsec:QJMC_algorithm}, once the QJMC dynamics jumps to an absorbing state it stops and yields exactly the same vector at all subsequent times. In other words, if we consider an individual trajectory $\tau$ $\tau$ which enters an absorbing GEP-like configuration at time $t_L$, the QJMC vector obeys
\be
    \ket{\psi(t_L, \tau)} = \ket{\psi(t, \tau)} \ \forall\, t \geq t_L
\ee
up to an irrelevant multiplicative phase which in the eQEP is trivial ($=1$). Taking the stationary (long-time) limit, this also implies
\be
    \ket{\psi(t_L, \tau)}= \ket{\psi(t \to +\infty, \tau)} \equiv \ket{\psi_{ss}(\tau)},
\ee
where the rightmost side is a shorthand for the first (and last) GEP-like absorbing configuration visited during the course of trajectory $\tau$.

%By calculating the contributions to our quantities of interest at $t_L$ for each trajectory we are therefore reconstructing their averages in the limit $t \to +\infty$. 
This means that, when considering a set of $M_{tr}>1$ trajectories, even though different trajectories $\tau_m$ "end" at different times $t_{L,m}$, for any observable $O$ we can approximate its stationary average as
\be
    \lim_{t\to \infty} \av{O(t)}_{\mal{L}} \approx  \frac{1}{M_{tr}} \sum_{m=1}^{M_{tr}} \bra{\psi(t_{L,m}, \tau_m)} O  \ket{\psi(t_{L,m}, \tau_m)},
\ee
which becomes an equality in the limit $M_{tr} \to \infty$.

\begin{figure}[h]
  \includegraphics[width=\columnwidth]{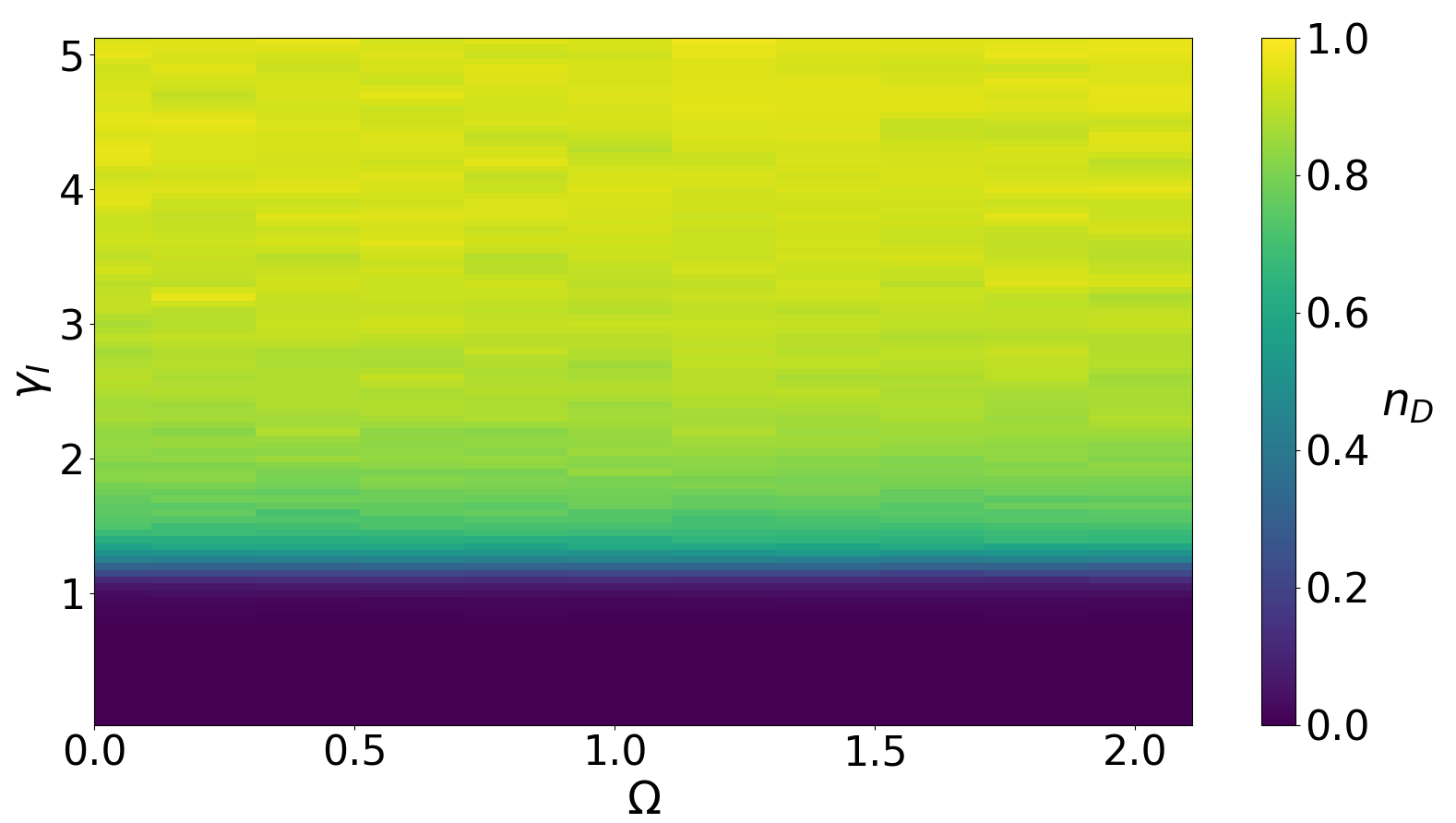}
  \caption{Stationary phase diagram of the eQEP. Here we have plotted, for each simulated parameter point ($\Omega, \gamma_I$), the stationary average DDS $n_D$. Brighter colors correspond to higher values of $n_D$, darker colors to lower ones. This plot shows a crossover between a phase with vanishing DDS at small $\gamma_I$ and one with finite DDS at larger $\gamma_I$s. } 
\label{fig:PD}
\end{figure}

Let us now work out expressions for the averages of the sigma operators $\sigma_k^{\mu\nu}$. To this end, we recall that the stationary vector $\ket{\psi_{ss}(\tau)}$ satisfies all the properties discussed in Secs.~\ref{subsec:weak_sym} and \ref{subsec:decoupling}, i.e., it can be written as
\be
    \ket{\psi_{ss} (\tau)} = \bigotimes_k \ket{\psi_{ss, k} (\tau)}_k \ \text{ with } \ket{\psi_{ss, k} (\tau)}_k \in \mal{H}_k^{(\zeta_k)}.
\ee
\begin{figure}[ht]
  \includegraphics[width=\columnwidth]{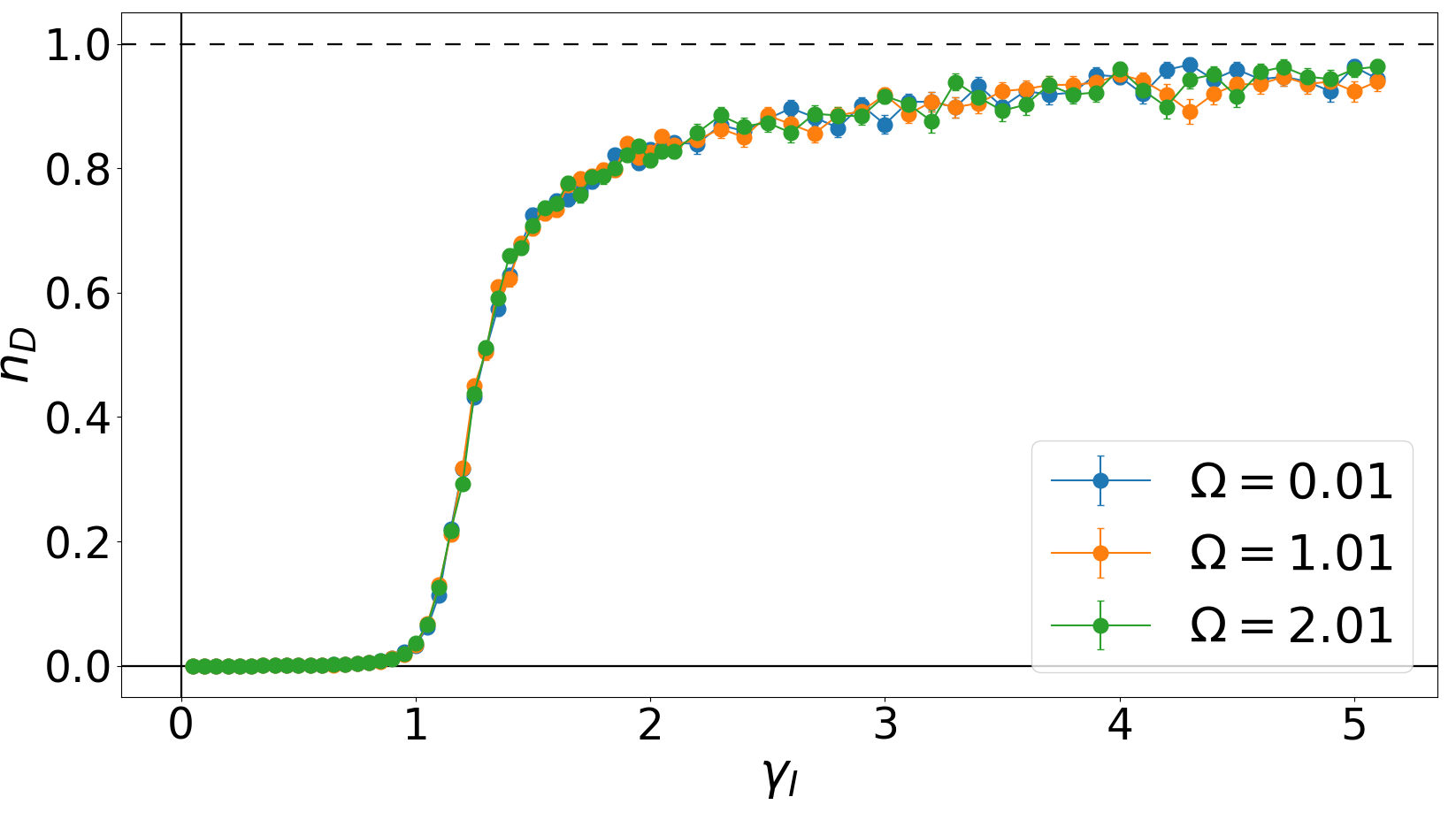}
  \caption{Three cross-sections of the phase diagram are plotted at constant $\Omega = 0.01$ (blue), $\Omega = 1.01$ (orange) and $\Omega = 2.01$ (green). The horizontal dashed line indicates the upper limit $n_D = 1$, to which all curves are expected to tend in the limit of prevailing infection $\gamma_I \to \infty)$. Curves for other values of $\Omega$ among those listed in \eqref{subeq:taken_Omegas} display exactly the same behavior, highlighting, in this parameter region, an independence of the stationary $n_D$ from the frequency $\Omega$.} 
\label{fig:slices}
\end{figure}
\begin{figure*}[ht]
  \includegraphics[width=0.95\textwidth]{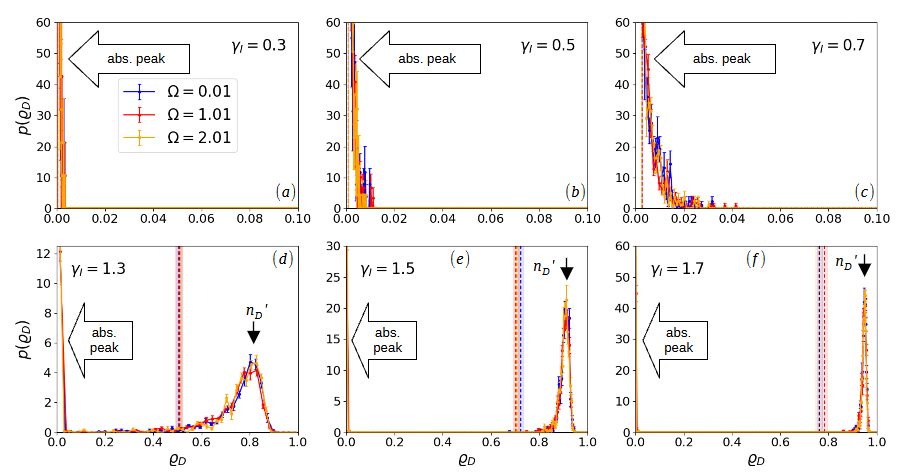}
  \caption{Each panel shows $p(\varrho_D)$ for a fixed $\gamma_I$ and three different frequencies $\Omega = 0.01$ (blue), $\Omega = 1.01$ (red) and $\Omega = 2.01$ (orange), see also the legend in panel (a). Panels are ordered such that $\gamma_I$ increases first from left to right, then from top to bottom. Vertical dashed lines highlight the position of the average $n_D$, while shaded areas the intervals $\lqq n_D - s_D, \, n_D + s_D \rqq$. The first row, panels (a)-(c), corresponds to values of $\gamma_I$ in the absorbing phase and consistently shows a single peak close to $\varrho_D = 0$. In panels (b) and (c) the peak develops a tail. The horizontal axis has been cut at $\rho_D = 0.1$ in order to make the widening of the tail more visible.  The second row, panels (d)-(f), shows instances of the active phase. The absorbing peak, still present at the left margin, is now accompanied by a second peak centered around some value $n_D' > n_D > 0$ (black downward arrows).  } 
\label{fig:nd_distr}
\end{figure*}

As shown in App.~\ref{app:eQEP_absorbing}, the absorbing states of the eQEP are in a one-to-one correspondence with the GEP ones, i.e. they are all the GEP-like configurations made up of only susceptible and dead sites ($\zeta_k \in \set{S, \, D}$ $\forall\, k$). Hence, using the relations \eqref{eqs:sigmas} we find
\be
    \bra{\psi_{ss}(\tau)} \sigma_k^{\mu\nu} \ket{\psi_{ss}(\tau)} = \delta_{\mu S} \,\delta_{\nu S}\, \delta_{\zeta_k S} + \delta_{\mu D}\, \delta_{\nu D}\, \delta_{\zeta_k D},
\ee
which leaves only two non-trivial stationary observables for site $k$: $\sigma_k^{SS}$ and $\sigma_k^{DD}$; additionally, these are not independent as their averages must sum up to $1$ and we can therefore study either one without loss of generality. 

Thus, the only relevant macroscopic order parameter we can build in the stationary state is the density of dead sites (DDS)
\be
    \varrho_D (\tau) =  \frac{1}{N} \sum_k \bra{\psi_{ss}(\tau)} \sigma_k^{DD} \ket{\psi_{ss}(\tau)},
    \label{eq:rand_varrho}
\ee
which is normalized here in such a way that $0 \leq \varrho_D \leq 1$. Correspondingly, we define its stochastic average
\be
    n_D = \overline{\varrho_D ( \tau)} \approx \frac{1}{M_{tr}} \sum_{m=1}^{M_{tr}}  \varrho_D (\tau_m),
\ee
where, again, in the r.h.s.~the sum runs over a set of $M_{tr}$ independent trajectories sampled by our algorithm.

The stationary phase diagram is shown in Fig.~\ref{fig:PD}. Close to $\Omega = 0$ the eQEP algorithm is merely an elaborate simulation of the GEP (see App.~\ref{app:eQEP_to_GEP}); unsurprisingly, therefore, moving vertically we observe a sharp crossover between an absorbing phase (very small stationary DDS) and an active one (non-negligible stationary DDS). The transition between these two regimes takes place around $\gamma_I \approx 1.2$.

Interestingly, the same behavior is identically replicated, within statistical error, at larger values of $\Omega$ and, in fact, throughout the entire range we have simulated.
Combining this observation with the previously stated fact that for $\Omega = 0$ the eQEP reduces to the GEP (see App.~\ref{app:eQEP_to_GEP}) and that the GEP is known to undergo a continuous phase transition (see Sec.~\ref{subsec:GEP}), it seems reasonable to assume that the smooth crossover in Fig.~\ref{fig:PD} is in fact signaling the presence of an underlying continuous phase transition for frequencies in the entire range $\Omega \in \lqq 0, \, 2.01 \rqq$.

In Fig.~\ref{fig:slices} we show some cross sections of the phase diagram, each taken at fixed $\Omega$. These curves display $n_D$ as a function of $\gamma_I$; clearly, they take sigmoidal shapes with a sharp increase and an inflection point around $\gamma_I \approx 1.2$. Error bars extend above and below each data point by the standard error of the mean $s_D$, estimated in the usual way
\be
    s_D^2 = \frac{1}{M_{tr} \lt M_{tr} - 1 \rt} \sum_{m=1}^{M_{tr}} \bigl(  \varrho_D (\tau_m)  - n_D \bigr)^2.
\ee
The comparison makes it apparent that the curves are statistically indistinguishable, which suggests that the stationary properties of the system do not depend on $\Omega$, at least up to its highest simulated value $\Omega_{max} = 2.01$.

\begin{figure*}[ht]
  \includegraphics[width=0.95\textwidth]{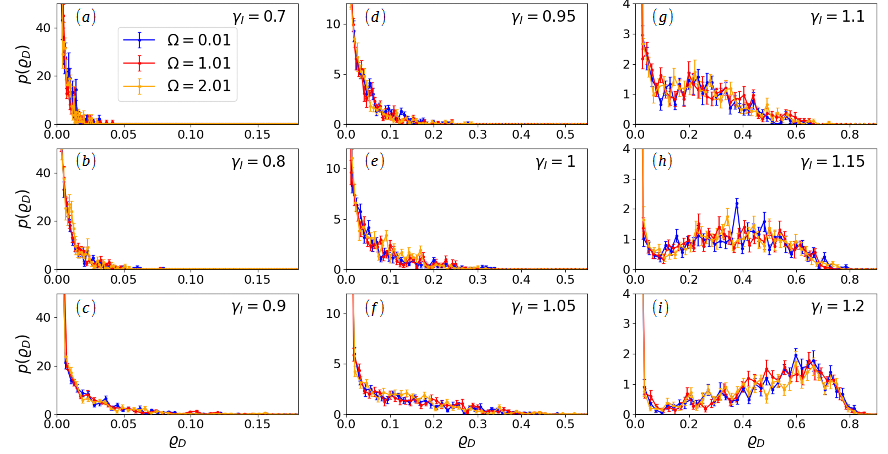}
  \caption{
  Behavior of the DDS's distribution $p(\varrho_D)$ close to the critical region. Each panel shows $p(\varrho_D)$ for three different values of $\Omega$, as in Fig.~\ref{fig:nd_distr}. In this case, however, we have transposed the ordering of the panels, i.e., $\gamma_I$ increases first from top to bottom, then from left to right. We have labeled them (a)-(i) following the same rule. In each column the horizontal axis is cut off at the same threshold to make the comparison easier. In panels (a)-(f) we see a tail developing from the absorbing peak close to $\varrho_D = 0$ towards finite values. In the last column, panels (g)-(i) we start seeing the second peak develop and quickly shift rightward. There is no significant difference to be found for different values of $\Omega$. 
}
\label{fig:nd_distr_crit}
\end{figure*}
To further test this, we have looked at the distribution $p(\varrho_D)$ of the random DDS in Eq.~\eqref{eq:rand_varrho}, which we have reconstructed numerically by simply subdividing the interval $\lqq 0,\,1\rqq$ in smaller "bins" and counting, for each, how many trajectories ended up producing a $\varrho_D$ within its bounds. We then divided these counts by the number of trajectories and the bin width, in such a way to reproduce the normalization
\be
    \int_0^1 \rmd \varrho_D \, p(\varrho_D) = 1.
\label{eq:ditr_norm}
\ee
Some examples are reported in Figs.~\ref{fig:nd_distr} and \ref{fig:nd_distr_crit}. To avoid confusion, let us remark that Eq.~\eqref{eq:ditr_norm} means that $p$ is normalized as a probability density, rather than a probability. Thus, it is not subject to the constraint $p \leq 1$.

Before discussing the plots, let us spend a few words on how we chose our "bins": whenever sharply-peaked distributions (with one or more peaks) are expected, it can be difficult to determine a meaningful bin width a priori, i.e., a width sufficiently large to capture several points per bin (at least close to the peaks), but small enough not to lose, under the effect of coarse-graining, the main features of the distribution.
To better visualize the peaks and their neighborhood, we have chosen to adapt the bin width in such a way to always have at least $O(10)$ non-empty bins; this is why the reader, upon attentive inspection, may notice that the spacing between points varies widely between different plots \footnote{For instance, for small values of the infection rate $\gamma_I < 0.7$ the distribution is non-zero only in a very narrow neighborhood of $\varrho_D = 0$. For $\gamma_I \approx 1$ its tail extends instead up to $\rho_D \approx 1$. If we were to pick the same bin width in both regimes, we would either have bins too large to see the (small, but present) tails of the absorbing peak at $\varrho_D = 0$ or so small that the curves at higher $\gamma_I $ would be dominated by stochastic noise. }.

%Since the distribution displays, for several parameter choices, a sharply-peaked structure (with either one or two distinct peaks), we have adapted the bin width so to have always at least $O(10)$ bins within its support (i.e.~at least $O(10)$ points where $p \neq 0$). Because of this, the spacing between points varies widely in our Figures \ref{fig:nd_distr} and \ref{fig:nd_distr_crit} further below\footnote{For instance, for small values of the infection rate $\gamma_I < 0.7$ the distribution is non-zero only in a very narrow neighborhood of $\varrho_D = 0$. For $\gamma_I \approx 1$ its tail extends instead up to $\rho_D \approx 1$. If we were to pick the same bin width in both regimes, we would either have bins too large to see the (small, but present) tails of the absorbing peak at $\varrho_D = 0$ or so small that the curves at higher $\gamma_I $ would be dominated by stochastic noise. }. 
We have roughly estimated uncertainties by subdividing every set of trajectories in $10$ subsets \footnote{As a matter of fact, we excluded some trajectories as we took the largest multiple of $10$ smaller than the available number; for instance, for parameter pairs with $1051$ trajectories we took, for error estimation, the first $1050$ and divided them into $10$ subsets of $105$ each. For parameters with $958$ trajectories we took the first $950$, and so on.}; for every parameter choice and every bin we thus obtained $10$ independent estimates of the count, which we then used to calculate a standard deviation.

Figure \ref{fig:nd_distr} illustrates the typical behavior of the distribution $p(\varrho_D)$ in the two phases. Each plot is associated to a given value of $\gamma_I$ and includes three curves, corresponding to three different values of the frequency $\Omega$. Panels (a)-(c) in the first row are taken in the absorbing phase and feature a single narrow peak around $0$. In the thermodynamic limit this "absorbing peak" would reduce to a Dirac delta (with an amplitude equal to the probability of not producing an outbreak in a trajectory). Because of this, the bins close to the origin $\varrho_D = 0$ are always very highly populated. To make the other features of the distribution (tails, other peaks, ...) visible, we always cut off the vertical range of our plots below the value taken by $p(0)$.

%The vertical range of our plots is thus consistently cut off, so that other features of the distribution (tails, other peaks, ...) become visible. 
With this choice we can observe that, as $\gamma_I$ increases, the tails of the peak become thicker and extend to the right. This means that, at least in some trajectories, the infection manages to propagate further and further away from the origin. In turn, this implies that in the steady state the system achieves a higher (albeit still sub-extensive) death count and, consequently, a larger DDS $\varrho_D$.

Panels (d)-(f), which make up the second row, show instead examples of the distribution in the active phase. We observe here the presence of a second peak around a finite DDS value $n_D' > 0$. This peak moves to the right and becomes higher as $\gamma_I$ is increased; at the same time, even though this this is not shown in the plots, the height of the absorbing peak decreases. This is analogous to what happens in the GEP: as $\gamma_I$ is increased, the probability of outbreak formation increases, which conversely means that the probability of a trajectory stopping close to the origin decreases by the same amount.  

In Fig.~\ref{fig:nd_distr_crit} we focus instead on the critical region. Again, each plot corresponds to a fixed value of $\gamma_I$ and includes three curves for the same values of $\Omega$ used in Fig.~\ref{fig:nd_distr}. The main difference lies in the infection rate being now taken in the interval $\lqq 0.7,\, 1.2 \rqq$. Note that, to make the comparison easier between curves that can be meaningfully plotted on the same $\varrho_D$ range (on the horizontal axis), we have ordered the plots from \emph{top to bottom first}, \emph{then left to right}. Each plot includes the entire support of the distribution, with plot (f) representing the only exception due to tiny statistical fluctuations at higher DDS.

%We remark that, to make the comparison between successive values of $\gamma_I$ a bit easier, we have ordered the panels first from top to bottom, then from left to right, meaning that instead of each column (instead of row) is the continuation of the previous one. Each column features the same restriction of the horizontal axis, chosen as to include most of the support of the distribution. 
Moving from panel (a) to (f) it is apparent that the absorbing peak develops a tail which extends further and further to the right as $\gamma_I$ is increased. The distribution appears to become non-monotonic in panel (g) and the second peak becomes visible in the last two plots. 

Analogously to the GEP's phenomenology, the finite peak seems to move out of the origin as $\gamma_I$ is increased, i.e., $n_D' \to 0$ at the critical point. As we discussed in Sec.~\ref{subsec:GEP} just before and after Eq.~\eqref{eq:greater_than_0}, the second peak starting to appear at a finite $n_D' \approx c_D' > 0$, instead of detaching from the absorbing one, could be a possible explanation of the discontinuities observed in Ref.~\cite{Espigares2017}. Hence, from the stationary order parameter's properties we find no discernible trace of discontinuity that could be regarded as any kind of  first-order phase transition. Furthermore, curves for different frequencies are not distinguishable, or more precisely their differences are not statistically significant. Even the distribution $p(\varrho_D)$ does not show any dependence on $\Omega$. Since the blue curves describe a very small $\Omega = 0.01$, we can consider them good approximations to the GEP's own $p(\varrho_D)$, and conclude that, in the eQEP, the stationary statistical properties of the order parameter are indistinguishable (in the range we have investigated) from the GEP's. 

We have been thus far unable to find a general, rigorous explanation for this. We can however provide some intuitive justification for it in the high frequency regime $\Omega \gg \gamma_I,\,\gamma_D$: the effective Hamiltonian $H_{eff}$ is, in this case, dominated by the actual Hamiltonian $H$ which, we recall, induces oscillations (precession) between the local GEP-like states $\vec{I}$ and $\vec{B}$. If the separation of the Hamiltonian and dissipative timescales is sufficiently large we can see the eQEP dynamics as an epidemic process where the active sites undergo extremely fast oscillations in an out the actual infected state, spending half of their time in it and half out of it. Consequently, infection and death both proceed at half their "nominal" speeds. In a sense, the Hamiltonian oscillations behave as a sort of annealed noise, in the sense that the dissipative part of the dynamics cannot respond to the very fast fluctuations of the coherent part and is thus only sensitive to the average behavior they produce.

In other words, fpr very large $\Omega$ we are looking at an effective GEP with rescaled rates
\be
    \gamma_I^{(eff)} = \gamma_I / 2 \ \ \text{ and } \ \ \gamma_D^{(eff)} = \gamma_D / 2.
\ee
This is equivalent to keeping the same rates and rescaling time $t \to t^{(eff)} =  2t$, i.e., slowing down time by half, which clearly cannot impact stationary properties in any way. Hence, we can reasonably expect the steady state to be approximately the same for $\Omega \approx 0$ and for (sufficiently) large $\Omega$. We provide some numerical evidence in support of this intuitive argument when discussing the dynamics in Sec.~\ref{subsec:dyn}.

In summary, at least in terms of ordinary one-point order parameters, the stationary properties of the eQEP reproduce exactly those of the GEP, as if the Hamiltonian were not present. 
In particular, this shows that the eQEP cannot substantiate the findings of \cite{Espigares2017}. The origin of the puzzling features of the RQEP's stationary phase diagram remains, unfortunately, elusive.

\subsection{Dynamical properties}
\label{subsec:dyn}
We now turn our attention to the dynamics. Being particularly interested in the formation and (potential) repetition of outbreaks, we focus first on the local density of infection $\overline{\av{\sigma_k^{II}(t)}}$. For this purpose, we have run simulations up to a fixed maximum time $t_{max}$, measuring this observable at regular steps $\Delta t_{step}$. These simulations are independent from the ones Fig.~\ref{fig:cells_and_trajs} alludes to but, as was the case before, we have taken a lower number of trajectories at higher values of the infection rate $\gamma_I$ and vice versa. We will specify the actual numbers when discussing each instance.
 
\begin{figure*}[ht]
  \includegraphics[width=0.98\textwidth]{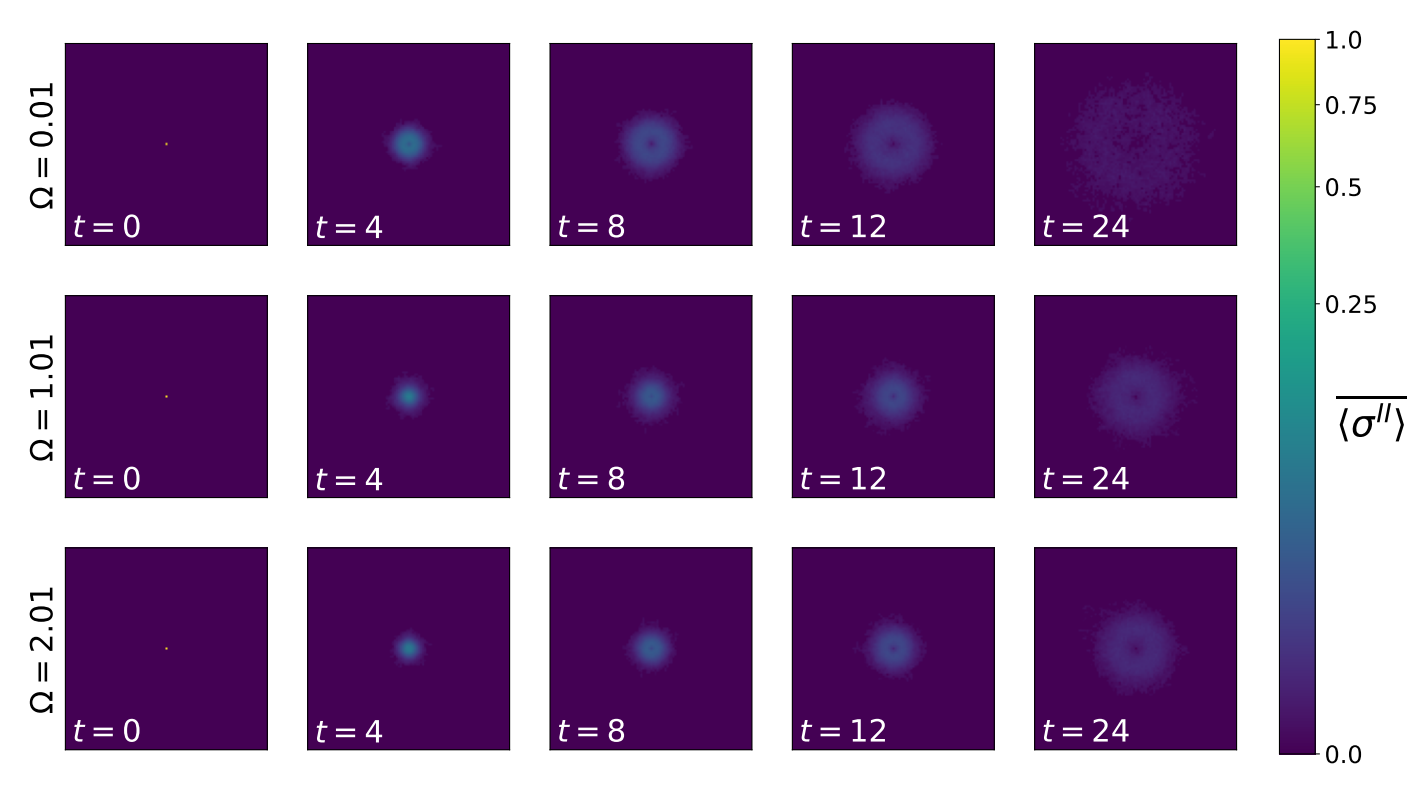}
  \caption{System evolution in the absorbing phase. Each of the panels above is a visualization of the infection density $\av{\sigma^{II}}$ on a lattice of $101 \times 101$ sites for $\gamma_I = 0.8$, averaged over $M_{tr} = 2 \times 10^4$ trajectories. Different columns correspond to different times (reported in the lower left corner of each panel), whereas different rows to different values of the frequency $\Omega$ (shown at the very left). From top to bottom, we have $\Omega = 0.01$, $\Omega = 1.01$ and $\Omega = 2.01$. As the infected density declines very rapidly in time, we have stretched the colormap ($c \to c^{1/3}$) to gain a higher resolution at lower values.
}
\label{fig:snaps_abs}
\end{figure*}

In Fig.~\ref{fig:snaps_abs} we display snapshots of the evolution of $\overline{\av{\sigma_k^{II}(t)}}$ on a $101 \times 101$ lattice in the absorbing phase ($\gamma_I = 0.8$). The first column corresponds to the initial state ($t = 0$) and shows a single, bright dot at the very center, which is nothing but the single initial infected site positioned at the origin.
%
%
%
% %
% %
% %
% \begin{figure*}[ht]
%   \includegraphics[width=\textwidth]{Movies_abs2.png}
%   \caption{Each of the panels above is a visualization of the infection density $\av{\sigma^{II}}$ on a lattice of $101 \times 101$ sites for $\gamma_I = 0.8$ (within the absorbing phase), averaged over $M_{tr} = 2 \times 10^4$ trajectories. Different columns correspond to different times, whereas different rows to different values of the frequency $\Omega$. From top to bottom, we have $\Omega = 0.01$, $\Omega = 1.01$ and $\Omega = 2.01$. As the infected density declines very rapidly in time, we have stretched the colormap ($c \to c^{1/3}$) to gain a higher resolution at lower values.
% }
% \label{fig:snaps_abs2}
% \end{figure*}
%
%
%
%
%
%
\begin{figure*}[ht]
  \includegraphics[width=\textwidth]{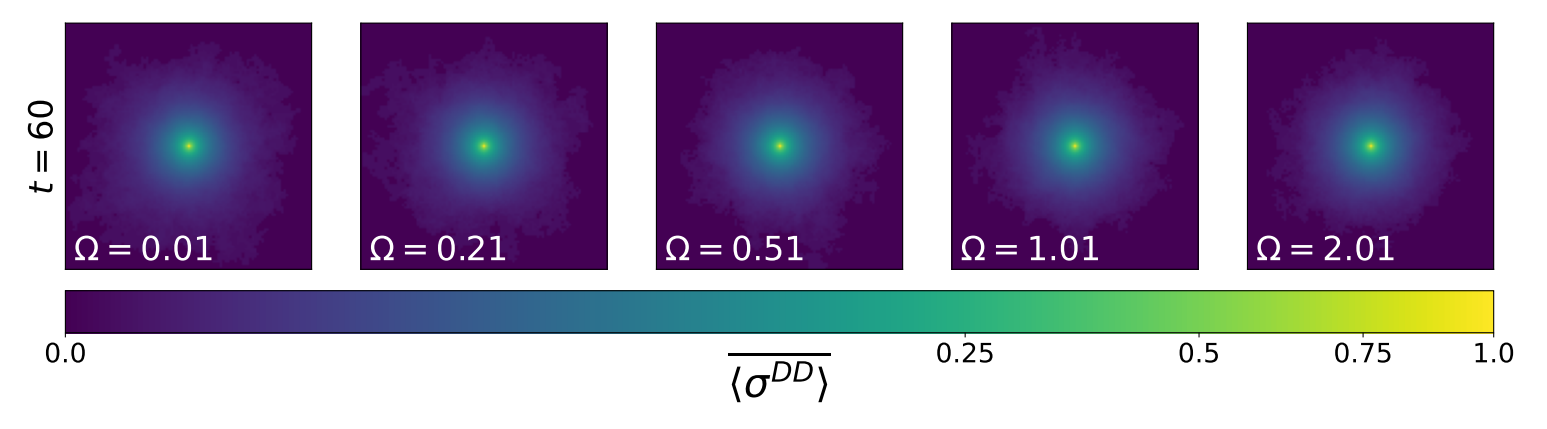}
  \caption{Long-time local DDS on a $101 \times 101$ lattice in the absorbing phase. The plots above are snapshots at $t = 60$ for the same infection rate of Fig.~\ref{fig:snaps_abs}, $\gamma_I = 0.8$. The colormap is also analogously stretched to highlight lower densities. In spite of the delay displayed by the dynamics at higher values of $\Omega$ in Fig.~\ref{fig:snaps_abs}, at long times the DDS shows the same profiles, which is consistent with our previous observations on the stationary phase diagram, Fig.~\ref{fig:PD}. 
}
\label{fig:snaps_final_D}
\end{figure*}
\begin{figure*}[ht]
  \includegraphics[width=\textwidth]{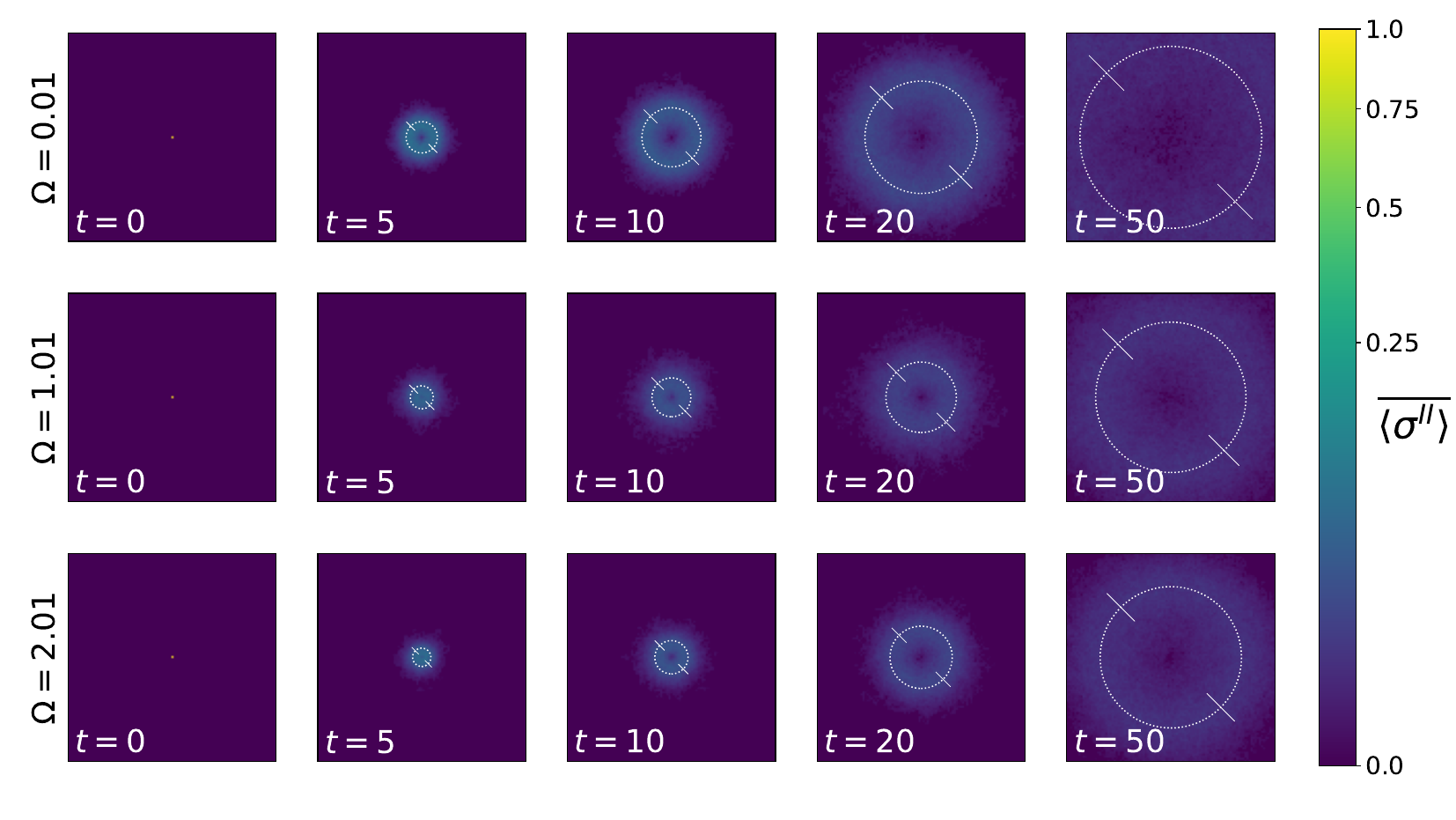}
  \caption{Close-to-critical evolution. The local infection density $\av{\sigma^{II}}$ is displayed here as in Fig.~\ref{fig:snaps_abs}; now, however, $\gamma_I = 1.2$, we average over a smaller set of trajectories ($M_{tr} = 2500$), and stretch the colormap according to a different law $c \to c^{2/5}$. Dotted white circles illustrate the average radius $R_{ring} (t)$ (see Eq.~\eqref{eq:ring_moments}), whereas the two superimposed white segments stretch both inward and outward from the dotted circle by an average ring width $W_{ring} (t)$.
}
\label{fig:snaps_crit}
\end{figure*}
%
%
%
%
% \begin{figure*}[ht]
%   \includegraphics[width=0.9\textwidth]{Final_DD_crit_ii.png}
%   \caption{Local DDS close to the critical point ($\gamma_I = 1.2$) at times $t = 50$ (first row) and $t = 100$ (second row). Note that, differently from Fig.~\ref{fig:snaps_crit}, the colormap here is linear.
%   }
% \label{fig:snaps_crit_D}
% \end{figure*}
%
%
\begin{figure*}[ht]
  \includegraphics[width=\textwidth]{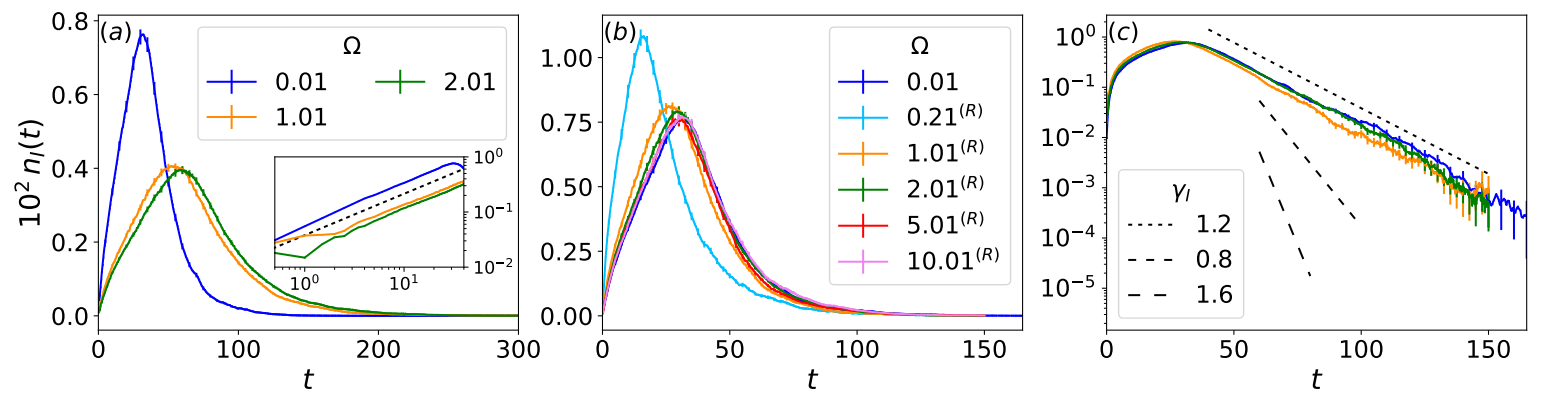}
  \caption{
  Evolution of the global infection density $n_I (t)$, Eq.~\eqref{eq:av_inf_density}, for the near-critical $\gamma_I = 1.2$ (same as Fig.~\ref{fig:snaps_crit}), averaged over $1520$ trajectories. We have introduced a $10^2$ magnification in all panels to make the vertical axes' labels more easily readable. Data points have been taken with a step of $\Delta t_{step} = 0.5$ and errorbars are shown once every $10$ points to avoid clutter. (a) Linear-scale plot for $\Omega = 0.01$ (blue), $1.01$ (orange) and $2.01$ (green). In the inset, the same curves are plotted in log-log scale on the interval $\lqq 0.5,\, 40  \rqq$, together with a pure power-law $\propto t^{0.75}$ (black, dashed line). (b) Same as panel (a), but with all curves but the blue one ($\Omega = 0.01$) rescaled according to Eq.~\eqref{eq:II_rescaling}, including three additional values of the frequency $\Omega = 0.21$ (cyan), $5.01$ (red) and $10.01$ (purple). Curves subject to rescaling  are identified by an $(R)$ in the legend. (c) Logplot of panel (b) with only the frequencies selected for panel (a). The dashed lines highlight the exponential decay extracted for $\gamma_I = 1.2$ (current plot), $0.8$ (subcritical) and $1.6$ (supercritical).
}
\label{fig:crit_multi}
\end{figure*}

As time progresses, the process quickly dies off: reading any row from left to right, we see a wavefront of infection becoming dimmer and dimmer as it expands outwards, signaling the fact that a smaller and smaller portion of trajectories supports infection up to that distance from the origin. Following infection of the immediate neighborhood of the origin ($t \approx 4$), the wavefront takes its typical ring-like structure. The outer dark area corresponds to susceptible sites the infection has not reached; the internal dark disk, instead, is where dead sites can be found. At $t \gtrsim 24$ we see that the process is essentially over (only a negligible portion of the trajectories has survived) and has not propagated to the edges. 

We encounter here the first actual effect of the Hamiltonian term: by comparing the three rows, we see that at larger values of $\Omega$ the infection seems to slow down. This is not completely unexpected, as the oscillations that infected sites undergo between $\ket{I}$ and $\ket{B}$ hinder propagation: remember, in fact, that the constraint strictly requires the infecting site to have component over $\ket{I}$; a site (mostly) in the bedridden state $\ket{B}$ \emph{cannot infect} any of its susceptible neighbors, but neither can it die. It will, after some time, oscillate back close to $\ket{I}$.

%At the same time, while in state $\ket{B}$ a site is prevented from dying and can "reactivate" later to participate again in the spreading of the infection. 
Interestingly, the delay induced by the presence of the bedridden state causes the process to eventually deposit the same DDS independently from the value of $\Omega$, as we have seen in Sec.~\ref{subsec:stat}. This is also illustrated in Fig.~\ref{fig:snaps_final_D}, where we compare the DDS over the entire lattice for different $\Omega$s, including the values used in Fig.~\ref{fig:snaps_abs}, at a time ($t = 60$) sufficiently large for the infection to have largely subsided. 

In Fig.~\ref{fig:snaps_crit} we move closer to the critical point, having set $\gamma_I = 1.2$. We observe in this case the formation of a single outbreak, meaning that infection manages to propagate all the way to the edges of the lattice. The progressive dimming of $\sigma^{II}$, although still observable, is now much less pronounced than in the absorbing phase. Because of the large differences in dimming at different values of $\gamma_I$, we have used different rescalings of our colormaps for different plots. The interested reader may find them specified in the figure captions and should generally keep in mind that lattice plots for different values of $\gamma_I$ cannot be easily compared in a quantitative way because of this.

It is difficult, on our lattice of $101 \time 101$ sites, to unambiguously determine whether the outbreak propagates outwards at constant speed: following Grassberger's example for the GEP \cite{Grassberger1983}, we introduce the infection radius moments
\be
    R_{ring}^{(m)} (t) = \frac{\sum_{k} \overline{\av{\sigma_k^{II} (t)}} \, d(k)^m }{ \sum_{k} \overline{\av{\sigma_k^{II} (t)} }},
    \label{eq:ring_moments}
    %= \frac{\sum_{k} \overline{\av{\sigma_k^{II} (t)}} \, d(k)^m }{N \, n_I (t)} ,
\ee
where $d(k)$ denotes the distance of site $k$ from the origin. We dub $R_{ring} \equiv R_{ring}^{(1)}$ the average infection radius and $W_{ring} = \sqrt{R_{ring}^{(2)} - \lt R_{ring} \rt^2}$ the average width of the ring-like outbreak. For a single outbreak, these quantities are reasonable descriptors of the outbreak's features. To illustrate this, in Fig.~\ref{fig:snaps_crit} we have drawn a white circumference at distance $R_{ring}$ from the origin in each panel; two small white radial segments have been superimposed (at angles $\pm 3\pi/4$) which stretch from $R_{ring} - W_{ring}$ to $R_{ring} + W_{ring}$. A power-law fit on the interval $t \in \lqq 5,\, 25 \rqq$ suggests $R_{ring}(t) \propto t^{0.9}$ for $\Omega = 0.01$, suggesting a mild deceleration. Analogously, fitting on $t \in \lqq 10,\, 50 \rqq$ yields the rough estimates $R_{ring}(t) \propto t^{0.85}$ for $\Omega = 1.01$ and $R_{ring}(t) \propto t^{0.9}$ for $\Omega = 2.01$. Interestingly, $W_{ring}$ also increases during the expansive phase of the outbreak, signaling the fact that the wavefront is widening as it expands outwards.

In Fig.~\ref{fig:crit_multi}(a) we look at the evolution of the global average infection density
\be
    n_I (t) = \frac{1}{N} \sum_k \overline{ \av{\sigma_k^{II} (t)}},
    \label{eq:av_inf_density}
\ee
magnified by a factor $100$, as a function of time.
Curves at higher $\Omega$ clearly lag behind the one at $\Omega = 0.01$, while showing the same qualitative structure: at short times ($t \sim O(1)$) there is a transient phase coinciding with the infection of the origin's immediate neighborhood. Subsequently, $n_I$ experiences a period of sub-linear growth as the outbreak travels outwards. The inset in panel (a) shows the curves of the parent plot in a log-log scale restricted to $t \in \lqq  0.5, \, 40 \rqq$, highlighting a common power-law increase $n_I(t) \propto t^{0.75}$ (dashed, black line).
After the outbreak has reached the lattice boundaries the growth of $n_I$ is stifled by the diminishing opportunities for new infections and the density peaks as the outbreak moves to the lattice corners. Following that, the outbreak has nowhere left to go and the dynamics becomes dominated by the progressive death of the remaining infected sites, yielding an exponential decay.

We can now test the assumption we made in Sec.~\ref{subsec:stat}, namely that under a large separation between the coherent and dissipative timescales ($\Omega \gg \gamma_I,\, \gamma_D$) the eQEP dynamics can be approximately seen as a classical GEP with half the infection and death rates (due to the bedridden state being inert) and producing half the infection density (again due to spending about half the time in I and half in B). Therefore, in Fig.~\ref{fig:crit_multi}(b) we take the $\Omega = 0.01$ curve (blue line) as a reference and rescale every other one doubling its vertical amplitude and shrinking it by half in the horizontal direction, i.e., instead of plotting $f(t) = \overline{\av{\sigma^{II}(t)}}$ we draw
\be
    f_{resc}(t) =  2 \,  \overline{\av{\sigma^{II}\lt 2t \rt}}.
    \label{eq:II_rescaling}
\ee
As predicted, we see the curves at higher frequency collapse (within statistical error) onto the one used as a reference. Unexpectedly, however, we observe good agreement already for $\Omega = 2.01$, i.e., less than twice the current infection rate $\gamma_I = 1.2$; even the curve for $\Omega = 1.01$ (orange line) gets rescaled rather close to the reference, suggesting that the effective "decoupling" between the coherent and dissipative dynamics takes hold very quickly as $\Omega$ is increased. For $\Omega \in \lqq 0, \, 1\rqq$, instead, curves do \emph{not} collapse onto the reference upon rescaling, as exemplified by the cyan line ($\Omega = 0.21$), marking the expected breakdown of our "separation of timescales" argument for small enough $\Omega$.   

In Fig.~\ref{fig:crit_multi}(c) we focus on long-time behavior and plot the evolution of the infected density in logarithmic scale; curves for $\Omega = 1.01$ and $2.01$ are still rescaled according to \eqref{eq:II_rescaling} to show that the aforementioned collapse continues to hold. Between $t \approx 40$ and $160$ all curves seem to assume a linear behavior, meaning that the evolution in this range is dominated by an exponential decay $\propto \rme{- t / T_{rel}}$ with a characteristic relaxation time $T_{rel} \approx 16.6$ extracted via linear fit. For comparison, the corresponding timescale at $\gamma_I = 0.8$ is $T_{rel} \approx 6.9$ and, at $\gamma_I = 1.6$, $T_{rel} \approx 3.5$; pieces of these two exponential decays are added for comparison as dashed lines.

Remember, still, that we are rescaling large-$\Omega$ curves according to \eqref{eq:II_rescaling}; hence, in order to retrieve the correct relaxation time for the $\Omega = 1.01$ and $2.01$ we need to multiply the aforementioned values by $2$. For instance, $T_{rel} \lt \gamma_I = 1.2,\, \Omega = 2.01 \rt \approx 2 \times 16.6 = 33.2$.

The large variations of $T_{rel}$ across the transition are hardly surprising: physical systems close to a continuous phase transition undergo "critical slowing down", i.e., they experience a delay in responding to perturbations and relaxing back to stationarity. Indeed, in the thermodynamic limit $T_{rel}$ is expected to diverge at the critical point, leading to algebraic, instead of exponential, relaxation of observables.

\begin{figure*}[ht]
    \includegraphics[width=\textwidth]{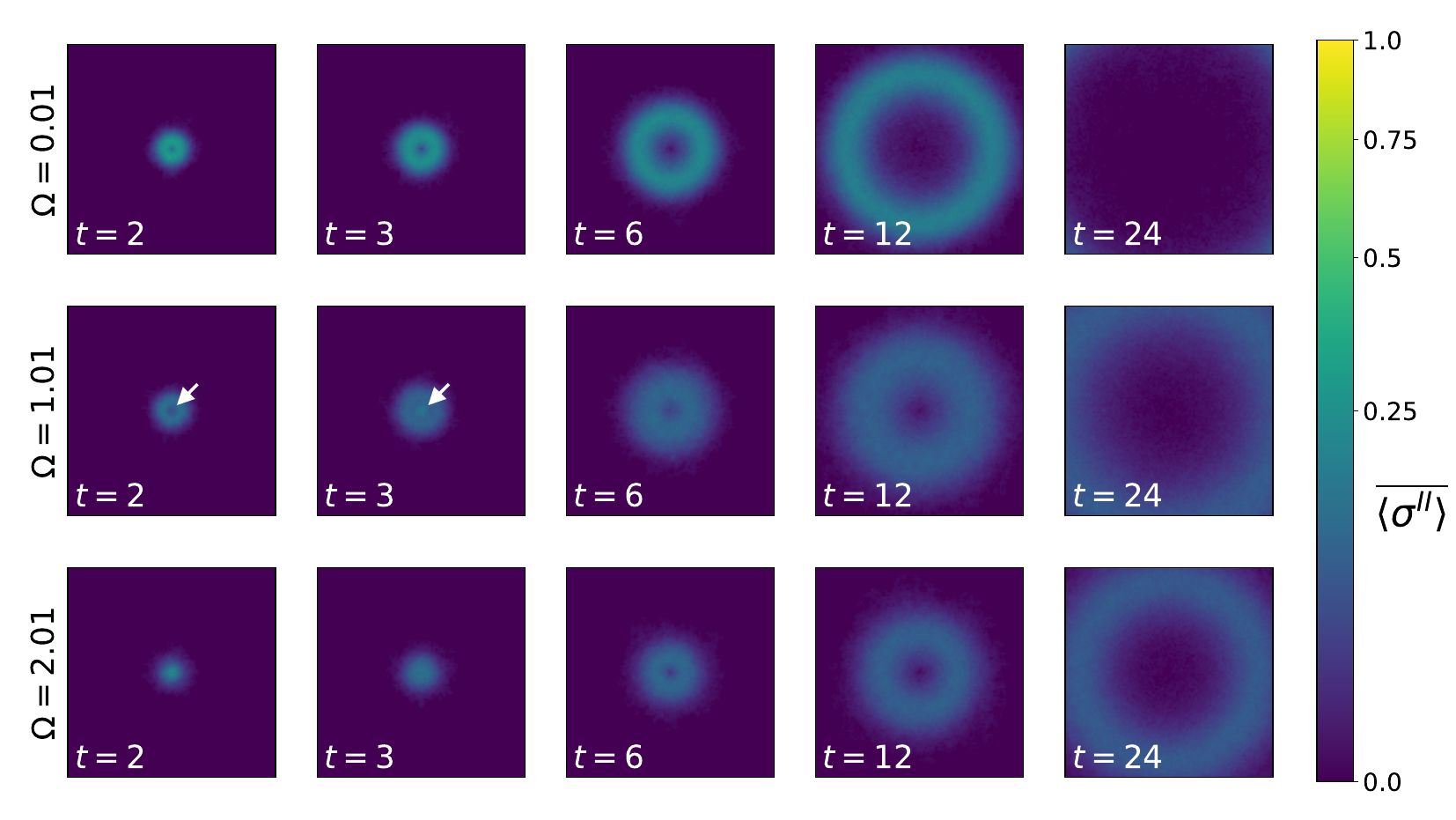}
  \caption{Snapshots of the local infection density for $\gamma_I = 2$, in the active phase, averaged over $1240$ trajectories, for the same three values of the frequency used in Figs.~\ref{fig:snaps_abs} and \ref{fig:snaps_crit}. The colormap rescaling has been changed again to $c \to c^{1/2}$. Two small white arrows in the first two panels of the second row ($\Omega = 1.01$) highlight the presence of oscillations close to the origin at short times.
}
\label{fig:snaps_act_low}
\end{figure*}
\begin{figure*}[ht]
    \includegraphics[width=\textwidth]{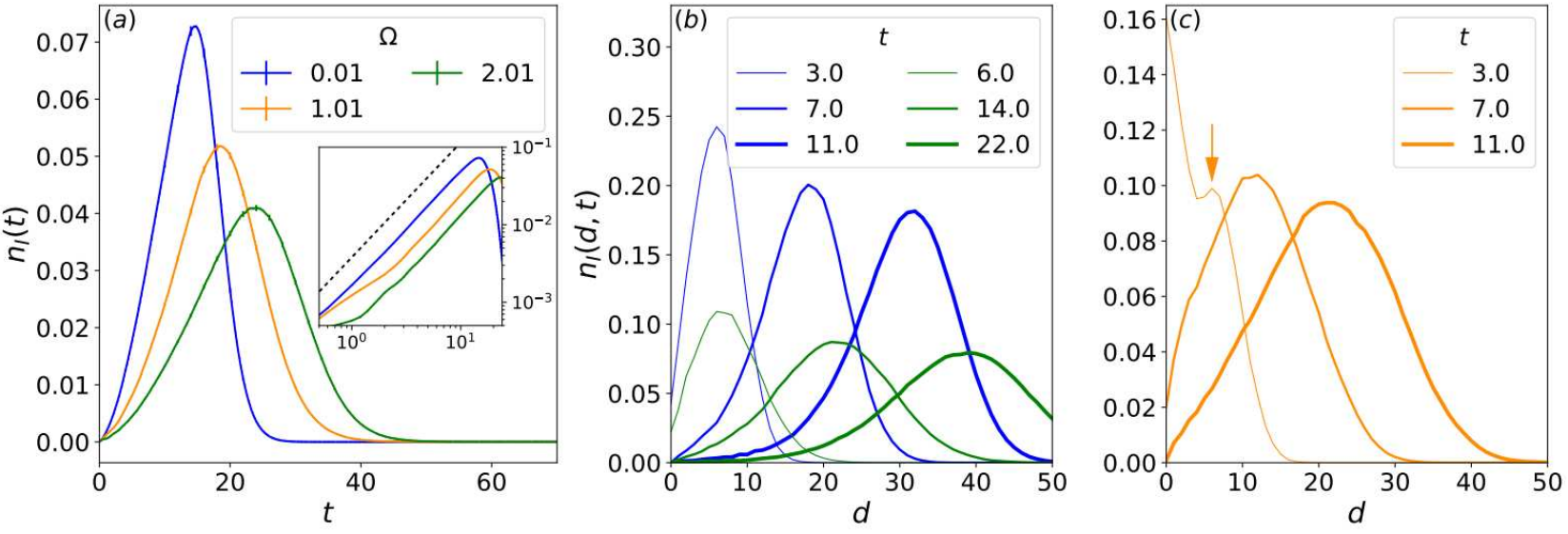}
  \caption{Infection density as a function of either time $t$ or distance $d$ from the origin, calculated from the same trajectories used for Fig.~\ref{fig:snaps_act_low}. (a) Evolution of the global $n_I(t)$, Eq.~\eqref{eq:av_inf_density}; the inset, in log-log scale, zooms onto the interval $t \in \lqq 0.6, \, 24 \rqq$ and superposes a power-law behavior $\propto t^{1.48}$ (dashed, black line) to the curves. (b) Profile of $n_I(d, t)$, with thicker lines corresponding to later times; the color scheme is the same as in panel (a): blue for $\Omega = 0.01$, green for $2.01$. (c) Same as (b), but for $\Omega = 1.01$. The arrow highlights a tiny second peak developing close to the origin at small times, which later disappears.
}
\label{fig:multi_act_low}
\end{figure*}
\begin{figure*}[ht]   %% 20
    \includegraphics[width=\textwidth]{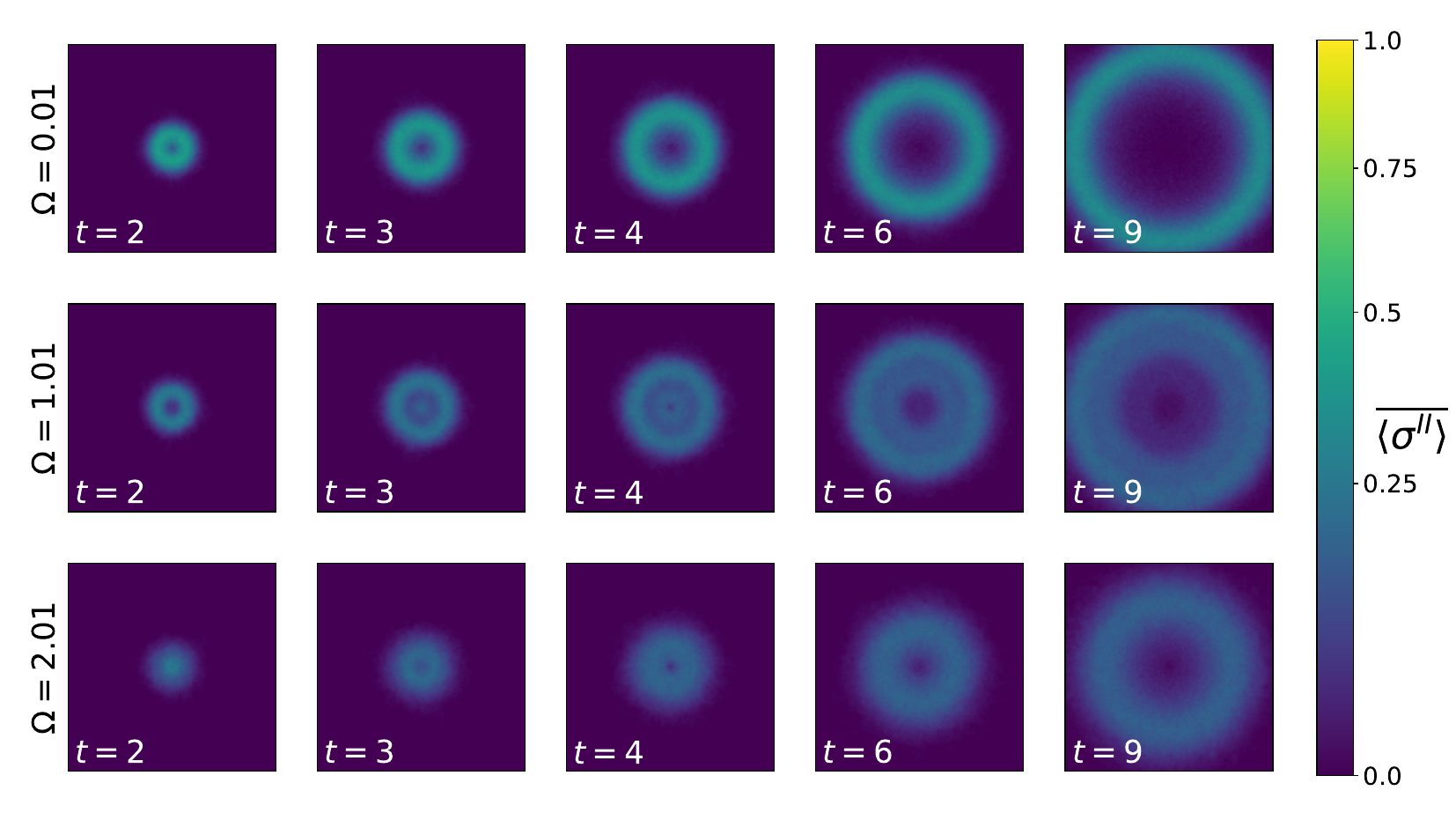}
  \caption{Local infection density for $\gamma_I = 3$ averaged over $M_{tr} = 800$ trajectories. The panels are organized as in Figs.~\ref{fig:snaps_abs}-\ref{fig:snaps_act_low}. The colormap rescaling is $c \to c^{2/3}$, chosen in this case to highlight the two rings in the second row.
}
\label{fig:snaps_act_mid}
\end{figure*}
\begin{figure*}[ht]
    \includegraphics[width=\textwidth]{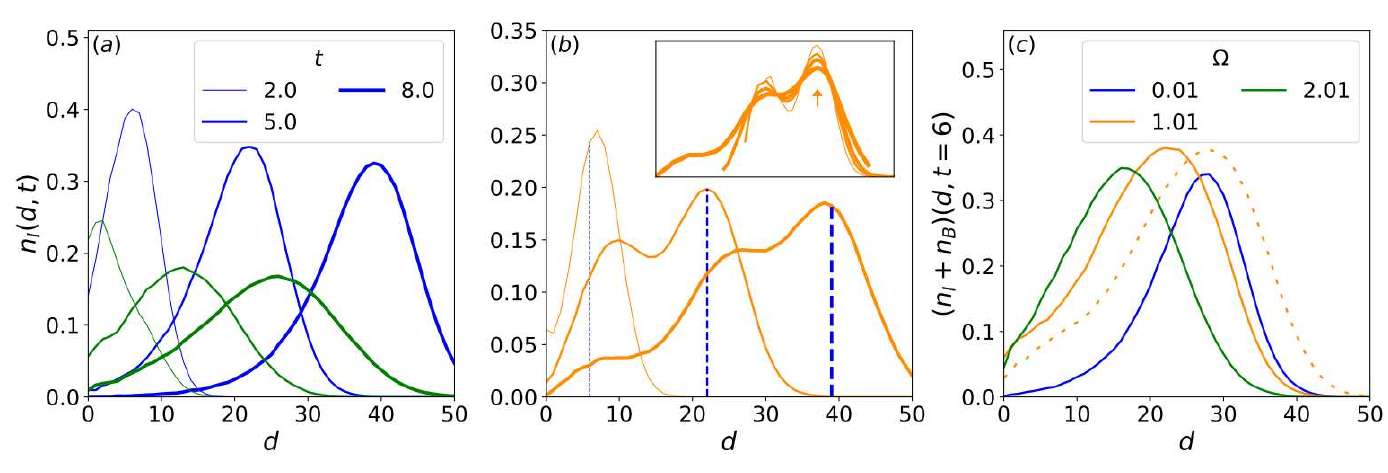}
  \caption{Density profiles against the distance $d$ from the origin, taken at different times and averaged over the same trajectory set used for Fig.~\ref{fig:snaps_act_mid}. (a) Shape of the infection wave for $\Omega = 0.01$ (blue lines) and $2.01$ (green ones) at three different times. (b) Same for $\Omega = 1.01$ (drawn in orange). The vertical, blue, dashed lines show the positions of the maxima for the $\Omega = 0.01$ curves at the corresponding times. In the inset we show $\Omega = 1.01$ profiles for a broader selection of times, shifted horizontally so that they reach their maxima (orange arrow) at the same point. (c) Profiles for the combined infected-bedridden density $n_I + n_B$ at $t = 6$. The dashed, orange line displays the $\Omega = 1.01$ curve later, at $t = 7$.
}
\label{fig:multi_act_mid}
\end{figure*}
%
%
%
%
%
% \begin{figure*}[ht]
%     \includegraphics[width=\textwidth]{Movies_act_high.pdf}
%   \caption{Local infection density on the same lattice deep in the active phase ($\gamma_I = 5$) for the same three values of the frequency $\Omega$ used before. Averages are calculated over a set of $M_{tr} = 650$ trajectories and we have further changed the colormap's rescaling to $c \to c^{3/4}$.
% }
% \label{fig:snaps_act_high}
% \end{figure*}
% %
% %
% %
% %
% %
% \begin{figure*}[ht]   %% 23
%     \includegraphics[width=\textwidth]{Act3_multiplot_ii.pdf}
%   \caption{Profiles of the I and D average densities as a function of the distance from the origin for a selection of times (later times correspond to thicker curves). (a) Restricted average density $n_I(d, t)$ for $\Omega = 1.01$. (b) Same for $\Omega = 2.01$. (c) Restricted DDS $n_D(d,t)$ for $\Omega = 1.01$. The dashed lines show the profiles at $t = t_{max} = 20$ for all the considered values of $\Omega$, with the same color code used before: $0.01$ (blue), $1.01$ (orange) and $2.01$ (green). 
% }
% \label{fig:multi_act_high}
% \end{figure*}

We now take a few steps into the active phase, starting with $\gamma_I = 2$; the corresponding evolution is shown in Fig.~\ref{fig:snaps_act_low}. Once more, we observe a single outbreak progressively covering the entire lattice. Differently from the previous case ($\gamma_I = 1.2$), however, we can spot a qualitative change in behavior at small times: for $\Omega = 1.01$ there are oscillations bringing about a revival of infection in the origin's neighborhood (small white arrows) when the outbreak has already formed its characteristic ring-like structure. These do not seem to cause any significant changes at longer times, though.

In Fig.~\ref{fig:multi_act_low}(a) we plot the evolution of the global average density of infection $n_I(t)$ against time. Following a brief initial transient, all curves grow algebraically (see inset), roughly $\propto t^{1.48}$. Unsurprisingly, the outbreak moves now significantly faster than in the previous, near-critical case, hitting the boundaries around $t \approx 15$ for $\Omega = 0.01$. Because of this, it is difficult to accurately study the propagation; our fitting attempts returned rough estimates $R_{ring} \propto t^{1.07}$ for $\Omega = 0.01$, $R_{ring} \propto t^{0.96}$ for $1.01$ and $t^{1.03}$ for $2.01$, compatible with the outbreak traveling at constant speed in the short window between the initial transient and the moment it reaches the boundaries. We still find that the ring's width $W_{ring}$ increases in time, whereas the wave's peak's height decreases.

To better visualize these qualitative features, we introduce the restricted infection density at distance $d \in \N$ from the origin, defined similarly to \eqref{eq:av_inf_density}, but including only sites $k$ which satisfy $d \leq d(k) < d + 1$; formally, if we call $\Lambda(d)$ the set of all such points, 
\be
    n_I(d, t) = \frac{1}{\abs{\Lambda(d)}} \sum_{k \in \Lambda(d)} \overline{\av{\sigma_k^{II}(t)}},
    \label{eq:restr_av_inf_dens}
\ee
where $\abs{\Lambda(d)}$ denotes here the number of elements in the set $\Lambda(d)$ (its cardinality). Considering that the snapshots in Fig.~\ref{fig:snaps_act_low} show, with good approximation, rotational symmetry, the restricted density \eqref{eq:restr_av_inf_dens} can be taken as a descriptor of the wavefront shape along any radial direction. In other words, it should reliably display the average behavior we would observe on any radial slice of the lattice. 

In Fig.~\ref{fig:multi_act_low}(b) we show $n_I(d,t)$ for the two cases $\Omega = 0.01$ and $2.01$ and observe that, in fact, the wavefront's profile widens (horizontally) and shortens (vertically) as it travels outwards. Note how the three representative instants chosen for the $\Omega = 2.01$ curves are twice the corresponding values for $\Omega = 0.01$. 
%HERE HERE HERE HERE
It is possible to discern by eye that the $\Omega = 2.01$ (green) curves are generally farther from the origin than the $\Omega = 0.01$ ones, showing that in the former case the outbreak travels at more than half the speed than in the GEP. This implies that the separation of timescales introduced earlier does not hold here, or, more precisely, $\Omega = 2.01$ is now not large enough.

In panel (c) we have plotted the restricted average density \eqref{eq:restr_av_inf_dens} for the remaining case $\Omega = 1.01$; note that the vertical range is about half of that of panel (b). At $t = 3$ (thinner curve), we observe the presence of two peaks: one around $d = 6$, which is approximately where the peak of the $\Omega = 0.01$ profile lies at the same time, and one at the origin. The former indicates the radius of the outbreak that has emerged from the initial transient, the latter is the "extra" oscillation we have observed in the second panel, second row of Fig.~\ref{fig:snaps_act_low}. 

Furthermore, by comparing panels (b) and (c) we see that in the former there is a clear tendency of the peak's height to decrease in time. In the latter, instead, notice how the tiny peak around $d = 6$ at $t = 3$ is shorter than the single peak which is formed by $t = 7$. Overall, it seems that the effect of the secondary oscillation is simply to strengthen the initially-produced outbreak, combining with it. Indeed, at later times the profile shows a single peak analogous to those observed at lower and higher frequencies $\Omega$.

We are now in the position to make some simple guesses on the behavior that can be expected at higher values of the infection rate. Consider the following: (A) the speed of $I$-$B$ oscillations, which generate the second peak in Fig.~\ref{fig:multi_act_low}(c), is mostly determined by the frequency $\Omega$. (B) The speed at which an outbreak leaves the origin can be 
augmented by increasing $\gamma_I$, as highlighted by the comparison between the first rows of Fig.~\ref{fig:snaps_crit} and \ref{fig:snaps_act_low}, and as can be intuitively understood since higher infection rates imply a faster local spread of infection.

Let us pretend for simplicity that new peaks are generated at the origin with frequency $\nu_P$, with comparable height $h_P$, width $w_P$ and traveling outwards at some constant speed $v_P$. By (A), $\nu_P = \nu_P (\Omega)$ is an increasing function of the parameter $\Omega$, whereas the speed $v_P = v_P (\gamma_I,\,\Omega)$ only weakly depends on it, so that
\be
    v_P (\gamma_I,\, \Omega \to \infty) = \ha v_P (\gamma_I, \,0),
\ee
and, by (B) is increasing in $\gamma_I$ for any fixed $\Omega$. We also ignore the time dependence of all these peak properties, focusing on the initial part of the dynamics. Then, we can say that the distance between two consecutive peaks is $d_P = v_P / \nu_P$. Generally, if $d_P \lesssim w_P$ it will be impossible to distinguish individual peaks and they will combine to form a unique, larger one. Conversely, if $d_P \gtrsim w_P$ we should be able to observe (some) distinct wavefronts leaving the origin one after the other. Note now that by increasing $\Omega$ we are both increasing $\nu_P$ and (weakly) decreasing $v_P$, making the distance $d_P$ smaller. Thus, we can expect there to be a (soft) threshold $\Omega_{max}$ beyond which the dynamics of the eQEP will be qualitatively similar to that of the GEP, with a single observable outbreak forming. At the same time, if $\Omega \ll \gamma_I, \gamma_D$ the stochastic dynamics would dominate over the oscillations and we should also expect a behavior very similar to the GEP's, up to tiny corrections. Thus, the only hope to observe something different is to choose a frequency $\Omega$ that is of the same order of $\gamma_I$, but at the same time is not too large. This way, every new peak will have enough time to travel away from the origin before another one is formed and follows in its tracks. If this intuition is correct, we should then be able to observe separate outbreaks as in Ref.~\cite{Espigares2017}.

%Still, this hints at the behavior we can expect at higher infection rate: in fact, while the speed at which a second peak can form at the origin is mostly controlled by the oscillation frequency $\Omega$, the speed at which it leaves the origin is instead mainly determined by how larger $\gamma_I$ is than its critical value. Hence, as we increase $\gamma_I$ further and further we will make it more difficult for any secondary peaks to be reabsorbed by their predecessors and we can reasonably expect them to give rise to separate outbreaks.

We start testing this simple hypothesis in Fig.~\ref{fig:snaps_act_mid}, where we plot the evolution of $\overline{\av{\sigma_k^{II} (t)}}$ for the usual $101 \times 101$ lattice and initial condition. Indeed, for the intermediate value $\Omega = 1.01$ we can distinguish two outbreaks, the first already formed around $t = 2$ and the second forming around $t \approx 3$. They can still be made out, albeit not as clearly, at $t = 9$, when the first one is at the lattice edges. 

In Fig.~\ref{fig:multi_act_mid} we again use the quantity $n_I \lt d,\, t\rt$ to compare the profile of the infection wavefront in the radial direction for (a) $\Omega = 0.01$ and $2.01$ and (b) $\Omega = 1.01$. In the former two cases we observe a single peak traveling outwards, whereas in the latter, after the emergence of the first peak, a second one develops and follows in its wake. In order to roughly compare the propagation speeds in the $\Omega = 1.01$ case and the "quasi-GEP" case $\Omega = 0.01$ we have added in panel (b) some vertical dashed lines indicating where the GEP peaks would be at each time (again, displayed at increasing level of thickness according to the legend in panel (a)). We can thus see that the first $\Omega = 1.01$ outbreak closely follows the GEP single one, albeit with a reduced amplitude (note that the vertical range of panel (b) is smaller than panel (a)'s).

Furthermore, the second peak appears to propagate outwards at the same speed of the first one; to better highlight this, in the inset of panel (b) we have taken the profiles of $n_I(d,\,t)$ at four different times ($t = 3$, $4$, $5$ and $8$, displayed at increasing line widths) and shifted each horizontally in such a way that their global maxima (the apices of the first peaks), indicated by a vertical arrow) are reached at the same point. For those who prefer a more mathematical approach, say we have a family of curves $f_i(x)$, indexed by $i$, that feature their global maxima at points $x_i$; then, we would be plotting $f_i (x + x_i)$ instead of $f_i(x)$, so that at $x = 0$ every shifted curve reaches its maximum.  

%it would appear that the second outbreak travels at the same speed of the first one: in the inset of panel (b) we have taken the profiles at four different times and shifted them horizontally so that each profile reaches its maximum at the same point (indicated by an orange arrow). In more mathematical terms, for a curve $f(x)$ that features its global maximum at $x_{max}$, we would be plotting $f(x - x_{max})$ instead of $f(x)$. 
The superimposed profiles in the inset show that, with good approximation, the position of the second peaks also coincide, which implies that the two peaks remain, at least roughly, at the same distance during the course of the evolution, up to the point when the first one hits the boundaries. This is in contrast to the (mean-field) physics of the RQEP, in which subsequent outbreaks behave as if they were produced under a reduced infection parameter, meaning that each would be slower than its predecessor.

In Fig.~\ref{fig:multi_act_mid}(c) we plot the combined $I$-$B$ density $n_I + n_B$, again as a function of the distance $d$ from the origin, for a single selected time $t = 6$; as one would naively expect, the "missing" part of the infection profile between different peaks for $\Omega = 1.01$, as well as the "missing" amplitude of the $\Omega = 2.01$ outbreaks, is made up of sites with non-negligible $B$ components. Adding their contributions, the single-peak structure is indeed recovered, with an amplitude comparable to the GEP (blue) case. 

At the same time, we see that the process at $\Omega = 1.01$ cannot be seen as a mere GEP with a "hidden portion" of infected, or at least this is not as straightforward as it was in our previous timescale separation argument: the $\Omega = 1.01$ profile is clearly broader and taller than the $\Omega = 0.01$ one. 
%Even for $\Omega = 2.01$ the peak is about as tall as it is for the GEP, showing, inter alia, that the rescaling \eqref{eq:II_rescaling} cannot work here. 
We cannot even reduce this to a mere retardation in the dynamics: at $t = 6$, the blue curve in Fig.~\ref{fig:multi_act_mid}(c) features its maximum at $d = 28$. When the orange curve's peak arrives at the same distance, around $t = 7$, its profile (dashed, orange line) is still both broader and taller than the blue curve; thus, simple time rescaling cannot bridge entirely between the different behaviors observed for $\Omega = 0.01$ and $\Omega = 1.01$. 

\begin{figure*}[ht]
    \includegraphics[width=\textwidth]{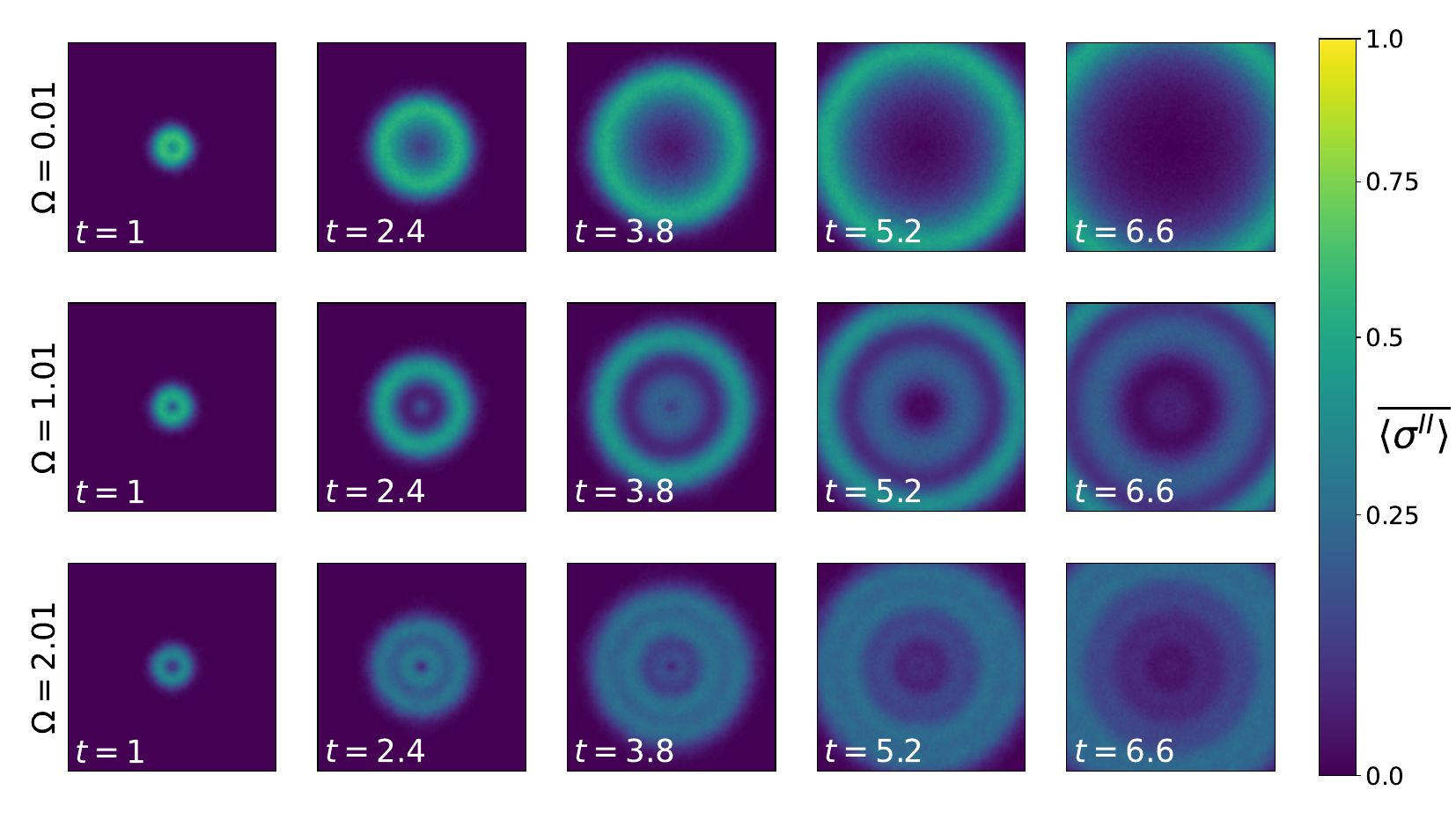}
  \caption{Local infection density on the same lattice deep in the active phase ($\gamma_I = 5$) for the same three values of the frequency $\Omega$ used before, listed at the very left. Averages are calculated over a set of $M_{tr} = 650$ trajectories and we have further changed the colormap's rescaling to $c \to c^{3/4}$.
}
\label{fig:snaps_act_high}
\end{figure*}
\begin{figure*}[ht]   %% 23
    \includegraphics[width=\textwidth]{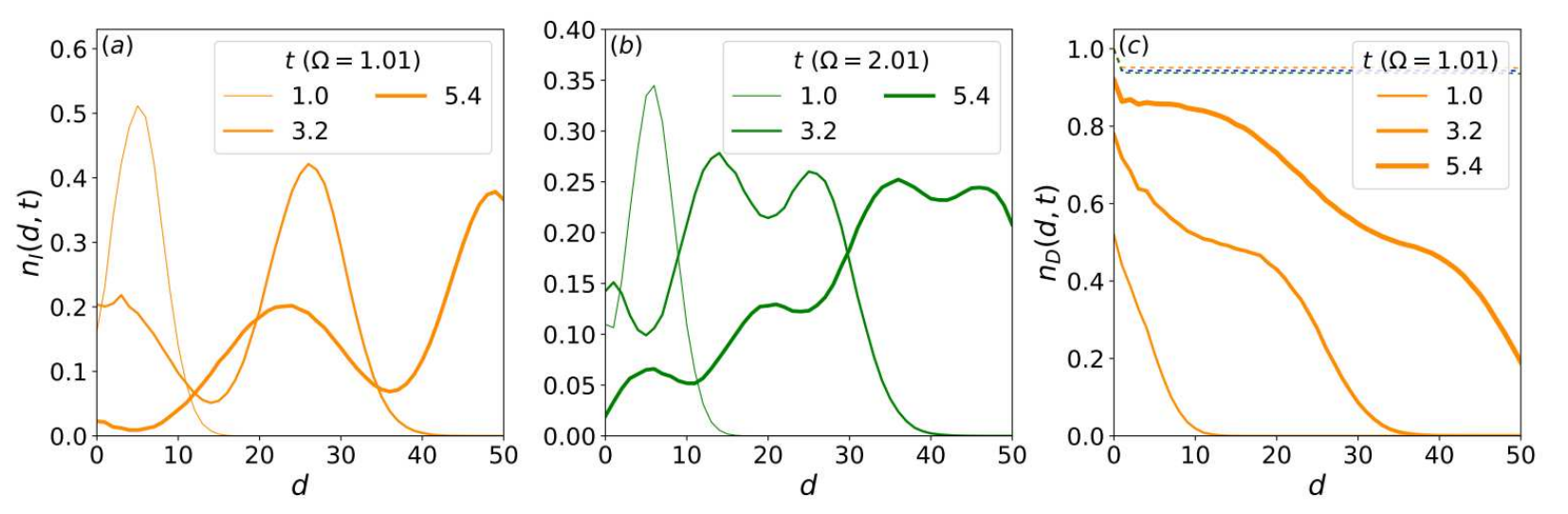}
  \caption{Profiles of the I and D average densities as a function of the distance from the origin for a selection of times (later times corresponding to thicker lines). (a) Restricted average density $n_I(d, t)$ for $\Omega = 1.01$. (b) Same for $\Omega = 2.01$. (c) Restricted DDS $n_D(d,t)$ for $\Omega = 1.01$. The dashed lines show the profiles at $t = t_{max} = 20$ for all the considered values of $\Omega$, with the same color code used before: $0.01$ (blue), $1.01$ (orange) and $2.01$ (green). 
}
\label{fig:multi_act_high}
\end{figure*}
In Fig.~\ref{fig:snaps_act_high} we plot the evolution of the local infection density for $\gamma_I = 5$, quite deep in the active phase of the model. Matching our intuition, the fast propagation of the infection allows the oscillations induced by the Hamiltonian term to produce several outbreaks. For $\Omega = 1.01$ we can distinguish, for instance, three different ones, the last being spawned at the origin right after the first one has hit the boundaries (see the rightmost panel at $t = 6.6$). As $\Omega$ is further increased more outbreaks appear, but, at the same time, they get squeezed closer together (see last row).

To highlight these features more clearly, in Fig.~\ref{fig:multi_act_high} we display some radial profiles of the infected density $n_I(d, t)$ and the corresponding DDS $n_D(d, t)$, the latter being defined analogously to Eq.~\eqref{eq:restr_av_inf_dens}:
\be
    n_D(d, t) = \frac{1}{\abs{\Lambda(d)}} \sum_{k \in \Lambda(d)} \overline{\av{\sigma_k^{DD}(t)}}.
    \label{eq:restr_av_death_dens}
\ee
Comparing panels (a) and (b) we can see that, as expected, oscillations are more frequent at $\Omega = 2.01$ than they are at $\Omega = 1.01$. Note that the speed of outward propagation is roughly the same for all three $\Omega$ values, as shown by Fig.~\ref{fig:snaps_act_high}; therefore, oscillations display both a smaller wavelength (in $d$) and a higher frequency (in $t$) as $\Omega$ is increased.

Again, the resolution of the individual peaks worsens as time passes, as each peak progressively broadens and shortens. This is particularly visible in panel (b), since the fast oscillations do not give enough time to each outbreak to move away before the next peak forms, resulting in a significant overlap from the very beginning. Panel (a) makes it also easier to observe the formation of a third, tiny wave when the first one reaches the edges of the lattice. 
%Again, as far as we can tell on a $101 \times 101$ lattice, subsequent outbreaks travel outwards at the same speed.

In Fig.~\ref{fig:multi_act_high}(c) we plot $n_D(d,t)$ for the $\Omega = 1.01$ case, showing the emergence of a "ladder-like" structure where each wavefront of panel (a) drives an increase of the local DDS wherever it passes, although it is far from being as clear-cut as the mean-field behavior of the RQEP, see Fig.~\ref{fig:old_one}(b). At long times the various outbreaks end up "depositing" a roughly uniform DDS (dashed lines in Fig.~\ref{fig:multi_act_high}(c)). Like for the global quantities of Sec.~\ref{subsec:stat}, stationary profiles obtained for different values of $\Omega$ are, with good approximation, the same.

% %
% %
% %
% \begin{figure*}[ht]
%     \includegraphics[width=\textwidth]{IDtimes_multiplot.pdf}
%   \caption{Qualitative distribution of infection ($t_I$) and death times ($t_D$) at distance $d = 26$, i.e., approximately halfway between the origin and lattice boundaries. Each column correspond to a given value of the infection rate $\gamma_I$, each row to a given frequency $\Omega$. Brighter colors correspond to higher counts; colormaps are not provided as they have been rescaled by different amounts in different panels to better highlight the structure in each.
% }
% \label{fig:IDtimes}
% \end{figure*}
% %
% %
% %

To summarize what we have learned from studying the eQEP dynamics thus far, let us extract a few key points:
\begin{itemize}
    \item[(i)] Our simulations have highlighted no qualitative change in the eQEP dynamics in the absorbing phase, except for a progressive slowdown as $\Omega$ is increased.
    \item[(ii)] If the frequency $\Omega$ is very small, the eQEP behaves essentially like a GEP with rates $\gamma_I$ and $\gamma_D$. 
    \item[(iii)] If $\Omega$ is instead sufficiently large, one can see the eQEP dynamics as an effective GEP with rates $\gamma_I / 2$ and $\gamma_D / 2$ and half the average density of infected sites at any given time.
    \item[(iv)] For intermediate values of $\Omega$ the eQEP features multiple outbreaks. 
    \item[(v)] Subsequent outbreaks move outwards from the origin at the same speed; the corresponding peaks become shorter and broader with time.
\end{itemize}
Point (v) above suggests that, for any given choice of parameters that does produce multiple outbreaks, one could find a sufficiently large lattice in which, in time, the peaks would overlap to the point of becoming indistinguishable from each other, merging into a very broad wave profile. If this were the case the emergence of multiple outbreaks would represent merely a short-time (or finite-size) feature of the eQEP.

In most cases, later peaks visibly appear less prominent than earlier ones (the one exception being the second peak in Fig.~\ref{fig:multi_act_high}(b)); this means that, at some point, an outbreak may start so small to be statistically indistinguishable from stochastic fluctuations upon the tails left behind by its predecessors. We might thus be unable to resolve those peaks unless we average over a much higher number of trajectories. Any estimate of the number of peaks, therefore, may be sensitive to changes in $M_{tr}$. In fact, we cannot even exclude that, for some parameter choices, the eQEP might produce an infinite succession of outbreaks, each suppressed with respect to its predecessor with their amplitudes, after a certain initial number of larger wavefronts, simply dipping underneath the typical size of fluctuations and getting thus concealed by the latter. 

Yet, in Figs.~\ref{fig:snaps_abs}-\ref{fig:snaps_act_high} the number of produced outbreaks is one of the most prominent qualitative features distinguishing the classical dynamics of the GEP ($\Omega \approx 0$) from that of the more general eQEP; hence, it seems to constitute an obvious choice for attempting to draw a dynamical phase diagram for the model. 

For this purpose, while running the trajectory sets displayed in Fig.~\ref{fig:cells_and_trajs} (those used for the stationary results of Sec.~\ref{subsec:stat}) we have collected a set of infection and death times according to the following procedure: we have divided our lattice in radial "slices", with the $d$-th one including all sites in $\Lambda (d)$, i.e., all the sites at distance $\geq d$ and $< d+1$ from the origin. For every $(\gamma_I, \, \Omega)$ point (see Fig.~\ref{fig:cells_and_trajs}) in parameter space and for every distance $d \in \set{0, 1, 2, \ldots, 50 }$ we have defined a "container" $\mathcal{T}_{ID}(d)$. If, in a trajectory $\tau$, a site $k \in \Lambda(d)$ was infected at time $t_I$ and died at time $t_D$ we recorded the pair $(t_I, t_D)$ in $\mathcal{T}_{ID}(d)$.

\begin{figure*}[ht]
    \includegraphics[width=\textwidth]{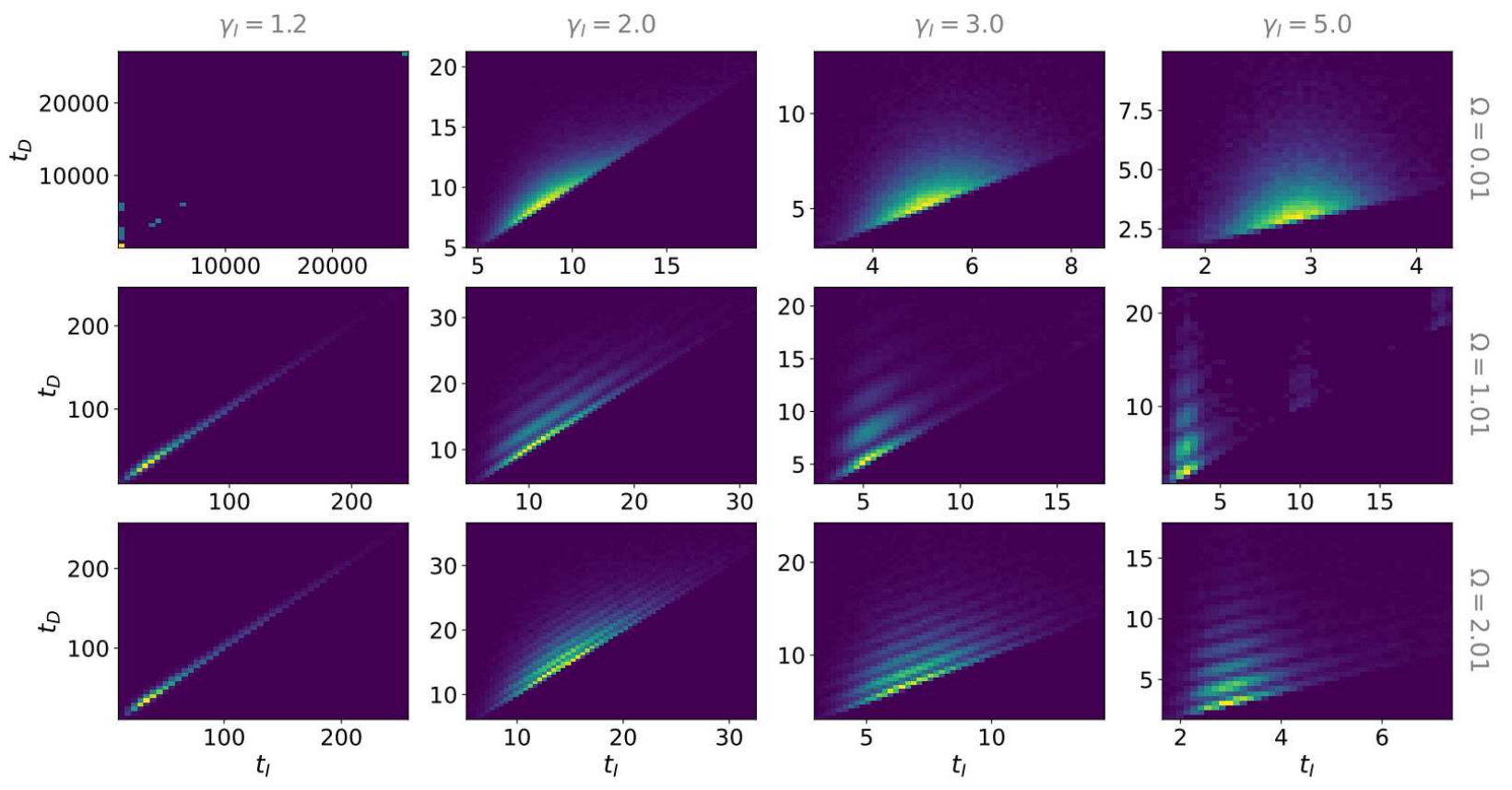}
  \caption{Qualitative distribution of infection ($t_I$) and death times ($t_D$) at distance $d = 26$, i.e., approximately halfway between the origin and lattice boundaries. Each column correspond to a given value of the infection rate $\gamma_I$, each row to a fixed frequency $\Omega$. Brighter colors correspond to higher counts; colormaps are not provided as they have been rescaled by different amounts in different panels to better highlight the structure in each.
}
\label{fig:IDtimes}
\end{figure*}

This allowed us to reconstruct, to an extent, the distribution of infection and death times. An example is provided in Fig.~\ref{fig:IDtimes}, where this distribution is qualitatively represented as a density plot (in this particular case, it may also be convenient to think of it as a two-dimensional histogram seen from above), for a fixed distance $d = 26$ about halfway between the origin and the edges; each row is associated to one of the three fixed values of the frequency $\Omega$: $0.01$ (first row), $1.01$ (second row) and $2.01$ (third row); each column to one of the infection rates used for the previous Figures, starting from the near critical $\gamma_I = 1.2$ (first column) and proceeding into the active phase: $2$ (second column), $3$ (third column) and $5$ (fourth and last column).

Each plot is subdivided in cells $\left[ t_I,\, t_I + \delta t_I \right) \times \lqq t_D,\, t_D + \delta t_D \rt$, with horizontal (vertical) bin width $\delta t_I$ ($\delta t_D$). These widths vary widely between different plots, as they have been automatically chosen by our algorithm in order to highlight the oscillations, if present, in the distribution. Each cell is brighter in proportion to the number of infection/death times recorded within its bounds; the colormap ranges and cutoffs have also been automatically fixed by our algorithm. Since we are only interested in the qualitative shape of these diagrams, we are not reporting here all these sundry values, and focus instead on the general properties we can recognize by eye.

As a consistency check, we remark that in every panel a portion at the bottom-right is empty, since a site \emph{cannot} die before being infected, and so we always have $t_D > t_I$; to avoid confusion note that, for the same reason, the vertical ($t_D$) axis typically stretches further than the horizontal ($t_I$) one and thus the line $t_I = t_D$ does not coincide with the panel diagonal.

%latter is qualitatively reconstructed as a two-dimensional histogram at distance $d = 26$ for three values of $\Omega$ ($0.01$ for the first row, $1.01$ for the second, $2.01$ for the third) and the same four values of $\gamma_I$ used for the previous Figures, from the near-critical $\gamma_I = 1.2$ (first column) to the last one $\gamma_I = 5$ (last column), positioned deep inside the active phase. In all panels, the bottom-right corner is empty as it should be, since a site \emph{cannot} die before being infected, and so we always have $t_D > t_I$; note that, for the same reason, the vertical ($t_D$) axis typically stretches further than the horizontal ($t_I$) one and thus the line $t_I = t_D$ does not coincide with the panel diagonal.

Second, we notice that the top-left panel, corresponding to parameters $\gamma_I = 1.2$ and $\Omega = 0.01$, is somewhat of an outlier, with its axes stretching up to the tens of thousands. It may be difficult to notice without magnification, but there is a bright spot at the very upper-right corner. This is a trace of the tiny deviations from the GEP induced by the very small, but finite, value of $\Omega$ in the eQEP. We propose the following interpretation: in a small portion of the trajectories the infection reached $d = 26$ extremely late, yielding $t_I \sim 3\times 10^4$; sites infected this late then died on much shorter, $O(1)$, timescales, as dictated by $\gamma_D = 1$. 

Let us now outline a perturbative argument (based on the fact that $\Omega \ll \gamma_I,\, \gamma_D$): 
for $\Omega = 0$, $\ket{I}$ and $\ket{B}$ are both eigenvectors of the effective Hamiltonian $H_{eff}$, each evolving on timescales dictated by the respective eigenvalue $\lambda_I = -i\gamma_{eff}/2$, $\lambda_B = 0$, where $\gamma_{eff}$ is the effective rate introduced in Eq.~\eqref{eq:gamma_eff}, whose definition we repeat below for the reader's convenience:
\be
    \gamma_{eff} = \gamma_D + \mal{N}_k^S \gamma_I \leq \gamma_D + 4 \gamma_I,
    % \label{eq:gamma_eff}
\ee
(remember that on a square lattice an infected site can have at most $4$ susceptible neighbors). With the introduction of a small frequency $\Omega$, the eigenvectors can be seen as slight perturbations of the infected and bedridden states:
\begin{subequations}
\begin{align}
    &\ket{I} \to \ket{\widetilde{I}} \approx \ket{I} + O\lt \frac{\Omega}{\gamma_{eff}} \rt  \ket{B} + \ldots, \\[0.2cm]
    &\ket{B} \to \ket{\widetilde{B}} \approx \ket{B} + O\lt \frac{\Omega}{\gamma_{eff}} \rt \ket{I} + \ldots, \label{subeq:perturb_B}
\end{align}
\end{subequations}
up to a higher-order normalization factor, and the same goes for the eigenvalues:
\begin{subequations}
\begin{align}
    &\lambda_{\widetilde{I}} = \lambda_I + O\lt \lt \frac{\Omega}{\gamma_{eff}} \rt^2 \rt, \\[0.2cm]
    &\lambda_{\widetilde{B}} = O\lt \lt \frac{\Omega}{\gamma_{eff}} \rt^2 \rt .
\end{align}
\end{subequations}
Consider now the initial condition at the origin: $\ket{I}$. By Eq.~\eqref{subeq:perturb_B}, the probability of finding the origin in state $\ket{\widetilde{B}}$ is $\abs{\bra{\widetilde{B}}I \Big\rangle}^2 = O \lt \lt \Omega / \gamma_{eff}  \rt^2  \rt$. Hence, in about an equivalent portion of the trajectories we may expect the evolution of the initial condition to take place on a timescale $1/\lambda_{\widetilde{B}} = O \lt \lt \gamma_{eff} / \Omega \rt^2  \rt$. With both rates $\gamma_I$ and $\gamma_D$ being $O(1)$, having $\Omega = 0.01$ implies that about one in $10^4$ trajectories will see its first jump at $t \sim 10^4$, which indeed matches what we observe. These "late" trajectories become rarer and rarer as $\gamma_I$ (and thus $\gamma_{eff}$) is further increased and, unsurprisingly, none has been captured in the remaining panels of the first row. We recall, after all, that the number of trajectories (see Fig.~\ref{fig:cells_and_trajs}) is $\approx 10^3$ for the first two columns and even lower for the third and fourth ones.

A similar argument can, in our opinion, explain the spots found at the very left in the plot, above the bottom-left corner: these are, we believe, due to trajectories where some sites at $d = 26$ get infected on short timescales ($\sim 26/\lambda_{\widetilde{I}}$), but have then, precisely as stated above for the origin, a probability of $O(10^{-4})$ to last for a time $O(10^4)$ before dying. Similar traces should be expected above the main diagonal of the plot, but would be suppressed by another factor $\sim 10^4$ and thus would require a proportionally larger sample of trajectories, which goes well beyond the scope of this study.

By far the most populated cell in this plot is, unsurprisingly, the bottom-left one, corresponding to $t_I,\, t_D = O(1-10)$ and recording the GEP-like trajectories. As we have discussed previously, the $\Omega = 0.01$ case can be considered almost equivalent to the GEP (and we used it on several occasions in this fashion), up to tiny corrections. Here we have encountered an example of such corrections.

Beyond the top-left plot, we can spot some additional traces of this perturbative structure in the last panel of the middle row, corresponding to $\gamma_I = 5$ and $\Omega = 1.01$. In it, albeit with some difficulty, one can see three distinct non-trivial infection intervals, the main (and most visible) one concentrated around $t_I \approx 2.5$, the second, much dimmer, one around $t_I \approx 10$ and the third, almost invisible without magnification, around $t_I \approx 20$.  

Generally, if we exclude the leftmost column, we observe oscillations emerging in the second and third row, absent in the first one. These take a form that we could describe as vertically-stacked, diagonal stripes. The absence of these structures in the $\gamma_I = 1.2$ column seems to suggest that oscillations can only be observed in the active phase, being absent in the critical region. Furthermore, by moving from top to bottom, i.e.~increasing $\Omega$, we notice the "stripes" being squeezed closer together.

The vertical stacking can be probably interpreted in the following way: once a site, in this case at distance $d = 26$ from the origin, has been infected, it starts oscillating between the I and B states and is most likely to jump to its death when, periodically, it gets a large component over $\ket{I}$.

% These oscillations are particularly pronounced moving in the vertical ($t_D$) direction; more precisely, we see the formation of approximately parallel, vertically-stacked "stripes". These stripes only seem to form once we move into the active phase and they tend to be squeezed closer together when $\Omega$ becomes sufficiently large. These stripes have a fairly intuitive interpretation: once a site (at this distance) has been infected, it starts oscillating between the I and B states and is most likely to jump to its death when, periodically, it gets a large component over $\ket{I}$. 

From the second column ($\gamma_I = 2$) we can glean how our "separation of timescales" argument takes shape for this distribution, i.e., how the GEP behavior is recovered as $\Omega$ starts dominating over the dissipative rates. If we compare the top and bottom panels, we see that their qualitative features are quite similar; in the latter, stripes are getting so packed that it becomes difficult to distinguish them from each other. Furthermore, note how the range of both the horizontal and vertical axes in the bottom panel approaches roughly twice the range of the top panel, displaying a dynamical slowdown by a factor of about $1/2$.
% The comparison between the top and bottom panels of the second column ($\gamma_I = 2$) shows how the GEP-like behavior is recovered when $\Omega$ dominates over the dissipative rates: the stripes get packed closer and closer together until they become undistinguishable from each other. Note how the range of both the horizontal and vertical axes in the bottom panel approaches roughly twice the range of the top panel, corresponding to the expected slowing down by approximately $1/2$. Additionally, were it not for the presence of alternating brighter and darker diagonal stripes, the qualitative structure (colors, proportions, shape) of the bottom panel would roughly match that of the top panel.

% %
% %
% \begin{figure}[h]
%   % \vspace{0cm}
%   \includegraphics[width=\columnwidth]{Npeaks_map.pdf}
%   \caption{Approximate count of the number of peaks in the distribution of death times $t_D$ at distance $d = 26$ from the origin, as a function of the parameters $\gamma_I$ and $\Omega$. The strip $\gamma_I \in \lqq 0,\, 1 \rqq$ has been expunged since in the absorbing phase there are no outbreaks; thereby, infection typically does not reach $d = 26$ and there are no (or very few) deaths to record. Despite the significant noisiness of the plot, we can see the emergence of a "wedge-like" structure, highlighted by the orange and red dashed lines: as one moves diagonally from the bottom-left corner the number of peaks seems to increase with some consistency. 
%   } 
% \label{fig:npeaks}
% \end{figure}
% %
% %
We recall that our current aim is to construct some approximate "dynamical phase diagram" by estimating the number of outbreaks observed for each of the parameter points that were used to compose the stationary phase diagram in Fig.~\ref{fig:PD}. To do so, we have employed the distribution introduced above, i.e.~the joint distribution of infection and death times at distance $d$; for brevity, let us express it as
\be
    p_{ID} \lt t_I, \, t_D\, | \, d   \rt;
\ee
with this notation, the quantity visualized in Fig.~\ref{fig:IDtimes} would read $p_{ID} \lt t_I,\, t_D \,|\, 26 \rt$.

Let us now consider a single outbreak in the active phase and in an idealized scenario without stochastic noise; we say that an outbreak reaches a radius $d$ away from the origin when the restricted average density of infection $n_I (d, \,t)$, defined in Eq.~\eqref{eq:restr_av_inf_dens}, features a local maximum in time. If we call $t(d)$ one such time, this is tantamount to asking that a wave profile like the ones depicted in Fig.~\ref{fig:multi_act_high}(a, b) at time $t(d)$ features a peak at distance $d$. Slightly before $t(d)$, the peak will be to the left of $d$; slightly after, to its right.

According to this definition, the average number of infected sites in $\Lambda(d)$ grows up to $t(d)$ and then starts decreasing. It is thus around $t(d)$ that we can reasonably expect there to be the more numerous opportunities for death to occur. Provided this intuitive picture is mostly correct, we can therefore expect the death times $t_D$ at distance $d$ to bunch around these "peak times" $t(d)$. Of course, as revealed by Fig.~\ref{fig:multi_act_mid}(c), not sites will immediately die, but some will instead rotate into the bedridden state to die at a later time, namely when a later outbreak passes at $d$.

If we now consider the marginal distribution
\be
    p_D \lt t_D \,|\, d \rt = \int \rmd t_I\,\, p_{ID} \lt t_I, \, t_D\, | \, d   \rt
\ee 
of death times at distance $d$, we derive from the argument above that it should present peaks close to the various times $t(d)$ when an outbreak "ring" reaches radius $d$. Hence, there should be as many peaks in $p_D$ as there are outbreaks in the course of the dynamics. We have therefore extracted $p_D$, up to a normalization, by summing the rows (see Fig.~\ref{fig:IDtimes}) of the previously reconstructed two-dimensional $p_{ID}$\footnote{Some readers may be induced by the diagonal-stripe structure highlighted in Fig.~\ref{fig:IDtimes} to think that a better quantity to study would be the marginal probability of time differences $p_{\Delta t} \lt t_D - t_I \,|\,d \rt$. It is true that for $p_{\Delta t}$ each stripe would contribute to a peak, reinforcing it and making the blurring due to stochastic noise less significant. However, the peaks of $p_{\Delta t}$ describe something different from what we are looking for: they tell us how likely it is for any site (at distance $d$ from the origin) to die some time after its original infection. We are instead interested in outbreaks, which are macroscopic phenomena (waves traveling through the system).}.

\begin{figure}[h]
  % \vspace{0cm}
  \includegraphics[width=\columnwidth]{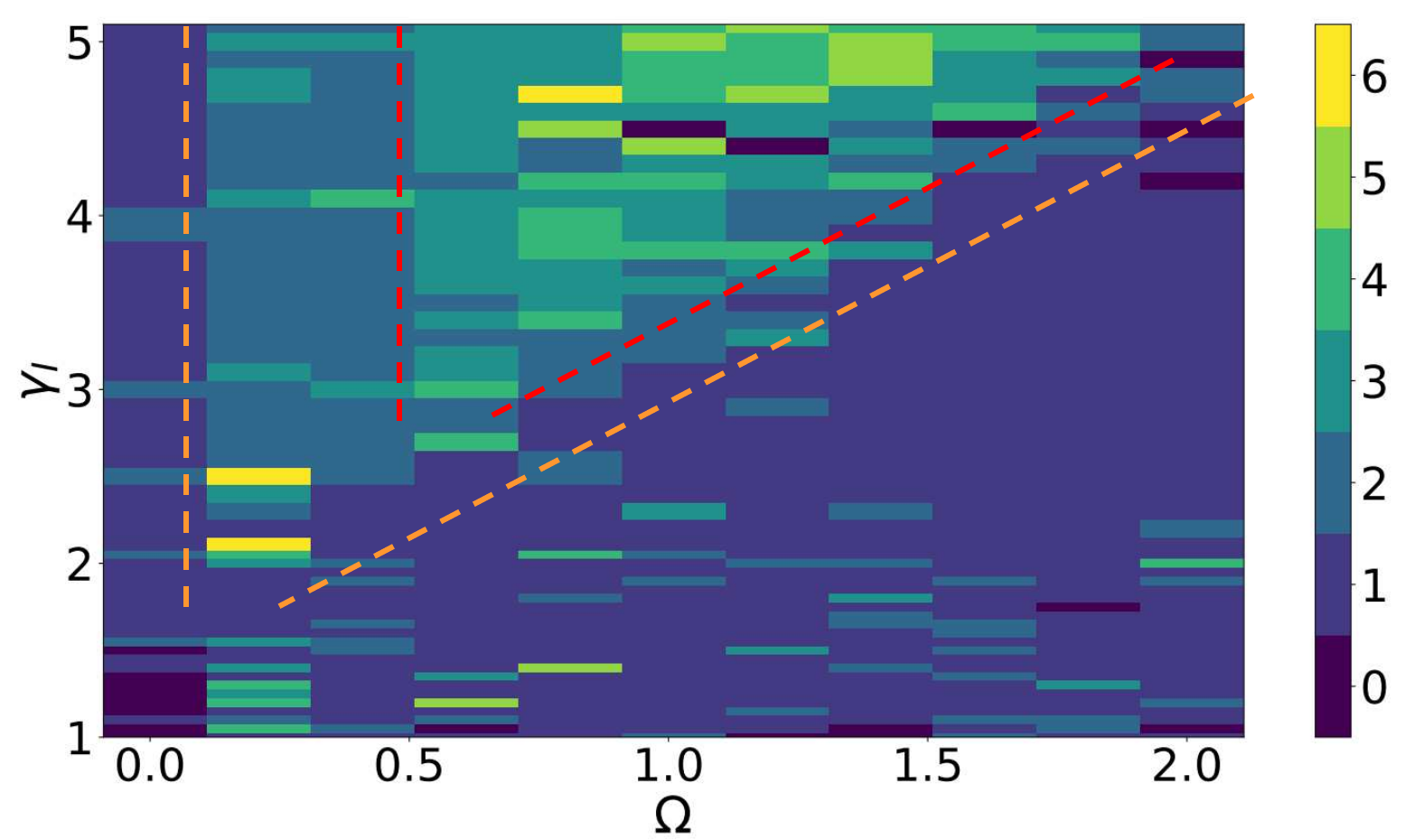}
  \caption{Approximate count of the number of peaks in the distribution $p_D$ of death times $t_D$ at distance $d = 26$ from the origin, as a function of the parameters $\gamma_I$ and $\Omega$. The strip $\gamma_I \in \lqq 0,\, 1 \rqq$ has been expunged since in the absorbing phase there are no outbreaks; thereby, infection typically does not reach $d = 26$ and there are no (or very few) deaths to record. Despite the significant noisiness of the plot, we can see the emergence of a "wedge-like" structure, highlighted by the orange and red dashed lines: as one moves diagonally from the bottom-left corner the number of peaks seems to increase with some consistency. 
  } 
\label{fig:npeaks}
\end{figure}

In Fig.~\ref{fig:npeaks} we show the results of a numerical counting of the peaks of $p_D \lt t_D \,|\, d = 26 \rt$. We have again chosen a distance about halfway between the origin and the boundaries in order to give time to outbreaks to fully form their ring-shaped structure and to avoid being affected by finite-size effects. 

Clearly, the counting operation is highly sensitive to noise, since stochastic fluctuations can generate a plethora of (physically meaningless) local maxima. We have thus applied a running average to smooth out the curves. The width of the averaging interval has been fixed by applying a discrete Fourier transform and looking for a clear separation of scales between low frequency modes (assumed to be the relevant oscillations) and high frequency ones (assumed to arise from stochastic noise). In some cases, mostly at larger values of $\gamma_I$, the algorithm has been unable to identify one such scale; those cases have been manually set to $0$ (black) in the plot and should be ignored. The whole area at $\gamma_I < 1$ is not displayed as it lies within the absorbing phase, where there are no (or very few) trajectories in which the infection manages to arrive at $d = 26$.

In spite of the heavy influence of noise, Fig.~\ref{fig:npeaks} does show a few interesting features: first, the leftmost vertical stripe at $\Omega = 0.01$ shows, in some cases, two peaks. This is a counterpart of the effect observed in the top-left panel of Fig.~\ref{fig:IDtimes}: for some parameter choices, a sufficient number of trajectories survived until very late times ($\sim \gamma_{eff}^2 / \Omega^2$) to create a statistically significant second peak in the distribution. We remark here, too, that the zeroes visible in the lower-left corner are not of the spurious type mentioned above, but are instead due to the very low number of trajectories in which infection continued up to distance $d = 26$. 

Second, the strip $\gamma_I \in \lqq  1, \, 2 \rqq$ seems to be particularly noisy. This is probably due to the fact that, although  more trajectories have been taken here than at higher $\gamma_I$s, see Fig.~\ref{fig:cells_and_trajs}, a smaller proportion thereof survives, i.e., not too deep in the active phase there is still a significant probability of infection stopping in the neighborhood of the origin. 

Third, for $\gamma_I > 2$ a wedge-like structure, which we tried to highlight by superposing  orange and red dashed lines, can be seen emerging, with higher peak counts encountered when moving diagonally, i.e., by increasing both $\gamma_I$ and $\Omega$. This too can be associated to features we have previously discussed: at small enough $\Omega$ we expect a GEP-like behavior with a single peak. As $\Omega$ is increased, we have seen the appearance of multiple wavefronts of infection and the peak count increases; however, at sufficiently large $\Omega$ a GEP-like behavior is recovered (although slowed down by a factor $1/2$) and the count regresses back to $1$. The larger $\gamma_I$ is, the larger this threshold value of $\Omega$ shall become.

Lastly, we mention a property that the eQEP shares with most classical epidemic models with a single-central-seed initial condition: as long as one takes a portion of the system far from the boundaries, there are no substantial finite-size effects. In the eQEP, in fact, there are no traces of any kind of "reflection" of the outbreaks at the edges of the lattice, nor is there any mechanism by which such a reflection could be achieved. Boundary effects can therefore be expected to affect the lattice up to some distance away from the edges themselves, a distance of the same order of the macroscopic correlation length $\xi$ of the system, which, away from the critical region of the phase diagram, remains typically quite small. Thus, farther away from the boundaries no appreciable differences would be observed when increasing the lattice size $N$. In other words, as long as one studies local properties far enough from the edges, the simulations can be virtually considered to be run in the thermodynamic limit $N \to \infty$.

On a more physically intuitive basis, think of being an observer of a GEP with an applied "black" filter on all sites beyond a distance $d = 26$ from the origin, so that nothing can be known about them. Whether the actual lattice is $101 \times 101$, as in our case, or $1000 \times 1000$, or $10^6\times 10^6$, the observer would always see the same behavior, up to extremely small corrections. The observer would thus not be able to tell what lattice size the model were being simulated on.

Since $d = 26$ is in our case quite far from the nearest edge at $d = 50$, we can safely suppose that, outside the near-critical region $1 \lesssim \gamma_I\lesssim 2$, the dynamical phase diagram in Fig.~\ref{fig:npeaks} is largely unaffected by finite-size effects and the noise is instead entirely stochastic in nature, due to a low number of trajectories $M_{tr}$. In the light of this and the other considerations above, we think it reasonable to hypothesize that the qualitative wedge-like features in the plot do represent a physical feature of the eQEP, and would survive, at least in some form, in the limit $M_{tr} \to \infty$.

%Of course, the evolution timescales of sites at the very edges are different than those in the bulk: remember, for instance, that the effective rate $\gamma_{eff}$ from Eq.~\eqref{eq:gamma_eff} depends on the number $\mal{N}_k^S$, which in turn is bounded by the number $\mal{N}_k$ of nearest neighbors, which is different at the boundaries.    

%In the light of these considerations, we expect the qualitative structure of this dynamical phase diagram to survive at much larger trajectory numbers $M_{tr}$ (much lower levels of stochastic noise) and to therefore represent a physical feature of the eQEP. 

All the results discussed in this Section suggest a simple interpretation of the eQEP's evolution in the active phase: if we think in terms of individual trajectories, the single most important consequence of the introduction of the Hamiltonian $H$ seems to be a delayed start of the infection's propagation. 

We have seen an example of this for $\Omega = 0.01$, when some trajectories are stuck for extremely long times $t \sim 10^4$. 
For larger values of $\Omega$, enough for the Hamiltonian oscillations not to be dampened ($\Omega > \gamma_{eff}/4$), consider for simplicity the earliest dynamical stage, before any jumps have occurred: the only active site in the whole lattice is the origin and both death and infection (of a neighbor) are suppressed when it has large component over $\ket{B}$. Thus, there is some probability that the first jump will only occur after an oscillations has been completed, some smaller probability that it will have two wait two cycles, and so on. 

Thus, most trajectories will "start" on timescales comparable to the GEP's; a smaller proportion, however, will only start after some time $t \approx 2\pi / \Omega$, or after some integer multiple thereof. After infection has first begun, it proceeds in a way qualitatively similar to infection in the GEP, i.e., it simply propagates outwards; since death and infection are affected by the oscillations exactly in the same way (they both require a site to be infected), we can roughly think about the process as a GEP in which periodically the rates are turned off and on again. 

Provided this interpretation holds over the entire active phase, the emergence of multiple outbreaks after averaging simply results from having statistically similar trajectories that simply do not start at the same point in time, but tend to start instead with some reciprocal delays. This would also explains why different outbreaks travel outward at the same speed: they are the same propagation delayed by some interval. Unfortunately, it would also imply that the eQEP cannot possibly capture the striking features of the RQEP, since those heavily rely on having each outbreak after the first one being affected by the DDS left behind by its predecessors, which can be seen as an effective reduction in the value of the infection parameter $\Omega_I$\cite{Espigares2017}.

\section{Conclusions}
\label{sec:Concl}

The non-equilibrium dynamics of open quantum systems still provides fertile ground for many interesting theoretical questions. This work was inspired by past research on the effects of quantum fluctuations on dynamical systems featuring absorbing states. Among many examples, the predicted emergence of a succession of discontinuous transitions in the RQEP \cite{Espigares2017} sparked our interest, as it constitutes a radical departure from the much simpler behavior of its closest classical relative, the GEP. As these predictions are the result of a non-uniform mean-field approximation, we think it is important to test them and, possibly, establish the results on steadier ground.

As a first attempt in this direction, we have combined ingredients from both the GEP and the RQEP to build a new model. From the former, we have taken the defining features of an epidemic process, namely the constrained nature of infection and the irreversibility of death. From the latter we extracted its most obvious difference with respect to the GEP, i.e., the presence of oscillations that periodically activate and deactivate sites. We have then tried to blend these building blocks in the most basic form we could think of, resulting in a new epidemic model dubbed the "elementary Quantum Epidemic Process", or eQEP.

In contrast to the GEP and RQEP, the eQEP features a fourth local state, the bedridden state ($B$). Thanks to this, however, it recovers the $S \rightarrow I \rightarrow D$ "directionality" of the GEP (violated by the RQEP), i.e., the fact that infected sites cannot become susceptible again (nor dead ones infected). 

Crucially, the eQEP features some weak symmetries that yield very significant simplifications, which essentially reduce QJMC simulations to Kinetic Monte Carlo ones with peculiar, time-dependent rates, expressed via a complicated survival probability. We have thus been able to simulate the eQEP at system sizes large enough to observe its epidemic dynamics with some clarity.

The introduction of oscillations between the $I$ and $B$ states alters the epidemic dynamics and leads to the emergence of multiple outbreaks, supporting the analogous prediction made for the RQEP: in the light of our findings, it seems reasonable to conclude that the RQEP, too, would display subsequent waves of infection. 

On the other hand, our stationary results are statistically indistinguishable from the GEP's. The stationary phase diagram, for instance, shows no discernible dependence on the coherent frequency $\Omega$. We are unable to provide a rigorous explanation of this, but our study of the eQEP's dynamics suggests that the most relevant effect caused by the introduction of the Hamiltonian is a periodic delay in the initial formation of the outbreaks: roughly speaking, most trajectories would qualitatively look the same once the infection starts propagating outward, but propagation would start at different times for different trajectories; additionally, these "starting times" would bunch close to points when the oscillating origin gets close to $\ket{I}$, i.e., integer multiples of the period of these oscillations.  

Provided our intuition is, at least roughly, correct, an individual outbreak in the eQEP cannot affect the properties of its successors, as is the case in the RQEP. In a sense, the eQEP is too elementary a model to capture this possibility. We cannot therefore draw any conclusions on the remarkable stationary phase structure of the RQEP. Whether the latter is a physical feature of some quantum epidemic models or a mere artifact of the mean-field approximation remains, to the best of our knowledge, unanswered.

On a different note, simplifications similar to the ones highlighted here for the eQEP may be possible in other open quantum systems. Potentially, they can also guide the construction of new "simulation-amenable" models aimed at tackling similar problems.

\section{Outlook}
\label{sec:outlook}

Clearly, our numerical analysis of the eQEP is not exhaustive: we have focused entirely on one-point observables, leaving out $n$-point and $n$-time correlations. We have not studied scaling in the critical region. Our current understanding of its dynamics, however, makes us doubt that there could be more complicated stationary order parameters showing a more interesting phase structure. In our opinion, further investigations of the eQEP are not warranted; instead, it would be a more fruitful endeavor to develop more complex quantum epidemic models while trying to retain some of the eQEP's intrinsic advantages.

A possible way forward is to close the distance with the RQEP, i.e., to include more of its features beside the presence of coherent oscillations. For instance, oscillations in the RQEP are generated by the same Hailtonian term which encodes infection and are, consequently, constrained on the presence of infected neighbors. This constraint can be included by replacing the eQEP Hamiltonian \eqref{eq:eQEP_H} with 
\be
    H = \sum_{\nnsum{k}{j}} \Omega \, \Pi_{j} \lt \sigma_k^{IB}  + \sigma_k^{BI} \rt
\ee
where the summation runs over neighboring site pairs and the projector $\Pi_{j}$ could be either
\begin{subequations}
\begin{align}
    &\Pi_{j} = \sigma_j^{II} + \sigma_j^{BB} \text{ or} \\
    &\Pi_j = \sigma_j^{II}. \label{subeq:nofact_projector}
\end{align}
\end{subequations}
For both choices, the Lindblad equation would still retain the same weak symmetries identified for the eQEP. In the former case, it would additionally retain the factorization of the effective Hamiltonian, as the restriction over a given eigensector would allow replacing $H$ with
\be
    H \to \sum_{k} \Omega \, \lt \mal{N}_k - \mal{N}_k^S - \mal{N}_k^D  \rt \lt \sigma_k^{IB}  + \sigma_k^{BI} \rt,
\ee
with $\mal{N}_k$ the number of nearest neighbors of site $k$ and $\mal{N}_k^{S(D)}$ the number of susceptible (dead) ones among them.

For the latter choice, instead, factorization would be lost. This would reintroduce the issue of the exponential growth of the computational complexity with the system size. However, the dimension of the effective vector space to be used for the deterministic part of the QJMC evolution would not be $\sim 2^N$, but rather $ = 2^{\chi_I}$ (see Eqs.~\eqref{eq:eigensp_dim} and \eqref{eq:chi_I}) with $\chi_I$ the number of active (I-B superposition) sites at any given time. As long as one does not push deep into the active phase, for individual trajectories we may expect $\chi_I$ to remain small, hopefully small enough to permit running simulations on bearable timescales.

% \begin{acknowledgments}
% \mm{Any acknowledgments?}
% \end{acknowledgments}

\appendix

\section{Legend of adopted terms and shorthands}
\label{App:Legend}
For the reader's convenience we briefly report here a list of the most relevant terms and shorthands used in the main text in alphabetic order:
\begin{itemize}
\item \underline{Absorbing:} the property of a local state or configuration that cannot be left in the course of the evolution. Once a system enters one such absorbing state, the dynamics effectively halts.
\item \underline{Activation:} see "Deactivation".
\item \underline{Active phase:} in the context of a process with absorbing states, a parameter region in which, in the thermodynamic limit, there is a finite probability of the dynamics carrying on indefinitely. At finite, but large, system size $\infty > N \gg 1$, the active phase is a parameter region in which typical trajectories appear in which the infection, starting from an initial seed, is able to cover a macroscopic ($O(N)$) portion of the system.  
\item \underline{Classical basis (local):} a particular orthonormal basis for the Hilbert subspace of a single site. Its constituent states $\ket{S/I/D}$ (with the addition of a fourth one $\ket{B}$ for the eQEP) are meant to act as analogues of the S/I/D classical states of the GEP. For this reason, we refer to said quantum states as the local "GEP-like" states. For simplicity, for the eQEP we include $\ket{B}$ among them, even though there is analog B state in the GEP. 
\item \underline{Classical basis (global):} the set of all product states over the entire system's Hilbert space $\ket{\vec{\mu}} = \bigotimes_{k = 1}^{N} \ket{\mu_k}_k$ in which every single factor $\ket{\mu_k}_k$ is taken from the local classical basis. These product states are also described as "GEP-like" in the main text, as they are analogous to the classical configurations of the GEP. For a $N$-site lattice, there are $3^N$ GEP-like configurations for the RQEP and $4^N$ for the eQEP.

\item \underline{Classical state:} in this work, we mean by this the state of a stochastic process. Global classical states are also referred to as "configurations". 
\item \underline{Configuration:} also called a "global classical state" or "system classical state", it is a complete specification of all the local classical states of a stochastic process. It uniquely defines the state of the (entire) system. See also "GEP-like configuration" for the term used in our main text for the RQEP and eQEP equivalent. 
\item \underline{Deactivation:} the new process introduced in the eQEP which makes sites that have been infected oscillate (precess) between states $\ket{I}$ and $\ket{B}$. It is associated to the Hamiltonian \eqref{eq:eQEP_H}. Its name derives from the fact that it can transform I sites into B sites which can neither infect nor die (they are "inactive" with respect to the epidemic terms). Since Hamiltonian terms generate a unitary evolution, one could refer to it as "Activation/Deactivation", due to the equivalent possibility of transforming Bs back to Is.
\item \underline{DDS:} density of dead sites (either local or averaged over the entire lattice). See Eq.~\eqref{eq:class_DDS} for the GEP or Eq.~\eqref{eq:rand_varrho} for the eQEP.
\item \underline{Epidemic process:} for the purposes of this work, a dynamical process featuring (i) an infectious state, i.e., a state that can replicate itself in sites nearby and (ii) a dead (or immune) state that is dynamically inert.
\item \underline{eQEP:} elementary Quantum Epidemic Process, the four-level-per-site open quantum system introduced in this paper.
\item \underline{GEP:} General Epidemic Process, a three-state-per-site classical stochastic process described in Section \ref{subsec:GEP}.
\item \underline{GEP-like configuration:} a global quantum state for the system of the form $\ket{\vec{\mu}} = \otimes_{k=1}^N \ket{\mu_k}_k$ with $\mu_k \in \set{S, I, D}$ for the RQEP or $\mu_k \in \set{S, I, B, D}$ for the eQEP. The $\mu_k$s can be fixed independently, so that there are $3^N$ different choices for the RQEP, $4^N$ for the eQEP. A GEP-like configuration constitutes the direct analog to a classical configuration of the GEP. Note that we use the same name to refer to the pure density matrices $\rho = \ket{\vec{\mu}} \bra{\vec{\mu}}$.
\item \underline{GEP-like state:} a quantum state directly associated to a classical GEP state. There are three local GEP-like states for the RQEP ($\ket{S/I/D}$), plus a fourth one ($\ket{B}$) for the eQEP. Global GEP-like states, or GEP-like configurations, are tensor products thereof (see also the entry for "classical basis, global" above).
\item \underline{Inactive phase:} in the context of a process with absorbing states, a parameter region in which it is certain that the dynamics will be trapped in one such state at long times, even in the thermodynamic limit.
%
% \item \underline{Jump:} A single, instantaneous stochastic event. It can be fully characterized by specifying unambiguously where, when, and what happened or, more precisely, the coordinates of the sites affected by a change of state (where), the time at which the change occurs (when) and the new states assigned to the affected sites (what). For the GEP, due to the local nature of death and infection, this triple takes the simple form $(\vec{r}, t, \chi)$ with $\vec{r}$ the position where, at time $t$, the state switches to $\chi \in \left\{ \text{I, D}\right\}$. 
%
\item \underline{Jump:} A single, instantaneous stochastic event, fully characterized by specifying its time and type. In the case of local events, the "type" accounts also for the location of the event. For instance, for the GEP, death at site $i$ and death at site $j \neq i$ would be considered different event types. We describe a jump via a pair $\lt t, \alpha \rt$, where $t$ is the instant at which an event takes place, and $\alpha$ labels the type. Given a system configuration, each jump may take place independently from all others at some rate $\gamma \geq 0$. If $\gamma = 0$ the specific jump has no chance of occurring in the present configuration.
\item \underline{Jump (quantum):} see entry for "Quantum Jump" below.
\item \underline{Kinetic constraint:} a constraint imposed on an elementary (typically local) dynamical process. If the constraint is not satisfied, a site (or multiple ones) cannot change its (their) state, or its (their) options are limited. In the GEP, infection is constrained because the transition $S \to I$ is only allowed when there is at least another infected site nearby. For the RQEP the constraint is even stricter, requiring the presence of one and only one infected site among the nearest neighbors.
\item \underline{No-jump probability:} The probability that the first stochastic jump will not occur before a certain time has elapsed, see also "TFJ" below. The moste frequent notation we have used for it is $\mathbb{P} ( TFJ \geq t \,|\, \ket{\psi_0}, t_0) $, reading as the probability that there are no jumps up to time $t$ conditioned on the fact that the state, or vector, was $\ket{\psi_0}$ at time $t_0$. When differentiated with respect to $t$ and multiplied by $-1$ it yields the distribution of the time of first jump $TFJ$: $-\partial_t \mathbb{P} = p_{TFJ}$. 
\item \underline{Origin:} the "central" site of a lattice where a single infected individual is placed at the start of the dynamics ($t = 0$). As we deal exclusively with square lattices of uniform width ($1,\, \ldots,\,\ell_x \in \N$) and height ($1,\, \ldots,\, \ell_y \in \N$), we simply fix the origin at dimensionless (integer) coordinates $\lt \floor{\frac{\ell_x + 1}{2}}, \,\floor{\frac{\ell_y + 1}{2}}  \rt$. Of course, if either $\ell_x$ or $\ell_y$ is even there is no exact center of the lattice and the origin is intended to lie at one of the best approximations thereof available.
\item \underline{Outbreak:} in the context of an epidemic process, a wave of infection propagating outward from the center (or any other initially infected site or region) to the boundaries of the lattice (or indefinitely, in the ideal case of infinite size). An outbreak leaves in its wake a finite DDS.
\item \underline{QJMC:} Quantum Jump Monte Carlo. A formalism that aims to reconstruct the dynamics of an open quantum system (which is a map $t \to \rho(t)$ on density matrices in Liouville space) via an effective stochastic process in Hilbert space $t \to \ket{\psi(t, \tau)}$, with $\tau$ the stochastic trajectory label. The evolved density matrix is recovered upon averaging over the trajectory distribution $p_{st}(\tau)$: $\rho(t) = \int \mal{D} \tau \, p_{st}(\tau) \ket{\psi(t, \tau)} \bra{\psi(t, \tau)}$.
\item \underline{Quantum Jump:} an instantaneous event produced in the QJMC formalism. It is the building block of the stochastic part of the QJMC evolution. Quantum jumps are in a one-to-one correspondence with the Lindblad jump operators $L_\alpha$ and can be given a physical interpretation: in quantum optics, for instance, they can be associated to external measurements performed on the system; for example, a decay event for a spin (or atom) may be associated to the detection of a photon by a photomultiplier. This measurement collapses the spin's state $\ket{\psi} \to L_\alpha \ket{\psi}$.
\item \underline{RQEP:} Rydberg Quantum Epidemic Process, a three-level-per-site open quantum system introduced in Ref.~\cite{Espigares2017}. Note that the RQEP denomination was not used in the original work; we introduced it here to more easily make the distinction with our own model, the eQEP.
\item \underline{Sector:} in the main text we employ this term to denote a global eigenspace $\mal{H}^{(\vec{\zeta})} \subseteq \mal{H}$, having specified the eigenvalues of $\sigma_k^{SS}$ and $\sigma_k^{DD}$ on all sites $k$. Notice the difference with local (one-site) Hilbert subspaces $\mal{H}_k$: the full space is a tensor product of the latter $\mal{H} = \bigotimes_k \mal{H}_k$, but a direct sum of sectors $\mal{H} = \bigoplus_{\vec{\zeta}} \mal{H}^{(\vec{\zeta})}$. Also note that we can introduce local sectors $\mal{H}^{(\zeta_k)}_k$ as well by fixing the eigenvalues of $\sigma_k^{SS}$ and $\sigma_k^{DD}$ on site $k$ only. Then, $\mal{H}_k = \bigoplus_{\zeta_k} \mal{H}^{(\zeta_k)}_k$ ($\zeta_k \in \set{S, I, D}$) and $\mal{H}^{(\vec{\zeta})} = \bigotimes_k \mal{H}^{(\zeta_k)}_k$. 
\item \underline{Stochastic trajectory:} see the entry for "Trajectory" further below.
\item \underline{Superoperator:} for the scope of this work, a linear operator acting on the space of density matrices (or a matrix space of equal dimension).
\item \underline{TFJ:} acronym for "Time of First Jump". In a QJMC, exactly as in a classical Kinetic Monte Carlo, the time of first jump is a random variable; its distribution is one of the ingredients that define a stochastic process. It can be described via the cumulative no-jump probability $\mathbb{P} \lt TFJ \geq t | \mal{S}, t' \rt$ conditioned on the knowledge that the system was in some state $\mal{S}$ at a previous time $t'$. The actual TFJ distribution can be obtained from this probability by straightforward differentiation: $p_{TFJ}(t\,|\,t') = - \partial_t \mathbb{P}$.
\item \underline{Trajectory:} a trajectory, or "stochastic trajectory", is a single random realization of (the dynamics of) a stochastic process. It can be described as a time-ordered list of "jumps" (see entry above) $\tau = \left\{ (t_z, \alpha_z)  \right\}_z$ plus, unless it has been uniquely fixed, an initial state. A trajectory has an associated probability density $p_{st}(\tau)$ determined by the stochastic process properties. In terms of Monte Carlo numerical simulations, a trajectory corresponds to a single run of the process; furthermore, standard Monte Carlo methods produce trajectories with a frequency $\propto p_{st}$ and therefore averages can be approximated with arithmetic averages:
\be
    \int \mal{D} \tau \, p_{st}(\tau)\, O(\tau) \approx \frac{1}{M} \sum_{i = 1}^M O\lt \tau_i \rt 
\ee
over a set of $M$ sample trajectories.
\end{itemize}

\section{Defining open quantum absorbing states}
\label{App:abs_dark}

We spend here a few words to explain why, in extending the notion of absorbing states to open quantum systems, we find stationarity ($\mal{L} \rho^{(a)} = 0$) insufficient and require additionally that absorbing states be \emph{dark}, i.e.,
\begin{itemize}
\item[(i)] $\rho^{(a)} = \ket{\psi^{(a)}} \bra{\psi^{(a)}}$;
\item[(ii)] $H\ket{\psi^{(a)}} = \lambda \ket{\psi^{(a)}}$;
\item[(iii)] $L_\alpha \ket{\psi^{(a)}} = 0 \,\,\, \forall\,\, \alpha$.
\end{itemize}

To understand this, we point out first that in the purely classical case, too, absorbing states are more narrowly defined than stationary states; after all, had it been otherwise any stochastic process with stationary states would be absorbing. To clarity, let us discuss an example: consider a symmetric random walk on a ring of $N$ sites; its stationary state corresponds to a uniform distribution over all lattice sites. Still, the random walker never stops moving, continuing to jump left and right indefinitely. However, add a second "death" process to the mix, whereby a random walker may disappear with a certain rate (or with some finite probability, if in discrete time). Then, the only stationary state of the system is the empty configuration, which is also absorbing: once reached there is nothing moving any longer.

The distinction can also be expressed in a statistical physics language: in the examples above, we are using the word "state" in two different senses, typically differentiated by adding the prefixes "macro-" and "micro-". The microstate corresponds to the configuration of the system; in the random walk example, the microstate could be described via a string $\vec{w}(t)$ of $N$ integers, whose component $w_k(t) \geq 0$ counts how many walkers occupy site $k$ at time $t$. The macrostate, on the other hand, is the probability distribution $p(\vec{w},t)$ of microstates (configurations) at a certain point in time. 

In the first half of our example, the undying walker on a ring, there are $N$ different microstates $\vec{w}$, each corresponding to the walker being in one of the $N$ sites of the ring. The stationary macrostate is
\be
    p_{SS} \lt \vec{w} \rt = \frac{1}{N}, 
\ee
where we assume $N$ odd for simplicity. This macrostate is indeed invariant under the stochastic dynamics. However, the microstate continues to change indefinitely as the walker jumps left and right. 

In the second half of our example, with the addition of "death", the empty microstate $\vec{w} = \vec{0}$ is absorbing, corresponding to the macrostate
\be
    p(\vec{w}) = \delta_{\vec{w}, \vec{0}} = \sysb{lc} 1 & \text{if $\vec{w} = \vec{0}$,}  \vspace{0.1cm} \\
                                               0 & \text{otherwise.}\syse
\ee

As a matter of fact, for classical stochastic processes we could have given the following definition: for a process acting on microstates (configurations) $\vec{w}$, a macrostate $p_{\vec{a}}$ is absorbing if
\begin{itemize}
    \item[(I)] $p_{\vec{a}} \lt \vec{w}  \rt = \delta_{\vec{w},\vec{a}}$;
    \item[(II)] None of the elementary moves (jumps, events) that define the stochastic process can take place over configuration $\vec{a}$.
\end{itemize}
We have simply opted for a less cumbersome (and, hopefully, clearer) account. Still, in the one just given, the similarity with the definition of open quantum absorbing states should be now apparent. Condition (i) mimics condition (I), while conditions (ii) and (iii) restate condition (II) in the open quantum case differentiating between the action of the Hamiltonian and that of the jump operators.

Let us verify that, with this definition of quantum absorbing states, there can be no jumps taking place once one has been reached. To this end, let us define the evolution under the effective Hamiltonian as
\be
    \ket{\psi^{(a)} (t)} = \rme{-iH_{eff}t} \ket{\psi^{(a)}}.
\ee
Then, we derive
\begin{align}
    &\partial_t  \ket{\psi^{(a)} (t)}  =  \nol
    & = \rme{-iH_{eff}t} \lt -iH_{eff} \rt \ket{\psi^{(a)}} = -i \lambda \ket{\psi^{(a)} (t)}
\end{align}
from Eq.~\eqref{subeq:abs_Heff}, which yields, upon integration,
\be
    \ket{\psi^{(a)} (t)} = \rme{-i\lambda t} \ket{\psi^{(a)}},
\ee
i.e., Eq.~\eqref{subeq:abs_exp}. Thus,
\begin{align}
    \mathbb{P} & \lt TFJ \geq t \,\,|\, \ket{\psi^{(a)}}, t'  \rt = \norm{\,\ket{\psi^{(a)}(t)}}^2 = \nol 
    & =\norm{\rme{-i\lambda t} \ket{\psi^{(a)} } }^2  = \norm{\,\ket{\psi^{(a)}}}^2 = 1
\end{align}
at all times. Literally, we are certain that $TFJ$ is greater than any possible time $t$; in other words, no jump will occur in finite time, no matter how large.

\section{Effects of the constraint on the dynamical connection between classical eigenstates}
\label{App:Afi}
We provide in this Appendix our reasoning for the right column of Fig.~\ref{fig:RQEP1}(b). For the reader's convenience, we recall that we say that two orthogonal states $\ket{\psi}$ and $\ket{\phi}$ are "connected" by the evolution under a Hamiltonian $H$ if $\exists$ a time $t$ at which
\be
	\bra{\phi} \rme{-iHt} \ket{\psi} \neq 0.
\ee
In other words, they are connected if, in time, $\ket{\psi}$ acquires a component over $\ket{\phi}$ and vice versa.

\begin{figure}[h]
  \vspace{0.5cm}
  \includegraphics[width=\columnwidth]{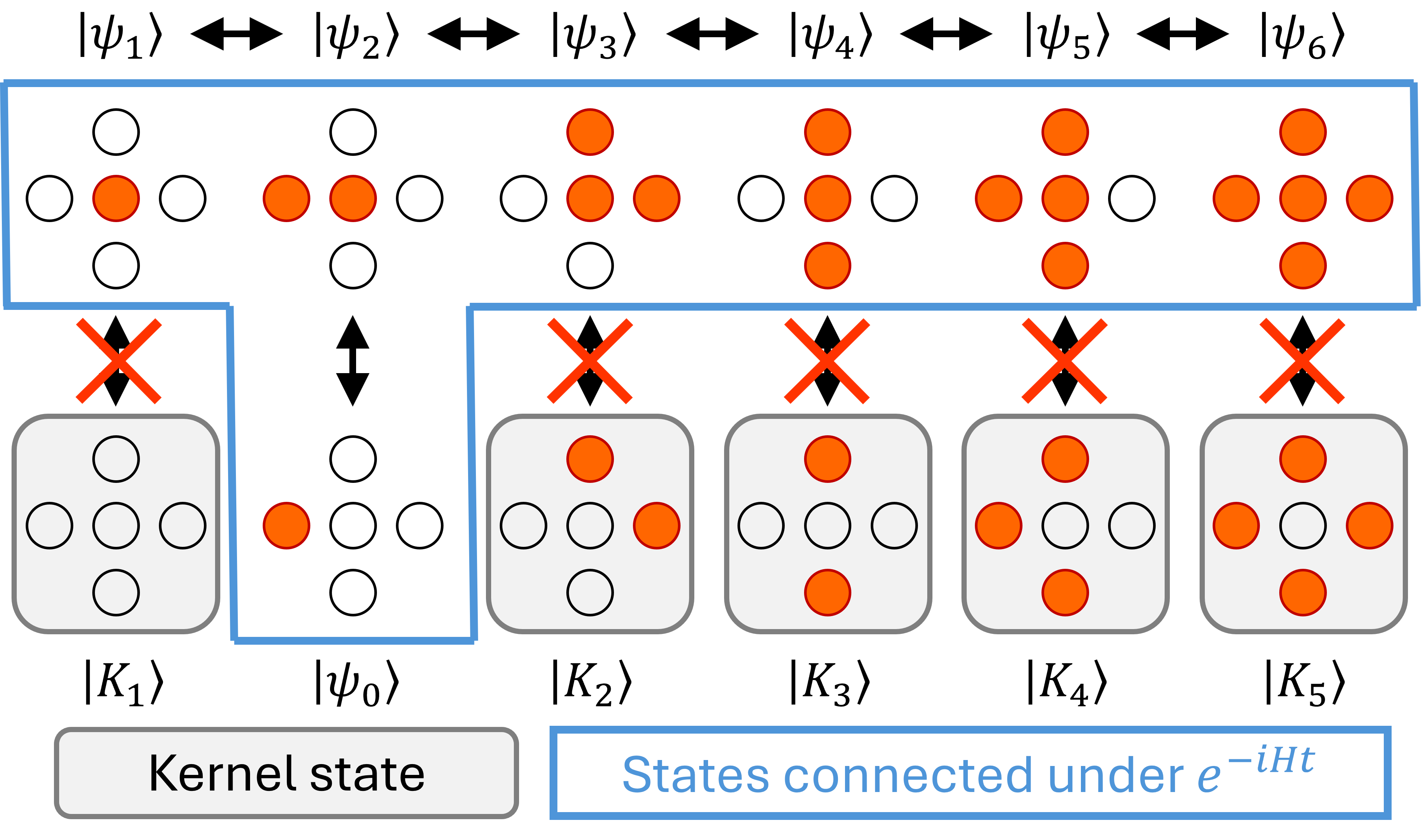}
  \caption{A collection of GEP-like states taken from panel (b) of Fig.~\ref{fig:RQEP1}, organized horizontally and slightly magnified. As in Fig.~\ref{fig:RQEP1}, black arrows indicate that configurations are dynamically connected, crossed-out arrows that they are not.
Configurations shown over a gray background feature no site satisfying the RQEP constraint. They thus belong to the kernel of the Hamiltonian and the corresponding GEP-like states are labeled $\ket{K_j}$. Conversely, the action of the RQEP Hamiltonian over configurations on a white background is non-trivial. To distinguish them from the kernel states, we have labeled them $\ket{\psi_j}$. As we argue in this Appendix, all the states displayed within the blue box are dynamically connected to each other. } 
\label{fig:RQEP_app}
\end{figure}

In the following, we take both $\ket{\psi}$ and $\ket{\phi}$ to be GEP-like states for five sites positioned in a cross-shaped format, as shown in Figs.~\ref{fig:RQEP1} and \ref{fig:RQEP_app}. We also recall that the system is isotropic, which on a square lattice means that if we rotate the GEP-like configurations by any multiple of $\pi / 2$ the results do not change. For brevity, we introduce shorthands for some selected mutually orthogonal states, as illustrated in Fig.~\ref{fig:RQEP_app}.

We also recall that the RQEP's constraint is such that a S site can only be infected if one \emph{and only one} of its nearest neighbors is infected; any GEP-like configuration featuring no S sites with a single I neighbor is annihilated by the projector \eqref{eq:RQEP_proj} and thus belongs to the kernel of the Hamiltonian. This is the case for all configurations $\ket{K_j}$, $j = 1, \ldots, 5$. Formally, $H \ket{K_j} = 0$ implies, for any $\ket{\phi} \perp \ket{K_j}$, 
\be
	\bra{\phi} \rme{-iHt} \ket{K_j} = \bracket{\phi}{K_j} = 0.
\ee

In general, whenever the evolution operator $\rme{-iHt}$ is analytic around $t = 0$, i.e.~the defining series
\be
	\rme{-iHt} = \suml{n = 0}{\infty} \frac{1}{n!} \lt -iHt \rt^n
\ee
has a non-zero convergence radius around $t = 0$, and in particular for all finite-dimensional cases, proving that two orthogonal states are dynamically connected can be reduced to showing that $\exists\, n \in \N$ such that $\bra{\phi} H^n \ket{\psi} \neq 0$. This is what we aim to prove in the following for the remaining GEP-like configurations $\ket{\psi_j}$.

For later convenience, let us define a counterclockwise "rotation" operator $R_{\pi/2}$ which, when applied to a configurations, returns its rotated counterpart; for instance,
\be
	R_{\pi/2}  \ket{\parbox[c]{0.1\linewidth}{ \includegraphics[scale=0.2]{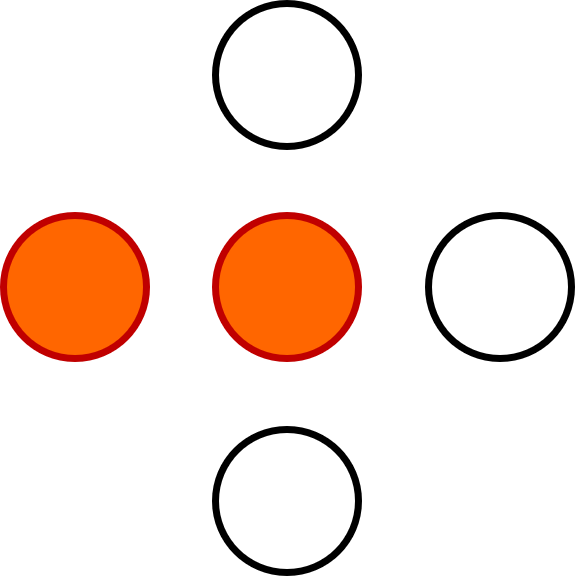} } } = \ket{\parbox[c]{0.1\linewidth}{ \includegraphics[scale=0.2]{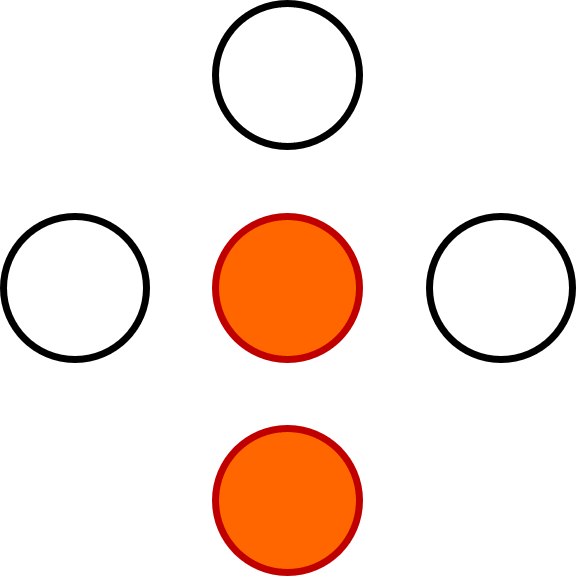} } }.
\ee
In a similar fashion, we introduce $R_{m\pi/2} = \lt R_{\pi/2}  \rt^m$, $m \in {0,\, 1,\, 2,\, 3}$, and the collective operator
\be
	R = \suml{m = 0}{3} R_{m\pi/2}
\ee
which makes any state rotationally invariant:
\begin{align}
    R_{\pi/2} R &= R_{\pi/2} \suml{m = 0}{3} \lt R_{\pi/2} \rt^m = \suml{m = 1}{4} \lt R_{\pi/2} \rt^m =  \nol 
    & = \suml{m = 0}{3} \lt R_{\pi/2} \rt^m = R,
\end{align}
having applied the identity $\lt R_{\pi/2} \rt^4 = R_{2\pi} = \id = R_0 = \lt R_{\pi/2} \rt^0$. 

Additionally, the Hamiltonian, being invariant under these rotations, commutes with all of these operators:
\be
    \comm{H}{R_{m\pi/2}} = \comm{H}{R} = 0.
\ee

Let us now analyze the action of $H$ over the GEP-like states by looking at two examples: first, consider $H \ket{\psi_1}$. The Hamiltonian \eqref{eq:RQEP_H} is as a sum of five terms, one per site; in each of these terms, the projector $\Pi_k$ is either $0$ or $1$ depending on whether the constraint is satisfied or not and controls whether the "precession" operator $\sigma_k^{SI} + \sigma_k^{IS}$ swaps $I \leftrightarrow S$ or leaves the local state untouched. $\ket{\psi_1}$ has a single infected site at the center, implying that the constraint (one infected neighbor) is satisfied for each of the peripheral sites of the cross. Thus, each corresponding addend in $H$ creates a peripheral infection and we can write
\be 
	H \ket{\psi_1} = \Omega_I R\ket{\psi_2}.
	\label{eq:Hpsi1}
\ee
This means that $\bra{\psi_2} H \ket{\psi_1} = \Omega_I \bra{\psi_2} R \ket{\psi_2} = \Omega_I \neq 0 $, i.e.~$\ket{\psi_1}$ and $\ket{\psi_2}$ are dynamically connected. 

We turn now to $H\ket{\psi_2}$; in this case all sites satisfy the constraint and thus the result will be a sum of five terms, each differing from the configuration in $\ket{\psi_2}$ by a single change:
\begin{align}
	\frac{1}{\Omega_I} H\ket{\psi_2} &= \ket{\psi_0} + \ket{\psi_1} + \nonumber\\
	& +\lt R_{\pi/2} + R_\pi \rt \ket{\psi_3} + R_{\pi/2} \ket{\psi_4} .
	\label{eq:Hpsi2}
\end{align}
Combining Eqs.~\eqref{eq:Hpsi1} and \eqref{eq:Hpsi2} and using the identity $R R_{m\pi/2} = R$ we find
\begin{align}
	\lt \frac{H}{\Omega_I} \rt^2 \ket{\psi_1} &= \frac{H}{\Omega_I}  R\ket{\psi_2} = R\, \frac{H}{\Omega_I} \ket{\psi_2} = \nol
	&= R \Bigl( \ket{\psi_0} + \ket{\psi_1} + 2\ket{\psi_3} + \ket{\psi_4} \Bigr).
	\label{eq:HHpsi1}
\end{align}
Notice the coefficient $2$ in front of $\ket{\psi_3}$. This counts the two ways in which the $\ket{\psi_3}$ configuration can be obtained from $\ket{\psi_1}$ in exactly two steps: (i) by infecting the top site first and then the rightmost one, or (ii) by infecting first the rightmost one and then the one at the top. Hence,
\be
	\bra{\psi_3} H^2 \ket{\psi_1} = 2 \lt \Omega_I \rt^2 \bra{\psi_3} R \ket{\psi_3} = 2  \lt \Omega_I \rt^2 \neq 0
\ee
and we conclude that $\ket{\psi_1}$ and $\ket{\psi_3}$ are dynamically connected.

As the reader might have inferred, the general rule is quite straightforward:
\be
	\bra{\psi_i} H^n \ket{\psi_j} = \lt \Omega_I \rt^n W_{ij}(n)
	\label{eq:Hn_conn}
\ee
with $W_{ij}(n)$ the number of ways to change configuration $j$ into $i$ via (exactly) $n$ individual site flips ($I \leftrightarrow S$). Of course, one has to take care to always satisfy the constraint (a flip is possible only in the presence of a single infected neighbor). For instance, take $i = 4$, $j = 1$ and $n = 2$. We need to change the configuration with a single central infected to what we could call a "vertical trimer". There are two ways to do this: (i) we first infect the topmost site, then the bottommost one, or (ii) the bottommost one first, then the topmost one; we thus expect $W_{41} (n=2) = 2$. From Eq.~\eqref{eq:HHpsi1}:
\be
	\bra{\psi_4} H^2 \ket{\psi_1} = \lt \Omega_I \rt^2 \bra{\psi_4} R \ket{\psi_4}
\ee
Now, $\bra{\psi_4} R \ket{\psi_4} = 2$ because $\ket{\psi_4}$ is invariant under rotations by $\pi$, i.e.~$R_\pi \ket{\psi_4} = \ket{\psi_4}$, and thus
\begin{align}
	R\ket{\psi_4} &= \ket{\psi_4} + R_{\pi/2} \ket{\psi_4} + R_\pi \lt \ket{\psi_4} + R_{\pi/2} \ket{\psi_4} \rt = \nol
	&= 2 \lt \ket{\psi_4} + R_{\pi/2} \ket{\psi_4} \rt.
\end{align}

The general rule \eqref{eq:Hn_conn} allows us to draw a very simple conclusion: two GEP-like states are dynamically connected under the action of the RQEP Hamiltonian if and only if the corresponding configurations can be obtained from one another via a sequence of allowed steps (single site flips $S \leftrightarrow I$ under the stricter RQEP constraint). Following this simple rule we can conclude that all the $\ket{\psi_i}$ states in Fig.~\ref{fig:RQEP_app} (and all their rotations) are dynamically connected to one another.

\section{The Quantum Jump Monte Carlo approach: a brief overview}
\label{App:QJMC}

As discussed in the main text, the QJMC algorithm is an effective dynamics taking place in the space of rays (unit vectors of the Hilbert space $\mal{H}$)
\be
	\widetilde{\mal{H}} = \left\{ \ket{\psi} \in \mal{H} \, | \, \bracket{\psi}{\psi} = 1    \right\}.
\ee
This space is in a one-to-one correspondence with the space of pure density matrices
\be
% \rho_A(t) \equiv
     \widetilde{\mal{D}} = \set{ \ket{\psi}\bra{\psi} \in \mal{H} \otimes \mal{H}^\ast \,|\, \ket{\psi} \in \widetilde{\mal{H}} }.
    \label{eq:matrices_space}
\ee
To understand why the QJMC works as intended, it is more convenient to work in the latter. Again, we point the reader interested in seeing the method discussed in more detail to Refs.~\cite{Qjumps1, Qjumps2, Breuer_book}. In this Appendix, we do not take a mathematically rigorous approach and sketch instead a proof that works for finite-dimensional Hilbert (and Liouville) spaces. 

Our aim here is to demonstrate the following: the matrix in Eq.~\eqref{eq:stoch_rho}, obtained as the stochastic average over QJMC trajectories, matches the one evolved via the Lindblad equation $\mal{L}$:
\be
    \rho_A(t) \equiv \overline{\bigl. \ket{\psi(t, \tau)} \bra{\psi(t, \tau)} \bigr.} = \rme{\mal{L}t} \rho_0.
    \label{eq:app_claim}
\ee
There are three preparatory steps for our proof: first, we need to translate the QJMC rules so that they act upon $\widetilde{\mal{D}}$ instead of $\widetilde{\mal{H}}$. Second, we will have to discuss how to combine and divide trajectories and how the associated probabilities behave under combination. Third, we need to discuss the integral over trajectories and subdivide it into subsets at fixed number of jumps.

\subsection{Translating to a dynamics over density matrices}
\label{subapp:translation}

We start by looking at the deterministic part of the QJMC evolution, i.e., the map
\be
    \ket{\psi} \to \ket{\psi_{eff}(t)} = \rme{-iH_{eff}t} \ket{\psi},
    \label{eq:Heff_action}
\ee
which at this stage still lacks normalization. This is straightforwardly generalized to 
\be
    \rho \to \rme{-iH_{eff}t} \rho \, \rme{iH_{eff}^\dag t} \equiv \rme{\mal{L}_0 t } \rho,
    \label{eq:L0_action}
\ee
where we have introduced the superoperator
\be
    \mal{L}_0 \rho = -iH_{eff} + iH_{eff}^\dag \, .
    \label{eq:first_L0}
\ee
The equality in \eqref{eq:L0_action} can be proved by simply differentiating both sides with respect to $t$. The deterministic evolution corresponds to the composition of the map \eqref{eq:Heff_action} with a normalization of the obtained vector, so that we remain in $\widetilde{\mal{H}}$. Hence, we need to combine \eqref{eq:L0_action} with a normalization ensuring that the matrix has unit trace:
\be
    \rho \to \frac{\rme{\mal{L}_0 t } \rho}{\trace{\rme{\mal{L}_0 t } \rho}}.
    \label{eq:det_with_normalization}
\ee

Before turning to the stochastic part, we take a closer look at the superoperator $\mal{L}_0$: if we substitute the definition \eqref{eq:Heff} of the effective Hamiltonian into \eqref{eq:first_L0} we can recast its expression as
\be
    \mal{L}_0 \rho = -i\comm{H}{\rho} - \ha \sum_\alpha \acomm{L_\alpha^\dag L_\alpha}{ \rho},
    \label{eq:L0}
\ee
which makes it apparent that $\mal{L}_0$ includes some of the terms of the Lindblad equation \eqref{eq:Lindblad}. For later convenience, we call $\mal{L}_1 = \mal{L} - \mal{L}_0$ the remaining portion:
\be
    \mal{L}_1 \rho = \sum_\alpha \mal{L}_1^{(\alpha)} \rho = \sum_{\alpha} L_\alpha \rho L_\alpha^\dag ,
    \label{eq:L1}
\ee
with
\be
    \mal{L}_1^{(\alpha)} \rho = L_\alpha \rho L_\alpha^\dag .
    \label{eq:L1alpha}
\ee
Inside a trace, the superoperators $\mal{L}_0$ and $\mal{L}_1$ can be exchanged (with a change of sign): 
\be
    \trace{\mal{L}_0 \, \rho} = - \trace{\mal{L}_1\, \rho} \ \ \forall\, \rho.
\ee
This is a consequence of the Lindblad equation being a trace-preserving generator:
\be
    \trace{\mal{L} \rho} = 0 \ \text{ and } \ \trace{\rme{\mal{L}t} \rho} = \trace{ \rho} \ \ \forall \, \rho, \, t,
\ee
which implies
\begin{align}
    0 &= \trace{\mal{L} \, \rho} =  \trace{\lt \mal{L}_0 + \mal{L}_1 \rt\, \rho} = \nol
    & = \trace{ \mal{L}_0\, \rho} + \trace{ \mal{L}_1 \, \rho}
\end{align}

Let us now turn to the probabilities to see how their expressions look like in terms of $\mal{L}_0$ and $\mal{L}_1$. The conditional no-jump probability or, more specifically, the probability of not having any jumps up to a time $t$ given that the (pure) matrix was some $\rho = \proj{\psi}$ at time $t'$, is
\begin{align}
    \mathbb{P} &\lt TFJ \geq t \,|\, \ket{\psi}, t'  \rt = \norm{\rme{-iH_{eff}(t-t')} \ket{\psi}}^2 = \nol
    & = \bra{\psi} \rme{i H_{eff}^\dag (t-t')} \rme{-iH_{eff}(t-t')} \ket{\psi}  = \nol
    & = \trace{\rme{-iH_{eff}(t-t')} \ket{\psi} \bra{\psi} \rme{i H_{eff}^{\dag} (t-t')}} = \nol
    & = \trace{\rme{\mal{L}_0 (t-t')} \rho},
    \label{eq:PP}
\end{align}
where we employed the general identity
\be
    \bracket{\phi}{\phi} = \trace{\proj{\phi}}.
\ee
The expression we have derived in \eqref{eq:PP} corresponds to the denominator in Eq.~\eqref{eq:det_with_normalization}. This should not come as a surprise: since we use $\sqrt{\mathbb{P}}$ to normalize the vector resulting from map \eqref{eq:Heff_action}, we should expect $\mathbb{P}$ to correctly normalize the evolved matrix in \eqref{eq:L0_action}. 

By differentiating Eq.~\eqref{eq:PP} we can obtain the probability density for the time of first jump $TFJ$:
\begin{align}
    p_{TFJ} & (t \,|\, \ket{\psi}, t')  = -\partial_t \,\mathbb{P} \lt TFJ \geq t \,|\, \ket{\psi}, t'  \rt = \nol
    & = -\partial_t \, \trace{\rme{\mal{L}_0 (t-t')} \rho} = - \trace{\mal{L}_0\, \rme{\mal{L}_0 (t-t')} \rho} = \nol
    & = \trace{\mal{L}_1 \, \rme{\mal{L}_0 (t-t')} \rho}.
    \label{eq:ptfj}
\end{align}
Next, we recall that the statistical weight of a jump type $\alpha$, introduced as part of step (II) in Sec.~\ref{subsec:QJMC_algorithm}, is 
\begin{align}
    Q_\alpha & \bigl( \ket{\psi} \bigr)  = \norm{L_\alpha \ket{\psi}}^2 = \bra{\psi} L_\alpha^\dag L_\alpha \ket{\psi} = \nol
    & = \trace{L_\alpha \proj{\psi} L_\alpha^\dag} = \trace{\mal{L}_1^{(\alpha)} \rho},
    \label{eq:Qalpha}
\end{align}
where we have made the dependence on the pre-jump vector explicit. The corresponding probability is
\be
    q_\alpha  \bigl( \ket{\psi} \bigr) = \frac{\trace{\mal{L}_1^{(\alpha)} \rho}}{\sum_\beta \trace{\mal{L}_1^{(\beta)} \rho}} = \frac{\trace{\mal{L}_1^{(\alpha)} \rho}}{ \trace{\mal{L}_1 \rho}}.
    \label{eq:qalpha_rho}
\ee

To be fair, we have up to this point assumed that the survival function $\mathbb{P}$ is a well-defined probability. We can now show that this assumption is indeed justified: from Eq.~\eqref{eq:Qalpha} we see that $Q_\alpha \lt \ket{\psi} \rt \geq 0$ for any vector $\ket{\psi} \in \mal{H}$. Combining Eqs.~\eqref{eq:Heff_action}, \eqref{eq:L0_action} and \eqref{eq:ptfj} we find that 
\begin{align}
    p_{TFJ} & (t \,|\, \ket{\psi}, t')  = \trace{\mal{L}_1 \, \rme{\mal{L}_0 (t-t')} \rho} = \nol
    & = \sum_\alpha \trace{\mal{L}_1^{(\alpha)} \, \rme{-iH_{eff} (t-t')} \rho\, \rme{iH^\dag_{eff} (t-t')}} \nol
    & = \sum_{\alpha} Q_{\alpha} \lt \ket{\psi_{eff}(t)}  \rt \geq 0.
    \label{eq:ptfj}
\end{align}
Hence, since this is minus the derivative of $\mathbb{P}$, we conclude that the survival function is non-increasing. Then, we just need to recall, from definition \eqref{eq:PP}, that $\mathbb{P} \geq 0$ since it is a squared norm and $\mathbb{P} (t=t')  = 1$ to arrive at our conclusion that $0 \leq \mathbb{P} \leq 1$ $\,\,\forall \,\,t \geq t'$. 

Finally, we recall that, by the QJMC rules, the $\alpha$-th jump transforms the matrix $\rho$ according to
\be
    \rho \to \frac{\mal{L}_1^{(\alpha)} \rho}{\trace{\mal{L}_1^{(\alpha)} \rho}}.
    \label{eq:jump_example}
\ee

To complete the transition to the $\widetilde{\mal{D}}$ picture, in the following we are going to substitute the dependence on vectors $\ket{\psi}$, used in the main text, with the (completely equivalent) dependence on pure density matrices $\rho = \proj{\psi}$, for instance
\be
    p_{TFJ}  (t \,|\, \ket{\psi}, t') \to p_{TFJ}  (t \,|\, \rho, t'), \ \ q_\alpha  \bigl( \ket{\psi} \bigr) \to q_\alpha  \bigl( \rho \bigr).
\ee
Additionally, since jumps formally introduce instantaneous discontinuities, for any jump time $t_z$ we introduce the notation $t_z^+$ for the time immediately after the jump and $t_z^-$ for the time immediately preceding it; as an example, we could restate Eq.~\eqref{eq:jump_example} as
\be
    \rho(t_z^-) = \rho \, \xRightarrow[\text{is } \alpha]{\text{if jump}} \,  \rho(t_z^+) =\frac{\mal{L}_1^{(\alpha)} \rho}{\trace{\mal{L}_1^{(\alpha)} \rho}}.
\ee

\subsection{Combining trajectories}
\label{subapp:traj_combination}

Having introduced the new notation over $\widetilde{\mal{D}}$, we turn out attention to the trajectories. In order to discuss how they can be combined and divided, we need to keep track of their initial conditions; we thus write
\be
    \tau \lt t \,|\, \rho', t' \rt = \set{\lt t_z,\, \alpha_z \rt_z;\, t \,|\, \rho', t'};
\ee
for a generic trajectory starting at time $t' \geq 0$ from $\rho' \in \widetilde{\mal{D}}$ and ending at time $t > t'$ after undergoing jumps at times $t_1 < t_2 < t_3 < \cdots$. A trajectory constitutes sufficient information to unambiguously reconstruct the matrix at any included time. In other words, 
\be
    \rho_\tau (t'') \equiv \ket{\psi \lt t'', \tau \rt} \bra{\psi \lt t'', \tau \rt} \in \widetilde{\mal{D}}
    \label{eq:unique_matrix}
\ee
is uniquely defined $\forall \, t'' \in \lqq t', \, t \rqq$. 

We can thus concatenate trajectories as long as we make sure that the final matrix for one coincides with the initial matrix for the next. Let us consider three times $t > t'' > t' $ and a trajectory 
\be
    \tau_< \lt t'' \,|\, \rho', t' \rt
\ee
defined from $t'$ to $t''$ and starting from matrix $\rho'$. By Eq.~\eqref{eq:unique_matrix}, the final matrix $\rho_{\tau_{<}}(t'')$ is uniquely defined. Thus, we can introduce a second trajectory
\be
    \tau_> \lt t \,|\, \rho_{\tau_{<}}(t'') \,,\, t'' \rt
\ee
which starts from the very same matrix at $t''$ and further evolves it into a new one at $t$. We can also look at the same evolution as a whole, i.e., recast it in terms of a single trajectory $\tau$ on the full interval $\lqq t',\, t \rqq$:
\be
    \tau \lt t \,|\, \rho', t' \rt = \tau_> \lt t \,|\, \rho_{\tau_{<}}(t'') \,,\, t'' \rt  \,\circ\, \tau_< \lt t'' \,|\, \rho', t' \rt,
\ee
where we use "$\circ$" to denote composition.

By construction, we have
\be
    \rho_\tau (s) = \sysb{lc}  \rho_{\tau_<} (s) & \text{if $s \in \lqq t',\, t''  \rqq$}  \\[0.2cm]
    \rho_{\tau_>} (s) & \text{if $s \in \lqq t'',\, t  \rqq$}\syse.
    \label{eq:equate_rho}
\ee
Thus, trajectory $\tau$ will feature all the jumps of trajectory $\tau_<$, followed by all those of $\tau_>$. Formally, if we call $\lt t^<_z, \alpha^<_z \rt$ ($z = 1, \ldots Z_<$) the former and $\lt t^>_z, \alpha^>_z \rt$ ($z = 1, \ldots Z_>$) the latter, then we will count $Z_< + Z_>$ jumps $\lt t_z, \alpha_z \rt$ in the combined trajectory, with
\be
    \lt t_z, \alpha_z \rt = \sysb{lc}   \lt t^<_z,\, \alpha^<_z \rt & \text{for $z \leq Z_<$}, \\[0.2cm]
                                        \lt t^>_{z - Z_<},\, \alpha^>_{z - Z_<} \rt & \text{for $z > Z_<$}.
                                        \syse
\ee

% \mm{Maybe this following example can be erased}
% For instance, if 
% \be
%     \tau \lt t \,|\, \rho', t' \rt = \set{\lt t_1, \, \alpha_1 \rt; \, t\,|\, \rho', t'}
%     \label{eq:concat_example}
% \ee
% is a one-jump trajectory and we fix $t'' = t_1^+$, then the earlier portion ends just after the jump and yields a matrix
% \begin{align}
%     \rho_{\tau_{<}}(t_1^+) & = \frac{\mal{L}_1^{(\alpha_1)} \rme{\mal{L}_0 \, \lt t_1^+ - t' \rt} \rho'}{\trace{\mal{L}_1^{(\alpha_1)} \rme{\mal{L}_0 \, \lt t_1^+ - t' \rt} \rho'}} = \nol
%     &= \frac{\mal{L}_1^{(\alpha_1)} \rme{\mal{L}_0 \, \lt t_1 - t' \rt} \rho'}{\trace{\mal{L}_1^{(\alpha_1)} \rme{\mal{L}_0 \, \lt t_1 - t' \rt} \rho'}}, 
%     \label{eq:rho<}
% \end{align}
% where the second line comes from the continuity of $\rme{\mal{L}_0 t}$ in $t$. The later portion of the trajectory (from $t_1^+$ to $t$) features no jumps, starts from the expression just above and yields
% \be
%     \rho_{\tau_{>}} (t) = \frac{\rme{\mal{L}_0 \, \lt t - t_1 \rt} \rho_{\tau_{<}}(t_1^+) }{\trace{\rme{\mal{L}_0 \, \lt t - t_1 \rt} \rho_{\tau_{<}}(t_1^+)}}.
%     \label{eq:rho>}
% \ee
% Substituting \eqref{eq:rho<} into \eqref{eq:rho>} we obtain the expression
% \be
%     \frac{\rme{\mal{L}_0 \, \lt t - t_1 \rt} \mal{L}_1^{(\alpha_1)} \rme{\mal{L}_0 \, \lt t_1 - t' \rt} \rho'}{\trace{\rme{\mal{L}_0 \, \lt t - t_1 \rt} \mal{L}_1^{(\alpha_1)} \rme{\mal{L}_0 \, \lt t_1 - t' \rt} \rho'}}
% \ee
% which is indeed the matrix $\rho_\tau (t)$ produced under $\tau$ in Eq.~\eqref{eq:concat_example}. \mm{End of previous comment}

Thankfully, the stochastic weight $p_{st}$ of a concatenation of smaller trajectories factorizes:
\be
    p_{st} \lt \tau \rt = p_{st} \lt \tau_> \rt \, p_{st} \lt \tau_< \rt.
    \label{eq:pst_sub}
\ee
This comes from the factorized structure of $p_{st}$. To see this in more detail, take the ratio between the two sides of Eq.~\eqref{eq:pst_sub} and perform the substitutions according to Eq.~\eqref{eq:pst_factors}. Most factors cancel out; the only survivors describe the time window between $t_< \equiv t_{Z_{<}}^+$ (just after the last jump of $\tau_{<}$) and $t_> \equiv t_{Z_{<} + 1}^-$ (just before the first jump of $\tau_{>}$):
\begin{align}
    & \frac{p_{st} \lt \tau \rt}{p_{st} \lt \tau_> \rt \, p_{st} \lt \tau_< \rt} = \nol 
    &=\frac{p_{TFJ}\lt t_> \,|\, \rho_{\tau }\lt t_< \rt, t_<  \rt}{ p_{TFJ}\lt t_> \,|\, \rho_{\tau_{<}} (t''), t''  \rt \mathbb{P} \lt TFJ \geq t'' \,|\, \rho_{\tau_{<}}\lt t_< \rt , t_< \rt  },
    \label{eq:pst_ratio}
\end{align}
with $t''$ being once again the intermediate time at which the two sub-trajectories meet.
This ratio is $1$ by the Markov property: the latter ensures that 
\begin{align}
    \mathbb{P} & \lt TFJ \geq t_> \,|\, \rho_{\tau_{<}}\lt t_< \rt , t_< \rt     = \nol
    & = \mathbb{P} \lt TFJ \geq t_> \,|\, \rho_{\tau_{<}}\lt t'' \rt , t'' \rt \times \nol
    &\times \mathbb{P} \lt TFJ \geq t'' \,|\, \rho_{\tau_{<}}\lt t_< \rt , t_< \rt  
\end{align}
and the required equality can be found by differentiation with respect to $t_>$.

However, it may not be immediately clear to all readers that the Markovianity of $\mal{L}$ transfers to the QJMC dynamics. Therefore, let us see in detail that the result of Eq.~\eqref{eq:pst_ratio} is indeed $1$: according to Eq.~\eqref{eq:equate_rho}, we can replace both $\rho_{\tau_<}$ in the denominator with $\rho_\tau$. Moreover, by construction there are no jumps between $t_<$ and $t_>$, which means that only the deterministic part of the evolution acts. From Eq.~\eqref{eq:det_with_normalization} we have
\be
    \rho_\tau (t'') = \frac{\rme{\mal{L}_0 \lt t'' - t_< \rt }\rho_\tau(t_<) }{\trace{\rme{\mal{L}_0 \lt t'' - t_< \rt }\rho_\tau(t_<)}}.
    \label{eq:rhot''}
\ee
Now it is just a matter of substituting expressions \eqref{eq:PP} and \eqref{eq:ptfj} in Eq.~\eqref{eq:pst_ratio}: the numerator reads
\be
    p_{TFJ} \bigl( t_> \,|\, \rho_{\tau }\lt t_< \rt, t_<  \bigr) = \trace{\mal{L}_1 \rme{\mal{L}_0 \lt t_> - t_< \rt} \rho_\tau(t_<)};
\ee
the denominator, on the other hand, 
\begin{align}
    &\lqq  \trace{\mal{L}_1 \rme{\mal{L}_0 \lt t_> - t'' \rt} \rho_\tau(t'')}   \rqq \, \lqq \trace{ \rme{\mal{L}_0 \lt t'' - t_< \rt} \rho_\tau(t_<)}  \rqq = \nol
    &= \lqq  \trace{\mal{L}_1 \rme{\mal{L}_0 \lt t_> - t'' \rt} \frac{\rme{\mal{L}_0 \lt t'' - t_< \rt }\rho_\tau(t_<) }{\trace{\rme{\mal{L}_0 \lt t'' - t_< \rt }\rho_\tau(t_<)}}}   \rqq \times \nol
    & \times \lqq \trace{ \rme{\mal{L}_0 \lt t'' - t_< \rt} \rho_\tau(t_<)}  \rqq = \nol
    &= \trace{\mal{L}_1 \rme{\mal{L}_0 \lt t_> - t'' \rt} \rme{\mal{L}_0 \lt t'' - t_< \rt }\rho_\tau(t_<)} = \nol
    &= \trace{\mal{L}_1 \rme{\mal{L}_0 \lt t_> - t_< \rt} \rho_\tau(t_<)},
\end{align}
having exploited for the first step Eq.~\eqref{eq:rhot''}.

\subsection{Dividing the integral over trajectories into parts at fixed number of jumps}
\label{subapp:submeasures}

The last operation we need to analyze is the integral over trajectories (see, e.g., Eq.~\eqref{eq:stoch_av_quantum}). We thus rewrite $\rho_A$ in Eq.~\eqref{eq:app_claim} as
\be
    \rho_A \lt t \rt = \int_0^t \mal{D} \tau \, \rho_{\tau} \! \lt t \rt  \, p_{st} \lt \tau \rt 
\ee
where we added "extrema" to the integral sign to make it explicit that we are averaging over trajectories starting at time $t_0 = 0$ and ending at $t$. This will prevent ambiguities further below.

Next, we subdivide the measure $\mal{D} \tau$ (acting on all possible trajectories) into sub-measures $\mal{D}_z \tau$, each acting only on trajectories undergoing exactly $z \in \N$ jumps. Clearly, trajectory subsets at different $z$ are disjoint (any trajectory has a defined number of jumps); additionally, their union gives back the full set. Hence, we can write the full measure as a sum
\be
    \mal{D} \tau = \sum_{z = 0}^{\infty} \mal{D}_z \tau.
\ee
For later convenience, we also introduce the notation
\be
    \mal{D}_{(>0)} \tau = \sum_{z = 1}^{\infty} \mal{D}_z \tau
\ee
for the measure restricted over trajectories jumping at least once; obviously, $\mal{D} \tau = \mal{D}_0 \tau + \mal{D}_{(>0)} \tau $.

To illustrate the rationale behind subdividing the measure in this fashion, let us look at a couple of examples: first, for $z = 0$, once we fix the initial time $t_0 = 0$, initial state $\rho_0$ and final time $t$, we find only one possible trajectory
\be
    \tau_0 \lt t\,|\, \rho_0, 0 \rt = \set{\emptyset ; \,  t\,|\, \rho_0, 0}
\ee
which produces the matrix
\be
    \rho_{\tau_0} \lt t \rt = \frac{\rme{\mal{L}_0 t} \rho_0}{\trace{\rme{\mal{L}_0 t} \rho_0}}
\ee
with finite probability
\be
    \mathbb{P} \lt TFJ \geq t \,|\, \rho_0 , 0   \rt =     \trace{\rme{\mal{L}_0 t} \rho_0}.
    \label{eq:prob_zero_jumps}
\ee
Therefore, the contribution to $\rho_A$ due to the zero-jump trajectory reads
\begin{align}
    \int_0^t \mal{D}_0 \tau \, & \rho_{\tau} (t) \, p_{st} \lt \tau \rt = \rho_{\tau_0} \lt t \rt \mathbb{P} \lt TFJ \geq t \,|\, \rho_0 , 0 \rt  \nol
    & = \rme{\mal{L}_0 t} \rho_0.
    \label{eq:zero_jump_traj}
\end{align}

Second, let us consider trajectories with a single jump ($z = 1$) of type, say, $\alpha_1$, occurring at some time $0 \leq t_1 \leq t$:
\be
    \tau_1 \lt t\,|\, \rho_0, 0 \rt = \set{ \lt t_1, \,\alpha_1 \rt ; \,  t\,|\, \rho_0, 0}
\ee
How do we express $\int_0^t \mal{D}_1 \tau$? We need to integrate over all possible values of $t_1$ and sum over all possible choices for $\alpha_1$, i.e.
\be
    \int_0^t \mal{D}_1 \tau \, \lt \cdot \rt = \sum_{\alpha_1} \int_0^t \rmd  t_1 \, \lt \cdot \rt .
\ee
Now, exploiting our previous discussion on the combination of trajectories, we subdivide $\tau_1$ into three portions around the jump time $t_1$:
\begin{align}
    \tau_1 \lt t\,|\, \rho_0, 0 \rt & = \tau_0' \lt t\,|\, \rho_1^+ , t_1^+ \rt \circ  \nol
    &\circ \mal{J}_{\alpha_1} \lt t_1 \rt \circ \tau_0'' \lt t_1^- \,|\, \rho_0, 0  \rt.
    \label{eq:decomposition}
\end{align}
with the shorthand 
\be
    \mal{J}_{\alpha_1} \lt t_1 \rt = \set{\lt t_1, \alpha_1 \rt;\, t_1^+ \,|\, \rho_1^- ,\, t_1^- }
\ee
for the part of infinitesimal duration including just the jump event, and
\be
    \rho_1^\pm \equiv \rho_{\tau_1} \lt t_1^{\pm} \rt.
\ee
The two remaining branches, $\tau_0'$ and $\tau_0''$ are both subtrajectories without jumps, describing the evolution after and before $t_1$, respectively.

Then,
\be
    p_{st} \lt \tau_1 \rt = p_{st} \lt \tau_0' \rt p_{st} \lt \tau_0'' \rt p_{st} \lt \mal{J}_{\alpha_1} \rt,
\ee
where the first two factors can be extracted from Eq.~\eqref{eq:prob_zero_jumps}, while the third can be reconstructed from Eqs.~\eqref{eq:pst_factors}, \eqref{eq:PP}, \eqref{eq:ptfj} and \eqref{eq:qalpha_rho}:
\begin{align}
    p_{st} & \lt \mal{J}_{\alpha_1} \rt = q_{\alpha_1} \lt \rho_1^-  \rt \, p_{TFJ} \lt t_1 \,|\, \rho_1^- , t_1  \rt \times \nol[0.2cm]
    & \times \mathbb{P} \lt TFJ \geq t_1 \,|\, \rho_1^+ ,\, t_1  \rt = \nol[0.2cm]
    & = \lqq \frac{\trace{\mal{L}_1^{(\alpha_1)} \rho_1^-}}{\trace{\mal{L}_1 \rho_1^-}} \rqq \, \lqq \trace{\mal{L}_1 \rho_1^-} \rqq \times \nol
    & \times \underbrace{\lqq \trace{\rho_1^+} \rqq}_{= 1} = \trace{\mal{L}_1^{(\alpha_1)} \rho_1^-} = Q_{\alpha_1} \lt \rho_1^-  \rt.
\end{align}
Thus, the probability density associated with the jump is, not very surprisingly, the jump's own statistical weight from Eq.~\eqref{eq:Qalpha}.
% \begin{align}
%     \tau_1 \lt t\,|\, \rho_0, 0 \rt & = \tau_0' \lt t\,|\, \rho_{\tau_1} \lt t_1^+  \rt, t_1^+ \rt \circ  \nol
%     &\circ \mal{J}_{\alpha_1} \lt t_1 \rt \circ \tau_0'' \lt t_1^- \,|\, \rho_0, 0  \rt.
% \end{align}
% with the shorthand 
% \be
%     \mal{J}_{\alpha_1} \lt t_1 \rt = \set{\lt t_1, \alpha_1 \rt;\, t_1^+ \,|\, \rho_{\tau_1} \lt t_1^-  \rt, t_1^- }
% \ee
% for the part of infinitesimal duration including just the jump event. Then,
% \be
%     p_{st} \lt \tau_1 \rt = p_{st} \lt \tau_0' \rt p_{st} \lt \tau_0'' \rt p_{st} \lt \mal{J}_{\alpha_1} \rt,
% \ee
% where the first two factors can be extracted from Eq.~\eqref{eq:prob_zero_jumps}, whereas the third can be reconstructed from Eqs.~\eqref{eq:pst_factors}, \eqref{eq:PP}, \eqref{eq:ptfj} and \eqref{eq:qalpha_rho}:
% \begin{align}
%     p_{st} & \lt \mal{J}_{\alpha_1} \rt = q_{\alpha_1} \lt \rho_{\tau_1} \lt t_1^-  \rt \rt \,\, p_{TFJ} \lt t_1 \,|\, \rho_{\tau_1} \lt t_1^-  \rt, t_1  \rt \times \nol
%     & \times \mathbb{P} \lt TFJ \geq t_1 \,|\, \rho_{\tau_1} \lt t_1^+ \rt  \rt = \nol
% \end{align}
The contribution coming from single-jump trajectories can be expressed is
\begin{align}
    \int_0^t & \mal{D}_1 \tau \,  \rho_{\tau} (t) \, p_{st} \lt \tau \rt = \nol
     & =  \sum_{\alpha_1} \int_0^t \rmd  t_1 \,  \rho_{\tau_1} (t) \,\, p_{st} \lt \tau_0' \rt p_{st} \lt \tau_0'' \rt p_{st} \lt \mal{J}_{\alpha_1} \rt .
\end{align}
For convenience, we express the matrix in the integrand as
\be
    \rho_{\tau_1} (t) = \frac{\rme{\mal{L}_0 \lt t - t_1 \rt} \mal{L}_1^{(\alpha_1)} \rho_1^- }{\trace{\rme{\mal{L}_0 \lt t - t_1 \rt} \mal{L}_1^{(\alpha_1)} \rho_1^-}}.
\ee
In this way, when combined with
\begin{align}
    p_{st} & \lt \tau_0' \rt p_{st} \lt \mal{J}_{\alpha_1} \rt = \trace{\rme{\mal{L}_0 \lt t - t_1 \rt} \rho_1^+} \, \trace{\mal{L}_1^{(\alpha_1)} \rho_1^-} = \nol
    & = \trace{\rme{\mal{L}_0 \lt t - t_1 \rt} \frac{\mal{L}_1^{(\alpha_1)} \rho_1^-}{\trace{\mal{L}_1^{(\alpha_1)} \rho_1^-}}}\, \trace{\mal{L}_1^{(\alpha_1)} \rho_1^-} = \nol
    & = \trace{\rme{\mal{L}_0 \lt t - t_1 \rt} \mal{L}_1^{(\alpha_1)} \rho_1^- },
\end{align}
yields the simplified expression
\begin{align}
    \int_0^t & \mal{D}_1 \tau \,  \rho_{\tau} (t) \, p_{st} \lt \tau \rt = \nol
     & =  \sum_{\alpha_1} \int_0^t \rmd  t_1 \,  \rme{\mal{L}_0 \lt t - t_1 \rt} \,\mal{L}_1^{(\alpha_1)}\, \rho_1^-  \, p_{st} \lt \tau_0'' \rt = \nol
     & = \int_0^t \rmd  t_1 \,  \rme{\mal{L}_0 \lt t - t_1 \rt} \,\mal{L}_1\, \rho_{\tau_0''} \lt t_1 \rt \,\, p_{st} \lt \tau_0'' \rt,
     \label{eq:one_jump_traj}
\end{align}
where we have used the fact that
\be
    \rho_1^- = \rho_{\tau_1} \lt t_1^- \rt = \rho_{\tau_0''} \lt t_1 \rt 
\ee
is the matrix evolved under subtrajectory $\tau_0''$ up to time $t_1$. 

Equation \eqref{eq:one_jump_traj} is now in the form of a time convolution between an integral kernel
\be
    \rme{\mal{L}_0 \lt t - t_1 \rt} \,\mal{L}_1
\ee
and
\be
    \rho_{\tau_0''} \lt t_1 \rt  p_{st} \lt \tau_0'' \rt = \int_0^{t_1} \mal{D}_0 \tau \rho_\tau \lt t_1 \rt p_{st} \lt \tau \rt,
\ee
i.e., the zero-jump contribution to $\rho_A \lt t_1 \rt$, see Eq.~\eqref{eq:zero_jump_traj}.
Following exactly the same steps, one can generalize this result to higher jump numbers $z$: 

\subsection{Recovering the Lindblad evolution}
\label{subapp:saving_private_Lindblad}

Let us consider the subset of trajectories for a fixed $z>0$ number of jumps and let $t_z$ denote the time at which the last ($z$-th) jump occurs (and $\alpha_z$ its type). Then, we can decompose any such trajectory in three as
\begin{align}
    \tau_z \lt t\,|\, \rho_0, 0 \rt & = \tau_0 \lt t\,|\, \rho_z^+ , t_z^+ \rt \circ  \nol
    &\circ \mal{J}_{\alpha_z} \lt t_z \rt \circ \tau_{z-1} \lt t_z^- \,|\, \rho_0, 0  \rt,
    \label{eq:decomposition_general}
\end{align}
with
\be
    \rho_z^{\pm} = \rho_{\tau_z} \lt  t_z^{\pm} \rt.
\ee
This is analogous to Eq.~\eqref{eq:decomposition} for the $z = 1$ case, i.e., we have divided $\tau_z$ in the part $\tau_0$ without jumps following $t_z$, the infinitesimal part including the last jump
\be
    \mal{J}_{\alpha_z} \lt t_z \rt = \set{\lt t_z, \alpha_z \rt;\, t_z^+ \,|\, \rho_z^- ,\, t_z^- } ,
\ee
and the portion $\tau_{z-1}$ from $t = 0$ up to $t_z$, which of course includes the remaining $z-1$ jumps. 

As seen above for the simpler $z = 1$ case, we can combine the matrix at time $t$ with the stochastic weights of the jump and last stretch of the trajectory
\be
    \sum_{\alpha_z} \, p_{st}(\tau_0) \,  p_{st} \lt \mal{J}_{\alpha_z}  \rt \rho_{\tau_z} (t) = \rme{\mal{L}_0 \lt t - t_z \rt} \mal{L}_1 \rho_z^-
\ee
to bring back $rho$ to $t_z$, immediately before the jump. Thus, the contribution coming from trajectories undergoing $z$ jumps can also be expressed as a time convolution between the already encountered integral kernel
\be
    \rme{\mal{L}_0 \lt t - t_z \rt} \mal{L}_1
\ee
and the contribution at time $t_z$ coming from trajectories undergoing exactly $z - 1$ jumps. Formally,
\begin{align}
    \int_0^t & \mal{D}_{z} \tau \, \rho_\tau (t) \, p_{st} \lt \tau \rt = \nol
    & = \int_0^t \rmd t_z \, \rme{\mal{L}_0 \lt t - t_z \rt}\, \mal{L}_1 \, \int_0^{t_z} \mal{D}_{z-1} \tau  \, \rho_\tau (t_z) \, p_{st} \lt \tau \rt.
\end{align}
Considering that $t_z$ (the last jump time) in the expression above is an integration variable, we can rename it $t_{L}$, which allows us to more easily show what happens when summing over $z \geq 1$:
\begin{align}
     \sum_{z=1}^\infty & \int_0^t  \mal{D}_{z} \tau \, \rho_\tau (t) \, p_{st} \lt \tau \rt = \int_0^t  \mal{D}_{(>0)} \tau \, \rho_\tau (t) \, p_{st} \lt \tau \rt = \nol
    & = \int_0^t \rmd t_{L} \, \rme{\mal{L}_0 \lt t - t_{L} \rt}\, \mal{L}_1 \, \int_0^{t_{L}} \mal{D} \tau  \, \rho_\tau (t_{L}) \, p_{st} \lt \tau \rt = \nol
    & = \int_0^t \rmd t_{L} \, \rme{\mal{L}_0 \lt t - t_{L} \rt}\, \mal{L}_1 \, \rho_A \lt t_L \rt .
\end{align}
Adding to both sides the zero-jump trajectory contribution to $\rho_A(t)$, Eq.~\eqref{eq:zero_jump_traj}, we finally arrive at
\begin{align}
    \rho_A(t) & = \rme{\mal{L}_0 t} \rho_0 + \int_0^t  \mal{D}_{(>0)} \tau \, \rho_\tau (t) \, p_{st} \lt \tau \rt = \nol
    & = \rme{\mal{L}_0 t} \rho_0 + \int_0^t \rmd t_{L} \, \rme{\mal{L}_0 \lt t - t_{L} \rt}\, \mal{L}_1 \, \rho_A \lt t_L \rt.
\end{align}
Differentiating both sides with respect to $t$ we find
\be
    \dot{\rho}_A(t) = \mal{L}_0 \rho_A(t) + \mal{L}_1 \rho_A(t) = \mal{L} \rho_A(t),
\ee
i.e., $\rho_A$ satisfies the Lindblad equation; with one last check on the initial condition
\be
    \rho_A (t = 0) = \rho_0
\ee
we get our desired result
\be
    \rho_A(t) = \rme{\mal{L}t} \rho_0.
\ee

\section{Strong and weak symmetries of the Lindblad equation}
\label{app:strong_weak}

In contrast to what we know for closed systems, there is a distinction in open quantum systems between "strong" and "weak" symmetries \cite{Buca2012, Albert2014}. For the reader's convenience, we provide here a brief overview of the difference.

Let us consider a unitary transformation $U$ and define the superoperator $\mal{U} \rho = U \rho U^\dag$. Intuitively, we say that $\mal{U}$ is a symmetry of the Lindblad equation $\mal{L}$ if it commutes with it: $\comm{\mal{U}}{\mal{L}} = 0$. Given that this commutator is expressed at the superoperatorial level, its implications in terms of the original $U$ may not be immediately clear. Let us take, then, one more step: if $\comm{\mal{U}}{\mal{L}} = 0$ then $\comm{\mal{U}}{\rme{\mal{L}t}} = 0$ at all times and
\begin{align}
   U \rho(t) U^\dagger &= \mal{U} \rho(t) = \mal{U} \rme{\mal{L} t} \rho(0) = \nol
  & = \rme{\mal{L} t} \mal{U} \rho(0) = \rme{\mal{L} t} \lt U \rho(0) U^\dagger \rt.
\end{align}
In other words, evolving the $U$-transformed state is equivalent to evolving the original state and then applying the unitary transformation. In this respect, symmetries act exactly as they would in a closed system.

\subsection{Strong symmetries}
\label{subapp:strong}

Strong symmetries additionally satisfy the stricter requirements 
\be
    \comm{U}{H} = 0 \text{  and  } \comm{U}{L_\alpha} = 0 \,\, \forall \, \alpha,
    \label{eq:Ucommutators}
\ee
i.e.~the Hamiltonian and all jump operators are individually invariant under the transformation $U$. Strong symmetries are associated to conserved quantities; for instance, if $U = \rme{iJ}$ for some Hermitian operator $J$, then $\av{J(t)}_{\mal{L}} = \av{J(0)}_{\mal{L}}$ for all times $t$. This can be seen by passing to the adjoint representation (essentially, the Heisenberg picture for open quantum system):
\be
    \av{J(t)}_{\mal{L}} = \trace{ J \lt \rme{\mal{L} t} \rho_0 \rt } = \trace{\lt \rme{\mal{L}^\ast t} J \rt \rho_0},
\ee
where 
\be
    \mal{L}^\ast J = i \comm{H}{J} + \sum_\alpha \lqq  L_\alpha^\dag J L_\alpha - \ha \acomm{L_\alpha^\dag L_\alpha}{J}  \rqq 
\ee
is the adjoint of $\mal{L}$. If $U$ satisfies the commutators \eqref{eq:Ucommutators}, then the same goes for $J$ and it is not difficult to see that $\mal{L}^\ast J = 0$, which implies $\rme{\mal{L}^\ast t} J = J$ $\forall \, t$. We were unable to find any non-trivial strong symmetries for the eQEP, except for the $\Omega = 0$ case discussed in App.~\ref{app:eQEP_to_GEP}.

\subsection{Weak symmetries}
\label{subapp:weak}

Conversely, weak symmetries do not guarantee the conservation of their generators. This does not mean that there are no conservation laws at all: any Lindblad equation featuring more than a single stationary state \emph{must} have conserved quantities, one per linearly-independent stationary state \cite{Albert2014}. More precisely, there is one less conserved quantity than there are stationary states, since one can always be chosen to be the identity operator, commonly considered to be trivially conserved (i.e., not providing any information). This is easily shown if we remember that the Liouville space ius itself a vector space: we can interpret $\rho$ as a vector $\rket{\rho}$, $\mal{L}$ as an operator and the trace
\be
    \trace{J^\dag \rho} = \rbracket{J}{\rho}
\ee
as a scalar product. Then, having $\mal{N}_{ss}$ independent stationary states translates to having $\mal{N}_{ss}$ linearly independent Liouville vectors $\rket{\rho_{ss}^{(n)}}$ that are right eigenvectors of $\mal{L}$ at eigenvalue $0$. Then, there must be an equal number of left kernel eigenvectors $\rbra{J_m}$ forming a reciprocal basis
\be
    \bigrbracket{J_m}{\rho_{ss}^{(n)}} = \delta_{mn}.
\ee
These quantities determine, for any initial state $\rho_0$, the component it takes on each stationary term:
\begin{align}
    \rho_0 &= \sum_{n = 1}^{\mal{N}_{ss}} c_n \rket{\rho_{ss}^{(n)}} + \lt \text{non-$ss$ terms} \rt \nol 
    & =\sum_{n = 1}^{\mal{N}_{ss}} \rbracket{J_n}{\rho_0} \rket{\rho_{ss}^{(n)}} + \lt \text{non-$ss$ terms} \rt.
\end{align}
Consequently, if known, they allow to unambiguously determine the stationary state for any given initial one:
\be
    \lim_{T\to \infty} \frac{1}{T}  \int_0^T \rmd t\, \rme{\mal{L}t} \rho_0 =  \sum_{n = 1}^{\mal{N}_{ss}} \rbracket{J_n}{\rho_0} \rket{\rho_{ss}^{(n)}}.
\ee
If $\mal{L}$ does not feature purely imaginary eigenvalues (giving rise to so-called "decoherence-free subspaces"), we can also conclude that
\be
    \lim_{t\to \infty}\rme{\mal{L}t} \rho_0 =  \sum_{n = 1}^{\mal{N}_{ss}} \rbracket{J_n}{\rho_0} \rket{\rho_{ss}^{(n)}}.
\ee

We know the eQEP has multiple absorbing states. Thus, it must feature a large number of conserved quantities $J_n$. For this work, we did not look for them. We remark that finding them would essentially amount to determining the stationary properties of the eQEP analytically, but would still not account for the dynamics.

\subsection{The eQEP case}
\label{subapp:eQEP_weak_syms}

Before discussing the weak symmetries introduced in the main text, we mention that they are not the only ones: for instance, on a square $N = n \times n$ lattice the model is invariant under all transformations belonging to the symmetry group of the square; if we expand our site index $k \to \lt k_x,\, k_y \rt$ with $k_o \to \lt 0,\,0 \rt$ the origin, we see for instance that a rotation by $\pi/2$, while leaving $H$ invariant, will map $L^D_{\lt 1,0 \rt}$ onto $L^D_{\lt 0,1 \rt} \neq L^D_{\lt 1,0 \rt}$, showing that these transformations do not commute with the jump operators. Still, the mapping of sites is one-to-one and, upon summation over all jump types, one recovers the same Lindblad equation. These are, in other words, also weak symmetries. 
The reason we did not include them in our discussion is that they do not commute with ours: a $\pi/2$ rotation, after all,  will map $\sigma^{DD}_{\lt 1,0 \rt}$ onto $\sigma^{DD}_{\lt 0,1 \rt} \neq \sigma^{DD}_{\lt 1,0 \rt}$. Out of the two symmetry sets, we have opted for the one which, to us, offered the greater simplification.

We now prove that the quantities $\sigma^{SS}_k$ and $\sigma^{DD}_k$, commuting with the eQEP's effective Hamiltonian, are indeed associated to weak symmetries for $\mal{L}$. With them being projectors, it is natural to introduce the corresponding parity transformations
\be
    U^S_k = \rme{i\pi \sigma_k^{SS}} \ \text{ and } \ U^D_k = \rme{i\pi \sigma_k^{DD}}.
\ee
These transformations merely change the signs of \emph{some} jump operators:
\begin{align}
    \lt U^D_k \rt^\dag & L_k^D \, U^D_k = \sqrt{\gamma_D} \, \rme{-i \pi \sigma_k^{DD}} \sigma_k^{DI} \rme{i\pi \sigma_k^{DD}} \nol 
    &=  \sqrt{\gamma_D} \, \rme{-i\pi} \, \sigma_k^{DI} = - L_k^D
\end{align}
and
\begin{align}
    \lt U^S_k \rt^\dag & L_{kj}^I \, U^S_k = \sqrt{\gamma_I} \sigma_j^{SS} \, \rme{-i \pi \sigma_k^{SS}} \sigma_k^{IS} \rme{i\pi \sigma_k^{SS}} \nol 
    &=  \sqrt{\gamma_I} \sigma_j^{SS} \, \sigma_k^{DI} \rme{i \pi} = - L_{kj}^I,
\end{align}
whereas
\begin{subequations}
\begin{align}
    &\lt U^D_j \rt^\dag  L_k^D \, U^D_j = L_k^D  & \forall \, j \neq k \\
    &\lt U^S_j \rt^\dag  L_k^D \, U^S_j = L_k^D  & \forall \, j \\
    &\lt U^D_j \rt^\dag  L_{kj'}^I \, U^D_j = L_{kj'}^I & \forall \, j \\
    &\lt U^S_j \rt^\dag  L_{kj'}^I \, U^S_j = L_{kj'}^I & \forall \, j \neq k.
\end{align}
\end{subequations}
where we have exploited the relations \eqref{eqs:sigmas}. Because the Lindblad equation only includes quadratic terms for any jump operator, a mere sign change like the ones above must leave $\mal{L}$ invariant. 

It is worth mentioning that, in the presence of a weak symmetry $U$, a representation of the Lindblad equation always exists in which the Hamiltonian is invariant while the jump operators only pick up a phase \cite{Kasia2021b}. In more mathematical terms, adopting the notation used in Eq.~\eqref{eq:Uinvariance}, if we have a generic Lindblad equation $\mal{L} \lqq H, \set{L_\alpha}_\alpha \rqq$ generated via a Hamiltonian term $H$ and jump operators $L_\alpha$ and if it satisfies
\be
    \mal{L} \lqq H, \set{L_\alpha}_\alpha \rqq = \mal{L} \lqq U^\dag H U, \set{U^\dag L_\alpha U}_\alpha \rqq,
\ee
then there exists a (possibly different) Hamiltonian $H$ and (possibly different) jump operators $L_\alpha$ such that
\be
    \mal{L} \lqq H, \set{L_\alpha}_\alpha \rqq = \mal{L} \lqq H', \set{L_\alpha'}_\alpha \rqq
\ee
and 
\begin{subequations}
\begin{align}
    &U^\dag H' U = H' ,\\
    &U^\dag L_\alpha' U = \rme{i \theta_\alpha} L_\alpha'
\end{align}
\end{subequations}
with $\theta_\alpha \in \R$. In this new representation, a simplification analogous to the one we discussed in Sec.~\ref{subsec:weak_sym} can be introduced, where the effective Hamiltonian leaves eigenspaces of $U$ invariant, whereas jump operators may swap the QJMC dynamics from one to another. The "convenient" representation can be found from the original one up to the diagonalization of a $m\times m$ matrix with $m$ the number of linearly independent jump operators $L_\alpha$. Ref.~\cite{Kasia2021b} contains a constructive proof and a few examples.

\section{Abosrbing states of the eQEP}
\label{app:eQEP_absorbing}

Here we provide our proof that all GEP-like states corresponding to absorbing GEP configurations are indeed absorbing for the eQEP; additionally, we consider whether there are additional absorbing states. 

We recall that absorbing GEP-like configuratiosn are described by factorized vectors $\ket{\psi} = \otimes_k \ket{\mu_k}_k$, where $\mu_k \in \left\{ S, D \right\}$. On such vectors we have, according to the rules \eqref{eqs:sigmas},
\be
    \sigma_k^{\bullet I} \ket{\mu_k}_k = \sigma_k^{\bullet B} \ket{\mu_k}_k = 0,  
\ee
which implies
\be
    H \ket{\psi} = 0 \ \text{, } \ L_k^D \ket{\psi} = 0 \ \text{ and } \ L_{jk}^I \ket{\psi} = 0 
\ee
for every site $k$ or pair of sites $k$, $j$. Thus, a density matrix $\rho = \ket{\psi} \bra{\psi}$ built on any such $\ket{\psi}$ satisfies conditions (i)-(iii) of the definition of open quantum absorbing states provided in Sec.~\ref{subsec:RQEP} and repeated in App.~\ref{App:abs_dark}.

We turn now to whether there could be any other absorbing states, taking for granted that $\gamma_D > 0$. To this end, consider a vector $\ket{\psi}$ and assume conditions (ii) and (iii) of the same definition. We first exploit the latter:
\be
    0 = L_k^D \ket{\psi} = \sqrt{\gamma_D} \, \sigma_k^{DI} \ket{\psi} = \sqrt{\gamma_D} \, \ket{D}_k \brak{I}{k} \left. \!  \psi  \right\rangle,
    \label{eq:LD_annihilation}
\ee
which implies 
\be
    \brak{I}{k} \left. \!  \psi  \right\rangle = 0,
    \label{eq:Ipsi}
\ee
with this denoting the partial scalar product
\be
    \brak{\varphi_1}{1}  \bigl( \ket{\psi_1}_1 \otimes \ket{\psi_2}_2  \bigr) = \brak{\varphi_1}{1} \left. \!  \psi_1  \right\rangle_1 \, \ket{\psi_2}_2.
\ee
Equation \eqref{eq:Ipsi} means that $\ket{\psi}$ cannot have an infected component on any site. Thus, the generic decomposition over the classical basis
\be
    \ket{\psi} = \sum_{\vec{\mu}} c_{\vec{\mu}} \bigotimes_{k=1}^N \ket{\mu_k}_k
    \label{eq:psi_components}
\ee
can now be restricted to a sum over $\vec{\mu}$s with $\mu_k \in \set{S, B, D}$ $\forall k$.  
%In fact, having already accounted for the absorbing GEP-like configurations, we can focus on the $3^N - 2^N$ combinations with at least one bedridden site. 

We also note that Eq.~\eqref{eq:Ipsi} is also a sufficient condition for
\be
    L_{jk}^I \ket{\psi} = 0,
\ee
which we obtain, in a sense, for free. This is nothing else but a reminder of the fact that a site can only die if it is infected; if it is not, however, it cannot infect a neighbor either.

We use now condition (ii), i.e., we request that $\ket{\psi}$ be an eigenvector of the Hamiltonian $H$ for some eigenvalue $\lambda$. For now, we set $\Omega \neq 0$. We recall that the eQEP Hamiltonian \eqref{eq:eQEP_H} is a sum of local terms
\be
    H = \sum_k H_k = \Omega \sum_k \lt \sigma_k^{IB} + \sigma_k^{BI}  \rt
\ee
which connect $\ket{I}_k$ with $\ket{B}_k$. For any $\ket{\psi}$ satisfying \eqref{eq:LD_annihilation} (i.e., without "I"s) we have
\begin{subequations}
\begin{align}
    &\bra{\psi} \sigma_k^{I\bullet}  =  \bracket{\psi}{I}_k \brak{\bullet}{k}  = 0, \label{subeq:psisigma}\\
    &\sigma_k^{\bullet I} \ket{\psi} = \ket{\bullet}_k \brak{I}{k} \left. \! \psi \right\rangle = 0. \label{subeq:sigmapsi}
\end{align}
\end{subequations}
In particular, this means that
\be
    \bra{\psi} \sigma_k^{IB} \ket{\psi} =  \bra{\psi} \sigma_k^{BI} \ket{\psi} = 0,
\ee
which in turn implies $\bra{\psi} H_k \ket{\psi} = 0$ $\forall \, k$ and thus $\bra{\psi} H \ket{\psi} = \lambda = 0$. We thus see that $\ket{\psi}$ must be a kernel vector for the Hamiltonian:
\be
    H\ket{\psi} = \sum_k  H_k \ket{\psi}  = 0.
    \label{eq:Hkernel}
\ee

Let now $k \neq j$ be two different lattice indices. Then, we claim that
\be
    \bra{\psi} H_j H_k \ket{\psi} = 0.
\ee
To prove it, we first notice that Eq.~\eqref{subeq:sigmapsi} implies
\begin{align}
    H_k &\ket{\psi}  = \Omega \, \sigma_k^{IB} \ket{\psi} = \Omega \, \sigma_k^{II} \sigma_k^{IB} \ket{\psi} = \nol
    & = \Omega \, \sigma_k^{II} \bigl( \sigma_k^{IB} + \sigma_k^{BI}  \bigr)  \ket{\psi} = \sigma_k^{II} H_k \ket{\psi}.
    \label{eq:H_k_comm}
\end{align}
Hence, 
\begin{align}
    \bra{\psi} H_j H_k \ket{\psi} &= \bra{\psi} H_j \sigma_k^{II} H_k \ket{\psi} =  \nol
    &= \bra{\psi} \sigma_k^{II} H_j H_k \ket{\psi} = 0,
\end{align}

the last equality coming from Eq.~\eqref{subeq:psisigma} and the previous one from the fact that we can always commute operators with disjoint support (in this case, acting on different sites). With this, we have proved that $H_k \ket{\psi} \perp H_j \ket{\psi}$ $\forall \, k \neq j$; therefore, the central expression in Eq.~\eqref{eq:Hkernel} is a sum of mutually orthogonal vectors. For their sum to vanish, every single one must be zero:
\be
    H_k \ket{\psi} = 0 \ \forall \ k.
\ee
Combining this with Eq.~\eqref{eq:H_k_comm} we find
\be
    0 = \Omega\, \sigma_k^{IB} \ket{\psi} = \Omega \ket{I}_k \brak{B}{k} \left. \! \psi \right\rangle \ \forall \,\, k,    
\ee
i.e.
\be
    \brak{B}{k} \left. \! \psi \right\rangle \ \forall \,\, k,
\ee
which places a further restriction on the non-zero components of $\ket{\psi}$, which can be thus written as in Eq.~\eqref{eq:psi_components} but with $\mu_k \in \set{S, D}$ $\forall k$. This identifies again the GEP-like absorbing configurations made up of only susceptible and dead sites.

This leaves us with just one remaining case, $\Omega = 0$; in the absence of a Hamiltonian term, the second restriction we have worked out above does not apply and we are left again with vectors as described in (and below) Eq.~\eqref{eq:psi_components}. In other words, any configuration of S, B, and D sites is absorbing. However, without a Hamiltonian $\ket{B}$ is completely decoupled from the other three states and thus, if one takes an initial state without B sites, none will appear in the course of the evolution. This is in particular true for our chosen initial state, meaning that, effectively, for $\Omega = 0$ we can forget about the existence of the bedridden component. This is discussed in more detail in App.~\ref{app:eQEP_to_GEP}, where we show that the case $\Omega = 0$ corresponds to simulating the GEP.

\section{Diagonalization of $h_k$}
\label{app:eigen}

For the reader's convenience we report below $h_k$ in matrix form (Eq.~\eqref{eq:hk_matrix}) in the $\ket{I}_k$, $\ket{B}_k$ basis: 
\be
    h_k = \matb{cc} - i \frac{\gamma_{eff}}{2} & \Omega \\ \Omega & 0    \mate.
\ee
For simplicity, since we focus on a single site and the spatial dependence only enters through $\gamma_{eff} = \gamma_D + \mal{N}_k^S \gamma_I$, for the rest of this Appendix we drop the lattice index $k$.

The eigenvalues of the matrix above (hereafter, $h$) are
\be
    \lambda_{\pm} = -i\frac{\gamma_{eff}}{4} \pm \sqrt{ \Omega^2 - \lt \frac{\gamma_{eff}}{4} \rt^2 }.
    \label{eq:eigenvalues}
\ee
The matrix not being Hermitian, we need to distinguish between right and left eigenvectors:
\be
    h \ket{r_\pm} = \lambda_\pm \ket{r_\pm} , \ \ \bra{l_\pm} h = \lambda_\pm \bra{l_\pm}, 
\ee
with the standard orthonormality relations
\be
    \bracket{l_\pm}{r_\pm} = 1 \ \text{ and } \ \bracket{l_\mp}{r_\pm} = 0,
\ee
with consistently chosen signs in each scalar product. They can be individually expressed as
\begin{subequations}
\begin{align}
    &\ket{r_\pm} = \pm \frac{1}{2\sqrt{ \Omega^2 - \lt \frac{\gamma_{eff}}{4} \rt^2 }} \matb{c} 1 \\ \frac{\Omega}{\lambda_\pm} \mate, \label{subeq:r_pm}\\
    &\ket{l_\pm} = \matb{c}  \lambda_{\pm}^\ast \\ \Omega \mate.
\end{align}
\end{subequations}
We see from Eq.~\eqref{subeq:r_pm} that the norm of $\ket{r_\pm}$ diverges when $\Omega = \gamma_{eff} / 4$, whereas $\ket{\l_\pm}$ remain finite. This highlights the presence of an exceptional point at which $\lambda_+ = \lambda_-$ and $h$ becomes non-diagonalizable. In principle, this is not a problem, since the evolution under one such $h$ just includes a secular term
\be
    \rme{-iht} \ket{\psi} = \rme{i \lambda_+ t} \lt \ket{\psi}  -it \ket{\phi}  \rt
\ee
for some $\ket{\phi}$ independent from $\ket{\psi}$. However, in our algorithm we opted to avoid these exceptional points altogether, since we can in any case get arbitrarily close to one. Hence, we have consistently used the decomposition
\begin{align}
    \ket{\psi(t)} &= \rme{-iht} \ket{\psi} = \nol
    & = \rme{-i\lambda_+ t} \bracket{l_+}{\psi} \ket{r_+} + \rme{-i\lambda_- t} \bracket{l_-}{\psi} \ket{r_-} .  
    \label{eq:eigendecomposition}
\end{align}

We use for a generic vector $\ket{\psi}$ the same parametrization introduced in the main text: 
\be
    \ket{\psi} = \ket{a,\, b} = a \ket{I} + ib\ket{B} = \matb{c} a \\ ib \mate,
\ee
which yields, for the right eigenvector components, 
\be
    \bracket{l_\pm}{\psi} = \lambda_\pm a + i\Omega b.
    \label{eq:rev_component}
\ee
Now, $\partial_t \ket{\psi(t)} = -ih \ket{\psi(t)}$ yields the system of equations
\be
    \sysb{l} \dot{a} = -\frac{\gamma_{eff}}{2} a + \Omega b   \\[0.2cm]
            \dot{b} = - \Omega a   \syse
    \label{eq:eq_of_motion}
\ee
which shows that we can work with $a$, $b \in \R$. 

We recall that the local probability of not jumping up to time $t$ is given, in our current notation, by
\be
    \mathbb{P}(TFJ \geq t | \ket{a,\,b}, 0) = \norm{\psi(t)}^2 = a(t)^2 + b(t)^2 ;
\ee
we can then apply Eqs.~\eqref{eq:eq_of_motion} to derive
\be
    p_{TFJ} \lt t | \ket{a,\, b}, 0 \rt = -\partial_t \lt a(t)^2 + b(t)^2  \rt = \gamma_{eff} \, a(t)^2,
\ee
which shows that when $a(t) = 0$ the probability distribution of jump times vanishes. Thus, no jump will ever be selected at a time when $a = 0$, which proves that in Eq.~\eqref{eq:qI} we are justified in simplifying the factors $\abs{a}^2$ appearing at the numerator and the denominator.

For step (IV) of the eQEP algorithm in Sec.~\ref{subsec:eQEP_algo} we need to find analytical expressions for $a$ and $b$. For this purpose, let us adopt the same notation (up to the reintroduction of the site index $k$): say that, at a certain time $t_z$, we have a vector $\ket{\psi^z} = \ket{a(t_z), \, b(t_z)}$ where $a(t_z)^2 + b(t_z)^2 = 1$. Then, we define
\be
    \ket{a(t_{z+1}), \, b(t_{z+1})} = \frac{\rme{-ih \lt t_{z+1} - t_z \rt} \ket{a(t_z), \, b(t_z)}}{ \norm{\rme{-ih \lt t_{z+1} - t_z \rt} \ket{a(t_z), \, b(t_z)}} }.
\ee
First, from Eq.~\eqref{eq:eigendecomposition} we can extract the yet unnormalized
\begin{subequations}
\label{eqs:unnormalized_ab}
\begin{align}
    a(t_{z+1}) &= \bracket{I}{\psi(t_{z+1})} = \nol
    & = \frac{\rme{-i\lambda_+ \delta t_z} \bracket{l_+}{\psi^z} - \rme{-i\lambda_- \delta t_z} \bracket{l_-}{\psi^z}}{2 \sqrt{ \Omega^2 - \lt \frac{\gamma_{eff}}{4} \rt^2 }  } , \\
    ib(t_{z+1}) &= \bracket{B}{\psi(t_{z+1})} = \nol
    & = \frac{\rme{-i\lambda_+ \delta t_z} \bracket{l_+}{\psi^z} / \lambda_+ - \rme{-i\lambda_- \delta t_z} \bracket{l_-}{\psi^z} / \lambda_-  }{2 \sqrt{ 1 - \lt \frac{\gamma_{eff}}{4\Omega} \rt^2 }  },
\end{align}
\end{subequations}
with $\delta t_z = t_{z+1}- t_z$ the time difference and the right eigenvector components calculated as in Eq.~\eqref{eq:rev_component}. Once numerical values for $a(t_{z+1})$ and $b(t_{z+1})$ have been extracted, normalizing the vector trivially reduces to the elementary operation
\be
    \sysb{l}  a(t_{z+1}) \to \frac{a(t_{z+1})}{\sqrt{a(t_{z+1})^2 + b(t_{z+1})^2}} , \\
    b(t_{z+1}) \to \frac{b(t_{z+1})}{\sqrt{a(t_{z+1})^2 + b(t_{z+1})^2}} . \syse
    \label{eq:being_normalized}
\ee

We now switch to the probability $\mathbb{P}(TFJ \geq t | \ket{\psi}, 0)$; for it, we go back to the eigenvector decomposition \eqref{eq:eigendecomposition}:
\begin{align}
    \mathbb{P}& (TFJ \geq t | \ket{\psi}, 0) = \nol
    &= \norm{\rme{-i\lambda_+ t} \bracket{l_+}{\psi} \ket{r_+} + \rme{-i\lambda_- t} \bracket{l_-}{\psi} \ket{r_-}}^2 = \nol
    & = \rme{i \lt \lambda_+^\ast - \lambda_+ \rt t}\, \abs{\bracket{l_+}{\psi}}^2 \,\bracket{r_+}{r_+} + \nol
    & + \rme{i \lt \lambda_-^\ast - \lambda_- \rt t}\, \abs{\bracket{l_-}{\psi}}^2 \,\bracket{r_-}{r_-} + \nol
    & + \rme{i \lt \lambda_+^\ast - \lambda_- \rt t}\, \bracket{\psi}{l_+} \bracket{l_-}{\psi} \,\bracket{r_+}{r_-} + \nol
    & + \rme{i \lt \lambda_-^\ast - \lambda_+ \rt t}\, \bracket{\psi}{l_-} \bracket{l_+}{\psi} \,\bracket{r_-}{r_+},
\end{align}
with the scalar products
\begin{subequations}
\label{eqs:scalar_prod}
\begin{align}
    % &\bracket{l_\pm}{\psi} = \lambda_\pm a + i\Omega b \\
    &\bracket{r_\pm}{r_\pm} = \frac{1}{\abs{4 \Omega^2 - \gamma_{eff}^2}} \lt 1 + \lt \frac{\Omega}{\abs{\lambda_{\pm}}} \rt^2 \rt \\
    &\bracket{r_\mp}{r_\pm} = - \frac{1}{\abs{4 \Omega^2 - \gamma_{eff}^2}} \lt 1 +  \frac{\Omega^2}{\lambda_{\mp}^\ast \lambda_{\pm}}  \rt.
\end{align}
\end{subequations}
Some further simplifications can be made, but require choosing which side of the exceptional point we are considering, i.e., which prevails between $\Omega$ and $\gamma_{eff} / 4$.

\subsection{Case $\Omega < \gamma_{eff} / 4$}
\label{subapp:small_Om}

In this case the square root in Eq.~\eqref{eq:eigenvalues} is imaginary:
\be
    \sqrt{\Omega^2 - \lt \frac{\gamma_{eff}}{4} \rt^2} = -i \sqrt{\lt \frac{\gamma_{eff}}{4} \rt^2 - \Omega^2} \equiv -i \Delta,
\ee
where without loss of generality we fix $\Delta > 0$.
The eigenvalues, too, are purely imaginary numbers $\lambda_\pm^\ast = - \lambda_{\pm} \in i\R$. Defining $\lambda_\pm = -i \eta_\pm$ with 
\be
    \eta_\pm = \frac{\gamma_{eff}}{4} \pm \Delta > 0
\ee
we find
\be
    i \lt \lambda_\pm^\ast - \lambda_\pm \rt = - 2\eta_\pm = -\frac{\gamma_{eff}}{2} \mp 2\Delta
\ee
and
\be
    i  \lt \lambda_\mp^\ast - \lambda_\pm \rt = -\eta_{\mp} - \eta_\pm = -\frac{\gamma_{eff}}{2}.
\ee

The right eigenvector components are now also purely imaginary
\be
   \bracket{l_\pm}{\psi} = i \lt \Omega b - \eta_\pm a \rt  \in i\R 
\ee
while the remaining cross products \eqref{eqs:scalar_prod} become
\begin{subequations}
\begin{align}
    &\bracket{r_\pm}{r_\pm} = \frac{\gamma}{8\Delta^2 \eta_{\pm}} \\
    &\bracket{r_\mp}{r_\pm} = - \frac{1}{2\Delta^2}.
\end{align}
\end{subequations}
Overall, the no-jump probability takes the form of a sum of three decaying exponential terms:
\begin{align}
    \mathbb{P}& (TFJ \geq t | \ket{\psi}, 0) = \nol[0.2cm]
    &= A_+ \rme{-\lt \frac{\gamma_{eff}}{2} + 2\Delta  \rt t} + 2A_0 \, \rme{-\frac{\gamma_{eff}}{2} t} + A_- \rme{-\lt \frac{\gamma_{eff}}{2} - 2\Delta  \rt t}
    \label{eq:three_exp}
\end{align}
with real coefficients
\begin{subequations}
\begin{align}
    A_\pm & = \abs{\bracket{l_\pm}{\psi}}^2 \,\bracket{r_\pm}{r_\pm} = \nol
    &=  \frac{\gamma_{eff} \lt \Omega b - \eta_\pm a  \rt^2}{8\Delta^2 \eta_\pm}  \geq 0   \\[0.2cm]
    A_0 & = \bracket{\psi}{l_\mp} \bracket{l_\pm}{\psi} \,\bracket{r_\mp}{r_\pm} = \nol
    & = -\frac{\Omega}{2\Delta^2} \lt \Omega \lt a^2 + b^2 \rt - \frac{\gamma_{eff}}{2} ab  \rt = \nol
    & = -\frac{\Omega}{2\Delta^2} \lt \Omega  - \frac{\gamma_{eff}}{2} ab  \rt
\end{align}
\end{subequations}
Note that $A_0$ is not guaranteed to be positive: immediately after the local site has been infected we have $a = 1$, $b = 0$ and $A_0 = - \Omega^2 / 2\Delta^2 < 0$. This means that we cannot interpret Eq.~\eqref{eq:three_exp} as a combination of three independent stochastic processes with three different rates. We have therefore chosen to solve the equation $\mathbb{P} \lt \cdots \rt = u$ numerically by bisection.

The yet unnormalized components $a(t_{z+1})$, $b(t_{z+1})$ in Eqs.~\eqref{eqs:unnormalized_ab} can be expressed as
\begin{subequations}
\begin{align}
    a(t_{z+1}) &= \frac{\rme{-\eta_- \delta t_z} \mal{E}_-(t_z) - \rme{-\eta_+ \delta t_z} \mal{E}_+(t_z)}{2 \Delta  } , \\[0.15cm]
    b(t_{z+1}) & = \frac{\Omega}{2\Delta} \lqq \frac{\rme{-\eta_- \delta t_z} \mal{E}_-(t_z)}{\eta_-} - \frac{\rme{-\eta_+ \delta t_z} \mal{E}_+(t_z)}{\eta_+} \rqq,
\end{align}
\end{subequations}
with shorthands
\begin{subequations}
\begin{align}
    \delta t_z &= t_{z+1} - t_z, \\
    \mal{E}_\pm (t_z) &= \Omega b(t_z) - \eta_\pm a(t_z).
\end{align}
\end{subequations}
%To obtain their normalized counterparts, Eq.~\eqref{eq:being_normalized}, we divide each by the expression in Eq.~\eqref{eq:three_exp}.

\subsection{Case $\Omega > \gamma_{eff} / 4$}
\label{subapp:large_Om}

When $\Omega$ prevails, the square root
\be
    \sqrt{\Omega^2 - \lt \frac{\gamma_{eff}}{4} \rt^2 } = \Phi > 0
\ee
is real and the eigenvalues obey the relations $\lambda_{\pm}^\ast = - \lambda_{\mp}$ and $\abs{\lambda_{\pm}} = \Omega$. The exponents in Eq.~\eqref{eq:eigendecomposition} can now be written as
\be
    i \lt \lambda_\pm^\ast - \lambda_\pm \rt =  -\frac{\gamma_{eff}}{2} 
\ee
and
\be
    i  \lt \lambda_\mp^\ast - \lambda_\pm \rt = -2i\lambda_{\pm} = -\frac{\gamma_{eff}}{2} \mp 2i\Phi.
\ee
The right eigenvector components are
\be
   \bracket{l_\pm}{\psi} = \pm a \Phi + i \lt \Omega b - \frac{\gamma_{eff}}{4} a \rt .  
\ee
In this regime the cross-products \eqref{eqs:scalar_prod} take the form
\begin{subequations}
\begin{align}
    &\bracket{r_\pm}{r_\pm} = \frac{1}{2\Phi^2} \\
    &\bracket{r_\mp}{r_\pm} =  \frac{i\gamma_{eff} }{8\Phi^2 \lambda_{\pm}}.
\end{align}
\end{subequations}
and the no-jump probability can be seen as the product between a single exponential decay and an oscillating part
\begin{align}
    \mathbb{P}& (TFJ \geq t | \ket{\psi}, 0) = \nol[0.2cm]
    &= \rme{-\frac{\gamma_{eff}}{2} t} \bigl[ 2 \mal{A}_0 + \mal{A}_C \cosa{ 2\Phi t} + \mal{A}_S \sina{ 2\Phi t}  \bigr]
    \label{eq:exp_oscill}
\end{align}
with real coefficients
\begin{subequations}
\begin{align}
    \mal{A}_0 & = \ha\abs{\bracket{l_+}{\psi}}^2 \,\bracket{r_+}{r_+} + \ha \abs{\bracket{l_-}{\psi}}^2 \,\bracket{r_-}{r_-} = \nol
    &=  \frac{\Omega}{2\Phi^2}   \lt \Omega \lt a^2 + b^2 \rt  - \frac{\gamma_{eff}}{2}ab \rt  = \nol 
    & = \frac{\Omega}{2\Phi^2}   \lt \Omega   - \frac{\gamma_{eff}}{2}ab \rt \geq 0, \\[0.2cm]
    \mal{A}_C & = 2\,\mathrm{Re} \, \bigl( \bracket{\psi}{l_+} \bracket{l_-}{\psi} \,\bracket{r_+}{r_-} \bigr) = \nol
    & = -\frac{\gamma_{eff}}{4\Phi^2} \lt \frac{\gamma_{eff}}{4} \lt a^2 + b^2 \rt - 2 \Omega ab  \rt = \nol
    & = \frac{\gamma_{eff}}{4\Phi^2} \lt 2 \Omega ab -  \frac{\gamma_{eff}}{4}  \rt, \\[0.2cm]
    \mal{A}_S & = -2\,\mathrm{Im}\, \bigl( \bracket{\psi}{l_+} \bracket{l_-}{\psi} \,\bracket{r_+}{r_-} \bigr) = \nol
    & = \frac{\gamma_{eff}}{4\Phi} \lt b^2  - a^2  \rt.
\end{align}
\end{subequations}

As for the unnormalized components $a(t_{z+1})$, $b(t_{z+1})$ in Eqs.~\eqref{eqs:unnormalized_ab}, we can express them as
\begin{subequations}
\label{eqs:oscillating_components}
\begin{align}
    a(t_{z+1}) &= \rme{-\frac{\gamma_{eff}}{4} \delta t_z}  \lqq  a(t_z) \cosa{\Phi \delta t_z} + \frac{\mal{F} (t_z)}{\Phi} \sina{\Phi \delta t_z} \rqq, \\
    b(t_{z+1}) & = \rme{-\frac{\gamma_{eff}}{4} \delta t_z}  \lqq  b(t_z) \cosa{\Phi \delta t_z} + \frac{\widetilde{\mal{F}} (t_z)}{\Phi} \sina{\Phi \delta t_z} \rqq,
\end{align}
\end{subequations}
with shorthands
\begin{subequations}
\begin{align}
    \delta t_z & = t_{z+1} - t_z, \\[0.15cm]
    \mal{F} (t_z) &= \Omega b(t_z) - \frac{\gamma_{eff}}{4} a(t_z), \\[0.15cm]
    \widetilde{\mal{F}} (t_z) &= \frac{\gamma_{eff}}{4} b(t_z) - \Omega a(t_z).
\end{align}
\end{subequations}
%In order to normalize them, one has to divide each by the expression in Eq.~\eqref{eq:exp_oscill}.

\subsection{Case $\Omega \gg \gamma_{eff}/4$}
\label{subapp:huge_Omega}

As mentioned in the main text, in the $\Omega \gg \gamma_{eff} / 4$ oscillations between GEP-like states $\ket{I}$ and $\ket{B}$ is are much faster than the dissipative processes driving infection and death. At intermediate timescales we can thus approximate our expressions with their time averages over a period $T = 2\pi / \Phi$, for instance
\begin{align}
    \mathbb{P}_{T}& (TFJ \geq t | \ket{\psi}, 0) =  \nol[0.2cm] = & \frac{1}{T} \int_{t - T/2}^{t + T/2} \rmd t'\,  \mathbb{P} (TFJ \geq  t' | \ket{\psi}, 0) .
\end{align}
The time integration of Eq.~\eqref{eq:exp_oscill} has the effect of eliminating the oscillating terms, whereas the exponential $\rme{-\gamma_{eff} t / 2}$, which by assumption varies little over a period, can be, with reasonable approximation, extracted from the integral, yielding
\be
    \mathbb{P}_{T} (TFJ \geq t | \ket{\psi}, 0) \approx 2 \mal{A}_0 \rme{-\frac{\gamma_{eff}}{2} t}. 
\ee
Additionally, since $\Phi \approx \Omega$ and $\Omega \gg \gamma_{eff} / 4 \geq \gamma_{eff}/ (2ab)$ we can approximate $\mal{A}_0 \approx 1$, which yields
\be
    \mathbb{P}_{T} (TFJ \geq t | \ket{\psi}, 0) \approx  \rme{-\frac{\gamma_{eff}}{2} t}. 
\ee
The expression above is identical to the no-jump probability of a purely classical process which jumps with rates $\gamma_I / 2$ (infection) and $\gamma_D / 2$ (death), i.e., a GEP which evolves twice as slow as the original rates would suggest.

Similarly, we can derive average occupations of states $\ket{I}$ and $\ket{B}$ by time averaging over the normalized squared vector components $a^2 / (a^2 + b^2)$ and $b^2 / (a^2 + b^2)$. It is convenient to rewrite here their expressions \eqref{eqs:oscillating_components} as
\begin{subequations}
\label{eqs:oscillating_components2}
\begin{align}
    a(t_z + t) &=  A_{amp} \, \rme{-\frac{\gamma_{eff}}{4} t} \cosa{\Phi t + \theta_a} , \\
    b(t_{z} + t) & = A_{amp} \, \rme{-\frac{\gamma_{eff}}{4} t} \sina{\Phi t + \theta_b},
\end{align}
\end{subequations}
with common amplitude
\begin{align}
    A_{amp}^2 & = a(t_z)^2 + \lt \frac{\mal{F}(t_z)}{\Phi} \rt^2 = b(t_z)^2 + \lt \frac{\widetilde{\mal{F}}(t_z)}{\Phi} \rt^2 = \nol
    & =  \frac{\Omega}{\Phi^2} \lt \Omega - \frac{\gamma_{eff}}{2} a(t_z) b(t_z)  \rt
\end{align}
and angles $\theta_a$, $\theta_b$ obeying
\begin{subequations}    
\begin{align}
    &\cosa{\theta_a} = \frac{a(t_z)}{A_{amp}}, \ \ &\sin{\theta_a} = - \frac{\mal{F}(t_z)}{\Phi A_{amp}}, \\[0.2cm]
    &\cosa{\theta_b} = \frac{\widetilde{\mal{F}}(t_z)}{\Phi A_{amp}}, \ \ &\sin{\theta_b} = \frac{b(t_z)}{A_{amp}}.
\end{align}
\end{subequations}
Introducing $\Delta \theta = \theta_b - \theta_a$ we find
\be
    \sina{\Delta \theta} = \frac{\gamma_{eff}}{4\Omega} \approx 0, 
\ee
which shows that for very large frequencies the relative phase $\Delta \theta$ vanishes; therefore, $a^2 + b^2 \approx  A_{amp} \, \rme{-\gamma_{eff}t / 4}$ and
\be
    \lt \frac{a^2}{a^2 + b^2} \rt_T \approx \frac{1}{T} \int_{t - T/2}^{t + T/2} \rmd t' \cossa{2}{\Phi t + \theta_a} = \ha
\ee
and the same goes for $b$ \footnote{In fact, this is more general and there is no need for approximation: for $\Omega > \gamma_{eff}/4$ the time averaging of $a^2 / (a^2 + b^2)$ invariably yields \emph{exactly} $1/2$.}. Hence, we can reasonably expect the average infected density $\overline{\av{\sigma^{II}(t)}}$ to be half of that we would expect from the relative GEP process at $\Omega \approx 0$, justifying our rescaling in Eq.~\eqref{eq:II_rescaling}.

\section{Local jump times: our extraction compared with the original survival probability}
\label{app:not_indep}

\begin{figure}[h]
  \includegraphics[width=\columnwidth]{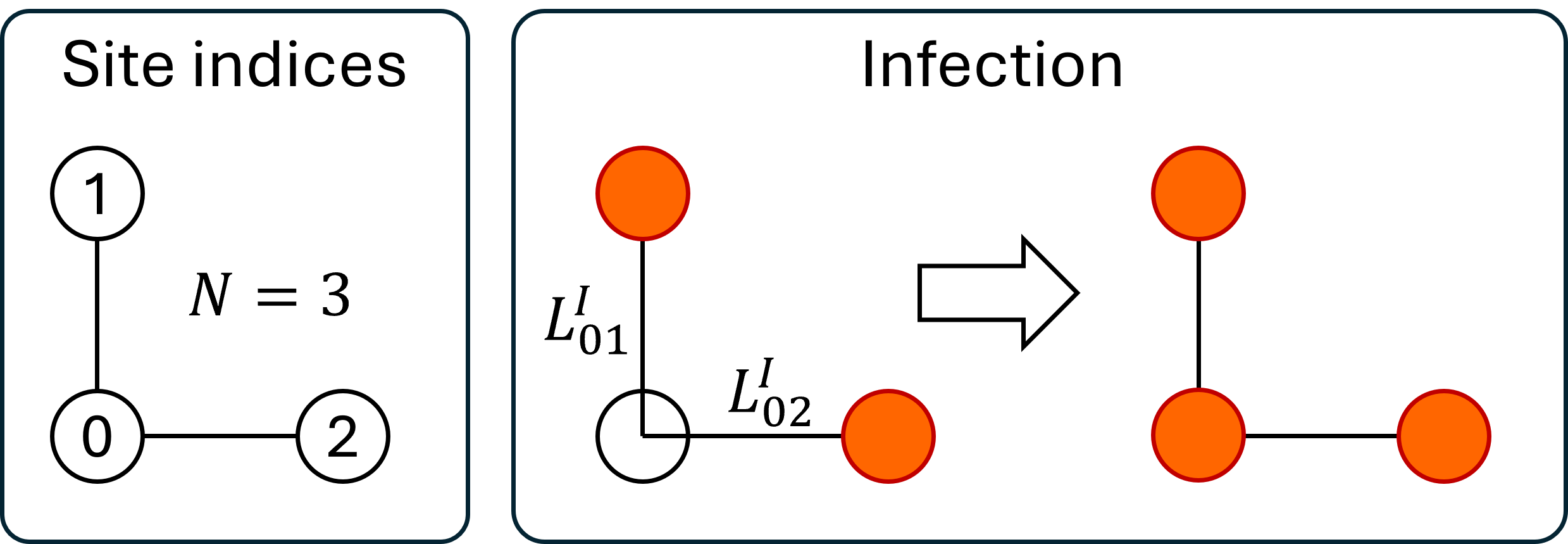}
  \caption{Example of a tiny $N = 3$ system with sites numbered as in the left panel. Setting for simplicity $\gamma_D = 0$ (no deaths), the first configuration in the right panel only admits two jumps, $L_{01}^I$ and $L_{01}^I$, i.e., infection of $0$ by $1$ or by $2$. Either jump transforms the configuration into the second one, but it is impossible for both to take place. }
\label{fig:dep_ex}
\end{figure}

In Sec.~\ref{subsec:decoupling}, we have seen that, thanks to the symmetries of $H_{eff}$, the no-jump probability $\mathbb{P} \lt TFJ \geq t | \ket{\psi}, t' \rt$ factorizes into local terms (see Eq.~\eqref{eq:Pfact}):
\begin{align}
    \mathbb{P} &\bigl( TFJ \geq t  \,\,|\, \ket{\psi}, t' \bigr)  = \norm{\rme{-ih (t-t')} \bigl( \otimes_k \ket{\psi_k}_k \bigr) }^2 \nol 
    & = \prod_k \norm{\rme{-ih_k (t-t')} \ket{\psi_k}_k}^2 \equiv \prod_k P_k (t).
    \label{eq:P_part_1}
\end{align}
We have then proceeded to treat the r.h.s.~as a product of probabilities $P_k(t)$ of independent events. Now, clearly jump times at different sites cannot be independent. To see this, consider the example in Fig.~\ref{fig:dep_ex}.
%    & = \prod_k \mathbb{P} \bigl( TFJ_k \geq t  \,\,|\, \ket{\psi_k}_k, t' \bigr) \equiv \prod_k P_k(t)
%
%
We take a tiny ($N=3$) system with sites numbered from $0$ to $2$ as shown in the left panel. We set for simplicity $\gamma_D = 0$ (no death jumps) and assume the configuration at $t = 0$ to be as the first one in the right panel, i.e., two infected sites connected to a middle, susceptible one. Then, there are only two jumps that can take place: infection of site $0$ due to site $1$, or due to site $2$, each with rate $\gamma_I$. Following our method, and in particular step (II) of Sec.~\ref{subsec:eQEP_algo}, we calculate two infection times, $t_1$ and $t_2$ respectively. Say, for instance, that $t_1 < t_2$; then $t_1$ is selected and site $1$ infects site $0$. After that, no other jump can take place, as there are no S sites remaining. Thus, $L^I_{02}$ will never act and the value $t_2$ holds no significance for the behavior of the system. 
%Note that the same considerations apply to the GEP.

Another way of approaching the problem is to recognize that the local probabilities are, effectively, conditioned not only on the time $t'$ and the QJMC vector $\ket{\psi_k}_k$ at $t'$, but also on $h_k$:
\be
    P_k(t) = P_k\lt t \,|\, \ket{\psi_k}, t', h_k  \rt.
\ee
We recall that $h_k$, as can be gleaned from Eq.~\eqref{eq:hk} or \eqref{eq:hk_matrix}, depends on $\gamma_{eff}$ and, in turn, on the number $\mal{N}^S_k$ of susceptible neighbors. Every infection changes that number for some (up to three) other sites. In doing so, it modifies their $h_k$, i.e.~one of the parameters the probability on those sites is conditioned upon. In the example of Fig.~\ref{fig:dep_ex}, the infection at time $t_1$ causes the number of susceptible neighbors of site $2$ to vanish:
\be
    \mal{N}_2^S = 1 \text{ at $t_1^-$} \to \mal{N}_2^S = 0 \text{ at $t_1^+$}
\ee
and the functional form of $P_2(t)$ therefore changes to $P_2(t) \equiv 1$ $\forall \, t \geq t_1$.

In what sense, then, can we say that events on different sites can be treated as independent? As a matter of fact, we are not making any physically meaningful statements on the nature of the system. Rather, we are introducing a mathematical trick that proves useful for our treatment, a trick that we will attempt to describe as making "copies", or "replicas" of the system. 

Remember that the actual aim is to extract a jump time from the global $\mathbb{P}$, corresponding to inverting the probability, i.e., solving $\mathbb{P} = u$ for a uniformly distributed $u$ over $\lqq 0,\, 1\rqq$. Whenever we reach step (II) of our algorithm (see Sec.~\ref{subsec:eQEP_algo}) we imagine that there are $N$ replicas of the system, each with an identical configuration. In the $k$-th copy, however, all jump operators but those generated by site $k$ are artificially set to $0$; by this we mean that only the jump operators $L_k^D$ and $L_{jk}^I$ are kept, corresponding to site $k$ dying or infecting one of its $\mal{N}_k^S$ susceptible neighbors. The same goes for the Hamiltonian sum: only the $k$-th addend is retained.

In this replica framework, we can extract $N$ jump times, one from each copy. As a matter of fact, a portion of these times can be easily discarded: if site $k$ is either susceptible or dead, the $k$-th replica is inactive and will never jump. Thus, we need only replicas corresponding to sites in a superposition of $\ket{I}_k$ and $\ket{B}_k$.

For the $k$-th replica the no-jump probability is, by construction,
\be
    \mathbb{P} \bigl( TFJ_k \geq t  \,\,|\, \ket{\psi_k}_k, t' \bigr) = \norm{\rme{-ih_k (t-t')} \ket{\psi_k}_k}^2
    \label{eq:P_part_2}
\ee
since $h_k$ is exactly the combination of all surviving terms for site $k$. By comparing Eqs.~\eqref{eq:P_part_1} and \eqref{eq:P_part_2} we have
\be
    \mathbb{P} \bigl( TFJ \geq t  \,\,|\, \ket{\psi}, t' \bigr) = \prod_k \mathbb{P} \bigl( TFJ_k \geq t  \,\,|\, \ket{\psi_k}_k, t' \bigr).
    \label{eq:P_part_12}
\ee
Since the random variables $TFJ_k$ are statistically independent (as they pertain to different replicas), we can recognize in the product on the r.h.s.~of Eq.~\eqref{eq:P_part_12} the probability of all $TFJ_k$ exceeding $t$ while conditioned on the entire vector $\ket{\psi}$:
\begin{align}
    \prod_k & \,\mathbb{P} \bigl( TFJ_k \geq t  \,\,|\, \ket{\psi_k}_k, t' \bigr) = \nol
     &= \mathbb{P} \Bigl( TFJ_k \geq t   \,\, \forall \, k\, \Bigr| \Bigl.\, \ket{\psi}, t' \Bigr).
\end{align}
As observed in the main text, asking that all $TFJ_k \geq t$ is equivalent to asking that $\min_k \left\{ TFJ_k \right\} \geq t$. Hence,
\be
    \mathbb{P} \bigl( TFJ \geq t  \,\,|\, \ket{\psi}, t' \bigr) = \mathbb{P} \bigl( \min_k\left\{ TFJ_k \right\} \geq t  \,\,|\, \ket{\psi}, t' \bigr),
    \label{eq:partial_result}
\ee
which proves that $TFJ$ and $\min_k\left\{ TFJ_k \right\}$ are equally distributed. Extracting a jump time from the latter yields values statistically equivalent to those that could be obtained from the former.

The attentive reader may object that Eq.~\eqref{eq:partial_result} is still not enough: when extracting the smallest $t_k^{(z+1)}$ in step (III) of our algorithm (see Sec.~\ref{subsec:eQEP_algo}) we also find out the site $\bar{k}$ at which the jump takes place; in the ordinary QJMC procedure (see Sec.~\ref{subsec:QJMC_algorithm}) the time is determined first, whereas the type (which for all models discussed in this work includes the position) is worked out in a separate step. 

To show that our approach is nonetheless valid, we start by stating that, as long as we look exclusively at jump times, Eq.~\eqref{eq:partial_result} ensures that values extracted via our method and values found by inverting the $\mathbb{P}$ at the l.h.s.~are identically distributed; for jump time estimation, the two methods are equivalent.

Then, the question becomes: does this equivalence hold when the determination of the position is taken into account? To answer this, we wish to compare the probability distributions of jumping at time $t$ and site $\bar{k}$ for both methods, starting from the "traditional" one. To avoid confusion, we need to introduce some new notation: let $\mathscr{A}$ be the set of all jump types, so that, for instance, the dissipator in the Lindblad equation \eqref{eq:Lindblad} can be written as
\be
    \sum_{\alpha \in \mathscr{A}} \lqq L_\alpha \rho L_\alpha^\dag - \ha \acomm{L_\alpha^\dag L_\alpha}{\rho(t)} \rqq.
\ee
Additionally, we call $\mathscr{A}_k \subseteq \mathscr{A}$ the subset of jump types pertaining to site $k$, i.e., the subset of jumps included in the $k$-th replica. Clearly, $\mathscr{A}_k \cap \mathscr{A}_j = \emptyset$ if $j \neq k$ and $\bigcup_k \mathscr{A}_k = \mathscr{A}$, which also implies
\be
    \sum_k \sum_{\alpha \in \mathscr{A}_k} = \sum_{\alpha \in \mathscr{A}} \equiv \sum_\alpha.
    \label{eq:sums_equiv}
\ee

\subsection{Distribution of jump times and positions with the standard algorithm}
\label{subapp:with_standard}

Working within the traditional framework, let $t_z$ be the current time, corresponding to a vector $\ket{\psi^z}$. Then, the distribution of jump times and types is, as we have seen,
\be
    p \lt t, \alpha \,|\, \ket{\psi^z}, t_z  \rt \equiv q_{\alpha}  \, p_{TFJ} \bigl( t \,|\, \ket{\psi^z}, t_z \bigr),
\ee
where 
\begin{align}
    q_{\alpha}  =  q_{\alpha} \bigl( \rme{-iH_{eff}\lt t - t_z \rt}\ket{\psi^z} \bigr)  
     = q_{\alpha} \bigl( t \,| \ket{\psi^z}, t_z \bigr)
\end{align}
is the probability of choosing jump type $\alpha$ introduced in step (II), Sec.~\ref{subsec:QJMC_algorithm}; for completeness, we have displayed here its dependence on the new jump time $t$, showing it has the same conditional structure of $p_{TFJ}$ and can be combined with it in a single object $p$ without issue. Now it is only a matter of calculating the corresponding aggregate probability of choosing a jump type among those relative to site $\bar{k}$:
\begin{align}
    d & \lt t, \bar{k} \,|\, \ket{\psi^z}, t_z  \rt = \sum_{\alpha \in \mathscr{A}_{\bar{k}}} p \lt t, \alpha \,|\, \ket{\psi^z}, t_z  \rt = \nol
    & = \lqq \sum_{\alpha \in \mathscr{A}_{\bar{k}}} q_{\alpha}  \rqq p_{TFJ} \bigl( t \,|\, \ket{\psi^z}, t_z \bigr)   .
    \label{eq:d_traditional}
\end{align}
This $d$ is the distribution of interest for us insofar as the traditional algorithm is concerned.

\subsection{Distribution of jump times and positions with our algorithm}
\label{subapp:with_ours}

We now switch to our own method. For brevity, we introduce the shorthands
\be
    p_k(t) \equiv p_{TFJ_k} \bigl( t_{k} \,|\, \ket{\psi^z_k}_k, \, t_z , \, h_k \bigr)
\ee
for the distributions relative to the individual replicas.
In order for site $\bar{k}$ to be selected, the time $t_{\bar{k}}$ must be the smallest one of all the $t_k$s. The probability to have simultaneously $\min_k \set{TFJ_k} \geq t$ and $TFJ_{\bar{k}} = \min_k \set{TFJ_k}$ can be expressed as
\begin{align}
    \mathbb{P} &\lt TFJ_k \geq TFJ_{\bar{k}} \geq t \,\, \forall k \,|\, \ket{\psi^z}, t_z  \rt = \nol
    & = \int_t^{\infty}  \rmd t_{\bar{k}} \,\, p_{\bar{k}} \lt t_{\bar{k}} \rt \, \prod_{k \neq \bar{k}}\, \int_{t_{\bar{k}}}^{\infty} \rmd t_k \,\, p_{k} \lt t_{k}  \rt
\end{align}
The distribution $\tilde{d}$ we are looking for, the equivalent of $d$ in Eq.~\eqref{eq:d_traditional}, can be found by differentiation:
\begin{align}
    \tilde{d} & \lt t, \bar{k} \,|\, \ket{\psi^z}, t_z  \rt = \nol[0.2cm]
     &= -\partial_t \, \mathbb{P} \bigl( TFJ_k \geq TFJ_{\bar{k}} \geq t \,\, \forall k \,|\, \ket{\psi^z}, t_z  \bigr) = \nol[0.2cm]
     & =  p_{\bar{k}} \lt t \rt \, \prod_{k \neq \bar{k}} \,\int_{t}^{\infty} \rmd t_k \,\, p_{k} \lt t_{k}  \rt
     \label{eq:tilde_d_p}
\end{align}
Recalling that, due to the equivalence \eqref{eq:partial_result},
\begin{align}
    \mathbb{P} &\lt TFJ  \geq t \,|\, \ket{\psi^z}, t_z  \rt = \prod_{k }\, \int_{t}^{\infty} \! \!\rmd t_k \,\, p_{k} \lt t_{k}  \rt,
\end{align}
we find
% \begin{align}
%     p_{TFJ} & \bigl( t \,|\, \ket{\psi^z}, t_z \bigr) = -\partial_t \, \mathbb{P} \lt TFJ  \geq t \,|\, \ket{\psi^z}, t_z  \rt = \nol
%     & = \sum_{\bar{k}}  p_{\bar{k}} (t) \, \prod_{k \neq \bar{k}} \,\int_{t}^{\infty} \rmd t_k \,\, p_{k} \lt t_{k}  \rt = \nol
%     & = \sum_{\bar{k}} \tilde{d}  \lt t, \bar{k} \,|\, \ket{\psi^z}, t_z  \rt.
%     \label{eq:CTON1}
% \end{align}
\begin{align}
    p_{TFJ} & \bigl( t \,|\, \ket{\psi^z}, t_z \bigr) = -\partial_t \, \mathbb{P} \lt TFJ  \geq t \,|\, \ket{\psi^z}, t_z  \rt = \nol
    & = \sum_{k}  p_{k} (t) \, \prod_{k' \neq k} \,\int_{t}^{\infty} \rmd t_{k'} \,\, p_{k'} \lt t_{k'}  \rt = \nol
    & = \sum_{k} \tilde{d}  \lt t, k \,|\, \ket{\psi^z}, t_z  \rt.
    \label{eq:CTON1}
\end{align}

\subsection{Comparing the distributions}
\label{subapp:comparing_distr}

Inserting Eq.~\eqref{eq:CTON1} into \eqref{eq:d_traditional} we see that the condition of equivalence of the two distributions $\tilde{d} \equiv d $ corresponds to
\be
    \tilde{d} \lt t, \bar{k} \,|\, \ket{\psi^z}, t_z  \rt =  \lqq \sum_{\alpha \in \mathscr{A}_{\bar{k}}} q_{\alpha}  \rqq \sum_k \tilde{d}  \lt t, k \,|\, \ket{\psi^z}, t_z  \rt.
    \label{eq:statement}
\ee
We now rewrite
\begin{align}
    \tilde{d} & \lt t, \bar{k} \,|\, \ket{\psi^z}, t_z  \rt = \nol[0.2cm]
     & =  \frac{p_{\bar{k}} \lt t \rt}{\int_t^{\infty}\rmd t_{\bar{k}} \, p_{\bar{k}} (t_{\bar{k}})}  \, \prod_{k} \,\int_{t}^{\infty} \rmd t_k \,\, p_{k} \lt t_{k}  \rt = \nol[0.2cm]
     & = \frac{p_{\bar{k}} \lt t \rt}{\mathbb{P} \lt TFJ_{\bar{k}} \geq t \,|\, \ket{\psi_z}, t_z \rt} \mathbb{P} \lt TFJ \geq t \,|\, \ket{\psi_z}, t_z \rt.
     \label{eq:statement2}
\end{align}
Hence, Eq.~\eqref{eq:statement} is equivalent to
\begin{align}
    & \frac{p_{\bar{k}} \lt t \rt}{\mathbb{P} \lt TFJ_{\bar{k}} \geq t \,|\, \ket{\psi_z}, t_z \rt} = \nol
    & \hspace{0.2cm} =\lqq \sum_{\alpha \in \mathscr{A}_{\bar{k}}} q_{\alpha}  \rqq \sum_k \frac{p_{k} \lt t \rt}{\mathbb{P} \lt TFJ_{k} \geq t \,|\, \ket{\psi_z}, t_z \rt}
    \label{eq:statement3}
\end{align}
To prove the equality above it is convenient to switch to a superoperatorial formalism, analogous to that introduced in App.~\ref{App:QJMC} but adapted to the system replicas introduced above. By construction, the $k$-th copy evolves under a Lindblad equation
\be
    \mal{L}_k = \mal{L} \lqq H_k, \set{L_\alpha}_{\alpha \in \mathscr{A}_{k}}   \rqq
\ee
with Hamiltonian 
\be
    H_k = \Omega \lt \sigma_k^{IB} + \sigma_k^{BI}  \rt,
\ee
a single death operator $L_k^D$ and all infection operators $L_{jk}^I$ acting on neighbors $j$. Clearly, $\sum_k \mal{L}_k = \mal{L}$. We now wish to separate each partial Lindbladian into its effective Hamiltonian part plus the remaining jump terms. We first notice that the effective Hamiltonian reads
\be
    H_{k, eff} = H_k - \frac{i}{2} \sum_{\alpha \in \mathscr{A}_{k}} L_{\alpha}^\dag L_{\alpha} = h_k,
    \label{eq:loc_and_glob_heff}
\ee
where $h_k$ is the local contribution to the full effective Hamiltonian $H_{eff}$ defined in Eq.~\eqref{eq:hk}. We recall that our replica trick acts at step (II) of our algorithm and therefore after step (I), which means we have identified at this stage all susceptible and dead sites in $\rho^z = \ket{\psi^z}\bra{\psi^z}$ and, in particular, worked out all the values $\mal{N}_k^S$ which enter the definition of $h_k$, which can be considered in the following unambiguously defined. We define
\be
    \mal{M}_0^{(k)} \rho = - iH_{k, eff} \, \rho + i \rho\, H_{k, eff}^\dag.  
\ee
Equation \eqref{eq:loc_and_glob_heff} ensures that
\be
    \sum_k \mal{M}_0^{(k)} = \mal{L}_0
    \label{eq:sum_M0}
\ee
with $\mal{L}_0$ in Eq.~\eqref{eq:first_L0}. 

Similarly, we take $\mal{L}_1^{(\alpha)}$ from \eqref{eq:L1alpha} and introduce
\be
    \mal{M}_1^{(k)} = \sum_{\alpha \in \mathscr{A}_{k}} \mal{L}_1^{(\alpha)}.
    \label{eq:def_M1k}
\ee
It is not difficult to see that summing over lattice indices returns the full $\mal{L}_1$ in Eq.~\eqref{eq:L1}:
\be
    \sum_k \mal{M}_1^{(k)} = \sum_k \sum_{\alpha \in \mathscr{A}_{k}} \mal{L}_1^{(\alpha)} = \sum_\alpha \mal{L}_1^{(\alpha)} \equiv \mal{L}_1.
    \label{eq:M1_to_L1}
\ee

The steps outlined in App.~\ref{App:QJMC} to translate the QJMC instructions in a superoperatorial formalism apply to any Lindblad equation; for the $k$-th replica, we can thus express the no-jump probability as
\be
    \mathbb{P} \lt TFJ_k \geq t \,|\, \rho^z , t_z \rt = \trace{ \rme{\mal{M}_0^{(k)} \lt t - t_z \rt}  \rho^z}
\ee
and the corresponding density as
\be
    p_k(t) = \trace{\mal{M}_1^{(k)} \rme{\mal{M}_0^{(k)} \lt t - t_z \rt}  \rho^z}.
\ee
We recall that we only need to consider sites $k$ that are neither S nor D, since otherwise $\mathbb{P} \equiv 1$ and $p_k \equiv 0$. Let us call $\Lambda^I$ the set of active lattice sites (sites in a I-B superposition) in $\rho^z$, $\Lambda^D$ the set of all dead sites, $\Lambda_k^S$ the set of susceptible neighbors of site $k$ and $\Lambda_{\not k}^S = \Lambda^S / \Lambda_k^S$ the set of all other susceptible sites. We further denote by
\be
    \xi_k^z = \rho_k^z \otimes \, \bigotimes_{j \in \Lambda_k^S} \rho_j^z = \rho_k^z \otimes \, \bigotimes_{j \in \Lambda_k^S} \ket{S}_j \brak{S}{j}
\ee
the local component of the density matrix covering the Hilbert subspace of site $k$ and its susceptible neighbors, and $\ptrace{kk}{\cdot}$ the partial trace over it. We will use the more conventional notation $\ptrace{k}{\cdot}$ for the partial trace on single-site subspaces.
With these definitions, exploiting the knowledge that our matrix remains factorized at all times, we have
\be
    \rho^z = : \xi_k^z \otimes \bigotimes_{j \in \Lambda_{\not k}^S } \rho_j \otimes \bigotimes_{l \in \Lambda^D } \rho_l \otimes \bigotimes_{m \in \Lambda^I / {k} } \rho_m :,
    \label{eq:factorized_rho}
\ee
where the notation $:\cdot:$ is just a reminder that the ordering of the tensor product is important and must be taken consistently throughout. 

We also need to recall that $\ket{S}_j$ and $\ket{D}_j$ are both kernel eigenvectors of $h_j$. Thus,
\be
    \mal{M}_0^{(j)} \ket{S}_j \brak{S}{j} = 0
\ee
and thus
\be
    \mal{M}_0^{(k)} \rho^z = \lqq  \mal{M}_0^{(k)} + \sum_{j \in \Lambda_k^S} \mal{M}_0^{(j)} \rqq \rho^z \equiv \mal{M}_0^{(kk)} \rho^z.
\ee
As we have seen for the matrix in Eq.~\eqref{eq:factorized_rho}, we have
\be
    \mal{L}_0 = \mal{M}_0^{(kk)} + \sum_{j\in \Lambda_{\not k}^S} \mal{M}_0^{(j)} + \sum_{l\in \Lambda^D} \mal{M}_0^{(l)} + \sum_{m\in \Lambda^I} \mal{M}_0^{(m)},
    \label{eq:separation_M0}
\ee
which is just a different grouping of the addends of \eqref{eq:sum_M0}.

By the local nature of the $\mal{M}$ superoperators, we can now recast the expressions for the replica probabilities in the forms
\begin{align}
    \mathbb{P} & \lt TFJ_k \geq t \,|\, \rho^z , t_z \rt = \ptrace{kk}{ \rme{\mal{M}_0^{(kk)} \lt t - t_z \rt}  \xi_k^z} = \nol
    & = \ptrace{k}{ \rme{\mal{M}_0^{(k)} \lt t - t_z \rt}  \rho_k^z}
\end{align}
and
\be
    p_k(t) = \ptrace{kk}{\mal{M}_1^{(k)} \rme{\mal{M}_0^{(kk)} \lt t - t_z \rt}  \xi_k^z}.
\ee
The strategy now is to take their ratio and complete both traces, exploiting again the factorized nature of our matrices:
\begin{align}
    & \frac{p_k(t)}{\mathbb{P}  \lt TFJ_k \geq t \,|\, \rho^z , t_z \rt} = \frac{\ptrace{kk}{\mal{M}_1^{(k)} \rme{\mal{M}_0^{(kk)} \lt t - t_z \rt}  \xi_k^z}}{\ptrace{kk}{ \rme{\mal{M}_0^{(kk)} \lt t - t_z \rt}  \xi_k^z}} = \nol
    & \hspace{0.2cm} = \frac{\ptrace{kk}{\mal{M}_1^{(k)} \rme{\mal{M}_0^{(kk)} \lt t - t_z \rt}  \xi_k^z}}{\ptrace{kk}{ \rme{\mal{M}_0^{(kk)} \lt t - t_z \rt}  \xi_k^z}} \times \nol
    & \hspace{0.2cm} \times \prod_{j \in \Lambda_{\not k}^S} \frac{\ptrace{j}{ \rme{\mal{M}_0^{(j)} \lt t - t_z \rt}  \rho_j^z}}{\ptrace{j}{ \rme{\mal{M}_0^{(j)} \lt t - t_z \rt}  \rho_j^z}} \, \prod_{l \in \Lambda^D} \frac{\ptrace{l}{ \rme{\mal{M}_0^{(l)} \lt t - t_z \rt}  \rho_l^z}}{\ptrace{l}{ \rme{\mal{M}_0^{(l)} \lt t - t_z \rt}  \rho_l^z}} \times \nol
    & \hspace{0.2cm} \times \prod_{m \in \Lambda^I} \frac{\ptrace{m}{ \rme{\mal{M}_0^{(m)} \lt t - t_z \rt}  \rho_m^z}}{\ptrace{m}{ \rme{\mal{M}_0^{(m)} \lt t - t_z \rt}  \rho_m^z}}.
\end{align}
Applying the combination rules \eqref{eq:factorized_rho} and \eqref{eq:separation_M0} to both numerator and denominator we finally find
\begin{align}
    & \frac{p_k(t)}{\mathbb{P}  \lt TFJ_k \geq t \,|\, \rho^z , t_z \rt} = \frac{\trace{\mal{M}_1^{(k)} \rme{\mal{L}_0 \lt t - t_z \rt}  \rho^z}}{\trace{ \rme{\mal{L}_0 \lt t - t_z \rt}  \rho^z}} = \nol
    & \hspace{0.2cm} = \sum_{\alpha \in \mathscr{A}_k} \frac{\trace{\mal{L}_1^{(\alpha)} \rme{\mal{L}_0 \lt t - t_z \rt}  \rho^z}}{\trace{ \rme{\mal{L}_0 \lt t - t_z \rt}  \rho^z}},
    \label{eq:pp_ratio}
\end{align}
having used definition \eqref{eq:def_M1k} in the second line. The sum over the entire lattice therefore produces
\be
    \sum_k \frac{p_k(t)}{\mathbb{P}  \lt TFJ_k \geq t \,|\, \rho^z , t_z \rt} = \frac{\trace{\mal{L}_1 \, \rme{\mal{L}_0 \lt t - t_z \rt}  \rho^z}}{\trace{ \rme{\mal{L}_0 \lt t - t_z \rt}  \rho^z}},
    \label{eq:pp_ratio_sum}
\ee
see Eq.~\eqref{eq:M1_to_L1}.

We now recall the expression of the probabilities $q_\alpha$ in terms of traces, Eq.~\eqref{eq:qalpha_rho}; with our current definitions,
\be
    q_\alpha = \frac{\trace{\mal{L}_1^{(\alpha)} \rme{\mal{L}_0 \lt t - t_z \rt}  \rho^z  }}{\trace{\mal{L}_1\, \rme{\mal{L}_0 \lt t - t_z \rt}  \rho^z }}
    \label{eq:qalpha_fact}
\ee
Substituting Eqs.~\eqref{eq:pp_ratio}, \eqref{eq:pp_ratio_sum} and \eqref{eq:qalpha_fact} into our desired result, Eq.~\eqref{eq:statement3}, we find
\begin{align}
    &\sum_{\alpha \in \mathscr{A}_k}  \frac{\trace{\mal{L}_1^{(\alpha)} \rme{\mal{L}_0 \lt t - t_z \rt}  \rho^z}}{\trace{ \rme{\mal{L}_0 \lt t - t_z \rt}  \rho^z}} = \nol
    & \hspace{0.2cm} = \lqq \sum_{\alpha \in \mathscr{A}_k} \frac{\trace{\mal{L}_1^{(\alpha)} \rme{\mal{L}_0 \lt t - t_z \rt}  \rho^z  }}{\trace{\mal{L}_1 \rme{\mal{L}_0 \lt t - t_z \rt}  \rho^z }}  \rqq \frac{\trace{\mal{L}_1 \, \rme{\mal{L}_0 \lt t - t_z \rt}  \rho^z}}{\trace{ \rme{\mal{L}_0 \lt t - t_z \rt}  \rho^z}}
\end{align}
which, after simplifying the common terms in the second line, is clearly true. In conclusion, the two distributions $d$ and $\tilde{d}$ are completely equivalent and our method of extracting jump time and site simultaneously is a perfectly valid alternative to following the "traditional" QJMC algorithm in Sec.~\ref{subsec:QJMC_algorithm}.

\section{Equivalence between the eQEP for $\Omega = 0$ and the GEP}
\label{app:eQEP_to_GEP}

In the main text we have claimed that the eQEP reduces to the GEP when $\Omega = 0$, i.e.~when the Hamiltonian vanishes. In this Appendix we provide a proof to this statement by showing that the QJMC algorithm in this case produces the stochastic rules of the GEP.

First, however, we discuss the "disappearance" of the fourth local state $\ket{B}$. For $\Omega = 0$ the density operators $\sigma_k^{BB}$ are all conserved quantities. As a matter of fact, each parity operator $U_k^B = \rme{i\pi \sigma_k^{BB}}$ becomes now a strong symmetry (see App.~\eqref{app:strong_weak}). Note that this is simply telling us that, in the absence of $H$, state $\ket{B}$ is dynamically disconnected from all the others. Hence, unless the initial state $\rho_0$ itself has some B components, none shall develop over the course of the evolution. Considering that our initial state is always chosen as a combination of $N-1$ susceptible sites and a single infected one, we can therefore completely disregard the presence of the bedridden state, as it will never play any role whatsoever. 

Combining this new fact with all our previous results derived from the weak symmetries of the model, we conclude that, in any given trajectory, at all times the vector is factorized $\ket{\psi} = \otimes_k\, \ket{\psi_k}$ and each component $\ket{\psi_k}_k$ is either $\ket{S}_k$, $\ket{I}_k$ or $\ket{D}_k$. We already see from this that superpositions do not develop in this case. The dynamics only performs jumps between different classical GEP-like configurations.

Let us take however a more rigorous approach: in the following, we will think of a purely classical stochastic process with three states per site, $S$, $I$, $D$. We will assume that site $k$ in this process is in state $\mu$ if the eQEP features component $\ket{\mu}_k$. When the eQEP performs the jump $\ket{\mu}_k \to \ket{\nu}_k$, this "parallel" classical process also switches from $\mu$ to $\nu$ at site $k$. The question now is: what are the stochastic rules of this classical process?

Let us look at the QJMC algorithm step by step. We start from the determination of the jump time. We have now
\be
    h_k = - \frac{i}{2} \lt \gamma_D + \mal{N}_k^S \gamma_I  \rt \sigma_k^{II} = -\frac{i}{2} \gamma_{eff}\, \sigma_k^{II}
\ee
and therefore
\begin{align}
    \mathbb{P} & \lt TFJ \geq t \,|\, \ket{\psi^z}, t_z  \rt = \prod_k \norm{\rme{-ih_k \lt t - t_z \rt} \ket{\psi^z_k}}^2 = \nol
    & =  \prod_k \rme{-\gamma_{eff}\,  c_k^I \lt t - t_z \rt} = \rme{ - \sum_k  \gamma_{eff}\, c_k^I \lt t - t_z \rt},
\end{align}
where
\be
    c_k^I = \sysb{lc} 1 & \text{if } \ket{\psi_k}_k = \ket{I}_k \\[0.2cm]
                        0 & \text{otherwise.}\syse
    \label{eq:def_ckI}
\ee
The probability above takes a very simple exponential form; the coefficient $\Gamma = \sum_k \gamma_{eff} \, c_k^I$ can be seen in this case as a collective rate (see also App.~\ref{app:KMC} for how this is used in classical Kinetic Monte Carlo). 

Let us now look at individual stochastic events. A death jump $L_k^D$ changes the $k$-th component (which we assume infected) only: $\ket{I}_k \to \ket{D}_k$; its statistical weight, i.e., the coefficient $Q_\alpha$, is in this case simply
\be
    \gamma_D \brak{I}{k} \sigma_k^{ID} \sigma_k^{DI} \ket{I}_k = \gamma_D.
\ee
For an infection jump $L_{kj}^I$ we assume the $k$-th component susceptible and the $j$-th one infected (in any other case the jump annihilates the vector and thus is never selected). Then, its effect is $\ket{S}_k \ket{I}_j \to \ket{I}_k \ket{I}_j$ with selection weight
\begin{align}
    \gamma_I \, \brak{I}{j} \brak{S}{k} \,\sigma_k^{SI} \sigma_j^{II} \sigma_j^{II} \sigma_k^{IS} \,\ket{S}_k \ket{I}_j = \gamma_I.  
\end{align}
When the parallel stochastic process jumps, it thus selects a death event $I \to D$ at site $k$ with probability
\be
    c_k^I \frac{\gamma_D}{ \sum_j \lt \gamma_D + \mal{N}_j^S \gamma_I \rt c_j^I}
\ee
or an infection event $S \to I$ at site $l$ from $k$ with probability
\be
    c_l^S c_k^I \frac{\gamma_I}{ \sum_j \lt \gamma_D + \mal{N}_j^S \gamma_I \rt c_j^I},
\ee
with $c_l^S$ defined analogously to \eqref{eq:def_ckI}.

The rules above precisely define a classical Kinetic Monte Carlo with rates $\gamma_D$ for events $I \to D$ and rates $\gamma_I$ for events $IS \to II$ involving neighboring pairs. In other words, infected sites die at rate $\gamma_D$, while infected sites can pass on the infection to neighboring susceptible sites at rate $\gamma_I$. These are precisely the rules of the GEP. 

Let us consider observables as well. The most generic one-time observable in the classical case is some function $f \lt \set{n_k^S,\, n_k^I, n_k^D}_k  \rt$ of the local counts 
\be
    n_k^\mu = \sysb{lc} 1 & \text{ if site $k$ is in state $\mu$} \\[0.15cm]
                        0 & \text{otherwise.}  \syse
                        \label{eq:parallel_rule}
\ee
Choosing as a quantum observable
\be
    O = f \bigl(   \set{\sigma_k^{SS}, \, \sigma_k^{II},\, \sigma_k^{DD}}   \bigr),
\ee
exploiting the commutation relations
\be
    \comm{\sigma_k^{\mu\mu}}{\sigma_j^{\nu\nu}} = 0 \ \ \forall \, k,\, j,\,\, \forall \, \mu,\, \nu,
\ee
and recalling that for $\Omega = 0$ the QJMC vector can be expressed at all times as
\be
    \ket{\psi} = \bigotimes \ket{\mu_k}_k
\ee
with $\mu_k \in \set{S, I, D}$, we see that the vector is always an eigenvector of $O$, whose arguments can be therefore replaced with the relative eigenvalues
\be
    \bra{\psi} O \ket{\psi} =  f \lt \set{c_k^S,\, c_k^I, c_k^D}_k  \rt,
\ee
with the $c_k^\mu$ defined analogously to Eq.~\eqref{eq:def_ckI}.
Stressing again that for any given trajectory $\tau$ the configuration of the parallel GEP matches that of $\ket{\psi}$, we have that
\be
    \bra{\psi} O \ket{\psi} =  f \lt \set{n_k^S,\, n_k^I, n_k^D}_k  \rt.
\ee
In other words, from each trajectory we get from the eQEP and from the GEP the same values for the observables they have in common; the same thus hold for the stochastic averages. In this sense, the eQEP at $\Omega = 0$ is equivalent to the GEP.

The case of two-time (or higher) correlations is more involved; discussing it goes beyond the scope of this work. 

% Similar considerations could be made for two-time (or higher) correlations, but this goes beyond the scope of this work.
%but these would require discussing in general the definitions taken in the quantum case (for open quantum systems not even $\av{A(t) B(t)} = \av{\lt AB \rt(t)}$ is guaranteed), which goes well beyond the scope of this work.  

\section{The classical KMC algorithm}
\label{app:KMC}

We report here for comparison the Kinetic Monte Carlo (KMC) \cite{VoterKMC} algorithm. Throughout this Appendix, we assume to be dealing with a classical system described by configurations $\mal{C}$. We shall denote again by $t_z$ the time of the $z$-th stochastic event, or "jump" for short, and by $\mal{C}^z$ the configuration at $t_z^+$ (immediately after the $z$-th jump). The goal is, again, the reconstruction of the probability distribution over configurations at some time $t_{end}$, with an initial condition set at time $t = 0$.

\subsection{Time-independent rates}
\label{subapp:time_indep_rates}

We start with the simpler case of time-independent rates. We therefore assume that the stochastic dynamics consists of a set of different events with the potential to change the configuration $\mal{C}$ of the system, each occurring at a rate $\gamma_\alpha = \gamma_\alpha \lt\mal{C} \rt$. The rates may generally depend on the configuration (for instance, when a constraint is present, if it is satisfied by the current configuration then $\gamma_\alpha \neq 0$, otherwise $\gamma_\alpha = 0$). We denote the collective rate by $\Gamma = \sum_\alpha \gamma_\alpha$, leading to a no-jump probability
\be
    \mathbb{P} \lt TFJ \geq t \,|\, \mal{C}, t'  \rt = \rme{-\Gamma \lt t - t'  \rt}.
\ee

The fundamental loop to repeat is then given by the following instructions:
\begin{itemize}
\item[(I)] \emph{\underline{Determine the time of the next jump:}} pick a random real $u$ uniformly on the interval $\lqq 0,\, 1 \rqq$ and set $t_{z+1} = t_z + \mathbb{P}^{-1} (u) = t_z - \lt 1 / \Gamma \rt \ln \lt u\rt $ as the new jump time. If $t_{z+1} > t_{end}$ return configuration $C^z$ and stop the procedure. Otherwise, continue.
\item[(II)] \emph{\underline{Find out which jump takes place:}} the $\alpha$-th jump has an associated probability $\gamma_\alpha / \Gamma$ of being selected. The extraction can be performed as follows: choose an ordering for the labels $\alpha$, then pick a random real number $v \in \lqq 0, \, \Gamma \rqq$ and find the label $\bar{\alpha}$ such that $\sum_{\alpha < \bar{\alpha}} \gamma_\alpha \leq v < \sum_{\alpha \leq \bar{\alpha}} \gamma_\alpha$; then $\bar{\alpha}$ is selected.
\item[(III)] \emph{\underline{Apply the selected jump:}} change the configuration $\mal{C}^z$ according to the selected $\bar{\alpha}$-th jump. The configuration thus obtained constitutes the new $\mal{C}^{z+1}$. Return to step (I). 
\end{itemize}

\subsection{Time-dependent rates}
\label{subapp:time_dep_rates}
We relax now our previous assumption of time-independence for the rates. There might be an underlying deterministic evolution on top of the stochastic one, or the rates may simply vary over time. Whatever the case may be, after the $z$-th jump and before the following one the rates do not exclusively depend on the configuration $\mal{C}^z$ at $t_z$:
\be
    \gamma_\alpha \equiv \gamma_\alpha \lt t \,|\, \mal{C}^z, t_z \rt.
\ee
Denoting once again the collective rate by 
\be
    \Gamma(t \,|\,  \mal{C}^z, t_z  ) = \sum_{\alpha} \gamma_\alpha \lt t \,|\, \mal{C}^z, t_z \rt
\ee
we have now a no-jump probability
\be
    \mathbb{P} \lt TFJ \geq t \,|\, \mal{C}^z, t_z  \rt = \rme{- \int_{t_z}^t \rmd t' \, \Gamma \lt t' \,|\,  \mal{C}^z, t_z  \rt  }
    \label{eq:P_from_Gamma}
\ee

The algorithm is then:
\begin{itemize}
\item[(I)] \emph{\underline{Determine the time of the next jump:}} pick a random real $u$ uniformly on the interval $\lqq 0,\, 1 \rqq$ and solve $\mathbb{P} \lt TFJ \geq t \,|\, \mal{C}^z, t_z  \rt = u $, numerically if necessary, to find a jump interval $\Delta t$; the jump time will then be $t_{z+1} = t_z + \Delta t$. If $t_{z+1} > t_{end}$ go to step (IIIb). Otherwise, continue.
\item[(II)] \emph{\underline{Find out which jump takes place:}} this is done precisely in the same way as for the time-independent case: the $\alpha$-th jump has an associated probability $\gamma_\alpha / \Gamma$ of being selected, with all rates calculated at time $t_{z+1}$. A label $\bar{\alpha}$ can be extracted as outlined in step (II) of the time-independent case.
\item[(IIIa)] \emph{\underline{Apply the selected jump:}} in the presence of an additional deterministic dynamics, evolve the configuration $\mal{C}^z$ for a time $t_{z+1} - t_z$ to obtain $C^z(t_{z+1})$; otherwise, $C^z(t_{z+1}) = C^z$. Change the configuration $\mal{C}^z(t_{z+1})$ according to the selected $\bar{\alpha}$-th jump. The configuration thus obtained constitutes the new $\mal{C}^{z+1}$. Return to step (I). 
\item[(IIIb)] \emph{\underline{Apply the deterministic evolution, if present:}} if there is an underlying deterministic evolution of the configuration, evolve the configuration $\mal{C}^z$ for a time $t_{end} - t_{z}$ and return the result. Otherwise, simply return $\mal{C}^{z}$. The trajectory is now over.
\end{itemize}

As a final remark, one could take an equivalent approach whereby, instead of providing the rates, the stochastic dynamics is defined via the no-jump probability $\mathbb{P} \lt TFJ \geq t \,|\, \mal{C}', t' \rt$ and individual jump probabilities $q_\alpha \lt \mal{C}, t  \rt$. The resulting evolution is statistically equivalent to one generated by rates
\begin{align}
    \gamma_\alpha & \lt t \,|\, \mal{C}^z, t_z \rt = q_\alpha \bigl( \mal{C}^z(t), t  \bigr) \times \nol
    & \times \Bigl[ -\partial_t \ln \mathbb{P} \bigl( TFJ \geq t \,|\, \mal{C}^z, t_z \bigr)  \Bigr].
    \label{eq:eff_rates}
\end{align}
We can show this to be the case by simply observing that, by definition, we have
\be
    \sum_\alpha q_\alpha \bigl( \mal{C}^z(t), t  \bigr) = 1 \ \ \forall \, t
\ee
and, therefore,
\begin{align}
    \Gamma & \lt t \,|\, \mal{C}^z, t_z \rt  \equiv \sum_\alpha \gamma_\alpha  \lt t \,|\, \mal{C}^z, t_z \rt = \nol
    & = -\partial_t \ln \mathbb{P} \bigl( TFJ \geq t \,|\, \mal{C}^z, t_z \bigr).
\end{align}
By integration and exponentiation of the equality above we recover Eq.~\eqref{eq:P_from_Gamma}, showing that the effective rates proposed in Eq.~\eqref{eq:eff_rates} give rise to the expected dynamics.

\section{Comparison with exact numerical diagonalization}
\label{app:comparison}

To check that our algorithm is, in fact, reproducing the Lindblad evolution for the eQEP we have performed some checks against exact numerical diagonalization of the Lindblad superoperator $\mal{L}$ for very small system sizes ($N = 2$). 
\begin{figure}[h]
  \includegraphics[width=\columnwidth]{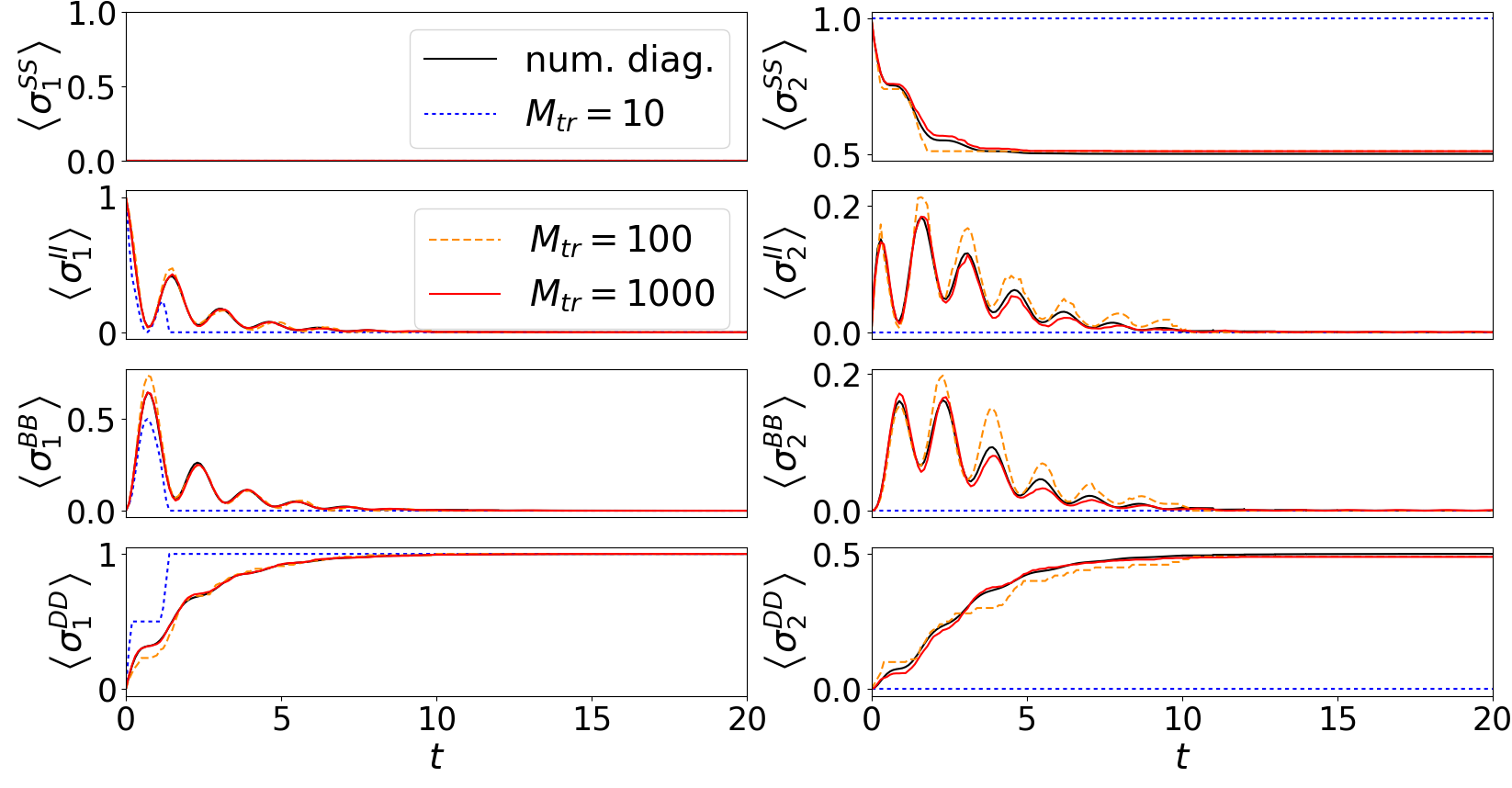}
  \caption{Evolution of the local S, I, B, and D populations for an eQEP on two sites ($N = 2$). We set $\gamma_D = 1$, or, equivalently, we measure time in units of $\gamma_D^{-1}$. Each panel includes four curves: one obtained from exact numerical diagonalization of the superoperator $\mal{L}$ (black, solid line) and three produced by our QJMC algorithm for different numbers of trajectories $M_{tr} = 10$ (blue, dotted line), $M_{tr} = 100$ (orange, dashed line) and $M_{tr} = 1000$ (red, solid line). The three latter trajectory sets are independent, i.e., a total of $1110$ distinct trajectories has been produced. As expected, as $M_{tr}$ increases the QJMC results approximate better and better the black lines. 
  In the top-right panel, showing the susceptible population of site $2$ the blue, dotted curve captures an extreme instance: in none of its $10$ trajectories site $2$ got infected }
\label{fig:comparison_ex_diag}
\end{figure}
One example is reported in Fig.~\ref{fig:comparison_ex_diag}, where we show the evolution of the means $\av{\sigma_k^{\mu\mu}}$ on either site ($k = 1$ or $2$) for parameters $\gamma_D = \gamma_I = 1$, $\Omega = 2$. In this case, we take the "origin" to be site $1$, which implies $\av{\sigma_1^{II}(t = 0)} = \av{\sigma_2^{SS}(t = 0)} = 1$ and
\be
    \av{\sigma_1^{SS}(t)} = 0 \ \forall \, t,
\ee
i.e., site $1$ can never become susceptible. Blue curves are obtained by averaging over a $10$ trajectory set and the discontinuous effects of jumps are still visible, especially for $\av{\sigma_1^{DD}}$. As more and more trajectories are included in the averages, these diconstinuous features are smoothed out and the corresponding curves (orange and red) more closely resemble the continuous ones obtained via numerical diagonalization. Other checks run for different parameter choices have displayed an analogous behavior.

\bibliography{biblio}

@article{Malossi2014,
  title = {Full Counting Statistics and Phase Diagram of a Dissipative Rydberg Gas},
  author = {Malossi, N. and Valado, M. M. and Scotto, S. and Huillery, P. and Pillet, P. and Ciampini, D. and Arimondo, E. and Morsch, O.},
  journal = {Phys. Rev. Lett.},
  volume = {113},
  issue = {2},
  pages = {023006},
  numpages = {5},
  year = {2014},
  month = {Jul},
  publisher = {American Physical Society},
  doi = {10.1103/PhysRevLett.113.023006},
  url = {https://link.aps.org/doi/10.1103/PhysRevLett.113.023006}
}

@article{Espigares2017,
  title = {Epidemic Dynamics in Open Quantum Spin Systems},
  author = {P\'erez-Espigares, Carlos and Marcuzzi, Matteo and Guti\'errez, Ricardo and Lesanovsky, Igor},
  journal = {Phys. Rev. Lett.},
  volume = {119},
  issue = {14},
  pages = {140401},
  numpages = {6},
  year = {2017},
  month = {Oct},
  publisher = {American Physical Society},
  doi = {10.1103/PhysRevLett.119.140401},
  url = {https://link.aps.org/doi/10.1103/PhysRevLett.119.140401}
}

@article{Struik1976,
author = {Struik, L. C. E.},
title = {PHYSICAL AGING IN AMORPHOUS GLASSY POLYMERS},
journal = {Annals of the New York Academy of Sciences},
volume = {279},
number = {1},
pages = {78-85},
doi = {https://doi.org/10.1111/j.1749-6632.1976.tb39695.x},
url = {https://nyaspubs.onlinelibrary.wiley.com/doi/abs/10.1111/j.1749-6632.1976.tb39695.x},
eprint = {https://nyaspubs.onlinelibrary.wiley.com/doi/pdf/10.1111/j.1749-6632.1976.tb39695.x},
year = {1976}
}

@article{Gillman2019,
doi = {10.1088/1367-2630/ab43b0},
url = {https://dx.doi.org/10.1088/1367-2630/ab43b0},
year = {2019},
month = {sep},
publisher = {IOP Publishing},
volume = {21},
number = {9},
pages = {093064},
author = {Edward Gillman and Federico Carollo and Igor Lesanovsky},
title = {Numerical simulation of critical dissipative non-equilibrium quantum systems with an absorbing state},
journal = {New Journal of Physics}
}

@article{Gillman2020,
  title = {Nonequilibrium Phase Transitions in ($1+1$)-Dimensional Quantum Cellular Automata with Controllable Quantum Correlations},
  author = {Gillman, Edward and Carollo, Federico and Lesanovsky, Igor},
  journal = {Phys. Rev. Lett.},
  volume = {125},
  issue = {10},
  pages = {100403},
  numpages = {6},
  year = {2020},
  month = {Sep},
  publisher = {American Physical Society},
  doi = {10.1103/PhysRevLett.125.100403},
  url = {https://link.aps.org/doi/10.1103/PhysRevLett.125.100403}
}

@article{Gillman2022,
  title = {Asynchronism and nonequilibrium phase transitions in $(1+1)$-dimensional quantum cellular automata},
  author = {Gillman, Edward and Carollo, Federico and Lesanovsky, Igor},
  journal = {Phys. Rev. E},
  volume = {106},
  issue = {3},
  pages = {L032103},
  numpages = {5},
  year = {2022},
  month = {Sep},
  publisher = {American Physical Society},
  doi = {10.1103/PhysRevE.106.L032103},
  url = {https://link.aps.org/doi/10.1103/PhysRevE.106.L032103}
}

@article{Brady2023,
  title = {Mean-field approach to Rydberg facilitation in a gas of atoms at high and low temperatures},
  author = {Brady, Daniel and Fleischhauer, Michael},
  journal = {Phys. Rev. A},
  volume = {108},
  issue = {5},
  pages = {052812},
  numpages = {7},
  year = {2023},
  month = {Nov},
  publisher = {American Physical Society},
  doi = {10.1103/PhysRevA.108.052812},
  url = {https://link.aps.org/doi/10.1103/PhysRevA.108.052812}
}

@ARTICLE{Brady2024,
       author = {{Brady}, Daniel and {Ohler}, Simon and {Otterbach}, Johannes and {Fleischhauer}, Michael},
        title = "{Anomalous Directed Percolation on a Dynamic Network using Rydberg Facilitation}",
      journal = {arXiv e-prints},
         year = 2024,
        month = apr,
          eid = {arXiv:2404.16523},
        pages = {arXiv:2404.16523},
          doi = {10.48550/arXiv.2404.16523},
archivePrefix = {arXiv},
       eprint = {2404.16523},
 primaryClass = {cond-mat.quant-gas},
       adsurl = {https://ui.adsabs.harvard.edu/abs/2024arXiv240416523B}
}

@article{Pancotti2020,
  title = {Quantum {East} Model: Localization, Nonthermal Eigenstates, and Slow Dynamics},
  author = {Pancotti, Nicola and Giudice, Giacomo and Cirac, J. Ignacio and Garrahan, Juan P. and Ba\~nuls, Mari Carmen},
  journal = {Phys. Rev. X},
  volume = {10},
  issue = {2},
  pages = {021051},
  numpages = {21},
  year = {2020},
  month = {Jun},
  publisher = {American Physical Society},
  doi = {10.1103/PhysRevX.10.021051},
  url = {https://link.aps.org/doi/10.1103/PhysRevX.10.021051}
}

@article{vanHorssen2015,
  title = {Dynamics of many-body localization in a translation-invariant quantum glass model},
  author = {van Horssen, Merlijn and Levi, Emanuele and Garrahan, Juan P.},
  journal = {Phys. Rev. B},
  volume = {92},
  issue = {10},
  pages = {100305},
  numpages = {5},
  year = {2015},
  month = {Sep},
  publisher = {American Physical Society},
  doi = {10.1103/PhysRevB.92.100305},
  url = {https://link.aps.org/doi/10.1103/PhysRevB.92.100305}
}

@ARTICLE{Causer2024,
       author = {{Causer}, Luke and {Ba{\~n}uls}, Mari Carmen and {Garrahan}, Juan P.},
        title = "{Dynamical heterogeneity and large deviations in the open quantum {East} glass model from tensor networks}",
      journal = {arXiv e-prints},
         year = 2024,
        month = apr,
          eid = {arXiv:2404.03750},
        pages = {arXiv:2404.03750},
          doi = {10.48550/arXiv.2404.03750},
archivePrefix = {arXiv},
       eprint = {2404.03750},
 primaryClass = {cond-mat.stat-mech},
       adsurl = {https://ui.adsabs.harvard.edu/abs/2024arXiv240403750C}
}

@ARTICLE{Brighi2024,
       author = {{Brighi}, Pietro and {Ljubotina}, Marko},
        title = "{Anomalous transport in the quantum East-West kinetically constrained model}",
      journal = {arXiv e-prints},
         year = 2024,
        month = may,
          eid = {arXiv:2405.02102},
        pages = {arXiv:2405.02102},
          doi = {10.48550/arXiv.2405.02102},
archivePrefix = {arXiv},
       eprint = {2405.02102},
 primaryClass = {quant-ph},
       adsurl = {https://ui.adsabs.harvard.edu/abs/2024arXiv240502102B}
}

@article{Makki2024,
  title = {Absorbing state phase transition with Clifford circuits},
  author = {Makki, Nastasia and Lang, Nicolai and B\"uchler, Hans Peter},
  journal = {Phys. Rev. Res.},
  volume = {6},
  issue = {1},
  pages = {013278},
  numpages = {11},
  year = {2024},
  month = {Mar},
  publisher = {American Physical Society},
  doi = {10.1103/PhysRevResearch.6.013278},
  url = {https://link.aps.org/doi/10.1103/PhysRevResearch.6.013278}
}

@article{Jo2021,
  title = {Absorbing phase transition with a continuously varying exponent in a quantum contact process: A neural network approach},
  author = {Jo, Minjae and Lee, Jongshin and Choi, K. and Kahng, B.},
  journal = {Phys. Rev. Res.},
  volume = {3},
  issue = {1},
  pages = {013238},
  numpages = {15},
  year = {2021},
  month = {Mar},
  publisher = {American Physical Society},
  doi = {10.1103/PhysRevResearch.3.013238},
  url = {https://link.aps.org/doi/10.1103/PhysRevResearch.3.013238}
}

@ARTICLE{Racz2002,
       author = {{Racz}, Zoltan},
        title = "{Nonequilibrium Phase Transitions}",
      journal = {arXiv e-prints},
         year = 2002,
        month = oct,
          eid = {cond-mat/0210435},
        pages = {cond-mat/0210435},
          doi = {10.48550/arXiv.cond-mat/0210435},
archivePrefix = {arXiv},
       eprint = {cond-mat/0210435},
 primaryClass = {cond-mat.stat-mech},
       adsurl = {https://ui.adsabs.harvard.edu/abs/2002cond.mat.10435R}
}

@article{Greissler2023,
  title = {Slow dynamics and nonergodicity of the bosonic quantum {East} model in the semiclassical limit},
  author = {Gei\ss{}ler, Andreas and Garrahan, Juan P.},
  journal = {Phys. Rev. E},
  volume = {108},
  issue = {3},
  pages = {034207},
  numpages = {7},
  year = {2023},
  month = {Sep},
  publisher = {American Physical Society},
  doi = {10.1103/PhysRevE.108.034207},
  url = {https://link.aps.org/doi/10.1103/PhysRevE.108.034207}
}

@article{Degenfeld2014,
  title = {Self-consistent projection operator theory for quantum many-body systems},
  author = {Degenfeld-Schonburg, Peter and Hartmann, Michael J.},
  journal = {Phys. Rev. B},
  volume = {89},
  issue = {24},
  pages = {245108},
  numpages = {12},
  year = {2014},
  month = {Jun},
  publisher = {American Physical Society},
  doi = {10.1103/PhysRevB.89.245108},
  url = {http://link.aps.org/doi/10.1103/PhysRevB.89.245108}
}

@article{Carollo2019,
  title = {Critical Behavior of the Quantum Contact Process in One Dimension},
  author = {Carollo, Federico and Gillman, Edward and Weimer, Hendrik and Lesanovsky, Igor},
  journal = {Phys. Rev. Lett.},
  volume = {123},
  issue = {10},
  pages = {100604},
  numpages = {6},
  year = {2019},
  month = {Sep},
  publisher = {American Physical Society},
  doi = {10.1103/PhysRevLett.123.100604},
  url = {https://link.aps.org/doi/10.1103/PhysRevLett.123.100604}
}

@book{Ma_book,
  title={Modern theory of critical phenomena},
  author={Ma, Shang-keng},
  url = {https://doi.org/10.4324/9780429498886},
  doi = {10.4324/9780429498886},
  year={2001},
  publisher={Routledge},
  address = "New York, NY",
  edition = {1st},
  series = {Advanced Book Classics}
}

@book{Gallagher_book,
  address = {Cambridge},
  author = {Gallagher, T F},
  publisher = {Cambridge University Press},
  title = {Rydberg Atoms},
  username = {jgurian},
  year = 1994
}

@article{Marcuzzi2014,
doi = {10.1088/1751-8113/47/48/482001},
url = {https://dx.doi.org/10.1088/1751-8113/47/48/482001},
year = {2014},
month = {nov},
publisher = {IOP Publishing},
volume = {47},
number = {48},
pages = {482001},
author = {M Marcuzzi and J Schick and B Olmos and I Lesanovsky},
title = {Effective dynamics of strongly dissipative {Rydberg} gases},
journal = {Journal of Physics A: Mathematical and Theoretical}
}

@article{Marcuzzi2015,
doi = {10.1088/1367-2630/17/7/072003},
url = {https://dx.doi.org/10.1088/1367-2630/17/7/072003},
year = {2015},
month = {jul},
publisher = {IOP Publishing},
volume = {17},
number = {7},
pages = {072003},
author = {M Marcuzzi and E Levi and W Li and J P Garrahan and B Olmos and I Lesanovsky},
title = {Non-equilibrium universality in the dynamics of dissipative cold atomic gases},
journal = {New Journal of Physics}
}

@article{Marcuzzi2016,
  title = {Absorbing State Phase Transition with Competing Quantum and Classical Fluctuations},
  author = {Marcuzzi, Matteo and Buchhold, Michael and Diehl, Sebastian and Lesanovsky, Igor},
  journal = {Phys. Rev. Lett.},
  volume = {116},
  issue = {24},
  pages = {245701},
  numpages = {7},
  year = {2016},
  month = {Jun},
  publisher = {American Physical Society},
  doi = {10.1103/PhysRevLett.116.245701},
  url = {http://link.aps.org/doi/10.1103/PhysRevLett.116.245701}
}

@article{Rigol2007,
  title = {Relaxation in a Completely Integrable Many-Body Quantum System: An Ab Initio Study of the Dynamics of the Highly Excited States of 1D Lattice Hard-Core Bosons},
  author = {Rigol, Marcos and Dunjko, Vanja and Yurovsky, Vladimir and Olshanii, Maxim},
  journal = {Phys. Rev. Lett.},
  volume = {98},
  issue = {5},
  pages = {050405},
  numpages = {4},
  year = {2007},
  month = {Feb},
  publisher = {American Physical Society},
  doi = {10.1103/PhysRevLett.98.050405},
  url = {https://link.aps.org/doi/10.1103/PhysRevLett.98.050405}
}

@article{Kollath2007,
  title = {Quench Dynamics and Nonequilibrium Phase Diagram of the Bose-Hubbard Model},
  author = {Kollath, Corinna and L\"auchli, Andreas M. and Altman, Ehud},
  journal = {Phys. Rev. Lett.},
  volume = {98},
  issue = {18},
  pages = {180601},
  numpages = {4},
  year = {2007},
  month = {Apr},
  publisher = {American Physical Society},
  doi = {10.1103/PhysRevLett.98.180601},
  url = {https://link.aps.org/doi/10.1103/PhysRevLett.98.180601}
}

@article{Bertini2015,
  title = {Prethermalization and Thermalization in Models with Weak Integrability Breaking},
  author = {Bertini, Bruno and Essler, Fabian H. L. and Groha, Stefan and Robinson, Neil J.},
  journal = {Phys. Rev. Lett.},
  volume = {115},
  issue = {18},
  pages = {180601},
  numpages = {6},
  year = {2015},
  month = {Oct},
  publisher = {American Physical Society},
  doi = {10.1103/PhysRevLett.115.180601},
  url = {https://link.aps.org/doi/10.1103/PhysRevLett.115.180601}
}

@article{Sieberer2013,
  title = {Dynamical Critical Phenomena in Driven-Dissipative Systems},
  author = {Sieberer, L. M. and Huber, S. D. and Altman, E. and Diehl, S.},
  journal = {Phys. Rev. Lett.},
  volume = {110},
  issue = {19},
  pages = {195301},
  numpages = {5},
  year = {2013},
  month = {May},
  publisher = {American Physical Society},
  doi = {10.1103/PhysRevLett.110.195301},
  url = {https://link.aps.org/doi/10.1103/PhysRevLett.110.195301}
}

@article{Sotiriadis2014,
doi = {10.1088/1742-5468/2014/07/P07024},
url = {https://dx.doi.org/10.1088/1742-5468/2014/07/P07024},
year = {2014},
month = {jul},
publisher = {IOP Publishing and SISSA},
volume = {2014},
number = {7},
pages = {P07024},
author = {Spyros Sotiriadis and Pasquale Calabrese},
title = {Validity of the GGE for quantum quenches from interacting to noninteracting models},
journal = {Journal of Statistical Mechanics: Theory and Experiment}
}

@article{Foini2017,
  title = {Measuring effective temperatures in a generalized Gibbs ensemble},
  author = {Foini, Laura and Gambassi, Andrea and Konik, Robert and Cugliandolo, Leticia F.},
  journal = {Phys. Rev. E},
  volume = {95},
  issue = {5},
  pages = {052116},
  numpages = {8},
  year = {2017},
  month = {May},
  publisher = {American Physical Society},
  doi = {10.1103/PhysRevE.95.052116},
  url = {https://link.aps.org/doi/10.1103/PhysRevE.95.052116}
}

@article{Gambassi2012,
  title = {Large Deviations and Universality in Quantum Quenches},
  author = {Gambassi, Andrea and Silva, Alessandro},
  journal = {Phys. Rev. Lett.},
  volume = {109},
  issue = {25},
  pages = {250602},
  numpages = {5},
  year = {2012},
  month = {Dec},
  publisher = {American Physical Society},
  doi = {10.1103/PhysRevLett.109.250602},
  url = {https://link.aps.org/doi/10.1103/PhysRevLett.109.250602}
}

@article{Heyl2013,
  title = {Dynamical Quantum Phase Transitions in the Transverse-Field Ising Model},
  author = {Heyl, M. and Polkovnikov, A. and Kehrein, S.},
  journal = {Phys. Rev. Lett.},
  volume = {110},
  issue = {13},
  pages = {135704},
  numpages = {5},
  year = {2013},
  month = {Mar},
  publisher = {American Physical Society},
  doi = {10.1103/PhysRevLett.110.135704},
  url = {https://link.aps.org/doi/10.1103/PhysRevLett.110.135704}
}

@article{Heyl2018,
doi = {10.1088/1361-6633/aaaf9a},
url = {https://dx.doi.org/10.1088/1361-6633/aaaf9a},
year = {2018},
month = {apr},
publisher = {IOP Publishing},
volume = {81},
number = {5},
pages = {054001},
author = {Markus Heyl},
title = {Dynamical quantum phase transitions: a review},
journal = {Reports on Progress in Physics}
}

@article{Hickey2014,
  title = {Dynamical phase transitions, time-integrated observables, and geometry of states},
  author = {Hickey, James M. and Genway, Sam and Garrahan, Juan P.},
  journal = {Phys. Rev. B},
  volume = {89},
  issue = {5},
  pages = {054301},
  numpages = {9},
  year = {2014},
  month = {Feb},
  publisher = {American Physical Society},
  doi = {10.1103/PhysRevB.89.054301},
  url = {https://link.aps.org/doi/10.1103/PhysRevB.89.054301}
}

@article{Chiocchetta2015,
  title = {Short-time universal scaling in an isolated quantum system after a quench},
  author = {Chiocchetta, Alessio and Tavora, Marco and Gambassi, Andrea and Mitra, Aditi},
  journal = {Phys. Rev. B},
  volume = {91},
  issue = {22},
  pages = {220302},
  numpages = {5},
  year = {2015},
  month = {Jun},
  publisher = {American Physical Society},
  doi = {10.1103/PhysRevB.91.220302},
  url = {https://link.aps.org/doi/10.1103/PhysRevB.91.220302}
}

@article{Caux2016,
doi = {10.1088/1742-5468/2016/06/064006},
url = {https://dx.doi.org/10.1088/1742-5468/2016/06/064006},
year = {2016},
month = {jun},
publisher = {IOP Publishing and SISSA},
volume = {2016},
number = {6},
pages = {064006},
author = {Jean-Sébastien Caux},
title = {The Quench Action},
journal = {Journal of Statistical Mechanics: Theory and Experiment}
}

@article{Marino2014,
  title = {Nonequilibrium dynamics of a noisy quantum Ising chain: Statistics of work and prethermalization after a sudden quench of the transverse field},
  author = {Marino, Jamir and Silva, Alessandro},
  journal = {Phys. Rev. B},
  volume = {89},
  issue = {2},
  pages = {024303},
  numpages = {16},
  year = {2014},
  month = {Jan},
  publisher = {American Physical Society},
  doi = {10.1103/PhysRevB.89.024303},
  url = {https://link.aps.org/doi/10.1103/PhysRevB.89.024303}
}

@article{Hinrichsen2000,
  title={Non-equilibrium critical phenomena and phase transitions into absorbing states},
  author={Hinrichsen, Haye},
  journal={Adv. Phys.},
  volume={49},
  number={7},
  pages={815--958},
  year={2000},
  publisher={Taylor \& Francis}
}

@article{Klocke2019,
  title = {Controlling excitation avalanches in driven {Rydberg} gases},
  author = {Klocke, Kai and Buchhold, Michael},
  journal = {Phys. Rev. A},
  volume = {99},
  issue = {5},
  pages = {053616},
  numpages = {8},
  year = {2019},
  month = {May},
  publisher = {American Physical Society},
  doi = {10.1103/PhysRevA.99.053616},
  url = {https://link.aps.org/doi/10.1103/PhysRevA.99.053616}
}

@article{Klocke2021,
  title = {Hydrodynamic Stabilization of Self-Organized Criticality in a Driven {Rydberg} Gas},
  author = {Klocke, K. and Wintermantel, T. M. and Lochead, G. and Whitlock, S. and Buchhold, M.},
  journal = {Phys. Rev. Lett.},
  volume = {126},
  issue = {12},
  pages = {123401},
  numpages = {6},
  year = {2021},
  month = {Mar},
  publisher = {American Physical Society},
  doi = {10.1103/PhysRevLett.126.123401},
  url = {https://link.aps.org/doi/10.1103/PhysRevLett.126.123401}
}

@article{Ates2007,
  title = {Antiblockade in {Rydberg} Excitation of an Ultracold Lattice Gas},
  author = {Ates, C. and Pohl, T. and Pattard, T. and Rost, J. M.},
  journal = {Phys. Rev. Lett.},
  volume = {98},
  issue = {2},
  pages = {023002},
  numpages = {4},
  year = {2007},
  month = {Jan},
  publisher = {American Physical Society},
  doi = {10.1103/PhysRevLett.98.023002},
  url = {https://link.aps.org/doi/10.1103/PhysRevLett.98.023002}
}

@article{Gutierrez2017,
  title = {Experimental signatures of an absorbing-state phase transition in an open driven many-body quantum system},
  author = {Guti\'errez, Ricardo and Simonelli, Cristiano and Archimi, Matteo and Castellucci, Francesco and Arimondo, Ennio and Ciampini, Donatella and Marcuzzi, Matteo and Lesanovsky, Igor and Morsch, Oliver},
  journal = {Phys. Rev. A},
  volume = {96},
  issue = {4},
  pages = {041602},
  numpages = {6},
  year = {2017},
  month = {Oct},
  publisher = {American Physical Society},
  doi = {10.1103/PhysRevA.96.041602},
  url = {https://link.aps.org/doi/10.1103/PhysRevA.96.041602}
}

@article{Grassberger1983,
title = "On the critical behavior of the general epidemic process and dynamical percolation",
journal = "Math. Biosci.",
volume = "63",
number = "2",
pages = "157 - 172",
year = "1983",
note = "",
issn = "0025-5564",
doi = "http://dx.doi.org/10.1016/0025-5564(82)90036-0",
url = "http://www.sciencedirect.com/science/article/pii/0025556482900360",
author = "P. Grassberger",
}

@article{Mattioli2015,
  author={Marco Mattioli and Alexander W Gl{\"a}tzle and Wolfgang Lechner},
  title={From classical to quantum non-equilibrium dynamics of {Rydberg} excitations in optical lattices},
  journal={New Journal of Physics},
  volume={17},
  number={11},
  pages={113039},
  url={http://stacks.iop.org/1367-2630/17/i=11/a=113039},
  year={2015}
}

@article{DallaTorre2013,
  title = {Keldysh approach for nonequilibrium phase transitions in quantum optics: Beyond the Dicke model in optical cavities},
  author = {Dalla Torre, Emanuele G. and Diehl, Sebastian and Lukin, Mikhail D. and Sachdev, Subir and Strack, Philipp},
  journal = {Phys. Rev. A},
  volume = {87},
  issue = {2},
  pages = {023831},
  numpages = {20},
  year = {2013},
  month = {Feb},
  publisher = {American Physical Society},
  doi = {10.1103/PhysRevA.87.023831},
  url = {https://link.aps.org/doi/10.1103/PhysRevA.87.023831}
}

@article{Lesanovsky2014,
  title = {Out-of-equilibrium structures in strongly interacting {Rydberg} gases with dissipation},
  author = {Lesanovsky, Igor and Garrahan, Juan P.},
  journal = {Phys. Rev. A},
  volume = {90},
  issue = {1},
  pages = {011603},
  numpages = {5},
  year = {2014},
  month = {Jul},
  publisher = {American Physical Society},
  doi = {10.1103/PhysRevA.90.011603},
  url = {http://link.aps.org/doi/10.1103/PhysRevA.90.011603}
}

@article{Lesanovsky2019,
doi = {10.1088/2058-9565/aaf831},
url = {https://dx.doi.org/10.1088/2058-9565/aaf831},
year = {2019},
month = {jan},
publisher = {IOP Publishing},
volume = {4},
number = {2},
pages = {02LT02},
author = {Igor Lesanovsky and Katarzyna Macieszczak and Juan P Garrahan},
title = {Non-equilibrium absorbing state phase transitions in discrete-time quantum cellular automaton dynamics on spin lattices},
journal = {Quantum Science and Technology}
}

@article{Nigmatullin2021,
  title = {Directed percolation in nonunitary quantum cellular automata},
  author = {Nigmatullin, Ramil and Wagner, Elisabeth and Brennen, Gavin K.},
  journal = {Phys. Rev. Res.},
  volume = {3},
  issue = {4},
  pages = {043167},
  numpages = {10},
  year = {2021},
  month = {Dec},
  publisher = {American Physical Society},
  doi = {10.1103/PhysRevResearch.3.043167},
  url = {https://link.aps.org/doi/10.1103/PhysRevResearch.3.043167}
}

@article{Buchhold2016,
  title = {Background field functional renormalization group for absorbing state phase transitions},
  author = {Buchhold, Michael and Diehl, Sebastian},
  journal = {Phys. Rev. E},
  volume = {94},
  issue = {1},
  pages = {012138},
  numpages = {7},
  year = {2016},
  month = {Jul},
  publisher = {American Physical Society},
  doi = {10.1103/PhysRevE.94.012138},
  url = {https://link.aps.org/doi/10.1103/PhysRevE.94.012138}
}

@article{Rose2022,
  title = {Hierarchical classical metastability in an open quantum {East} model},
  author = {Rose, Dominic C. and Macieszczak, Katarzyna and Lesanovsky, Igor and Garrahan, Juan P.},
  journal = {Phys. Rev. E},
  volume = {105},
  issue = {4},
  pages = {044121},
  numpages = {28},
  year = {2022},
  month = {Apr},
  publisher = {American Physical Society},
  doi = {10.1103/PhysRevE.105.044121},
  url = {https://link.aps.org/doi/10.1103/PhysRevE.105.044121}
}

@article{Kasia2021b,
  title = {Quantum jump {Monte Carlo} approach simplified: Abelian symmetries},
  author = {Macieszczak, Katarzyna and Rose, Dominic C.},
  journal = {Phys. Rev. A},
  volume = {103},
  issue = {4},
  pages = {042204},
  numpages = {12},
  year = {2021},
  month = {Apr},
  publisher = {American Physical Society},
  doi = {10.1103/PhysRevA.103.042204},
  url = {https://link.aps.org/doi/10.1103/PhysRevA.103.042204}
}

@article{Valado2016,
  title = {Experimental observation of controllable kinetic constraints in a cold atomic gas},
  author = {Valado, M. M. and Simonelli, C. and Hoogerland, M. D. and Lesanovsky, I. and Garrahan, J. P. and Arimondo, E. and Ciampini, D. and Morsch, O.},
  journal = {Phys. Rev. A},
  volume = {93},
  issue = {4},
  pages = {040701},
  numpages = {5},
  year = {2016},
  month = {Apr},
  publisher = {American Physical Society},
  doi = {10.1103/PhysRevA.93.040701},
  url = {http://link.aps.org/doi/10.1103/PhysRevA.93.040701}
}

@article{Lindblad1976,
	author = "Lindblad, G.",
	journal = "Comm. Math. Phys",
	pages = "119",
	title = "{On the generators of quantum dynamical semigroups}",
	volume = "48",
	year = "1976"
}

@article{Gorini1976,
    author = {Gorini, Vittorio and Kossakowski, Andrzej and Sudarshan, E. C. G.},
    title = "{Completely positive dynamical semigroups of N‐level systems}",
    journal = {Journal of Mathematical Physics},
    volume = {17},
    number = {5},
    pages = {821-825},
    year = {1976},
    month = {05},
    issn = {0022-2488},
    doi = {10.1063/1.522979},
    url = {https://doi.org/10.1063/1.522979},
    eprint = {https://pubs.aip.org/aip/jmp/article-pdf/17/5/821/19090720/821\_1\_online.pdf},
}

@book{Gardiner2004handbook,
  address = {Berlin},
  author = {Gardiner, C. W.},
  edition = {Third},
  isbn = {3-540-20882-8},
  mrclass = {00A69 (60-01 60Hxx 60Jxx 82C31)},
  mrnumber = {2053476 (2004m:00008)},
  pages = {xviii+415},
  publisher = {Springer-Verlag},
  series = {Springer Series in Synergetics},
  title = {Handbook of stochastic methods for physics, chemistry and the
              natural sciences},
  volume = 13,
  year = 2004,
  chapter = 5,
}

@article{Rigol2008,
	author = "Rigol, Marcos and Dunjko, Vanja and Olshanii, Maxim",
	doi = "10.1038/nature06838",
	issn = "1476-4687",
	journal = "Nature",
	month = apr,
	number = "7189",
	pages = "854--8",
	pmid = "18421349",
	publisher = "Nature Publishing Group",
	title = "{Thermalization and its mechanism for generic isolated quantum systems.}",
	url = "http://www.ncbi.nlm.nih.gov/pubmed/18421349",
	volume = "452",
	year = "2008"
}

@article{Langen2016,
doi = {10.1088/1742-5468/2016/06/064009},
url = {https://dx.doi.org/10.1088/1742-5468/2016/06/064009},
year = {2016},
month = {jun},
publisher = {IOP Publishing and SISSA},
volume = {2016},
number = {6},
pages = {064009},
author = {Tim Langen and Thomas Gasenzer and Jörg Schmiedmayer},
title = {Prethermalization and universal dynamics in near-integrable quantum systems},
journal = {Journal of Statistical Mechanics: Theory and Experiment}
}

@article{Calabrese2006,
  title = {Time Dependence of Correlation Functions Following a Quantum Quench},
  author = {Calabrese, Pasquale and Cardy, John},
  journal = {Phys. Rev. Lett.},
  volume = {96},
  issue = {13},
  pages = {136801},
  numpages = {4},
  year = {2006},
  month = {Apr},
  publisher = {American Physical Society},
  doi = {10.1103/PhysRevLett.96.136801},
  url = {https://link.aps.org/doi/10.1103/PhysRevLett.96.136801}
}

@article{Calabrese2007,
doi = {10.1088/1742-5468/2007/06/P06008},
url = {https://dx.doi.org/10.1088/1742-5468/2007/06/P06008},
year = {2007},
month = {jun},
publisher = {},
volume = {2007},
number = {06},
pages = {P06008},
author = {Pasquale Calabrese and John Cardy},
title = {Quantum quenches in extended systems},
journal = {Journal of Statistical Mechanics: Theory and Experiment}
}

@article{Calabrese2012,
doi = {10.1088/1742-5468/2012/07/P07016},
url = {https://dx.doi.org/10.1088/1742-5468/2012/07/P07016},
year = {2012},
month = {jul},
publisher = {IOP Publishing and SISSA},
volume = {2012},
number = {07},
pages = {P07016},
author = {Pasquale Calabrese and Fabian H L Essler and Maurizio Fagotti},
title = {Quantum quench in the transverse field Ising chain: I. Time evolution of order parameter correlators},
journal = {Journal of Statistical Mechanics: Theory and Experiment}
}

@article{Calabrese2012b,
doi = {10.1088/1742-5468/2012/07/P07022},
url = {https://dx.doi.org/10.1088/1742-5468/2012/07/P07022},
year = {2012},
month = {jul},
publisher = {IOP Publishing and SISSA},
volume = {2012},
number = {07},
pages = {P07022},
author = {Pasquale Calabrese and Fabian H L Essler and Maurizio Fagotti},
title = {Quantum quenches in the transverse field Ising chain: II. Stationary state properties},
journal = {Journal of Statistical Mechanics: Theory and Experiment}
}

@article{Calabrese2016,
doi = {10.1088/1742-5468/2016/06/064003},
url = {https://dx.doi.org/10.1088/1742-5468/2016/06/064003},
year = {2016},
month = {jun},
publisher = {IOP Publishing and SISSA},
volume = {2016},
number = {6},
pages = {064003},
author = {Pasquale Calabrese and John Cardy},
title = {Quantum quenches in $1 + 1$ dimensional conformal field theories},
journal = {Journal of Statistical Mechanics: Theory and Experiment}
}

@ARTICLE{Agarwal1973,
       author = {{Agarwal}, G.~S.},
        title = "{Open quantum Markovian systems and the microreversibility}",
      journal = {Zeitschrift fur Physik},
         year = 1973,
        month = oct,
       volume = {258},
       number = {5},
        pages = {409-422},
          doi = {10.1007/BF01391504},
       adsurl = {https://ui.adsabs.harvard.edu/abs/1973ZPhy..258..409A}
}

@article{Garrahan2011,
  title={Kinetically constrained models},
  author={Garrahan, Juan P and Sollich, Peter and Toninelli, Cristina},
  journal={Dynamical heterogeneities in glasses, colloids, and granular media},
  volume={150},
  pages={111--137},
  year={2011},
  publisher={International Series of Monographs on Physics},
  eprint = {arXiv:1009.6113},
}

@article{PolkovnikovRMP,
  title = {Colloquium: Nonequilibrium dynamics of closed interacting quantum systems},
  author = {Polkovnikov, Anatoli and Sengupta, Krishnendu and Silva, Alessandro and Vengalattore, Mukund},
  journal = {Rev. Mod. Phys.},
  volume = {83},
  issue = {3},
  pages = {863--883},
  numpages = {0},
  year = {2011},
  month = {Aug},
  publisher = {American Physical Society},
  doi = {10.1103/RevModPhys.83.863},
  url = {https://link.aps.org/doi/10.1103/RevModPhys.83.863}
}

@article{Mitra_quench,
   author = "Mitra, Aditi",
   title = "Quantum Quench Dynamics", 
   journal= "Annual Review of Condensed Matter Physics",
   year = "2018",
   volume = "9",
   number = "Volume 9, 2018",
   pages = "245-259",
   doi = "https://doi.org/10.1146/annurev-conmatphys-031016-025451",
   url = "https://www.annualreviews.org/content/journals/10.1146/annurev-conmatphys-031016-025451",
   publisher = "Annual Reviews",
   issn = "1947-5462",
   type = "Journal Article",
  }

@article{Rossini2009,
  title = {Effective Thermal Dynamics Following a Quantum Quench in a Spin Chain},
  author = {Rossini, Davide and Silva, Alessandro and Mussardo, Giuseppe and Santoro, Giuseppe E.},
  journal = {Phys. Rev. Lett.},
  volume = {102},
  issue = {12},
  pages = {127204},
  numpages = {4},
  year = {2009},
  month = {Mar},
  publisher = {American Physical Society},
  doi = {10.1103/PhysRevLett.102.127204},
  url = {https://link.aps.org/doi/10.1103/PhysRevLett.102.127204}
}

@article{Rossini2010,
  title = {Long time dynamics following a quench in an integrable quantum spin chain: Local versus nonlocal operators and effective thermal behavior},
  author = {Rossini, Davide and Suzuki, Sei and Mussardo, Giuseppe and Santoro, Giuseppe E. and Silva, Alessandro},
  journal = {Phys. Rev. B},
  volume = {82},
  issue = {14},
  pages = {144302},
  numpages = {17},
  year = {2010},
  month = {Oct},
  publisher = {American Physical Society},
  doi = {10.1103/PhysRevB.82.144302},
  url = {https://link.aps.org/doi/10.1103/PhysRevB.82.144302}
}

@book{Mussardo_book,
      author        = "Mussardo, Giuseppe",
      title         = "{Statistical field theory: an introduction to exactly
                       solved models in statistical physics; 1st ed.}",
      publisher     = "Oxford Univ. Press",
      address       = "New York, NY",
      series        = "Oxford graduate texts",
      year          = "2010",
      url           = "https://cds.cern.ch/record/1281256",
}

@book{Huang_book,
  author = {Huang, Kerson},
  edition = 2,
  publisher = {John Wiley \& Sons},
  title = {Statistical Mechanics},
  year = 1987
}

@article{Pelissetto2002,
title = {Critical phenomena and renormalization-group theory},
journal = {Physics Reports},
volume = {368},
number = {6},
pages = {549-727},
year = {2002},
issn = {0370-1573},
doi = {https://doi.org/10.1016/S0370-1573(02)00219-3},
url = {https://www.sciencedirect.com/science/article/pii/S0370157302002193},
author = {Andrea Pelissetto and Ettore Vicari}
}

@book{ZinnJustin_book,
  asin = {0198509235},
  author = {Zinn-Justin, Jean},
  description = {Quantum Field Theory and Critical Phenomena…Amazon.co.uk: Jean Zinn-Justin: Books},
  dewey = {530.143},
  ean = {9780198509233},
  edition = 4,
  isbn = {0198509235},
  publisher = {Clarendon Press},
  title = {Quantum Field Theory and Critical Phenomena (International Series of Monographs on Physics)},
  url = {http://www.amazon.co.uk/Quantum-Critical-Phenomena-International-Monographs/dp/0198509235%3FSubscriptionId%3D13CT5CVB80YFWJEPWS02%26tag%3Dws%26linkCode%3Dxm2%26camp%3D2025%26creative%3D165953%26creativeASIN%3D0198509235},
  year = 2002
}

@book{Sachdev_book,
  author = {Sachdev, Subir},
  publisher = {Wiley Online Library},
  title = {Quantum phase transitions},
  year = 2007
}

@article{Chanda2020,
  title = {Confinement and Lack of Thermalization after Quenches in the Bosonic Schwinger Model},
  author = {Chanda, Titas and Zakrzewski, Jakub and Lewenstein, Maciej and Tagliacozzo, Luca},
  journal = {Phys. Rev. Lett.},
  volume = {124},
  issue = {18},
  pages = {180602},
  numpages = {7},
  year = {2020},
  month = {May},
  publisher = {American Physical Society},
  doi = {10.1103/PhysRevLett.124.180602},
  url = {https://link.aps.org/doi/10.1103/PhysRevLett.124.180602}
}

@article{Kinoshita2006,
  author = {Kinoshita, Toshiya and Wenger, Trevor and Weiss, David S.},
  day = 13,
  doi = {10.1038/nature04693},
  issn = {0028-0836},
  journal = {Nature},
  month = apr,
  number = 7086,
  pages = {900--903},
  publisher = {Nature Publishing Group},
  title = {{A quantum Newton's cradle}},
  url = {http://dx.doi.org/10.1038/nature04693},
  volume = 440,
  year = 2006
}

@article{Gring2012,
author = {M. Gring  and M. Kuhnert  and T. Langen  and T. Kitagawa  and B. Rauer  and M. Schreitl  and I. Mazets  and D. Adu Smith  and E. Demler  and J. Schmiedmayer },
title = {Relaxation and Prethermalization in an Isolated Quantum System},
journal = {Science},
volume = {337},
number = {6100},
pages = {1318-1322},
year = {2012},
doi = {10.1126/science.1224953},
URL = {https://www.science.org/doi/abs/10.1126/science.1224953},
eprint = {https://www.science.org/doi/pdf/10.1126/science.1224953}}

@article{AduSmith2013,
doi = {10.1088/1367-2630/15/7/075011},
url = {https://dx.doi.org/10.1088/1367-2630/15/7/075011},
year = {2013},
month = {jul},
publisher = {IOP Publishing},
volume = {15},
number = {7},
pages = {075011},
author = {D Adu Smith and M Gring and T Langen and M Kuhnert and B Rauer and R Geiger and T Kitagawa and I Mazets and E Demler and J Schmiedmayer},
title = {Prethermalization revealed by the relaxation dynamics of full distribution functions},
journal = {New Journal of Physics}
}

@article{Greiner2002,
  author = {Greiner, Markus and Mandel, Olaf and Hansch, Theodor W. and Bloch, Immanuel},
  day = 05,
  doi = {10.1038/nature00968},
  issn = {0028-0836},
  journal = {Nature},
  month = sep,
  number = 6902,
  pages = {51--54},
  publisher = {Nature Publishing Group},
  title = {{Collapse and revival of the matter wave field of a Bose–Einstein condensate}},
  url = {http://dx.doi.org/10.1038/nature00968},
  volume = 419,
  year = 2002
}

@article{Ritort2003,
	author = "Ritort, F. and Sollich, P.",
	doi = "10.1080/0001873031000093582",
	journal = "Adv. Phys.",
	number = "4",
	pages = "219--342",
	title = "{Glassy dynamics of kinetically constrained models}",
	volume = "52",
	year = "2003"
}

@book{Breuer_book,
  address = {Great Clarendon Street},
  author = {Breuer, H. P. and Petruccione, F.},
  nota = {GK: excellent textbook!},
  publisher = {Oxford University Press},
  title = {The theory of open quantum systems},
  year = 2002
}

@article{Qjumps1,
	author = "M{\o}lmer, Klaus and Castin, Yvan and Dalibard, Jean",
	doi = "10.1364/JOSAB.10.000524",
	journal = "J. Opt. Soc. Am. B",
	month = "Mar",
	number = "3",
	pages = "524--538",
	publisher = "OSA",
	title = "{Monte Carlo wave-function method in quantum optics}",
	url = "http://josab.osa.org/abstract.cfm?URI=josab-10-3-524",
	volume = "10",
	year = "1993"
}

@article{Qjumps2,
	author = "Plenio, M. B. and Knight, P. L.",
	doi = "10.1103/RevModPhys.70.101",
	issue = "1",
	journal = "Rev. Mod. Phys.",
	month = "Jan",
	numpages = "0",
	pages = "101--144",
	publisher = "American Physical Society",
	title = "{The quantum-jump approach to dissipative dynamics in quantum optics}",
	url = "http://link.aps.org/doi/10.1103/RevModPhys.70.101",
	volume = "70",
	year = "1998"
}

@article{DMRG1,
  title = {Density matrix formulation for quantum renormalization groups},
  author = {White, Steven R.},
  journal = {Phys. Rev. Lett.},
  volume = {69},
  issue = {19},
  pages = {2863--2866},
  numpages = {0},
  year = {1992},
  month = {Nov},
  publisher = {American Physical Society},
  doi = {10.1103/PhysRevLett.69.2863},
  url = {https://link.aps.org/doi/10.1103/PhysRevLett.69.2863}
}

@article{DMRG2,
  title = {The density-matrix renormalization group},
  author = {Schollw\"ock, U.},
  journal = {Rev. Mod. Phys.},
  volume = {77},
  issue = {1},
  pages = {259--315},
  numpages = {0},
  year = {2005},
  month = {Apr},
  publisher = {American Physical Society},
  doi = {10.1103/RevModPhys.77.259},
  url = {https://link.aps.org/doi/10.1103/RevModPhys.77.259}
}

@article{Albert2014,
  title = {Symmetries and conserved quantities in Lindblad master equations},
  author = {Albert, Victor V. and Jiang, Liang},
  journal = {Phys. Rev. A},
  volume = {89},
  issue = {2},
  pages = {022118},
  numpages = {14},
  year = {2014},
  month = {Feb},
  publisher = {American Physical Society},
  doi = {10.1103/PhysRevA.89.022118},
  url = {https://link.aps.org/doi/10.1103/PhysRevA.89.022118}
}

@article{Buca2012,
doi = {10.1088/1367-2630/14/7/073007},
url = {https://dx.doi.org/10.1088/1367-2630/14/7/073007},
year = {2012},
month = {jul},
publisher = {IOP Publishing},
volume = {14},
number = {7},
pages = {073007},
author = {Buča, Berislav and Prosen, Tomaž},
title = {A note on symmetry reductions of the Lindblad equation: transport in constrained open spin chains},
journal = {New Journal of Physics}
}

@InProceedings{VoterKMC,
author="Voter, Arthur F.",
editor="Sickafus, Kurt E.
and Kotomin, Eugene A.
and Uberuaga, Blas P.",
title="INTRODUCTION TO THE KINETIC MONTE CARLO METHOD",
booktitle="Radiation Effects in Solids",
year="2007",
publisher="Springer Netherlands",
address="Dordrecht",
pages="1--23",
isbn="978-1-4020-5295-8"
}

\end{document}